\documentclass[11pt,twoside]{mybook}

\usepackage{amsmath,amssymb,graphicx,thesis,rotating}

\renewcommand{\baselinestretch} {1.5}
\begin{document}

\frontmatter

\pagestyle{empty}

\begin{center}
\Huge{\bf Particle Splitting: A New Method for SPH Star Formation Simulations}
\vspace*{40mm}

\large{by}
\vspace*{10mm}

\Large{\bf Spyridon Kitsionas}
\vspace*{60mm}

\normalsize{A thesis submitted to the \\
University of Wales\\
for the degree of \\
Doctor of Philosophy}
\vspace*{10mm}

\large{\bf September, 2000}
\end{center}
\cleardoublepage

\noindent DECLARATION
\vspace{2mm}

\noindent This work has not previously been accepted in substance for any degree and is 
not being concurrently submitted in candidature for any degree.
\vspace{6mm}

\noindent Signed ........................................ (candidate)
\vspace{2mm}

\noindent Date ...........................................
\vspace{15mm}

\noindent STATEMENT 1
\vspace{2mm}

\noindent This thesis is the result of my own investigations, except where otherwise stated. 

\noindent Other sources are acknowledged by footnotes giving explicit 
references. A bibliography is appended.
\vspace{6mm}

\noindent Signed ........................................ (candidate)
\vspace{2mm}

\noindent Date ...........................................
\vspace{15mm}

\noindent STATEMENT 2
\vspace{2mm}

\noindent I hereby give consent for my thesis, if accepted, to be available for 
photocopying and for inter-library loan, and for the title and summary to be
made available to outside organisations.
\vspace{1mm}

\noindent Signed ........................................ (candidate)
\vspace{2mm}

\noindent Date ...........................................
\cleardoublepage

\renewcommand{\baselinestretch} {1.0}

\begin{center}
\LARGE{\bf Acknowledgements}
\end{center}

\normalsize{
\noindent I would 
like to thank my family for their love and support in all possible ways, during the period of my PhD studies. I still do not know how could I thank $M \upsilon \rho \tau \acute{\omega}$ for her endless 
love, support and patience during the difficult three years of me living in a different country 
and the ``strange'' period of writing up. I thank my supervisor, 
Prof. Anthony Whitworth, for guiding me through the PhD project with his methodic work and enthusiasm. I also thank him for correcting the awful first drafts of the thesis with great patience and for making useful comments on the language and the style. I am grateful to Neil Francis for his great help.
\vspace{3mm}

\noindent It has been a pleasure working with Dr. D. Ward-Thompson, 
Dr. A.S. Bhattal, Dr. S.J. Watkins, Philip Gladwin, Seung-Hoon 
Cha, Glynn Hosking and Jason Kirk. I would like to make a special mention to Dr. 
Henri Boffin for his assistance. I thank the computing personnel of the department, 
Phil, Rodney and Roger, for the help, patience and skills they kindly offered me. I acknowledge the use 
of the Sun E4000 computer of the Cardiff Centre for Computational Science and Engineering. I also 
acknowledge the use of Starlink and Departmental software and equipment. I would also like to thank the head of the department, Prof. 
P. Blood, and my supervisor for offering me a short term research assistanceship that has been a great financial relief for me.
\vspace{3mm}

\noindent I thank my housemates $\Pi \alpha \nu \alpha \gamma \iota \acute{\omega} \tau \eta$ and Simone. Living with 
Simone has been a great experience not just astronomically but also gastronomically! I will never forget all the friends from Italy, Spain, Germany, France, Portugal, Greece, South and Central America, India and of course 
Britain who have been excellent dinner, pub and party partners. I thank the 
five-a-side departmental team for the weekly exercise. I should also mention my friends from the BBC National Chorus of 
Wales as they played a special part in my social life. I thank the Akkizidis brothers and Vivian as well as Sabela and Benny for their hospitality, and my cousin $H \acute{\omega}$ for lending me her laptop.
}
\cleardoublepage

\begin{center}
\Huge{\bf Particle Splitting: A New Method for SPH Star Formation Simulations}
\end{center}
\vspace*{40mm}

\begin{center}
\LARGE{\bf Abstract}
\end{center}

\normalsize{
\noindent Particle Splitting is a new algorithm invented for use with self-gravitating SPH  
codes. It is designed to enable numerical simulations to obey the Jeans condition at all times (but it could be used in other contexts, to satisfy other conditions which require high resolution locally). With particle splitting, all coarse particles in 
regions of interest, are erased and replaced by sets of fine particles, increasing the numerical resolution of the simulations. A new algorithm for calculating smoothing lengths was added to our code, to accommodate the mixing of different mass particles. Our particle splitting SPH code reproduced the results of standard test simulations. 
\vspace{3mm}

\noindent Simulations of rotating clouds with m=2 density perturbations produce a binary and a bar. We confirm that fragmentation of the bar should be attributed to inadequate resolution. By applying Particle Splitting to such simulations, we reproduce the results of high resolution finite difference simulations (Truelove {\it et al}. and Klein {\it et al}.), 
where bar fragmentation is absent. We obtain these results with great computational economy.
\vspace{3mm}

\noindent We apply Particle Splitting to simulations of clump-clump collisions. We investigate the influence of clump mass, clump velocity and collision impact parameter on the structures formed. We show that such collisions lead to the formation of shocked layers. Networks of filaments or spindles, and groups of protostellar discs form within the layers. In all collisions, fragmentation of the filaments was the common mechanism producing the groups of protostars. Low-mass secondary companions to the protostars may form subsequently by instabilities in the discs and/or dynamical interaction between the protostars. However, due to time-step requirements, we cannot follow the disc evolution for long enough to explore the formation of secondaries. 
\vspace{3mm}

\noindent We show that all the protostars formed have mass accretion rates of $\sim$5 x 10$^{-5}$ M$_{\odot}$ yr$^{-1}$ for the first 10-20 thousand years after their formation. This mechanism shows 10-20\% Star Formation efficiency. The efficiency increases with increasing clumps mass. Collisions with impact parameter $b < $0.4 and Mach number $\mathcal{M} \geqslant$10 give the highest efficiency. We predict the existence of filaments with $n_{H_{2}} \gtrsim$5 x 10$^{5}$ cm$^{-3}$ in sites of dynamical Star Formation. These filaments could be observed in NH$_{3}$ or CS line radiation, in star forming regions lying within 1 kpc.
}
\cleardoublepage

\renewcommand{\baselinestretch} {1.5}

\normalsize{

\pagestyle{headings}

\tableofcontents

\listoffigures

\listoftables

\mainmatter

\chapter{Introduction}

Star Formation is a field relying on theoretical work since it is difficult to obtain 
observationally a picture of the way stars form. Numerical simulation of Star Formation has lately become a very 
popular theoretical tool due to the rapid increase in computing power. In this thesis, 
we deal with simulations of low-mass star formation and in particular, we establish a method for increasing the resolution and dynamic range of simulations. In this chapter, we briefly review the main theoretical and observational 
constraints on Star Formation, so that we can put our work in context with the other research in this field. 

\section{Stars: the end-product of Star Formation}

Stars are a fundamental constituent of the Universe. Stars are hot massive dense gas 
spheres emitting radiation produced in their centres from nuclear
reactions and/or gravitational contraction. Stars are very well studied. Analysing their spectra provides information not 
only on their outer layers that can be seen emitting, but also on their interiors. In particular, 
we can infer their chemical composition and from known nuclear reaction cycles we can produce 
models for their structure and time evolution. Core hydrogen burning stars of different age, mass and chemical 
composition define a locus on the surface-temperature {\it vs}. luminosity 
diagram: the Main Sequence of the Hertzsprung-Russell diagram. A star reaches the Main 
Sequence, following a period of gravitational contraction, as soon as the conditions at its centre become hot and dense enough to start burning hydrogen. 
The gravitational contraction phase of a forming star is qualitatively understood and there are 
theoretical models predicting its evolution towards the main-sequence (pre-main-sequence tracks). 
What is relatively unknown, and constitutes one of the fundamental questions for Star Formation 
theory, is the mechanism that brings star-forming gas together in the first place. At this point, a review 
of the properties of the interstellar medium, which provides the ingredients for Star Formation, 
can help us put this question into perspective. 

\section{Properties of the ISM}

The Interstellar Medium (ISM) consists of gas in all states (atoms, molecules, ions) and dust 
grains. The dust component accounts only for 1-2\% of the mass of the interstellar medium. The 
gas consists mainly of hydrogen and helium. A small percentage of the gas mass is in the form of heavier elements.

The ISM is far from being uniform, as it consists of dense clouds of gas ($n \gtrsim 100$ cm$^{-3}$). In these clouds, the gas is mainly molecular and cold ($T \lesssim 100$ K). Apart form H$_{2}$, the clouds contain quite a few other molecules, such as 
CO, OH, CH, H$_{2}$O, CS, NH$_{3}$ as well as more massive and more complex carbon chains 
\cite{vanDishoeckLADA1999}. Such clouds are called Giant Molecular Clouds (GMCs) 
because of their size ($R \sim$ 10-20 pc and $M \sim 10^{3}$-$10^{4}$ M$_{\odot}$). These clouds 
are immersed in a warm ($T \gtrsim 10^{4}$ K) and rarefied ($n \gtrsim 1$ cm$^{-3}$) interclump medium, 
partly atomic and partly ionised. 

The structure of the ISM is evolving with time. Parts of a GMC in proximity to massive stars can be 
ionised by expanding nebulae (HII regions, 
supernova remnants) and become part of the interclump medium, or be squashed and forced to form more stars. 
GMCs form in galactic spiral arms possibly through condensation of HI clouds and /or collisional 
agglomeration of lower mass clouds. GMCs live for about 10$^{7}$ yr.

GMCs are self-gravitating and they are supported by supersonic MHD turbulence. Some of the turbulent motions are generated by stars within these clouds. In particular, 
stellar winds and outflows, HII regions and supernova remnants are 
mechanisms for exciting turbulent motion. There is also evidence for magnetic 
fields threading the clouds (observed via polarimetry and Zeeman splitting measurements, e.g. Ward-Thompson 
{\it et al.} \shortcite{Ward-ThompsonApJ2000}). 

Molecular Clouds are the birthplaces of all known young 
stars. They provide the initial conditions for Star Formation. ``They host Young Stellar Objects (YSOs) in a 
wide range of evolutionary states; from Class 0 protostars some 10$^{-2}$ Myr old, deriving most of their 
luminosity from gravitational infall, to T-Tauri stars a few Myr old, deriving their luminosity from quasi-static 
contraction. They also host stars in a wide range of spatial groupings; from isolated single stars as in Taurus, 
having no known neighbours within a few pc, to rich star clusters as in Orion, having a few thousand stars 
within a few pc. The masses of the stars in GMCs range from 0.1-30 M$_{\odot}$, nearly the whole range 
of known stellar masses. Indeed the Initial Mass Function (IMF), or the distribution of stellar masses at birth, 
is indistinguishable between stars in molecular clouds and field stars.'' \cite{MyersLADA1999} 

GMCs have structure of their own. Dense clumps can be observed within GMCs, typically with densities above 
10$^{3}$ cm$^{-3}$, via molecular line observations using tracers such as CO and its isotopomers. These clumps 
have masses between 10 and 100 M$_{\odot}$. In sites of high-mass star formation, these clumps tend to be 
associated with groups and clusters of young stars \cite{LadaApJ1991a}. Within these clumps, 
dense cores can be observed using molecules such as CS and NH$_{3}$ which trace densities of 
10$^{4}$ cm$^{-3}$ and above \cite{MyersApJ1983a,MyersApJ1983b,MyersApJ1983c,BensonApJ1983}. In sites of 
low-mass star formation, such cores (with masses of $\sim$1 M$_{\odot}$) are directly associated with the 
formation of single, or at the most 
binary, stars. It is, therefore, of great importance to understand how these cores and clumps can evolve 
to produce young stars. Before we list some of the available models for Star Formation, let us summarise 
the properties of young stars.

\section{Observations of YSOs}{\label{sec:collapse}}

Some of the cores observed in NH$_{3}$ are associated with IR sources which are identified as protostars. There
are also starless cores or pre-stellar cores \cite{Ward-ThompsonMNRAS1994}, which are believed to be the 
precursors of protostars. They are more extended than cores containing IR sources. Some show double-peaked velocity profiles with a
stronger blue-shifted peak suggesting that they are collapsing \cite{Ward-ThompsonMNRAS1996}. Pre-stellar cores are believed to be supported by magnetic fields. 
The magnetic support is removed by ambipolar diffusion \cite{ShuARAnA1987} or by reconnection of magnetic field lines 
\cite{LubowMNRAS1996}. Without this support the pre-stellar cores start collapsing.

The cores associated with IR sources also show signatures of collapse. 
The YSOs within these cores are not all at the same evolutionary stage. The evolution of YSOs can be divided 
into two major phases: the embedded phase and the revealed phase \cite{LadaLADA1999a}. There are two different 
classes of objects in each phase. 

In the embedded phase the protostars are collapsing fast. Their luminosity is produced by gravitational contraction and accretion. 
They are characterised either as Class 0 or as Class I objects depending on their Spectral Energy 
Distribution (SED). Class 0 objects have SEDs that can be fitted with single temperature black body functions. 
The Class 0 SEDs peak at sub-mm wavelengths and the objects are not observed below 20 $\mu$m. Class 0 protostars 
are highly obscured (A$_{V} \sim$ 1000) and of extremely low temperature (20-30 K). Class 0 SEDs are 
attributable to emission from dust grains in the infalling envelope. The dust absorbs all radiation coming from the protostar and then re-emits it in the sub-mm spectral range
\cite{AndreApJ1993}. The Class 0 phase lasts for about 1-5 x 10$^{4}$ 
yr. During this period, Class 0 protostars accrete matter with a mass accretion rate of $\gtrsim 10^{-5}$ M$_{\odot}$ yr$^{-1}$. 
These objects are associated with very powerful outflows.

The SED of a Class I object is broader than a single temperature blackbody function. Class I SEDs peak at sub-mm 
wavelengths but they also show a characteristic 
excess of infrared emission. This IR emission is associated with large amounts of circumstellar dust. Class I 
sources are deeply embedded and there is no emission at optical wavelengths. However, there is NIR emission (e.g. 
2.2 $\mu$m) and a significant fraction of this NIR emission is from scattered light. Class I 
objects are also associated with outflows, but these are not so powerful as those of Class 0 objects. The lifetime of a Class I 
object is 1-5 x 10$^{5}$ yr. During this time, the mass accretion rate is $\lesssim 10^{-6}$ M$_{\odot}$ yr$^{-1}$. There 
is some evidence that Class I objects have come through a Class 0 phase.

In the revealed phase the protostars have evolved into pre-main sequence stars. The infalling envelope of gas and dust has been removed. The luminosity in this phase is produced by Kelvin-Helmholtz 
contraction and deuterium burning. Again, there are two classes of 
objects: Class II  and Class III. Classification of pre-main-sequence stars into these two classes is based on 
their SEDs. The Class II SEDs 
are broad like those of Class I objects but they peak at IR wavelengths. For wavelengths longer than 2.2 $\mu$m 
they decrease in a power-law fashion. The IR excess of Class II objects is smaller than that of Class I 
objects. Again, it indicates the presence of circumstellar dust. These YSOs can be observed in optical 
wavelengths, and therefore are better known than the protostars of the embedded phase. They show H$_{\alpha}$ 
emission which is associated with accretion onto the protostar. Optical identification of 
accretion discs silhouetted against a bright background has been 
possible with HST \cite{O'DellApJ1994}. Some of these objects have weak outflows associated with their 
accretion discs. Class II objects have variable emission lines; they are also known as Classical T-Tauri Stars (CTTS). From pre-main-sequence tracks, their age is estimated to be 1-4 x 10$^{6}$ yr. 
Their accretion rates are estimated to be $\sim 10^{-8}$ M$_{\odot}$ yr$^{-1}$. 

During the infall of circumstellar gas onto a protostar, its angular momentum increases. Above a critical value, the excess in angular momentum, produced by more infalling mass, has to be removed. The discs provide a natural mechanism for angular momentum removal  \cite{Lynden-BellMNRAS1974}. Mass is accumulated on the plane perpendicular to the angular momentum vector and angular momentum is transfered to the mass at the outer edge of this disc.

Class III objects are visible at both optical and IR wavelengths. Their SEDs can be fitted with single temperature black body functions. At wavelengths longer than 2.2 $\mu$m their emission decreases more steeply than that of Class II objects. They show no excess IR emission. Their SEDs are thought to be arising directly from their photospheres. However, there is still considerable amount of dust around these objects (remnant of the infalling envelope of previous protostellar phases) which produces a substantial extinction and reddening. These objects could not be distinguished from main-sequence stars, if they were not emitting hydrogen lines and X-rays. The H$_{\alpha}$ emission is not so strong as in the Class II stage, but it indicates traces of a disc. Class III objects are also known as Weak-line T-Tauri Stars (WTTS). Their positions on the H-R diagram can be directly compared to predictions of pre-main-sequence tracks (they are positioned on top and to the right of the main-sequence). From these tracks we obtain their ages of 10$^{6}$ - 10$^{7}$ yr. 

The discs around Classical T-Tauri stars are rather extended (100-1000 AU) but not very massive (0.001-0.01 M$_{\odot}$). The central protostar, which has accreted most of its mass by this time, has mass 0.5-1.5 M$_{\odot}$ \cite{BeckwithLADA1999}. At the end of their evolution (Class III stage), pre-main-sequence stars have their discs stripped of their gaseous component. What little gas remains, together with the dust, is believed to form planetary systems \cite{RudenLADA1999}.

T-Tauri stars were first observed in the Taurus molecular cloud. Taurus is a low-mass Star Formation Region (SFR) on the northern sky at a distance of $\sim$140 pc. Taurus is the best example of distributed star formation, where forming stars are not part of dense groups or clusters \cite{GomezAJ1993,GladwinMNRAS1999}. Young stars in Taurus have masses between 0.5 and 1.0 M$_{\odot}$ and are in small clusters of $\lesssim$10 members.

The Orion molecular cloud (also on the northern sky, at a distance of $\sim$440 pc) is the most well-studied example of clustered star formation. It has been shown that in the center of Orion (close to the Trapezium OB stars) there are more than a thousand stars within one or two pc \cite{HillenbrandAJ1997,HillenbrandApJ1998}. In Orion B, almost all stars ($\sim$700) have formed in just 4 clusters \cite{LadaApj1991b}. Discs around young stars in Orion (proplyds) have smaller radii than those in Taurus possibly due to photo evaporation of gas by the radiation field of the nearby massive stars \cite{HollenbachApJ1994}. This illustrates the strong effect that the molecular cloud environment has on Star Formation \cite{LadaLADA1999b}.

The more massive of the YSOs in Orion (M $\gtrsim$2 M$_{\odot}$) are Herbig Ae-Be stars and are the precursors of main-sequence stars of type A or earlier. These YSOs are not so well studied as T-Tauri stars because they are fewer of them and they are further away. Despite having luminosities 
large enough to drive stellar winds, these YSOs are highly obscured by their infalling envelopes. Like T-Tauri stars, they are associated with violent phenomena like jets and outflows \cite{ChurchwellLADA1999}.

It has been shown that the IMF, $\phi(M)$, can be fitted with

\[ \phi(\mathrm{M}) \, \mathrm{dM} \simeq 50 \exp \left[ - \log_{10}^{2} \left(\frac{\mathrm{M}}{0.01 \mathrm{M}_{\odot}} \right) \right], \; \; \; \; 0.2 \mathrm{M}_{\odot} \leqslant \mathrm{M} \leqslant 50 \mathrm{M}_{\odot} \]
\cite{MillerApJSS1979}. This IMF extends the previous Salpeter \shortcite{SalpeterApJ1955} IMF 

\[ \phi(\mathrm{M}) \, \mathrm{dM} \simeq K \mathrm{M}^{-2.35} \mathrm{dM}, \; \; \; \; 0.4 \mathrm{M}_{\odot} \leqslant \mathrm{M} \leqslant 10 \mathrm{M}_{\odot} \]
to smaller and larger masses. In the low-mass range the Miller \& Scalo IMF is flatter than the -2.35 power law, whereas in the high mass range it is steeper than the -2.35 power law. This IMF suggests that low-mass stars are greater in number than high-mass stars, and account for most of the mass. 

The observed IMF puts a strong constraint to Star Formation theories, as these theories must predict a similar mass distribution. Another statistical constraint is set by the observed spatial distribution of stars. 

It is well known that almost 50\% of all solar-type main-sequence stars are part of a binary or higher multiple system \cite{DuquennoyAnA1991}. Duquennoy \& Mayor have measured the distributions of periods, eccentricities and mass ratios for field binary systems. They have shown that the distribution of periods can be fitted with a Gaussian that peaks at $\sim$200 years, corresponding to a separation $\sim$30 AU. 
Pre-main-sequence binaries are observed in the same range of separations as field star binaries (from a few to 10$^{4}$ AU, e.g. Ghez {\it et al}. \shortcite{GhezApJ1997}). Their distribution of separations is similar to that of the field stars (e.g. Prosser {\it et al}. \shortcite{ProsserApJ1994}). This indicates that stars often form in binaries. 

The binary frequency may depend on the environment of the SFR \cite{BrandnerApJ1998}. It appears that closer binaries are formed in regions near massive stars. However, there is evidence that universal processes determine the multiplicity of young stars, since the surface density of companions for pre-main-sequence stars can be fitted with a single power law, $\bar{N}(\theta) \propto \theta^{-2.0 \pm 0.1}$ for 0.0001$^{\circ} \leqslant \theta \leqslant 0.01^{\circ}$, in many different regions, e.g. Chamaeleon I, Taurus, Orion, $\rho$ Ophiuchus, Lupus, Vela (Larson \shortcite{LarsonMNRAS1995}; Simon \shortcite{SimonApJ1997}; Nakajima {\it et al}. \shortcite{NakajimaApJ1998}; for a review see Gladwin {\it et al}. \shortcite{GladwinMNRAS1999}).

We have seen that protostars, which are very young but not very bright sources, are heavily obscured by circumstellar material for a significant period of their formation. It is, therefore, rather difficult to obtain detailed observational information for the physical mechanisms that convert the ISM into stars. Theoretical work therefore has an important role to play in Star Formation research. Modelling the complex gas dynamics in the ISM, including several orders of magnitude change in density and linear scale, is achieved by means of numerical simulations. In this thesis, we model cloud-cloud collisions. In the next section, we review the most important theories of Star Formation and previous cloud-cloud collision simulations. 

\section{Theories and Simulations of Star Formation}{\label{sec:Simulations}}

``Every piece, or part, of the whole of nature is always merely an {\it approximation} 
to the complete truth, or the complete truth so far as we know it. In fact, everything 
we know is only some kind of approximation, because {\it we know that we do not know 
all the laws} as yet. Therefore, things must be learned only to be unlearned again or, 
more likely, to be corrected.'' \cite{FeynmanBOOK1995}.

Star Formation theories should predict the formation of groups of protostars with the above distributions of masses and separations. One of the oldest mechanisms proposed for the production of binary stars was fission, the splitting of a single object into two due to excess rotation. However, both finite difference and SPH simulations of Durisen {\it et al}. \shortcite{DurisenApJ1986} have shown that spiral arms remove angular momentum efficiently from the rotating object, and fission does not occur.

Capture predicts that an initially unbound system of two or more protostars will become bound when it loses energy in a close encounter. Star-disc capture \cite{HallMNRAS1996,BoffinMNRAS1998} is a mechanism sufficient to reproduce the binary frequency in very small dense stellar clusters, but not in larger looser clusters or the field \cite{ClarkeMNRAS1991a}. 

The well known model of low-mass star formation of Shu, Adams \& Lizano \shortcite{ShuARAnA1987} involves the collapse of an isothermal sphere producing a single protostar. The isothermal sphere loses its magnetic support via ambipolar diffusion, where neutrals slowly decouple from the ions and the field to produce a $\rho \propto r^{-2}$ density profile. Collapse of a sphere with such a profile gives a constant accretion rate. However, Whitworth {\it et al}. \shortcite{WhitworthMNRAS1996} have argued that such a profile is unlikely to arise in nature. In addition, Class 0 protostars are observed to be undergoing collapse with a less centrally condensed profile \cite{Ward-ThompsonMNRAS1994}.

Fragmentation is a process where separate parts of a cloud, clump or core become gravitationally unstable and contract. Groups of protostars are usually produced by this process. Fragmentation requires a mechanism to induce gravitational instabilities. Several numerical simulations of fragmentation have been conducted involving clouds of different geometries or applying different mechanisms to produce the instabilities (for a review on fragmentation simulations see Bonnell \shortcite{BonnellLADA1999}). Simulations of rotating spherical clouds \cite{BossApJ1979} or cylindrical clouds \cite{BonnellApJ1992b,BonnellApJ1991a} produce binaries and/or bars that break up into lumps. In disc fragmentation simulations, spiral arm rotational instabilities in the rotating disc produce companions to the central objects \cite{BonnellMNRAS1994a,BonnellMNRAS1994b}.

Mechanisms of a dynamic nature that can induce gravitational instabilities and binary formation via fragmentation, include the creation of shock compressed layers triggered by the interaction of stellar winds with the ambient gas, or by cloud-cloud collisions.

Early simulations of cloud-cloud collisions produced coalescence of clouds \cite{StoneApJ1970a,StoneApJ1970b,GildenApJ1984a,LattanzioMNRAS1985}. Fragmentation of a shocked layer produced at the interface of the collision was found in simulations with equations of state that included cooling of the shocked gas \cite{NagasawaPTP1987,HabePASJ1992}.

Subsequent SPH simulations of cloud-cloud collisions \cite{ChapmanNATURE1992,PongracicMNRAS1992,TurnerMNRAS1995,WhitworthMNRAS1995,BhattalMNRAS1998} have been able to resolve several orders of magnitude change in density and linear scale. These simulations produced large rotationally supported protostellar discs (200-4000AU) of high mass (5-60 M$_{\odot}$). The simulations of Chapman {\it et al}. \shortcite{ChapmanNATURE1992} involved higher mass clouds and produced filaments that fragmented, producing several discs. Rotational instabilities in the discs produced secondary companions to the protostars \cite{TurnerMNRAS1995,WhitworthMNRAS1995}. Increasing the impact parameter of the collision \cite{PongracicMNRAS1992,BhattalMNRAS1998} assisted disc fragmentation as the angular momentum was increased. However, such simulations did not always obey the Jeans condition \cite{TrueloveApJ1997,BateMNRAS1997} and as a result they are suspect, i.e. the protostars may have formed numerically.

In this thesis, we apply a new method (particle splitting) that enables the Jeans condition to 
be obeyed throughout our simulations with minimum computational cost. With 
this method, the resolution of 
the simulations can be increased at certain times and places, creating 
simulations with increasingly fine resolution, nested inside an 
initial coarse simulation. We apply particle splitting to numerical 
simulations of cloud-cloud collisions.

\section{Thesis plan} 

In chapter \ref{sec:SPH}, we describe in detail the numerical code we use (Smoothed Particle Hydrodynamics with Tree-Code-Gravity). Most of the features of this code have been discussed in previous theses 
\cite{ChapmanPhD1992,TurnerPhD1993,BhattalPhD1996,WatkinsPhD1996} and in Turner {\it et al}. \shortcite{TurnerMNRAS1995}, but 
they are presented here in detail in order to make the thesis self-contained.
At the end of the chapter, we 
apply our code to standard tests for a self-gravitating hydrodynamical code, to demonstrate that it is appropriate for the problems we study.

In chapter \ref{sec:rotating}, we present simulations of rotating clouds with 
an m=2 density perturbation. With these simulations, we try to 
reproduce previous results \cite{BateMNRAS1997,TrueloveApJ1997,TrueloveApJ1998,KleinTOKYO1998}, and we 
investigate the dependence of the code on numerical parameters such as the 
number of particles, the number of neighbours, the choice of the 
interpolating function, the initial spatial distribution of the particles. We consider this chapter to be
an examination of the efficiency of the code when applied to more realistic problems.

In chapter \ref{sec:part-split}, we describe particle splitting in detail. 
The method is then tested thoroughly. 
From the tests, it is shown that the noise introduced to the simulations 
when the method is applied, is not significant. We also apply particle splitting 
to simulations of rotating clouds with m=2 density perturbations: results of finite difference simulations are reproduced with great computational efficiency.

In chapter \ref{sec:results}, we apply particle splitting to simulations of cloud-cloud 
collisions. We repeat the simulations of Bhattal {\it et al}.
\shortcite{BhattalMNRAS1998} 
but now satisfying the Jeans condition. 
These simulations involve collisions between intermediate-mass clouds (75 M$_{\odot}$). By comparing our 
results to those of Bhattal {\it et al}., we can estimate the efficiency 
of the new method and quantify its benefits. We then present simulations of 
low-mass clump collisions (10 M$_{\odot}$) that have not been studied before. We investigate the influence of cloud mass, collision impact parameter and relative velocity on the filamentary structures formed in the shocked layers, on the protostellar discs formed within the filaments and on the Star Formation efficiency. Finally, we compare the properties of the stellar systems produced in our 
simulations and those obtained from observations of YSOs. 

In chapter \ref{sec:conclusions}, we summarise our main conclusions emphasising the computational efficiency achieved with particle 
splitting. Finally, we make suggestions for improvements to our code as well as for further applications of particle splitting.

Appendix A gives a derivation of the Jeans criterion for fragmentation. 


\chapter{Self-Gravitating Smoothed Particle Hydrodynamics}{\label{sec:SPH}}

\section{Self-gravitating Hydrodynamics}

The gas in the interstellar medium is highly compressible. In star formation the self-gravity of the gas is also important. In order to describe the evolution of a self-gravitating, inviscid, compressible, non-magnetic fluid, we need to solve a system of four equations -- the continuity equation, Euler's equation, energy equation and equation of state \cite{LandauBOOK1966,ShuBOOK1992} -- with four unknowns, namely the velocity {\bf \em v}, pressure $P$, specific internal energy $u$, and density $\rho$, at each position ${\bf r}$ in the fluid. It is implicit that these four quantities are also functions of time. The four equations read as follows:

\begin{itemize}

\item Continuity equation 

\begin{equation}
\label{equa:continuity}
\frac{\mathrm{d}\rho({\bf r})}{\mathrm{d}t} \; = \; - \rho({\bf r}) \; {\bf \nabla} \cdot \mbox{\bf \em v}({\bf r}),
\end{equation}

\noindent where we have used the Lagrangian time-derivative 
$\mathrm{d} \, / \, \mathrm{d}t \; = \; \partial \, / \, \partial t \; + \; \mbox{\bf \em v} \cdot {\bf \nabla}$ and the fact that $\partial \rho \, / \, \partial t \; = \; - {\bf \nabla} \cdot (\rho \mbox{\bf \em v})$. The continuity equation expresses the conservation of mass.

\item Euler's  equation

\begin{equation}
\label{equa:Euler}
\frac{\mathrm{d}\mbox{\bf \em v}({\bf r})}{\mathrm{d}t} \; = \; - \frac{1}{\rho({\bf r})} {\bf \nabla} P({\bf r}) + {\bf a}_{grav}({\bf r}) + {\bf a}_{visc}({\bf r}),
\end{equation}

\noindent where ${\bf a}_{grav}$ is the self-gravitational acceleration, given by 

\begin{equation}
\label{equa:self-gravity}
{\bf a}_{grav}({\bf r}) \; = \; G \int_{\mathrm{all} \; \mathrm{space}} \frac{\rho({\bf r'}) ({\bf r'}- {\bf r}) \mathrm{d}^{3}{\bf r'}}{|{\bf r'}- {\bf r}|^{3}},
\end{equation}

\noindent and ${\bf a}_{visc}$ is the artificial viscous acceleration (see discussion in \S \ref{sec:artvisc}). Euler's equation expresses the conservation of momentum.

\item Energy equation and equation of state

In general, the equation for the rate of change of the specific internal energy reads

\[
\rho({\bf r}) \frac{\mathrm{d}u({\bf r})}{\mathrm{d}t} \; = \, - \, P({\bf r}) \; {\bf \nabla} \cdot \mbox{\bf \em v}({\bf r}) \, + \,(\Gamma - \Lambda),
\]

\noindent where $\Gamma$ and $\Lambda$ are the radiative heating and cooling rates per unit volume respectively. This equation expresses the conservation of energy. The pressure is then given by the ideal gas equation of state \[ P = (\gamma - 1) \rho u, \] where $\gamma$ is the ratio of specific heats.

\end{itemize}

However, it is well established \cite{TohlineCANUTO1982} that prestellar gas at low densities ($\rho \ll \rho_{0} \simeq 10^{-14}$ g cm$^{-3}$) is approximately isothermal at $T \sim 10$K (this is mainly due to the strong temperature sensitivity of $\Lambda$). At high densities ($\rho \gg 10^{-14}$ g cm$^{-3}$), the gas is approximately adiabatic. Therefore, we can reduce the system of equations we have to solve, by using a barotropic equation of state instead of the two equations involving the specific internal energy. The barotropic equation of state which we use reads

\begin{equation}
\label{equa:state}
\frac{P({\bf r})}{\rho({\bf r})} \; = \; c_{0}^{2} \left[1 + \left( \frac{\rho({\bf r})}{\rho_{0}} \right) ^{4/3} \right]^{1/2}, 
\end{equation}

\noindent where $c_{0}$ is the isothermal sound speed of the gas at low densities and $\rho_{0}$ is the density above which the gas becomes adiabatic. For $\rho \ll \rho_{0}$ Eqn. \ref{equa:state} gives  $P \sim \rho \, c_{0}^{2}$, while for $\rho \gg \rho_{0}$, $P \propto \rho^{5/3}$.

At high densities, the gas is approximated with a polytrope, having an index of 3/2 \footnote{The two rotational degrees of freedom that bring Tohline's \shortcite{TohlineCANUTO1982} polytropic index to 5/2 are frozen out while the gas is below $\sim$400K \cite{WinklerApJ1980}. Above this temperature, the rotational degrees of freedom are excited, and the 5/2 index switches on for one or two orders of magnitude increase in density, before dissociation happens and the gas finally becomes atomic with an index of 3/2.}, only for a few orders of magnitude. For further evolution toward stellar densities detailed radiation transport calculations are necessary. To-date, no simulation in the literature has advanced that far. Bate \shortcite{BateApJ1998} has presented results approaching stellar densities, but with an approximate barotropic (piecewise polytropic) equation of state.

With our code, we do not attempt to model interactions between the gas and interstellar magnetic fields that thread molecular clouds. This would require a larger set of governing equations (e.g. see the MHD equations in Vazquez-Semadeni, Canto \& Lizano \shortcite{Vazquez-SemadeniApJ1998}), but more importantly it would require a two-, or possibly a three-, component fluid. Such a fluid has never been modelled with an SPH code and it is unknown what the resolution requirements for each fluid component would be. Exploring this is beyond the scope of the present thesis; and in any case, it is not clear that the magnetic fields observed in molecular clouds are sufficiently strong to greatly influence the dynamics of the cloud-cloud collisions studied in this thesis.

To solve the above system of equations, we have used the numerical method 
described in Turner {\it et al.} \shortcite{TurnerMNRAS1995} with some later 
improvements. The method consists of two numerical techniques: SPH, which 
provides values for the hydrodynamical properties of the fluid (\S 
\ref{sec:equations}), and Tree-Code-Gravity (TCG), which provides values for 
the gravitational accelerations (\S \ref{sec:TCG}). In this chapter, we shall 
describe in detail the method as well as the integration scheme with which we 
follow the evolution of the fluid in time (\S \ref{sec:integrate}). To 
summarise the chapter, we will give a schematic picture of a complete cycle of 
the integration scheme (\S \ref{sec:step}). Finally, we will briefly describe 
the performance of the numerical method when applied to some standard tests 
for a self-gravitating hydrodynamical code (\S \ref{sec:SPHtests}). The 
novel modifications to this numerical method associated with particle splitting are developed in detail in chapter \ref{sec:part-split}. Additional 
numerical techniques used for the purposes of special applications of this 
numerical method, are described in the relevant chapters, e.g. radiative 
cooling of shocked layers in chapter \ref{sec:results}, sink particles in 
chapter \ref{sec:part-split}.

\section{Fundamentals of SPH}{\label{sec:fundamentals}}

Smoothed Particle Hydrodynamics (SPH) \cite{GingoldMNRAS1977,LucyAJ1977}, is a Lagrangian numerical method that assumes no symmetries or imposed grids. It is, therefore, very efficient in describing problems which involve complex 3-dimensional geometries. SPH represents the fluid by $N$ discrete but extended/smoothed particles (i.e. Lagrangian sample points). The particles are overlapping, so that all the physical quantities involved can be treated as continuous functions both in time and space. To implement this, a smoothing function (kernel) with compact support is used. This smoothing function describes the strength and extent of a particle's influence. Bhattal \shortcite{BhattalPhD1996} has shown that the M4-kernel, that has been frequently used in SPH \cite{MonaghanAnA1985}, gives good results. The 3-D M4-kernel is a polynomial

\begin{equation}
\label{equa:M4}
W_{M4}(s) \; = \; \frac{1}{\pi} \left \{ 
       \begin{array}{ll}
          1  -  3s^{2}/2  +  3s^{3}/4, & 0 \leqslant s \leqslant 1; \\
          (2  -  s)^{3}/4, & 1 \leqslant s \leqslant 2; \\
          0, & s \geqslant 2,
       \end{array} \right .
\end{equation}

\noindent which smoothes the mass of a particle out to two smoothing lengths. In SPH, the value of any quantity $A$ at a position {\bf r} is evaluated using

\begin{equation}
\label{equa:alpha}
A({\bf r}) \; = \; \sum_{i} m_{i} \frac{A_{i}}{\rho_{i}}h_{i}^{-3} W \left( \frac{|{\bf r} - {\bf r}_{i}|}{h_{i}} \right),
\end{equation}

\noindent where ${\bf r}_{i}$ is the position, $m_{i}$ the mass and $h_{i}$ is the smoothing length of particle $i$ \cite{MonaghanAnA1985,MonaghanCPC1988,MonaghanARAnA1992}. $A_{i}$ and $\rho_{i}$ are the values of $A$ and $\rho$ at ${\bf r}_{i}$. Note the normalising term $h_{i}^{-3}$ which is introduced when we use Eqn. \ref{equa:M4} for the kernel and substitute $s=|{\bf r}|/h$. In Eqn. \ref{equa:alpha} the summation is finite due to the fact that the kernel function has compact support. This implies that contributions from only a few close neighbouring particles are taken into account. The fact that the summation does not need to be over all particles greatly reduces the computational cost of all SPH calculations. The benefit of this fact must be balanced against the need to obtain accurate results. With $h$ being large enough we can reduce sampling errors. The value of $h$ is, therefore, of great importance. It is discussed in detail later in this chapter.

\begin{figure}
\resizebox{7.75cm}{!}{\includegraphics{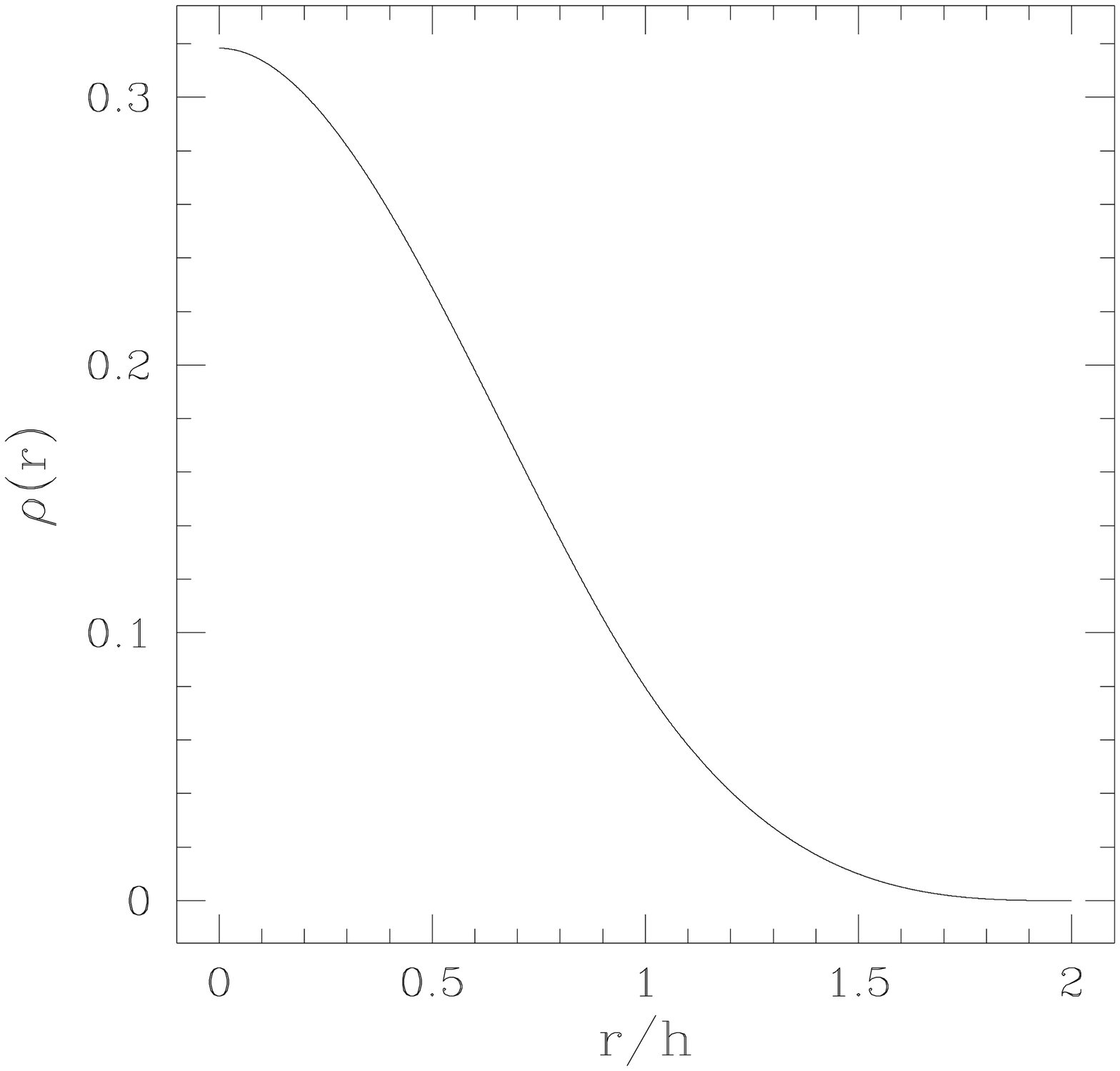}}
\resizebox{7.75cm}{!}{\includegraphics{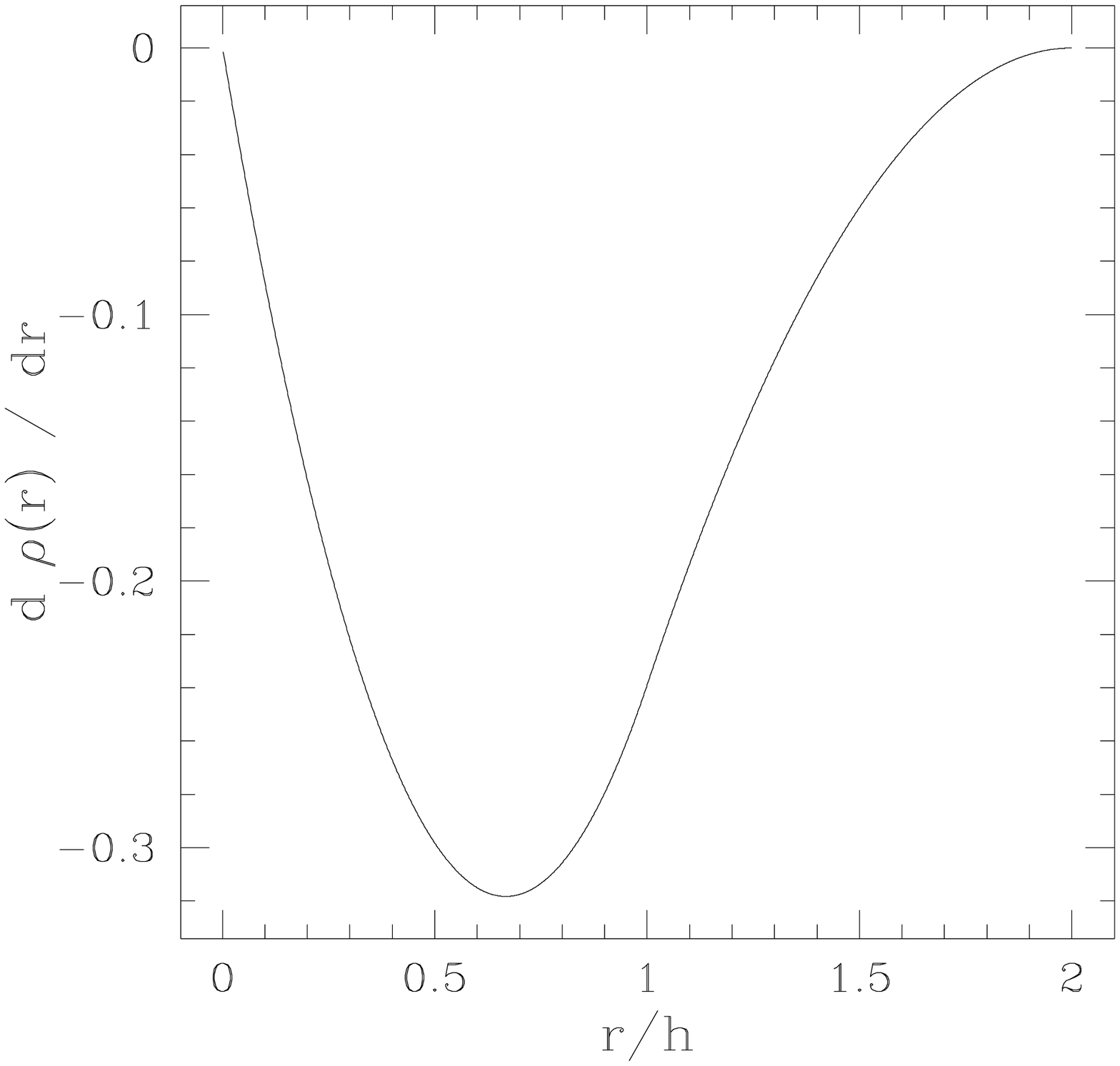}}  
      \caption[Density and its gradient for an isolated particle]{\underline{Left}: Radial density profile of an isolated particle of unit mass and smoothing length $h$. \underline{Right}: The gradient of the radial density profile of an isolated particle of unit mass and smoothing length $h$.}
      \label{fig:sph_profile}
\end{figure}

For example, the density at position {\bf r}, is given by

\begin{equation}
\label{equa:density}
\rho({\bf r}) \; = \; \sum_{i} m_{i} h_{i}^{-3} W \left( \frac{|{\bf r} - {\bf r}_{i}|}{h_{i}} \right).
\end{equation}

\noindent The density profile of an isolated particle of unit mass and smoothing length $h$, calculated with Eqn. \ref{equa:density}, is shown on the left panel of Fig. \ref{fig:sph_profile}.

The gradient of any quantity $A$ at a position {\bf r} is evaluated using

\begin{equation}
\label{equa:gradalpha}
{\bf \nabla} A({\bf r}) \; = \; \sum_{i} m_{i} \frac{A_{i}}{\rho_{i}}h_{i}^{-4} W^{'} \left( \frac{|{\bf r} - {\bf r}_{i}|}{h_{i}} \right) \frac{{\bf r} - {\bf r}_{i}}{|{\bf r} - {\bf r}_{i}|},
\end{equation}

\noindent where $W^{'}(s) \equiv \mathrm{d}W/\mathrm{d}s$ \cite{MonaghanAnA1985,MonaghanCPC1988,MonaghanARAnA1992}. Note that in such an expression, a second order truncation error\footnote{These errors are introduced because we substitute the integral expressions with sums \cite{MonaghanARAnA1992}.}, $O(h^{2})$, is introduced \cite{MoorePhD1995}. For the M4-kernel, the gradient is given by

\begin{equation}
\label{equa:gradM4}
W^{'}_{M4}(s) \; = \; - \frac{1}{\pi} \left \{
       \begin{array}{ll}
          3s  -  9s^{2}/4, & 0 \leqslant s \leqslant 1; \\
          3(2  -  s)^{2}/4, & 1 \leqslant s \leqslant 2; \\
          0, & s \geqslant 2.
       \end{array} \right .
\end{equation}

\noindent The gradient of the density is therefore given by

\begin{equation}
\label{equa:graddensity}
{\bf \nabla} \rho({\bf r}) \; = \; \sum_{i} m_{i} h_{i}^{-4} W^{'} \left( \frac{|{\bf r} - {\bf r}_{i}|}{h_{i}} \right) \frac{{\bf r} - {\bf r}_{i}}{|{\bf r} - {\bf r}_{i}|}.
\end{equation}

\noindent The gradient of the density profile of an isolated particle of unit mass and smoothing length $h$, calculated with Eqn. \ref{equa:graddensity}, is shown on the right panel of Fig. \ref{fig:sph_profile}.

\section{Basic SPH equations}{\label{sec:equations}}

Following Eqn. \ref{equa:gradalpha}, the first two of the set of three hydrodynamical equations that we need to solve (namely Eqns. \ref{equa:continuity} \& \ref{equa:Euler}) should read as follows:

\begin{equation}
\label{equa:SPHdrhodt}
\frac{\mathrm{d}\rho({\bf r})}{\mathrm{d}t} \; = \; - \rho({\bf r}) \sum_{i} \frac{m_{i}}{\rho_{i}} h_{i}^{-4} \mbox{\bf \em v}_{i} \cdot W^{'} \left( \frac{|{\bf r} - {\bf r}_{i}|}{h_{i}} \right) \frac{{\bf r} - {\bf r}_{i}}{|{\bf r} - {\bf r}_{i}|}
\end{equation}

\noindent and

\begin{equation}
\label{equa:SPHdvdt}
\frac{\mathrm{d}\mbox{\bf \em v}({\bf r})}{\mathrm{d}t} \; = \; - \frac{1}{\rho({\bf r})} \sum_{i} m_{i} h_{i}^{-4} \frac{P_{i}}{\rho_{i}} W^{'} \left( \frac{|{\bf r} - {\bf r}_{i}|}{h_{i}} \right) \frac{{\bf r} - {\bf r}_{i}}{|{\bf r} - {\bf r}_{i}|} + {\bf a}_{grav}({\bf r}) + {\bf a}_{visc}({\bf r}),
\end{equation}

\noindent where ${\bf a}_{grav}$ is given by Eqn. \ref{equa:self-gravity} and ${\bf a}_{visc}$ is discussed in \S \ref{sec:artvisc}.

However, in such a case, we would have to use the value of $\rho({\bf r})$, numerically estimated using Eqn. \ref{equa:alpha}, which would introduce an error greater than the second order truncation error \cite{MoorePhD1995}. Instead, we use a formulation that includes density in the gradient of the function. This formulation derives from the identities 
\[
\rho \nabla A \; = \; \nabla(\rho A) \, - \, A \nabla \rho,
\] and
\[
\frac{\nabla A}{\rho} \; = \; \nabla(\frac{A}{\rho}) \, + \, \frac{A}{\rho^{2}} \nabla \rho.
\] Substituting for $\mbox{\bf \em v}$ and $P$ respectively, we obtain

\begin{equation}
\label{equa:golden1}
\rho \nabla \cdot \mbox{\bf \em v}\; = \; \nabla \cdot (\rho \mbox{\bf \em v}) \, - \, \mbox{\bf \em v} \cdot \nabla \rho,
\end{equation}

\noindent and

\begin{equation}
\label{equa:golden2}
\frac{\nabla P}{\rho} \; = \; \nabla(\frac{P}{\rho}) \, + \, \frac{P}{\rho^{2}} \nabla \rho.
\end{equation}

Eqns. \ref{equa:SPHdrhodt} \& \ref{equa:SPHdvdt} become

\begin{equation}
\label{equa:SPHdrhodtfinal}
\frac{\mathrm{d}\rho({\bf r})}{\mathrm{d}t} \; = \; \sum_{i} m_{i} h_{i}^{-4} \left( \mbox{\bf \em v} ({\bf r}) - \mbox{\bf \em v}_{i} \right) \cdot W^{'} \left( \frac{|{\bf r} - {\bf r}_{i}|}{h_{i}} \right) \frac{{\bf r} - {\bf r}_{i}}{|{\bf r} - {\bf r}_{i}|}
\end{equation}

\noindent and

\begin{equation}
\label{equa:SPHdvdtfinal}
\frac{\mathrm{d}\mbox{\bf \em v}({\bf r})}{\mathrm{d}t} \; = \; - \sum_{i} m_{i} h_{i}^{-4} \left( \frac{P_{i}}{\rho_{i}^{2}} + \frac{P({\bf r})}{\rho({\bf r})^{2}} \right) W^{'} \left( \frac{|{\bf r} - {\bf r}_{i}|}{h_{i}} \right) \frac{{\bf r} - {\bf r}_{i}}{|{\bf r} - {\bf r}_{i}|} + {\bf a}_{grav}({\bf r}) + {\bf a}_{visc}({\bf r}).
\end{equation}

As mentioned at the beginning of the previous section, SPH follows the evolution of the hydrodynamical properties of a fluid represented by a system of particles -- sample points. These particles are allowed to move with the fluid, as  they trace elements of constant mass. The particle motion will be discussed further in \S \ref{sec:smoothing}. A consequence of this fact is that we only need to estimate the fluid hydrodynamical properties at the particle positions, $j$. Therefore, Eqns. \ref{equa:SPHdrhodtfinal} \& \ref{equa:SPHdvdtfinal} reduce to

\begin{equation}
\label{equa:drhodtSPH}
\frac{\mathrm{d}\rho_{j}}{\mathrm{d}t} \; = \; \sum_{i} m_{i} \bar{h}_{ij}^{-4} \mbox{\bf \em v}_{ij} \cdot W^{'} \left( \frac{|{\bf r}_{ij}|}{\bar{h}_{ij}} \right) \frac{{\bf r}_{ij}}{|{\bf r}_{ij}|}
\end{equation}

\noindent and

\begin{equation}
\label{equa:dvdtSPH}
\frac{\mathrm{d}\mbox{\bf \em v}_{j}}{\mathrm{d}t} \; = \; - \sum_{i} m_{i} \bar{h}_{ij}^{-4} \left( \frac{P_{i}}{\rho_{i}^{2}} + \frac{P_{j}}{\rho_{j}^{2}} \right) W^{'} \left( \frac{|{\bf r}_{ij}|}{\bar{h}_{ij}} \right) \frac{{\bf r}_{ij}}{|{\bf r}_{ij}|} + {\bf a}_{grav,j} + {\bf a}_{visc,j},
\end{equation}

\noindent where $\mbox{\bf \em v}_{ij} = \mbox{\bf \em v}_{j} -  \mbox{\bf \em v}_{i}$, ${\bf r}_{ij} = {\bf r}_{j} - {\bf r}_{i}$ and $\bar{h}_{ij} \, = \,  0.5 (h_{i} + h_{j})$. The latter is used in order to utilise the symmetrical forms that Eqns. \ref{equa:drhodtSPH} \& \ref{equa:dvdtSPH} have obtained with the use of Eqns. \ref{equa:golden1} \& \ref{equa:golden2}. This way we make sure that every interaction between pairs of particles is symmetric and hence we ensure that linear and angular momentum are conserved.

To calculate density we use Eqn. \ref{equa:density} instead of the continuity Eqn. \ref{equa:drhodtSPH}. This reads as

\begin{equation}
\label{equa:densityfinal}
\rho_{j} \; = \; \sum_{i} m_{i} \bar{h}_{ij}^{-3} W \left( \frac{|{\bf r}_{ij}|}{\bar{h}_{ij}} \right).
\end{equation}

\noindent This formulation conserves mass very accurately because of the fact that the kernel function has compact support and it is appropriately normalised. It is therefore prefered to Eqn. \ref{equa:drhodtSPH}. It is also simpler and faster to calculate. In our integration scheme (\S \ref{sec:integrate}), we use Eqn. \ref{equa:dvdtSPH} to calculate the acceleration. 

The equation for the rate of change of the specific internal energy can also be formulated in a symmetric form \cite{MonaghanARAnA1992}, 
\[ 
\frac{\mathrm{d}u_{j}}{\mathrm{d}t} = \frac{1}{2} \sum_{i} m_{i} \bar{h}_{ij}^{-4} \left( \frac{P_{i}}{\rho_{i}^{2}} + \frac{P_{j}}{\rho_{j}^{2}} \right) \mbox{\bf \em v}_{ij} \cdot W^{'} \left( \frac{|{\bf r}_{ij}|}{\bar{h}_{ij}} \right) \frac{{\bf r}_{ij}}{|{\bf r}_{ij}|} + \frac{1}{\rho_{j}} \sum_{i} m_{i} \bar{h}_{ij}^{-3} \frac{(\Gamma_{i} - \Lambda_{i})}{\rho_{i}} W \left( \frac{|{\bf r}_{ij}|}{\bar{h}_{ij}} \right).
\] However, here we are using a barotropic equation of state (Eqn. \ref{equa:state}) to obtain the pressure. In our formulation, the barotropic equation of state reads as follows:

\begin{equation}
\label{equa:statefinal}
\frac{P_{j}}{\rho_{j}} \; = \; c_{0}^{2} \left[1 + \left( \frac{\rho_{j}}{\rho_{0}} \right) ^{4/3} \right]^{1/2}.
\end{equation}

\section{Artificial viscosity}{\label{sec:artvisc}}

In regions where particle streams collide interpenetration may occur. This is undesirable since colliding fluid elements should retain their relative positions; the colliding streams should be decelerated by shocks.

In order to make sure that particle interpenetration is inhibited and that we obtain well-defined (but not necessarily well-resolved) shocks, we have
to include an `artificial viscosity'. We use the artificial viscosity
described in Monaghan \shortcite{MonaghanARAnA1992}. This includes a linear
bulk viscosity component that prevents interpenetration as well as a Von
Neumann-Richtmeyer type viscosity component.
The artificial viscous acceleration that acts on a particle $j$ is given by

\begin{equation}
\label{equa:viscosity}
{\bf a}_{visc,j} \; = \; - \sum_{i} m_{i} \bar{h}_{ij}^{-4} \Pi_{ij}
W^{'} \left( \frac{|{\bf r}_{ij}|}{\bar{h}_{ij}} \right)
\frac{{\bf r}_{ij}}{|{\bf r}_{ij}|},
\end{equation}

\noindent where 

\begin{equation}
\label{equa:Piartvisc}
\Pi_{ij} \; = \; \left \{
\begin{array}{ll}
\frac{- \alpha \mu_{ij} \bar{c}_{ij} + \beta \mu_{ij}^{2}}
{\bar{\rho}_{ij}}, & (\mbox{\bf \em v}_{ij} \cdot {\bf r}_{ij}) < 0; \\
0, & (\mbox{\bf \em v}_{ij} \cdot {\bf r}_{ij}) > 0,
\end{array} \right .
\end{equation}

\noindent and

\begin{equation}
\label{equa:mu}
\mu_{ij} \, = \, \frac{(\mbox{\bf \em v}_{ij} \cdot {\bf r}_{ij})
\bar{h}_{ij}}{|{\bf r}_{ij}|^{2} \, + \, 0.01 \bar{h}_{ij}^{2}} 
\end{equation}

\noindent while ${\bar{\rho}_{ij}} = 0.5 (\rho_{i} + \rho_{j})$ and ${\bar{c}_{ij}} = 0.5 (c_{i} + c_{j})$
(the average sound speed). The $\mbox{\bf \em v}_{ij} \cdot
{\bf r}_{ij} < 0$ condition makes sure that only particles that are
approaching particle $j$ will contribute to its artificial
viscous acceleration. The above formula is symmetric in order to make
sure that particle $j$ will in turn exert an equal and opposite artificial viscous acceleration on any of the neighbouring particles, $i$, that are approaching.

$\Pi_{ij}$ has dimensions of velocity squared over density. The $\alpha$-term
gives an artificial viscous acceleration similar to the hydrodynamical acceleration if
sound speed squared is replaced by the product of sound speed times the
relative velocity. It can treat a bulk colliding flow, but in high
Mach number shocks the $\beta$-term is needed, as in such cases, the relative
velocity of the colliding flows is large compared to the local sound speed.
$\alpha$ and $\beta$ are tunable parameters and should take appropriately
large values to prevent interpenetration. If $\alpha=\beta=0$ then the
artificial viscous acceleration is zero, particles penetrate each other and
shocks cannot be modelled. If $\alpha$ and $\beta$ are very large then the
shock is very broad and it ends up not very well resolved; even mild sound waves are rapidly damped. We have adopted the value $\alpha=\beta=1$. It has been shown that these values give good results (\S \ref{sec:tube}, also Patsis (1999 private communication)).

Watkins {\it et al.} \shortcite{WatkinsAnASS1996} have shown that one can use the Navier-Stokes equations for viscous compressible flows to derive a viscous
acceleration similar to the $\alpha$ term, but the $\beta$ term is still
needed to model accurately high Mach number shocks. The Navier-Stokes viscosity was derived to give a more realistic treatment of shear viscosity (e.g. in problems involving the evolution of protostellar discs). The formulation of Watkins {\it et al.} \shortcite{WatkinsAnASS1996} is derived from the standard SPH
cross product expressions \cite{MonaghanARAnA1992}.

The artificial viscous acceleration of Eqn. \ref{equa:viscosity} should be added
to the hydrodynamical acceleration of Eqn. \ref{equa:dvdtSPH}, as we shall see in \S \ref{sec:final}.

\section{Tree Code Gravity}{\label{sec:TCG}}

We now have to define the $a_{grav,j}$ term of Eqn. \ref{equa:dvdtSPH}. In the
case of point masses, the gravitational acceleration on each particle $j$ should
be calculated as

\begin{equation}
\label{eq:N2grav}
{\bf a}_{grav,j} \; = \; - \sum_{i, i \ne j} m_{i} \frac{{\bf r}_{ij}}
{|{\bf r}_{ij}|^{3}},
\end{equation}

\noindent where we have used units so that $G = 1$. However, for a large number
of particles, $N$, such a formulation becomes very expensive, as the computational cost scales as
$O(N^{2})$. Instead, we implement Tree-Code-Gravity (TCG) \cite{BarnesNATURE1986,HernquistApJSS1989}, which scales as $O(N log N)$.

With this method, a tree is constructed containing spatial information on 
individual particles and the centres of mass of groups of particles. This way,
for distant interactions we can substitute individual particles with groups 
of particles. At every integration we have to construct 
the tree, walk up the tree to calculate the centres of mass and walk down the 
tree to calculate the gravitational acceleration.

For the construction of the tree, we use the whole computational domain as 
the rootcell, the top level of the tree. We then divide the rootcell into 
8 subcells. These subcells define the first level down the tree. Each of 
them is subsequently divided into another 8 subcells, i.e. the next level 
down in the tree, etc. A cell at any level is not divided further only 
when it contains either a single particle or no particle at all.

If a cell contains more than one particle, its centre of mass is calculated
using all its subcells in all lower levels. This way, for the rootcell we
calculate the centre of mass of the whole computational domain. For each cell, 
we save the following information: cell centre, linear dimensions, pointers 
to its subcells, total mass and centre of mass, pointers to particles.

We calculate the gravitational acceleration at particle $j$ using the centre 
of mass of a cell unless the cell is so close that we must use its 
subcells instead. We apply a geometrical criterion in order to decide whether 
we shall use a cell or its subcells. Specifically, a cell is used if 

\[
\frac{l}{D} <  \theta,
\]

\noindent where $l$ is the linear size of the cell under consideration, $D$ is
the distance between particle $j$ and this cell and $\theta$ is the maximum opening 
angle, an accuracy parameter. If the criterion is not satisfied then the cell 
is `opened' and its subcells are examined. If a subcell contains a single 
particle then a particle-particle interaction is calculated. The value of 
$\theta$ should be sufficiently small for close interactions to be calculated as
particle-particle. Salmon, Warren \& Winckelmans \shortcite{SalmonJSA1994} examined different values for $\theta$ and they determined that accurate calculations of the gravitational acceleration were obtained for $\theta < 0.577$. We have chosen the value of $\theta = 0.576$ again driven by the need to balance accuracy against speed of calculation.

Therefore, the gravitational acceleration of particle $j$ is the sum of 
contributions from other particles and cells. For a particle-particle 
interaction with particle $i$ we use

\begin{equation}
\label{equa:part-part}
{\bf a}_{grav,ij} \; = \; - m_{i} \frac{{\bf r}_{ij}}{|{\bf r}_{ij}|^{3}},
\end{equation}

\noindent with the potential energy given by

\begin{equation}
\label{equa:potepart-part}
m_{i} \Phi_{ij} \; = \; - m_{i} m_{j} \frac{1}{|{\bf r}_{ij}|}.
\end{equation}

\noindent For the interaction with cell $k$ we write 
\cite{GoldsteinBOOK1980}

\begin{equation}
\label{equa:part-cell}
{\bf a}_{grav,kj} \; = \; - m_{k} \frac{{\bf r}_{kj}}{|{\bf r}_{kj}|^{3}} + \frac{{\bf Q} \cdot {\bf r}_{kj}}
{|{\bf r}_{kj}|^{5}} - \frac{5}{2} \left( {\bf r}_{kj} \cdot {\bf Q} \cdot {\bf r}_{kj} \right) \frac{{\bf r}_{kj}}{|{\bf r}_{kj}|^{7}},
\end{equation}

\noindent and

\begin{equation}
\label{equa:potepart-cell}
m_{k} \Phi_{kj} \; = \; - m_{k} m_{j} \frac{1}{|{\bf r}_{kj}|} - 
\frac{1}{2} \left( {\bf r}_{kj} \cdot {\bf Q} \cdot
{\bf r}_{kj} \right) \frac{m_{k}}{|{\bf r}_{kj}|^{5}}.
\end{equation}

\noindent Here ${\bf Q}$ is the traceless quadrupole tensor about the centre of mass. It is defined as \cite{GoldsteinBOOK1980}

\begin{equation}
\label{equa:Q}
Q_{ab} = \sum_{p=1}^{N_{ptcls}} m_{p} (3 x_{a,p} \, x_{b,p} - r_{p}^{2} 
\delta_{ab}),
\end{equation}

\noindent where $a,b$ run from 1 to 3 for each direction in Euclidean space 
and $\delta_{ab}$ is the Kronecker delta, e.g. $Q_{11} = \sum m_{p}
(2x_{p}^{2} - y_{p}^{2} - z_{p}^{2})$ and $Q_{12} = 3 \sum m_{p} x_{p} y_{p}$, etc.

The quadrupole moment of a cell in the tree is based on the quadrupole moments of 
its subcells:

\begin{equation}
\label{equa:Qcell}
Q_{ab} = \sum_{p=1}^{N_{subcell}} (Q_{ab})_{p} + \sum_{p=1}^{N_{subcell}} 
m_{p} (3 x_{a,p} \, x_{b,p} - r_{p}^{2} 
\delta_{ab}),
\end{equation}

\noindent where $x_{a,p}, x_{b,p}$ are now the coordinates of the subcells 
in the reference frame of the parent cell.

With SPH we try to describe all quantities involved as continuous functions  
both in space and time. Therefore, we have to reduce the magnitude of close  
particle interactions in order to avoid unduly large gravitational accelerations
($\propto 1/r^{2}$). The particles are smoothed similarly to the way they are  
treated for hydrodynamics, i.e. as spherically symmetric and finite in extent  
with a radius of $2\epsilon$. If two particles overlap, the mass involved 
in the calculation of the mutual gravitational acceleration is calculated  
from Gauss' gravitational theorem, otherwise the particles are treated as point masses. 
 
For particle $i$ the mass interior to radius $s \epsilon_{i}$ is  
given by $m(s) = m_{i} W^{*}(s)$ using $\rho_{i}(s) \, = \, m_{i}  
\epsilon_{i}^{-3} W(s)$ and hence 
 
\[ 
W^{*}(s) = \int_{0}^{s} 4 \pi u^{2} W(u) \mathrm{d}u. 
\] 
 
If particle $i$ is less than $2\bar{\epsilon}_{ij} = \epsilon_{i} + \epsilon_{j}$ away from particle $j$ then the gravitational acceleration at particle $j$ will be 
 
\begin{equation} 
\label{equa:realacc} 
{\bf a}_{grav,ij} \; = \; - m_{i} W^{*} \left( \frac{|{\bf r}_{ij}|}{\bar{\epsilon}_{ij}} \right) \frac{{\bf r}_{ij}}{|{\bf r}_{ij}|^{3}}, 
\end{equation} 

\noindent which means that the mass of particle $i$ outside $r_{ij} =  
|{\bf r}_{ij}|$ is not being taken into account. With this symmetric description of the gravitational interaction between particles $i$ and $j$, we do not have to specify to which of the two particles we have applied Gauss' gravitational theorem. Then the total gravitational acceleration at particle $j$ is

\begin{equation} 
\label{equa:totalrealacc} 
{\bf a}_{grav,j} \; = \; - \sum_{i,i \ne j} m_{i} W^{*} \left( \frac{|{\bf r}_{ij}|}{\bar{\epsilon}_{ij}} \right) \frac{{\bf r}_{ij}}{|{\bf r}_{ij}|^{3}}.
\end{equation}

\noindent The potential at distance $r_{ij}$ away from particle $j$ is then given by  
 
\[ 
\Phi_{ij} = - \int_{r_{ij}}^{\infty} \frac{m_{j}}{r^{2}} W^{*}(r/\bar{\epsilon}_{ij}) 
\mathrm{d}r. 
\] 
 
Integrating by parts we obtain the expression for the mutual potential energy of two 
particles $i$ and $j$:
 
\begin{equation} 
\label{equa:potenergy} 
m_{i} \Phi_{ij} \; = \; - \frac{m_{i} m_{j}}{r_{ij}} \left( W^{*}(s) +  
W^{**}(s) \right), 
\end{equation} 
 
\noindent where  
 
\[ 
W^{**}(s) = s \int_{s}^{\infty} 4 \pi u W(u) \mathrm{d}u. 
\] 
 
For the M4-kernel we obtain 
 
\begin{equation} 
\label{equa:wstar} 
W_{M4}^{*}(s) \; = \; \frac{1}{30} \left \{
       \begin{array}{ll}
          40s^{3}  -  36s^{5}  +  15s^{6}, & 0 \leqslant s \leqslant 1; \\
          80s^{3}  -  90s^{4}  +  36s^{5}  -  5s^{6}  -  2, & 1 \leqslant s \leqslant 2; \\
          30, & s \geqslant 2,
       \end{array} \right .
\end{equation} 
 
\noindent and  
 
\begin{equation} 
\label{equa:wstarsq} 
W_{M4}^{**}(s) \; = \; \frac{s}{10} \left \{
       \begin{array}{ll}
          14  -  20s^{2}  +  15s^{4}  -  6s^{5}, & 0 \leqslant s \leqslant 1; \\
          (2s  +  1)(2  -  s)^{4}, & 1 \leqslant s \leqslant 2; \\
          0, & s \geqslant 2,
       \end{array} \right .
\end{equation} 
 
\noindent $\epsilon$ can take either a small constant value ($\epsilon$- softening)  
which is not very good when the distance between particles changes  
considerably during the simulation; or it can take the value of the hydrodynamical  
smoothing length $h$, which has the benefit that it smoothes the gravitational  
and hydrodynamical forces by the same amount \cite{BateMNRAS1997}. We use the latter choice, i.e. $\epsilon_{ij} = h_{ij}$. 

\section{Smoothing length}{\label{sec:smoothing}}

In this section we shall discuss the importance of the hydrodynamical smoothing length, $h$, and we shall define the range of values that it should take. SPH is a Lagrangian particle method, with the particles -- sample points moving with the fluid, containing constant mass within their radius of influence. In particular, the mass of the SPH particles is considered to be smoothed over a finite volume. As mentioned in \S \ref{sec:fundamentals}, at any time, only a few neighbouring particles overlap with the smoothing radius of any particle. Therefore, we choose the radius of influence of the smoothing kernel function, $2h$, so that for any particle and at any time, it contains an approximately constant number of neighbouring particles. For this reason, we will use a time-varying non-universal $h$, as each particle requires its own smoothing length (e.g. in a dense region a smaller $h$ is required, than in a rarefied region, in order to contain this constant number of neighbours). Nelson \& Papaloizou \shortcite{NelsonMNRAS1993,NelsonMNRAS1994} have shown that with adaptive $h$ the energy is conserved quite accurately.

Because we have to treat all the physical quantities involved as continuous functions both in time and space, we need to take into account a large number of interactions to reduce sampling errors. This needs to be balanced by the fact that for too big a smoothing length SPH will under-estimate the self-density of the particles and will over-smooth all the properties of the fluid. Therefore the problem of finding the correct value for $h$ is reduced to finding the right value for the number of neighbours, $N_{want}$, for each particle. Tests on known distributions have shown that $N_{want} \sim 50$ neighbours within $2h$ gives good results with the 3-D M4-kernel.

We allow a 10\% fluctuation in the value of the number of neighbours, $N$,  
for each particle at any time. Thus, we allow $N$ to be between $N_{min}=45$ 
and $N_{max}=55$. If $N_{j}^{n}$ is the number of neighbours of particle $j$ at time-step $n$, then a trial value for $h_{j}$ at time-step ($n+1$) is obtained by its value at the $n$th time-step according to 
 
\begin{equation} 
\label{equa:hupdate} 
h_{j}^{n+1} \; = \; h_{j}^{n} \left( \frac{N_{want}}{N_{j}^{n}} \right) ^{1/3}.
\end{equation} 
 
\noindent We then count the number of neighbours within $2h_{j}^{n+1}$, $N_{j}^{n+1}$. If $N_{j}^{n+1}$ is within the above limits then the value of $h_{j}^{n+1}$ is accepted. However, if the value of $N_{j}^{n+1}$ is not within the limits, we iterate over Eqn. \ref{equa:hupdate}, finding a new trial value for $h$ using $N_{j}^{n+1}$, obtaining a new number of neighbours according to this new trial value of $h$, and so on. We stop when $N$ is acceptable or when the fractional change of two successive trial values for $h$ becomes less than 1\%. After we have obtained the $h^{n+1}$ values for all particles, we substitute these values to the current value of $h$. To set the initial values of $h$ for all particles, we repeat the above procedure until either all particles have acceptable value for $N$, or we exceed a finite number of iterations. Numerical tests indicate that after 20 iterations most particles have an acceptable $N$.

\begin{figure}
\resizebox{7.75cm}{!}{\includegraphics{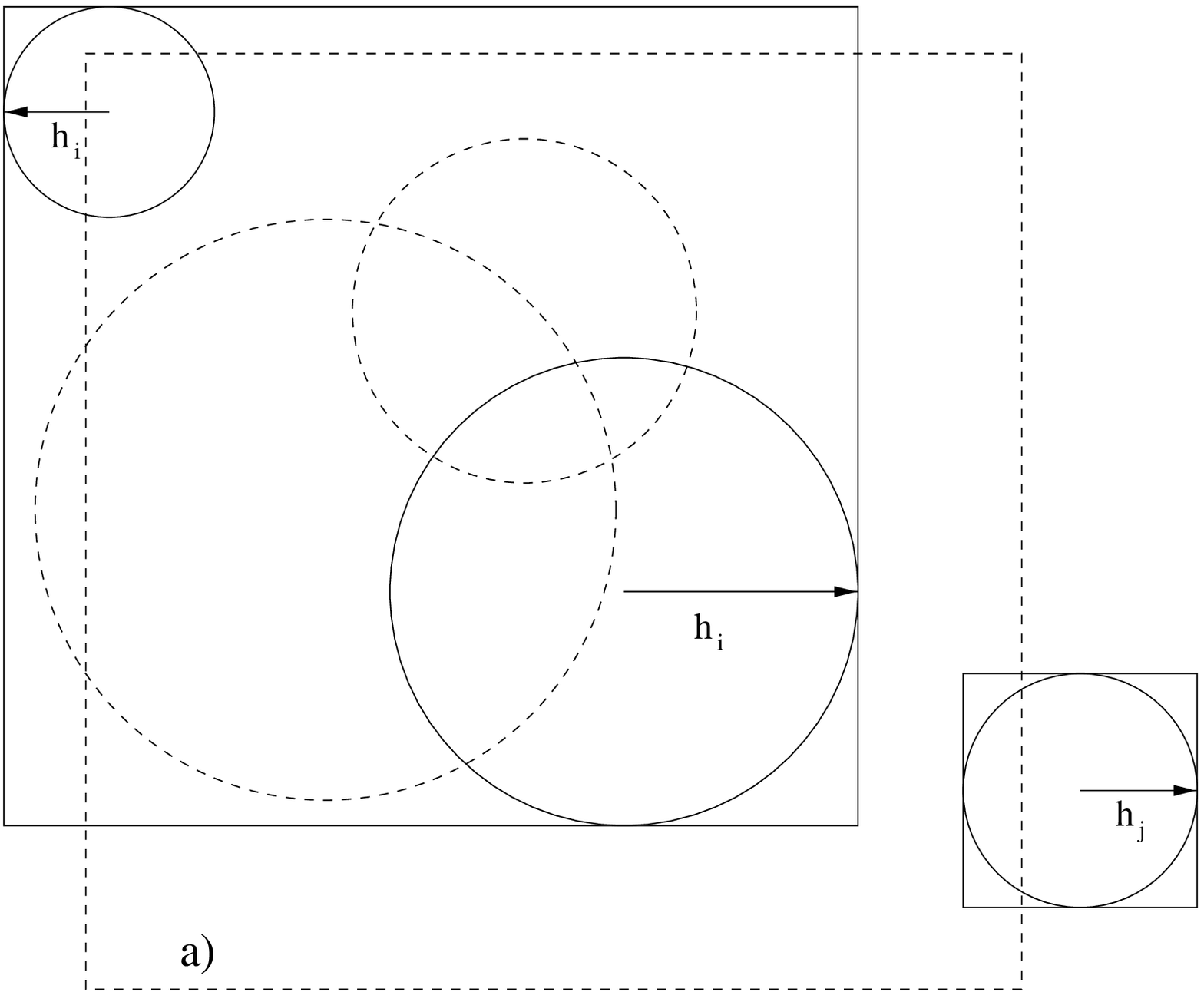}}
\resizebox{7.75cm}{!}{\includegraphics{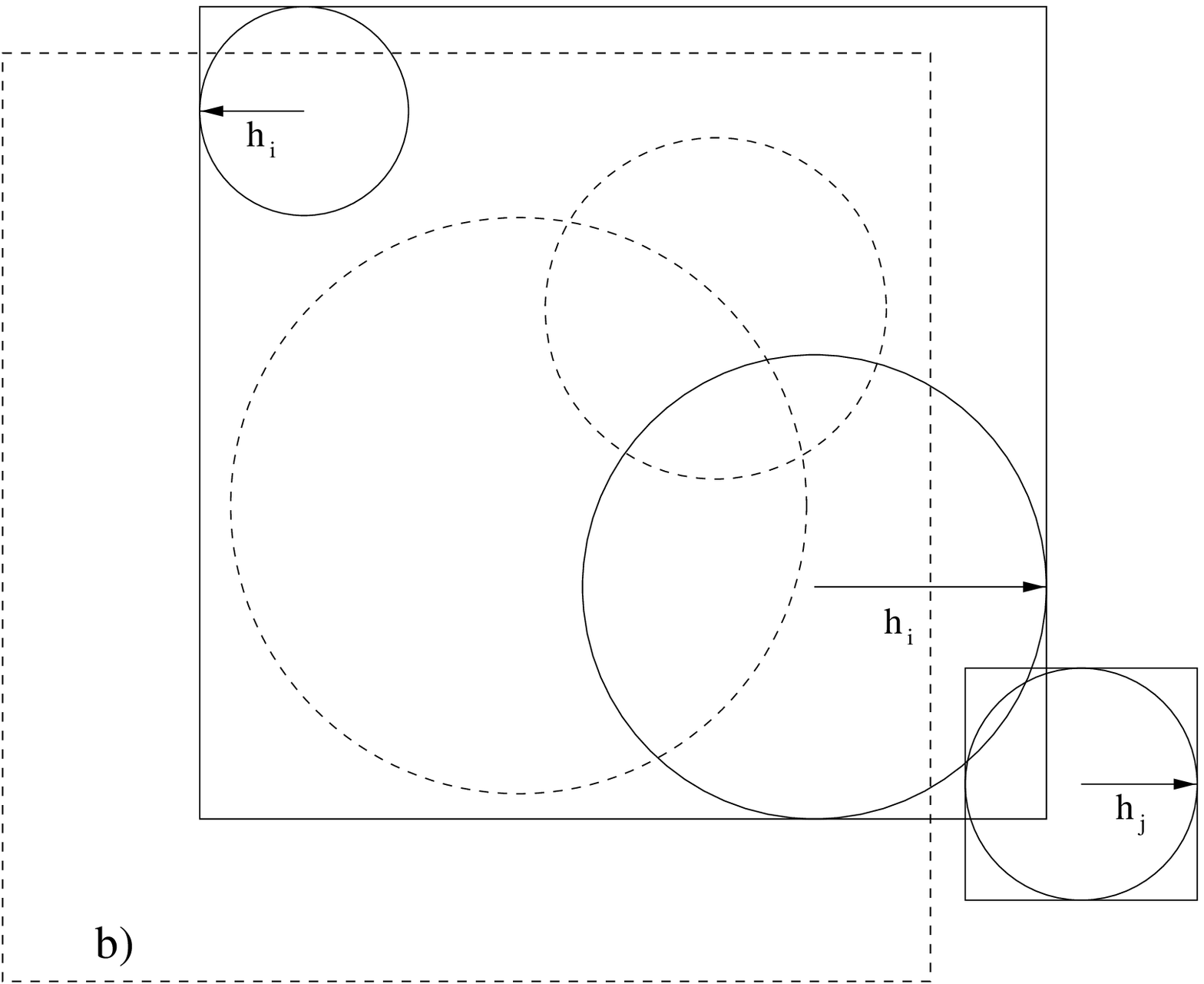}}  
      \caption[Finding SPH neighbours with the kernel box method]{Example of the kernel box method for finding SPH neighbours. We use the minimum box that contains all the particles in the cell, when the particles are taken each to extend to its smoothing radius $h_{i}$, instead of the cell itself (dashed box). a) Avoiding to open a cell unnecessarily: Particle $j$ would interact with the cell but now it does not interact with the kernel box. b) Obtaining an interaction we would have missed: Particle $j$ overlaps with particle $i$, but it does not overlap with the cell.}
      \label{fig:kernel-box}
\end{figure} 
 
We will now discuss the way we identify neighbours. A direct search would be an $N^{2}$ procedure, but using the spatial information stored in the gravity tree we can speed it up considerably. The original idea \cite{HernquistApJSS1989} for looking up in the tree for neighbours involved constructing a trial cube of side $4h_{j}$ for particle $j$ and then identifying which cells this cube overlaps, always starting with the rootcell. If these cells contain subcells they are then examined in a hierarchical fashion until we open subcells containing single particles. We then compare the inter-particle separation with $2h_{j}$ and decide accordingly. 

We use a slightly modified method in order to avoid opening cells unnecessarily \cite{BhattalPhD1996}. We store in the tree the values of $h$ for each particle contained in a cell. This way we can construct a bounding box, or `kernel box', for each cell. This represents the minimum box that contains all the particles in the cell, when each  particle is taken to extend to its smoothing radius $h_{i}$ (see Fig. \ref{fig:kernel-box}). The search process is as before but with a trial cube of $2h_{j}$ this time, using the kernel box for each cell instead of the cell itself. With this technique, we find all neighbours, missing no interactions within $2 \bar{h}_{ij}, \forall i,j : i \ne j$, as $\bar{h}_{ij}$ is the value we are actually using for our SPH equations (Eqn. \ref{equa:densityfinal} \& \ref{equa:dvdtSPH}). We only open a few cells unnecessarily and we can formulate this technique together with the calculation of the centres of mass for each cell so that we do not have to perform unnecessary walks of the tree.

\section{SPH equations}{\label{sec:final}}

We can now present the final forms of the three equations determining the evolution of the simulation. These are 

\begin{equation}
\label{equa:rhofinal}
\rho_{j} \; = \; \sum_{i} m_{i} \bar{h}_{ij}^{-3} W \left( \frac{|{\bf r}_{ij}|}{\bar{h}_{ij}} \right),
\end{equation}

\begin{equation}
\label{equa:dvdtfinal}
\frac{\mathrm{d}\mbox{\bf \em v}_{j}}{\mathrm{d}t} \; = \; - \sum_{i} m_{i} \frac{{\bf r}_{ij}}{|{\bf r}_{ij}|} \left( \bar{h}_{ij}^{-4} \left[ \left( \frac{P_{i}}{\rho_{i}^{2}} + \frac{P_{j}}{\rho_{j}^{2}} + \Pi_{ij} \right) W^{'} \left( \frac{|{\bf r}_{ij}|}{\bar{h}_{ij}} \right) \right] + \frac{1}{|{\bf r}_{ij}|^{2}} W^{*} \left( \frac{|{\bf r}_{ij}|}{\bar{h}_{ij}} \right) \right),
\end{equation}

\begin{equation}
\label{equa:eosfinal}
\frac{P_{j}}{\rho_{j}} \; = \; c_{0}^{2} \left[1 + \left( \frac{\rho_{j}}{\rho_{0}} \right) ^{4/3} \right]^{1/2}.
\end{equation}

\noindent where $\Pi_{ij}$ is given by Eqn. \ref{equa:Piartvisc}. For Eqn. \ref{equa:dvdtfinal} we have combined Eqns. \ref{equa:dvdtSPH}, \ref{equa:viscosity} \& \ref{equa:totalrealacc}. $W$, $W^{'}$ and $W^{*}$ are given by Eqns. \ref{equa:M4}, \ref{equa:gradM4} \& \ref{equa:wstar}, respectively.

Parenthetically, we note that if we were solving for $u$, the equation for the rate of change of the specific internal energy, after including heating from the artificial viscous forces, should read as

\[
\frac{\mathrm{d}u_{j}}{\mathrm{d}t} = \frac{1}{2} \sum_{i} m_{i} \bar{h}_{ij}^{-4} \left( \frac{P_{i}}{\rho_{i}^{2}} + \frac{P_{j}}{\rho_{j}^{2}} + \Pi_{ij} \right) \mbox{\bf \em v}_{ij} \cdot W^{'} \left( \frac{|{\bf r}_{ij}|}{\bar{h}_{ij}} \right) \frac{{\bf r}_{ij}}{|{\bf r}_{ij}|} + \frac{1}{\rho_{j}} \sum_{i} m_{i} \bar{h}_{ij}^{-3} \frac{(\Gamma_{i} - \Lambda_{i})}{\rho_{i}} W \left( \frac{|{\bf r}_{ij}|}{\bar{h}_{ij}} \right).
\] 

\section{Integration scheme}{\label{sec:integrate}}

We advance the positions and velocities of all particles in time using the second order Runge-Kutta integration scheme. This means that to advance particle $j$ from the $n$th to the $(n+1)$th step, first we need to calculate its position and velocity at the midpoint. This is given by

\begin{equation}
\label{equa:rmidposition}
{\bf r}_{j}^{n+1/2} \; = \; {\bf r}_{j}^{n} + \mbox{\bf \em v}_{j}^{n} \; \Delta t/2
\end{equation}

\begin{equation}
\label{equa:vmidposition}
\mbox{\bf \em v}_{j}^{n+1/2} \; = \; \mbox{\bf \em v}_{j}^{n} + {\bf a}_{j}^{n} \; \Delta t/2,
\end{equation}

\noindent where ${\bf a}_{j}^{n}$ is the total acceleration of particle $j$ at the $n$th step and $\Delta t$ is the discrete time-step with which all particles will be advanced \cite{PressBOOK1990}. We calculate ${\bf a}_{j}^{n}$ from Eqn. \ref{equa:dvdtfinal}. It is then easy to obtain the position of particle $j$ at the $(n+1)$th step from

\begin{equation}
\label{equa:rfinalposition}
{\bf r}_{j}^{n+1} \; = \; {\bf r}_{j}^{n} + \mbox{\bf \em v}_{j}^{n+1/2} \; \Delta t.
\end{equation}

\noindent However, in the meantime, we must calculate the total acceleration for $j$ at the midpoint, since its velocity at the $(n+1)$th step is given by

\begin{equation}
\label{equa:vfinalposition}
\mbox{\bf \em v}_{j}^{n+1} \; = \; \mbox{\bf \em v}_{j}^{n} + {\bf a}_{j}^{n+1/2} \; \Delta t.
\end{equation}

The selection of the time-step $\Delta t$ is of great importance. There are several time scales that can be defined locally in systems like the ones we follow. Firstly, the inverse of the local velocity divergence. Secondly, the ratio of the local length scale to the velocity at this scale. Thirdly, the square root of the ratio of the local length scale to the acceleration at this scale. And finally, a time scale similar to the second one, except that it involves the local sound speed instead of the total local velocity (separating the hydrodynamical properties of the fluid from its overall behaviour). For each particle, $i$,  we calculate the smallest of these time scales using its smoothing radius (radius of influence) as a local length scale, i.e. 

\begin{equation}
\label{equa:deltatime1}
\Delta t_{i} \; = \; \gamma \; MIN \left[ \frac{1}{|\nabla \cdot \mbox{\bf \em v}|_{i}}, \frac{h_{i}}{|\mbox{\bf \em v}_{i}|}, \left( \frac{h_{i}}{|{\bf a}_{i}|} \right)^{1/2}, \frac{h_{i}}{\sigma_{i}} \right],
\end{equation}

\noindent where 

\begin{equation}
\label{equa:sigma}
\sigma_{i} \; = \; c_{i} + \zeta \; (\alpha \, c_{i} + \beta \; \underset{j}{MAX} \left \{ \mu_{ij} \right \})
\end{equation}

\noindent is a modified sound speed which includes the effect of artificial viscosity. $\zeta$ is a parameter usually taken equal to 1.2. $\alpha$ and $\beta$ are the viscosity parameters (\S \ref{sec:artvisc}). $\underset{j}{MAX} \left \{ \mu_{ij} \right \}$ gives the largest contribution by a neighbour of particle $i$ to its viscous acceleration. The value of $\gamma$ is usually referred to as the Courant number and is given a sufficiently small value for the simulation to be well behaved as well as to conserve the total energy. We have adopted the value of $\gamma = 0.3$.

By choosing the smallest of these scales, we ensure that we do not evolve each particle for a time longer than any of the time scales dictated by the local dynamical/hydrodynamical properties. For the same reasons, we select the smallest of these minimum particle time scales for the value of the global time-step $\Delta t$ at any step $n$. Formally this is given by

\begin{equation}
\label{equa:deltatime2}
\Delta t \; = \; \overset{N}{\underset{i=1}{MIN}} \left \{ \gamma \; MIN \left[ \frac{1}{|\nabla \cdot \mbox{\bf \em v}|_{i}}, \frac{h_{i}}{|\mbox{\bf \em v}_{i}|}, \left( \frac{h_{i}}{|{\bf a}_{i}|} \right)^{1/2}, \frac{h_{i}}{\sigma_{i}} \right] \right \}.
\end{equation} 

\section{Multiple time-steps}{\label{sec:timesteps}}

\begin{figure}
\begin{center}
\resizebox{12cm}{!}{\includegraphics{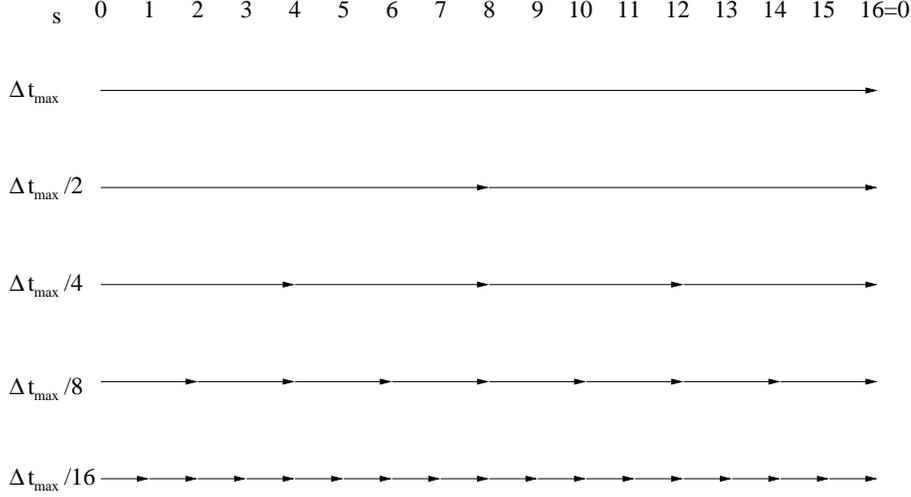}}
\end{center}  
      \caption[Multiple time-steps]{Graphical representation of multiple time-steps for the example of $N=5$ time bins. The steps of particles in different time bins for a period of $\Delta t_{max}$. A time-step $\Delta t = \Delta t_{max}/4$ (time bin $n = 2$) is accepted for a particle, only at $s=0, 4, 8, 12$.}
      \label{fig:arrows}
\end{figure}

If Eqn. \ref{equa:deltatime2} gives the value of the global time-step, it is obvious that there should be particles whose minimum local time scale is much larger than $\Delta t$. This means that if we can avoid evolving these particles with the minimum global time-step but with a time-step closer to the value they require, then there is a great gain in the speed of computation. This is the basic reason why we use the method of multiple time-steps \cite{BhattalPhD1996}. This method is ideal for simulations where there exist within the computational domain both dense regions (requiring small time-steps) and rarefied regions (not requiring small time-steps). The particles are assigned individual time-steps which are allowed to vary from step to step according to their need.

Taking into account the fact that at each step there are two half steps, the method creates a hierarchy of time-step bins, each containing particles that take one half step while the particles in the immediately lower bin take one full step. Therefore, the time-steps at any bin are twice as large as the ones at the immediately lower bin. The fact that the particles are not allowed to move with arbitrary time-steps but under this hierarchy of time bins, is dictated by the need to know the positions of all particles every time we calculate the accelerations and to therefore keep the system synchronised at regular intervals.

The values of the discrete times-steps used by particles in different time bins, are therefore calculated as fractions of a maximum time-step, $\Delta t_{max}$. In particular, the time-steps can take the following values: $\Delta t_{max}$, $\Delta t_{max}/2$, $\Delta t_{max}/4$, $\Delta t_{max}/8$, $\dots$ , $\Delta t_{max}/2^{N_{bins} - 1}$. We choose the total number of available time bins, $N_{bins}$, sufficiently large in order not to put any constraint on the time evolution of the simulation. Since at any time during a simulation $\Delta t_{max} = \Delta t_{min} 2^{N_{min} - 1}$, 
where $\Delta t_{min}$ is the minimum time-step from the ones used at this time (corresponding to the $N_{min}$th time bin), we can express the current position along the largest time-step as $s \Delta t_{min}$, where $s = 0, 1, 2, 3, \dots , 2^{N_{min} - 1}$. If $s=0$ the system is at the start of the maximum step and all particles are in-synch. 

For any particle, we calculate the ideal value of its time-step, $\Delta t_{ideal}$, from Eqn. \ref{equa:deltatime1}. We then put this particle into the next smaller time bin, $n$, with time-step $\Delta t = \Delta t_{max} / 2^{n}$, where $n$ is defined as

\begin{equation}
\label{equa:deltatimeideal3}
n \; = \; INT \left \{ \frac{\mathrm{ln} (\Delta t_{max} / \Delta t_{ideal})}{\mathrm{ln}2} \right \} + 1, \; \; \; \; n = 0, 1, 2, 3, \dots , (N_{bins} - 1).
\end{equation}

\noindent To ensure that at the end of $\Delta t_{max}$ all particles are in-synch, we have to check whether we can accept this bin, $n$, or not. A time bin is accepted only if the time from the beginning of the $\Delta t_{max}$ period is a multiple of the time this bin represents. Otherwise we choose the immediately lower acceptable time bin. For example, for N=5 we have $\Delta t_{max} = 16 \Delta t_{min}$. If the time-step we are checking is $\Delta t = \Delta t_{max}/4$ ($n = 2$), we can accept it only if $s=0, 4, 8, 12$. Otherwise we will assign to this particle a time-step $\Delta t = \Delta t_{max}/8$ ($n = 3$) if $s=0, 2, 4, 6, 8, 10, 12, 14$, or the lowest available time-step (bin $n = 4$) if $s$ is odd (Fig. \ref{fig:arrows}). With this test we ensure that a particle can move down to a lower time bin at any time, but can only move up the time ladder at times which allow the system to remain in-synch. 

As mentioned above, we need the positions and velocities of all particles to calculate the acceleration of the particles in the minimum time bin. To minimise errors we must update the positions and velocities of the particles in the upper bins for the intervening times. This is achieved by extrapolating the positions and velocities of these particles. Finally, at the end of every $\Delta t_{max}$ period the system is synchronised and the particles in all time bins have their positions and velocities updated by the integration scheme.

\section{Going through the code `step by step'}{\label{sec:step}}

\begin{figure}
\resizebox{15cm}{!}{\includegraphics{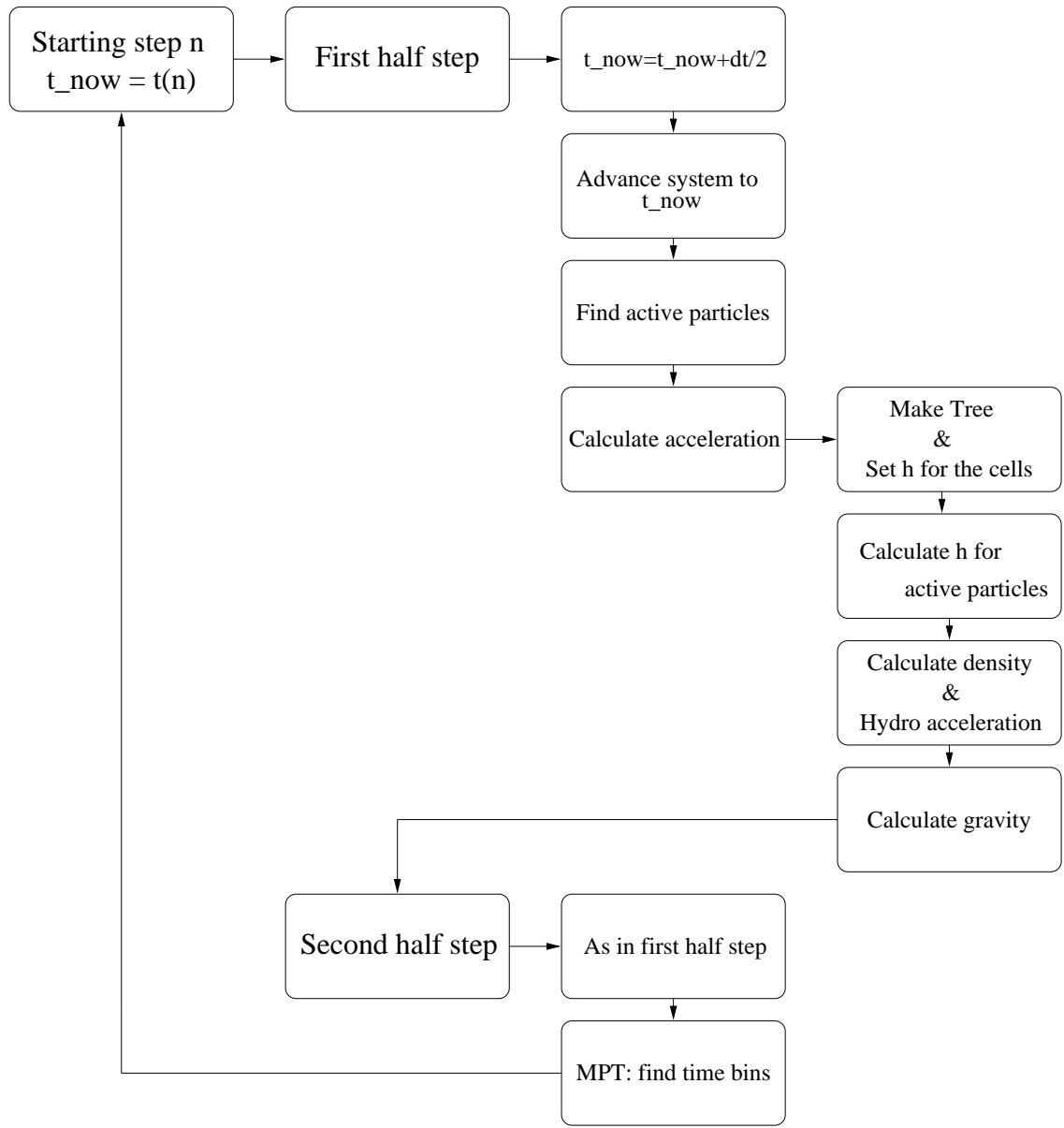}}  
      \caption[The code `step by step']{Flow-chart of the algorithm that dictates the evolution of the fluid in time. It represents the cycle, $n$, of the integration scheme, when the system advances with the time-step in the minimum time bin, $dt = \Delta t_{min}$.}
      \label{fig:step-by-step}
\end{figure}

Fig. \ref{fig:step-by-step} shows a flow-chart of the algorithm that dictates the evolution of the fluid in time. It demonstrates the way we assemble all the previous features of the numerical method. It represents the cycle $n$ of the integration scheme, when the system advances with the time-step in the minimum time bin, $dt = \Delta t_{min}$.

As mentioned in \S \ref{sec:integrate}, each step consists of two half steps. The first half step starts by advancing the system by $\Delta t_{min}/2$. Active particles have their positions and velocities updated by the integration scheme (Eqns. \ref{equa:rmidposition} \& \ref{equa:vmidposition}), while all other particles have their positions and velocities updated by extrapolation. We then need to calculate the acceleration at the half time-step for the particles in the minimum time bin. We proceed as follows: we make a new tree since the particles have just moved (\S \ref{sec:TCG}). Subsequently, we update the value of $h$ for the particles in the minimum time bin (\S \ref{sec:smoothing}). We can then calculate the hydrodynamical (\S \ref{sec:equations}), viscous (\S \ref{sec:artvisc}) and gravitational (\S \ref{sec:TCG}) acceleration for these particles. The total acceleration at the half time-step is given by Eqn. \ref{equa:dvdtfinal}.

We then proceed to the second half step, when tasks similar to those of the first step are performed. Specifically, the second half step starts by advancing the system by another 
$\Delta t_{min}/2$ (Eqns. \ref{equa:rfinalposition} \& \ref{equa:vfinalposition}), and thus completing a full time-step of the minimum bin $\Delta t_{min}$. Once again, particles in the minimum time bin have their positions and velocities updated by the integration scheme (using the acceleration at the midpoint calculated during the first half step), while all other particles have their positions and velocities updated by extrapolation. 
We then find the active particles for the next time-step, i.e. the particles that are going to be updated by the integration scheme and therefore need their acceleration calculated. This time the active particles are not just those in the minimum time bin, but all the particles for which the total time of the simulation is a multiple of, either the whole time-step, or half the time-step of the time bin they are in. 
For example, for $s=4$ in Fig. \ref{fig:arrows}, all the particles in the three smaller time bins are active as they start a new step, as well as the particles in the second larger time bin as they have just completed their first half step. We will then calculate the acceleration for all the active particles, following exactly the same steps as in the first half step. Finally, before starting time-step $(n+1)$ we re-distribute the particles into time bins. In order to make sure that at the end of a $\Delta t_{max}$ period the system is in-synch, we allow particles to move up the time ladder only if $\Delta t_{max}$ is a multiple of the time-step of their time bin, which in our example means the particles in the three smaller time bins. All particles that have just completed a full step are allowed to move to smaller time bins, in our example again the particles in the three smaller time bins.
                                                                                                                                                                                                      
Having calculated the accelerations for all the active particles, we can then proceed to the next time-step, $(n+1)$. 

\section{Tests}{\label{sec:SPHtests}}

We shall now describe the performance of the above numerical method on some standard tests for a self-gravitating hydrodynamical code. In particular, first we simulate the interaction of two colliding flows in order to test the efficiency of the code's treatment to hydrodynamics. In particular, we would like to quantify the efficiency of artificial viscosity (\S \ref{sec:artvisc}). We then allow a uniform sphere to collapse freely to test the efficiency of the TCG method. 
Finally, we combine both gravity and hydrodynamics to follow the evolution of a stable isothermal sphere in order to test the overall performance of the code. For each test we compare our results with the relevant analytic or semi-analytic solutions. We conclude that our numerical method treats effectively both the gravitational and hydrodynamical properties of the fluids we are going to simulate in chapters \ref{sec:rotating} \& \ref{sec:results}.

In quantifying the results of our tests, we need to be able to associate these results only with the performance of our numerical code. In order to decrease the numerical noise input by the initial distribution of particles, we perform all tests using clouds whose particles are taken initially to be on a lattice.

\subsection{Colliding Flows}{\label{sec:tube}}

\begin{figure}
\resizebox{7.75cm}{!}{\includegraphics{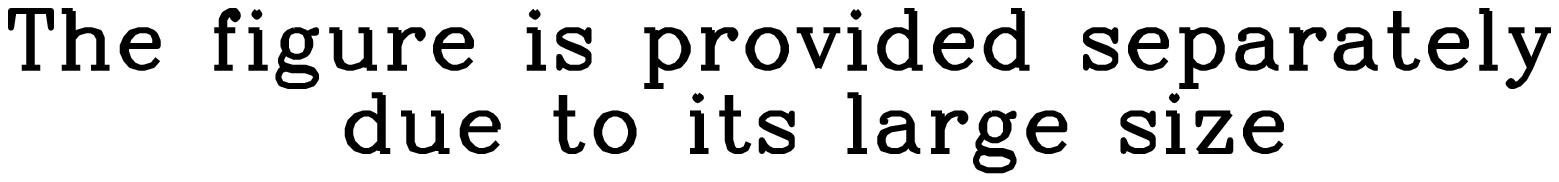}}
\resizebox{7.75cm}{!}{\includegraphics{./figs/size.eps}}  
      \caption[SPH tests: Colliding flows]{The density (left) and the velocity (right) of the simulated shock (points). The analytic solution is given by the solid lines (Dyson \& Williams 1980). The shock is broader by less than $2h$ compared to the analytic solution. It is very well resolved (contains $\sim$ 20 post-shock $h$).}
      \label{fig:colliding}
\end{figure}

In order to demonstrate the efficiency of the code's treatment of hydrodynamics, we have chosen to simulate the interaction between two colliding flows instead of a Riemann shock tube. 
We believe that the former test is more relevant to the problems we will investigate later in this thesis, i.e. high Mach number isothermal shocks with high compression factors. The typical Riemann shock tube test only involves a compression factor of $\sim 2$. Nevertheless, Hosking \shortcite{HoskingREPORT1999} has shown that the numerical method we are using, gives very good results for the Riemann shock tube test.

Two colliding isothermal flows will produce a strong shock, provided the Mach number of the collision, $\mathcal{M}$, is high. If $u_{0}$ is the pre-shock velocity of each flow in the reference frame of the shock and $c_{0}$ is the isothermal sound speed, then the Mach number is given by $\mathcal{M} = u_{0} / c_{0}$ and we obtain the post-shock velocity of each flow, $u_{2}$, from

\begin{equation}
\label{equa:post-shock}
u_{2} = \frac{c_{0}^{2}}{u_{0}}
\end{equation}

\noindent \cite{DysonBOOK1980}. From Eqn. \ref{equa:post-shock} we obtain the value for the velocity of the shock, $V_{s}$, with respect to the frame of the simulation, knowing the pre-shock and post-shock velocities of each flow in this frame of reference, $v_{0}$ and $v_{2}$ respectively. Specifically, because $u_{0} = v_{0} - V_{s}$ and  $u_{2} = - V_{s}$ (as $v_{2} = 0$), Eqn. \ref{equa:post-shock} gives \[- V_{s} = \frac{c_{0}^{2}}{v_{0} - V_{s}}.\] After some algebra we obtain

\begin{equation}
\label{equa:shockvel}
V_{s} = \frac{v_{0}}{2} - \frac{\surd{(v_{0}^{2} + 4 c_{0}^{2})}}{2}. 
\end{equation}

From the continuity equation we can write \[\rho_{2} = \frac{u_{0}}{u_{2}} \rho_{0}\] where $\rho_{0}$ and $\rho_{2}$ are the pre-shock and post-shock density for each flow, respectively. Finally, we can obtain an analytic estimate for the post-shock density for each flow

\begin{equation}
\label{equa:shockden}
\rho_{2} = \frac{V_{s} - v_{0}}{V_{s}} \rho_{0}. 
\end{equation}

The simulation involves two colliding isothermal flows ($T = 10$K) each of unit length in the direction of the collision. Each flow has a velocity of 1 km s$^{-1}$ in this direction. The isothermal sound speed at $10$K is $\sim 0.2$ km s$^{-1}$. Both flows have unit pre-shock densities. There are $\sim 10,000$ particles in total, taken initially on a lattice. The other two dimensions have lengths of $\sim 4h$. Eqn. \ref{equa:shockvel} gives $V_{s} = 0.039$ km s$^{-1}$. Therefore, the Mach number of the shock is $\cal{M} \sim$ 5.20. This is typical of the values we shall later model. Eqn. \ref{equa:shockden} gives a post-shock density of $\rho_{2} = 26.64$.

We evolve this collision with our 3-D SPH code, using periodic boundary conditions. We do not include the TCG method as it is not relevant for this test. Fig. \ref{fig:colliding} shows our results at the point where the rarefaction waves at the opposite ends of the tube have propagated to one tenth of the initial length for each flow. The mean value for the post-shock density, $\rho_{2}$, is $\sim$ 26 having a $\sim$ 11\% dispersion around the mean. The mean value for the post-shock velocity, $v_{2}$, is 0 to the 3rd significant figure, with a dispersion of $\sim$ 0.19 km s$^{-1}$ around the mean. The shocked layer is not broader than the analytic solution. In fact, the wings of the shocked layer are extended to less than $2h$. The layer is well resolved as its width is $\sim$ 20 times larger than the post-shock $h$.

The simulation reproduces the analytic compression factor accurately. In fact, it is slightly lower than the analytic value, as some small fraction of the pre-shock kinetic energy is transformed to post-shock random velocity dispersion instead of work done in compression. However, the biggest contribution to this random velocity dispersion comes from the breaking of the symmetry of the lattice. The particles in the shocked layer are not on a lattice any more and have random motions in all 3 dimensions. The artificial viscosity is very efficient though, as at the centre of the shocked layer the random motions are damped very effectively in all directions. Very little interpenetration is observed in the layer. 

We conclude that the results of this test are encouraging, and thus the choice of the values for the artificial viscosity parameters $\alpha = \beta = 1$ is justified. The simulation has produced a well resolved, not broad shock with a compression factor very close to the analytic value. Comparison of this test with realistic simulations will follow in chapter \ref{sec:results}.

\subsection{Free-Fall collapse}{\label{sec:ffcollapse}}

For the free-fall collapse simulation, we let a uniform density sphere of mass $M_{0}$ and radius $R_{0}$ collapse under its self-gravity. For this test we do not calculate the SPH equations, since the aim of this test is to verify the efficiency of the TCG method. Therefore, the acceleration in the integration scheme (\S \ref{sec:integrate}) is given only by Eqn. \ref{equa:totalrealacc}. 

\begin{figure}
\begin{center}
\resizebox{7.75cm}{!}{\includegraphics{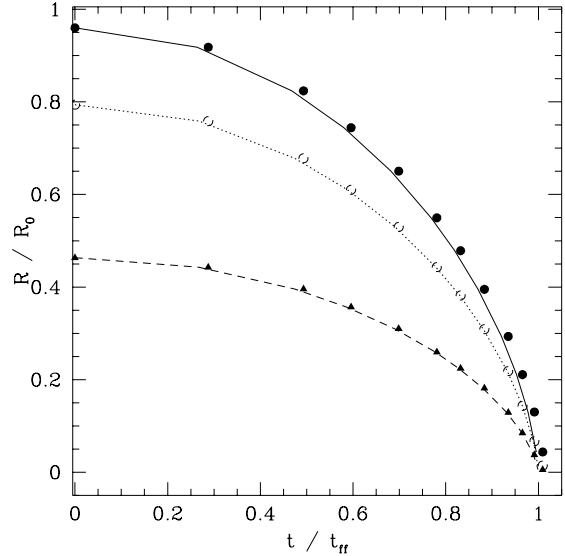}}
\end{center}
      \caption[SPH tests: Free-fall collapse]{Evolution of the 90\%, 50\% and 10\% mass radii of a uniform sphere in free-fall collapse. The points show the values obtained from the simulation every 50 time-steps, while the solid lines indicate the analytic solutions.}
      \label{fig:ff_collapse}
\end{figure}

The sphere should collapse homologously to a point after a free-fall time, $t_{ff}$. The analytic solution for the free-fall collapse of a uniform sphere states that any fluid element initially at radius $r_{0}$ will arrive at radius $r$ at time $t$ given by

\begin{equation}
\label{equa: free-fall}
\frac{t}{t_{ff}} \; = \; 1 - \frac{2}{\pi} \left\{ \sin^{-1} \left[ \left( \frac{r}{r_{0}} \right)^{1/2} \right] - \left( \frac{r}{r_{0}} \right)^{1/2} \left(1 - \frac{r}{r_{0}} \right)^{1/2} \right\}
\end{equation}

\noindent where $0 \leqslant t \leqslant t_{ff}$ \cite{SpitzerBOOK1978}. The free-fall time for a uniform density sphere is defined as 

\begin{equation}
\label{equa:freefalltime}
t_{ff} = \frac{\pi}{2} \left( \frac{R_{0}^{3}}{2 G M_{0}} \right)^{1/2}.
\end{equation}

We construct a $M_{0} = 1$ M$_{\odot}$ uniform density sphere with $\sim 10,000$ SPH particles on a lattice. The sphere is then let to collapse for $\sim$ 1 $t_{ff}$. As particles come closer due to collapse, the gravity softening $\epsilon = h$ is becoming smaller. However, $h$ decreases slower for the particles close to the edge of the sphere than for those at the centre. This happens because particles close to the edge can find their $\sim 50$ neighbours only from the inner side of the outer boundary of the sphere. Having a larger $h$ than they should, the gravitational acceleration for these particles is over-softened. This reduces the rate with which they collapse. To overcome this problem, and only for this test, we assign to all particles the same $h = \bar{h}$, the mean value of $h$. 

Fig. \ref{fig:ff_collapse} compares the analytic solution (solid lines) with the values obtained from the simulations every 50 time-steps (points), for the 90\%, 50\% and 10\% mass radii. The above edge effects, although reduced due to our treatment of $h$, give a small over-estimation for the 90\% mass radius. The inner two radii agree closely with the analytical solutions. 

\subsection{Stable Isothermal Sphere}{\label{sec:test_stable}}

\begin{figure}
\resizebox{7.75cm}{!}{\includegraphics{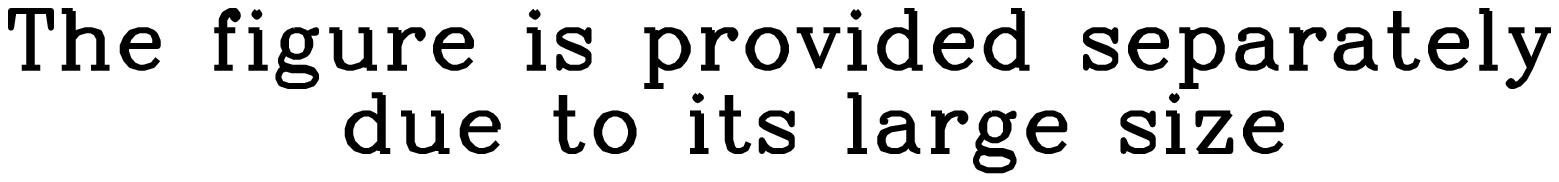}}
\resizebox{7.75cm}{!}{\includegraphics{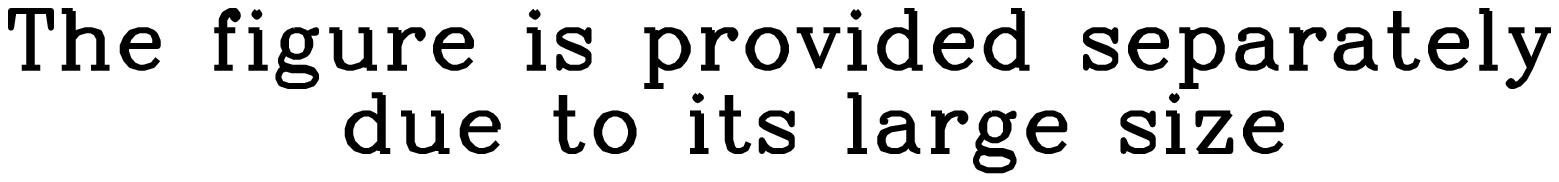}}  
      \caption[SPH tests: Stable isothermal sphere]{Initial radial density profile (left panel) and the radial density profile after $\sim$ 25 $t_{ff}$ (right panel). The solution to Eqn. \ref{equa:hydrobalancedimeless} is given by the solid lines.}
      \label{fig:iso_sphere}
\end{figure}

Finally, we test both the SPH and TCG methods in evolving a stable isothermal sphere. We construct the sphere using a combination of the hydrostatic balance and mass conservation equations

\begin{equation}
\label{equa:hydrobalance}
\frac{1}{r^{2}} \frac{\mathrm{d}}{\mathrm{d} r} \left( \frac{r^{2}}{\rho} \frac{\mathrm{d} P}{\mathrm{d} r} \right) + 4 \pi G \rho = 0
\end{equation}

\noindent with initial conditions \[ \rho(0) = \rho_{0}, \; \; \; \frac{\mathrm{d} \rho}{\mathrm{d} r}(0) = 0. \]
 
Using the Chandrasekhar \shortcite{ChandraBOOK1939} dimensionless expression for Eqn. \ref{equa:hydrobalance} we obtain

\begin{equation}
\label{equa:hydrobalancedimeless}
\frac{1}{\xi^{2}} \frac{\mathrm{d}}{\mathrm{d} \xi} \left( \xi^{2} \frac{\mathrm{d} \psi}{\mathrm{d} \xi} \right) - \mathrm{e}^{- \psi} = 0, \; \; \; \psi(0) = \frac{\mathrm{d} \psi}{\mathrm{d} \xi}(0) = 0,
\end{equation}

\noindent where we have substituted $P = c_{0}^{2} \, \rho$, $\rho = \rho_{0} \, \mathrm{e}^{- \psi}$ and $r = (4 \pi G \rho_{0})^{-1/2}  c_{0} \xi$. The solution of Eqn. \ref{equa:hydrobalancedimeless} formally extends to $\xi \rightarrow \infty$. We can truncate the sphere at a finite radius, $\Xi$, provided that the pressure at the boundary of the sphere is balanced by an external pressure. For $\Xi < 6.45$ the sphere is stable. 

We then solve Eqn. \ref{equa:hydrobalancedimeless} numerically. We tabulate the values for $\xi$, $\psi$ and $\mathrm{d} \psi / \mathrm{d} \xi$. Giving values to the total mass of the sphere $M_{0}$, the sound speed $c_{0}$ and  $\Xi$ we can substitute back to find $r$, $\rho(r)$ and M$(r)$. We have chosen M$_{0} = 1$ M$_{\odot}$, $c_{0}=0.17$ km s$^{-1}$ (corresponding to $T=7.9$K) and $\Xi = 3$. 

We then move $\sim$ 10,000 particles taken from a uniform density lattice to reproduce the tabulated values for $r$, $\rho(r)$ and M$(r)$. The solution to Eqn. \ref{equa:hydrobalancedimeless} is given by the solid lines in Fig. \ref{fig:iso_sphere}. We have also plotted the initial radial density profile (left panel) and the radial density profile after $\sim$ 25 $t_{ff}$ (right panel). We use crosses to present our initial data due to the fact that the initial distribution comes from a lattice and the radial profile consists of too few different points. We do not need crosses in the right panel of Fig. \ref{fig:iso_sphere} as the particles have moved from their initial positions.  However, the simulated radial density profile obeys the analytic solution very well. Note that at radii larger than $\sim$ 0.04 pc, there is an under-estimation of the density due to boundary effects similar to those described in \S \ref{sec:ffcollapse}. In particular, particles close to the edge of the cloud have larger $h$ than they should.

\chapter[Rotating Clouds with m=2 Density Perturbations]{Simulations of Rotating Clouds with m=2 Density Perturbations}{\label{sec:rotating}} 

Simulations of fragmentation are only reliable if the Jeans condition is 
obeyed (Truelove {\it et al}. \shortcite{TrueloveApJ1997,TrueloveApJ1998}, 
Klein {\it et al}. \shortcite{KleinTOKYO1998}, Boss {\it et al.} 
\shortcite{BossApJ2000}, Bate \& Burkert \shortcite{BateMNRAS1997}, 
Whitworth \shortcite{WhitworthMNRAS1998a}; for a review see \S 
\ref{sec:jeans_cond}). In this chapter we perform the standard test 
simulation first proposed by Boss \& Bodenheimer \shortcite{BossApJ1979}. 
We show that our SPH code faithfully reproduces the results obtained by 
Truelove {\it et al}. \shortcite{TrueloveApJ1997,TrueloveApJ1998} using an 
adaptive finite difference code, and by Bate \& Burkert 
\shortcite{BateMNRAS1997} using their SPH code (\S \ref{sec:solution}).
This way, we test our code on a more realistic application and find that the 
results are consistent with those of Eulerian codes and other SPH codes. We 
also draw conclusions on the significance of the Jeans condition for 
fragmentation simulations. We note that the Bate \& Burkert 
\shortcite{BateMNRAS1997} SPH code has been developed independently 
from our code and differs from our code in several fundamental regards.

We perform a series of simulations by 
changing the density above which adiabatic heating operates (\S 
\ref{sec:heating}). This way we can make a direct comparison between 
the results of our code and those of Bate \& Burkert 
\shortcite{BateMNRAS1997}. 

We also perform a series of simulations by 
gradually increasing the number of neighbours as a means of testing 
convergence to the known solutions (\S \ref{sec:neighbours}). We perform both isothermal 
simulations and simulations which include 
adiabatic heating. For each case, there are 2 subsets of simulations: low resolution and 
high resolution, depending on the total number of particles used. 

Knowing 
the solution to this problem, we can also explore some other numerical 
parameters of the code. In particular, we verify the choice of the 
interpolating kernel (M4) by repeating a few simulations using a different 
choice of kernel (\S \ref{sec:kernel}). We also test whether our results 
are corrupted by the choice of the initial spatial distribution of particles 
(\S \ref{sec:settle}). But first, let us define the initial conditions for 
the simulations described in this chapter.

\section{Initial Conditions}{\label{sec:rot-initial}}

The standard test simulation for a Star Formation code is the evolution of a 
rotating, spherical, uniform-density, isothermal cloud with an m = 2 perturbation. 
The initial conditions we have used for this simulation are taken from 
Boss \& Bodenheimer \shortcite{BossApJ1979}. In particular, we have 
constructed a uniform-density, isothermal spherical cloud of mass M = 1 M$_{\odot}$ 
and radius R $\approx$ 0.02 pc. The sound speed is $c_{0}=0.17$ km s$^{-1}$ (corresponding to $T=7.9$K), 
which assigns to the cloud a ratio of thermal to gravitational potential 
energy $\alpha \approx$ 0.26. The spherical cloud is constructed with 
particles cut either from a settled uniform distribution or from a uniform 
face-centred cubic lattice. The particles are then given an m = 2 azimuthal 
perturbation with amplitude $A$ = 10$\%$, by adjusting their spherical polar 
azimuthal coordinate, $\phi$, to a value $\phi^{*}$ given by 
\[ \phi \, = \, \phi^{*} + \frac{A \sin(m \phi^{*})}{m}. \] 
Finally, the cloud is given a uniform rotation (angular velocity $\Omega$ = 7.2 x 
10$^{-13}$ rad s$^{-1}$), so that the ratio of rotational energy to gravitational 
potential energy is $\beta \approx$ 0.16. We use clouds having different numbers of 
particles depending on the problem. In all cases, we apply our full SPH code as 
given in chapter \ref{sec:SPH}. When a cloud has to be evolved isothermally, 
we use \[ \frac{P_{j}}{\rho_{j}} \; = \; c_{0}^{2} \] instead of Eqn. 
\ref{equa:eosfinal}. 

In order to decrease the numerical noise input by the initial distribution of 
particles, we perform most simulations using clouds whose particles are taken 
initially to be on a lattice. To verify that the results of such simulations 
are not biased due to some preferred orientation of the initial lattice, we 
also perform one simulation where particles are taken initially from a 
``settled'' distribution (\S \ref{sec:settle}). Such a distribution of 
particles is produced when the particles are taken in random positions and 
then they are relaxed to uniform density, using the SPH formulation described 
in Whitworth {\it et al}. \shortcite{WhitworthAnA1995}.

All figures presented here are column density plots viewed along the rotation 
axis. The geometry of the problem (fragmentation happens on a flattened disc) 
supports the use of such plots, since 
projection effects are small. It also allows comparison with density contour 
plots or density equatorial slices, used by other workers, as most of the 
mass of the system ends up in the disc. The figure captions indicate the 
units of the colour tables. They also give the linear size of the figure and 
the time of the simulation. 

\section{The solution in Eulerian and Lagrangian formulations}{\label{sec:solution}}

The evolution of a rotating, spherical, uniform-density, isothermal cloud with an m = 2 
perturbation predicts that the cloud forms a flattened structure due to 
rotation, and that the inner region of the cloud initially expands; at all times, there are 
two over-dense zones due to the perturbation. After $t$ $\sim$1 $t_{ff}$ the inner 
region starts collapsing and forms an elongated high-density structure at 
$t$ $\sim$1.15 $t_{ff}$. At the two ends of the elongated structure material falls 
faster towards the over-dense regions. The two ends finally become self-gravitating at 
$t$ $\sim$1.20 $t_{ff}$ and form a protostellar binary at $t$ $\sim$1.25 $t_{ff}$. The 
elongated structure between the binary components increases in density and forms 
a uniform density bar at $t$ $\sim$1.30 $t_{ff}$. 

The adaptive finite difference code of Truelove {\it et al.} 
\shortcite{TrueloveApJ1997,TrueloveApJ1998}, implemented to obey the Jeans 
condition, shows that the bar does not fragment 
while the gas remains isothermal. In particular, Truelove {\it et al.} 
\shortcite{TrueloveApJ1998} find that after $t$ $\sim$1.32 $t_{ff}$ the binary 
fragments are elongated and collapse to 
filamentary singularities (their Fig. 13), as suggested by Inutsuka \& Miyama 
\shortcite{InutsukaApJ1992}. Moreover, with their Fig. 12 they show that the bar 
between the binary does not fragment, contrary to what had been suggested by simulations 
using other grid codes \cite{BurkertMNRAS1993}. They prove that fragmentation of 
the bar in these other simulations was a consequence of inadequate resolution. 
Therefore, since in their convergence tests, 
the resolution of their code can grow infinitely while resolving the local Jeans 
length and since with their highest resolution the bar does not fragment, they 
conclude that the bar should not fragment. Their results were subsequently confirmed by 
Boss {\it et al.} \shortcite{BossApJ2000}.

In Klein {\it et al}. \shortcite{KleinTOKYO1998} the simulation is repeated 
and extended to higher densities with an equation of state that includes 
adiabatic heating with $\rho_{0}$ = 5 x 10$^{-14}$ g cm$^{-3}$. Again, the bar does 
not fragment, but due to adiabatic heating and thermal support the binary 
fragments now become 
spherical at $t$ $\sim$1.35 $t_{ff}$ (their Fig. 2). They follow this binary for a 
few revolutions around the centre of the domain. The binary fragments accrete 
material from the bar and they finally form a detached binary at 
$t$ $\sim$1.45 $t_{ff}$ (their Fig. 5). The remains of the bar form spiral arms 
around the rotating fragments (their Fig. 6).

Bate \& Burkert \shortcite{BateMNRAS1997} repeat both the isothermal 
simulation and the simulation with adiabatic heating using their SPH code. 
They, like Truelove {\it et al.} \shortcite{TrueloveApJ1998}, perform a 
convergence test. They increase the numerical resolution of their code by 
increasing the total number of particles in the simulations. In the 
isothermal regime, they obtain the expected evolution described above, with the 
simulation having the highest resolution (80,000 
particles). They find that convergence appears to occur from the simulation 
with 40,000 particles. To prevent the simulation advancing with very small 
time-steps they use a minimum smoothing length of 10$^{14}$ cm. Their 
isothermal simulations progress until $t$ $\sim$1.29 $t_{ff}$, where the Jeans 
condition is violated even for the simulation with the highest resolution 
(80,000 particles). Up to this point, the bar between the binary has not 
fragmented. They continue the simulation with the highest resolution (80,000 
particles) including adiabatic heating\footnote{They use a polytropic index 
of 5/2, implicitly including the two rotational degrees of freedom of H$_{2}$. With their 
formulation, collapse is decelerated more slowly than with our code.} initiating at 
$\rho_{0}$ $\sim$ 10$^{-13}$ g cm$^{-3}$. This way the Jeans condition 
continues to be obeyed, as the Jeans mass increases with increasing density, 
due to the increase in temperature and thus in sound 
speed ({\it cf.} with Eqn. \ref{equa:jeans8}). They find that adiabatic 
heating provides the bar with extra support against collapse 
towards a filamentary singularity, and -- in contrast to the finite difference simulations of Klein {\it et al}. \shortcite{KleinTOKYO1998} -- the bar fragments at 
$t$ $\sim$1.31 $t_{ff}$. This fragmentation is attributable to particle noise.

They find that with higher $\rho_{0}$ the bar becomes thinner and produces 
more fragments as the ratio of the length of the bar to its thickness 
increases, in accordance with the results of the grid code of Burkert \& 
Bodenheimer \shortcite{BurkertMNRAS1993}. In particular, they show that fewer 
fragments are produced when more heating is applied, i.e. when adiabatic 
heating initiates at a lower density. They find that at $t$ $\sim$1.315 
$t_{ff}$ five fragments are 
produced when $\rho_{0}$ = 10$^{-13}$ g cm$^{-3}$, only one for $\rho_{0}$ 
= 3 x 10$^{-14}$ g cm$^{-3}$, and none for $\rho_{0}$ = 10$^{-14}$ 
g cm$^{-3}$. The survival or merger of the fragments depend on the chaotic 
dynamics of the system of these protostellar fragments.

\section{Changing $\rho_{0}$}{\label{sec:heating_changerho}}

We have conducted the simulation for the evolution of a rotating, spherical, 
uniform-density, isothermal cloud with an m = 2 perturbation using our SPH code 
(\S \ref{sec:SPH}). We have used a cloud of 80,000 particles with particles 
initially taken on lattice. We have included adiabatic heating. We have tried 
different values for $\rho_{0}$, the density at which heating starts, in an 
effort to explore whether convergence to the known solutions can be achieved 
with an equation of state that includes an adiabatic heating regime.
Like Bate \& Burkert \shortcite{BateMNRAS1997}, we start with a value of 
$\rho_{0}$ = 10$^{-13}$ g cm$^{-3}$. We then 
repeat the simulations with smaller (\S \ref{sec:heating_lrho}) and larger 
(\S \ref{sec:heating_mrho}) values of $\rho_{0}$. We also set up a simulation using 600,000 
particles that always obeys the Jeans condition and has the highest value for 
$\rho_{0}$ = 5 x 10$^{-12}$ g cm$^{-3}$ (\S \ref{sec:600kptcls}). With this, we 
aim to use the least 
possible heating and let the simulation evolve isothermally as long as possible. 
Its high resolution enables us to draw conclusions on whether our previous 
results are resolution dependent. Table \ref{tab:changerho} shows a summary of 
our findings.

Fig. \ref{fig:initial_80} shows the 
density projected on the x-y plane initially (at $t=0$), where the density 
enhancements indicate the m=2 perturbation.

\begin{figure}
\begin{center}
\resizebox{8cm}{!}{\includegraphics{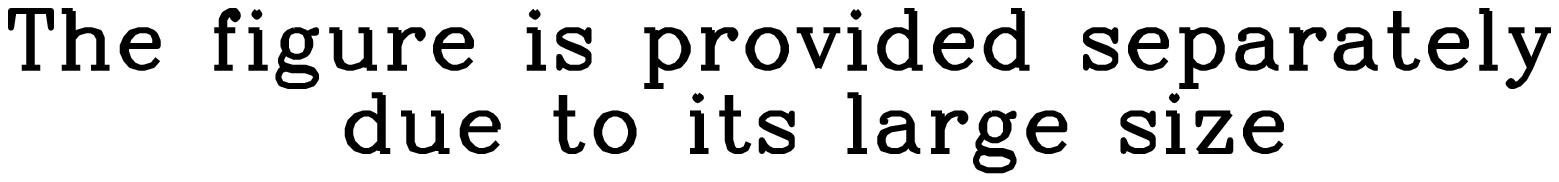}}
\end{center}    
      \caption[Column density plot of the initial sphere using 80,000 particles]
      {Column density plot of the initial sphere using 80,000 particles. The 
      linear size 
      of this plot is 0.04 pc. The colour table has units of 1.18 x 10$^{6}$ 
      g cm$^{-2}$.}
      \label{fig:initial_80}
\end{figure}

\begin{figure}
\begin{center}
\resizebox{8cm}{!}{\includegraphics{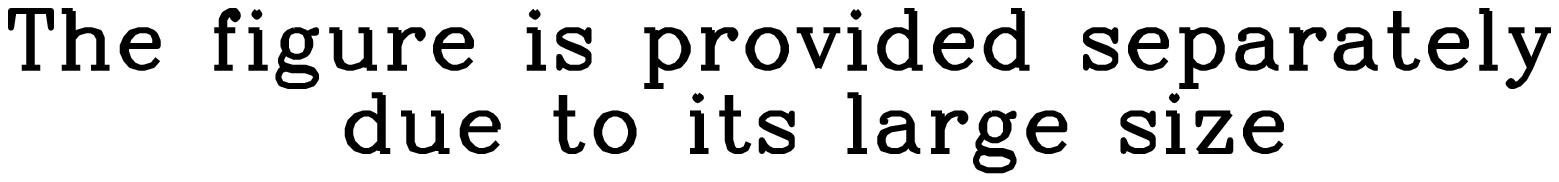}}
\end{center}  
\begin{center}
\resizebox{8cm}{!}{\includegraphics{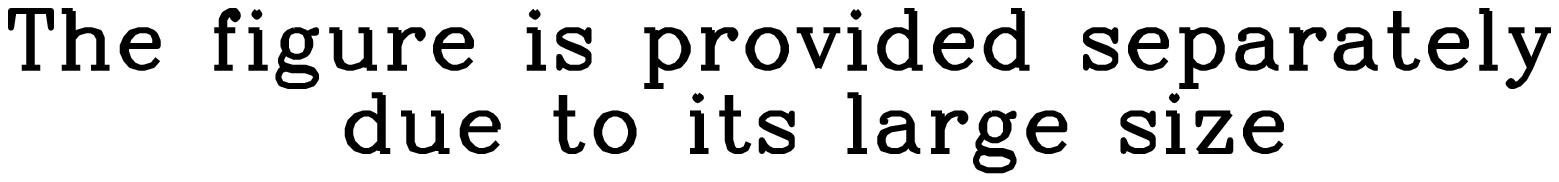}}
\end{center}
      \caption[Column density plots for a cloud of 80,000 particles before 
      heating starts and at the end ($\rho_{0}$ = 10$^{-13}$ g cm$^{-3}$)]
      {Column density plots for a cloud of 80,000 particles before heating 
      starts and at the end ($\rho_{0}$ = 10$^{-13}$ g cm$^{-3}$). The linear 
      size of these plots is 0.004 pc. The colour table has units of 1.18 x 
      10$^{8}$ g cm$^{-2}$. 
      \underline{Top}: Column density plot before heating starts ($t$ = 1.25 
      $t_{ff}$). 
      \underline{Bottom}: Column density plot at the end ($t$ = 1.30 $t_{ff}$).}
      \label{fig:13_heat}
\end{figure}

\begin{figure}
\begin{center}
\resizebox{8cm}{!}{\includegraphics{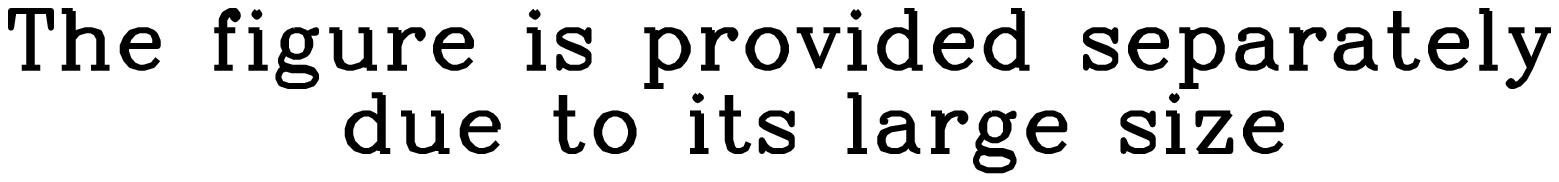}}
\end{center}    
      \caption[Column density plot for a cloud of 80,000 particles at the end ($\rho_{0}$ = 5 x 10$^{-14}$ g cm$^{-3}$)]
      {Column density plot for a cloud of 80,000 particles at the end 
      ($\rho_{0}$ = 5 x 10$^{-14}$ g cm$^{-3}$). The time is $t$ = 1.31 
      $t_{ff}$. The linear size of this plot is 0.004 pc. The colour table has 
      units of 1.18 x 10$^{8}$ g cm$^{-2}$.}
      \label{fig:5x14_end}
\end{figure}

\begin{figure}
\begin{center}
\resizebox{8cm}{!}{\includegraphics{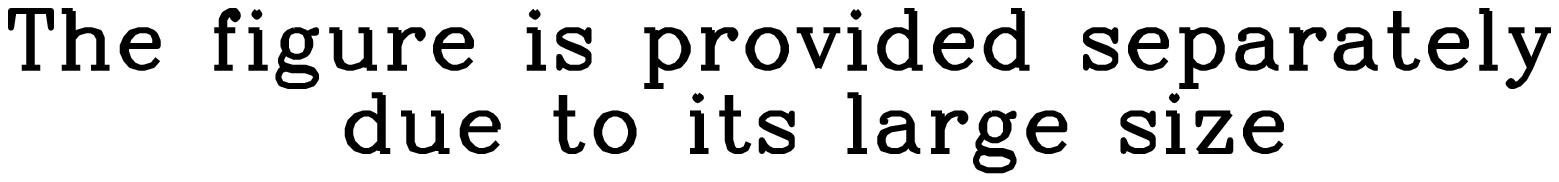}}
\end{center}    
      \caption[Column density plot for a cloud of 80,000 particles at the end 
      ($\rho_{0}$ = 10$^{-14}$ g cm$^{-3}$)]
      {Column density plot for a cloud of 80,000 particles at the end 
      ($\rho_{0}$ = 10$^{-14}$ g cm$^{-3}$). The time is $t$ = 1.33 $t_{ff}$. 
      The linear size of this plot is 0.004 pc. The colour table has units of 
      1.18 x 10$^{8}$ g cm$^{-2}$.}
      \label{fig:14_end}
\end{figure}

\subsection{$\rho_{0}$ = 10$^{-13}$ g cm$^{-3}$}{\label{sec:rho0=-13}}

The results are similar to those of Bate \& Burkert 
\shortcite{BateMNRAS1997}. The simulation evolves isothermally until $t$ 
$\sim$1.25 $t_{ff}$ producing a binary and a thin bar. At this stage, the 
binary components are elongated (top panel of Fig. \ref{fig:13_heat}).

After this point adiabatic heating initiates. Subsequent collapse onto the 
binary components is decelerated. Each of these objects has thermal 
support. They are forced into solid-body rotation by the high effective shear 
viscosity. As the rotating 
elongated objects interact with the slowly infalling envelope, they form 
larger prolate objects with thermal support. At a later stage, the binary 
components become spherical. The spherical 
objects are followed by thin spiral tails. At $t$ $\sim$1.28 $t_{ff}$, the
bar -- which also has thermal support -- fragments. At the end of the simulation 
(bottom panel of Fig. \ref{fig:13_heat}), at $t$ $\sim$1.30 $t_{ff}$, the bar 
has produced 3 fragments, one 
at the bar's centre and two closer to the binary components. The binary 
components are likely to merge with the bar fragments. There are also 2 
smaller fragments in the spiral tails, one in each. The mass for each of the 
binary components is 0.1 M$_{\odot}$ 
and the radius is 39 AU. Their separation is 330 AU. The total mass of the bar 
fragments is 0.02 M$_{\odot}$. The simulation has reached a peak 
density of $\rho_{peak}$ = 2.1 x 10$^{-11}$ g cm$^{-3}$. 

The simulation could 
resolve fragmentation up to a density of $\rho_{max}$ = 9.6 x 10$^{-14}$ 
g cm$^{-3}$. Therefore, fragmentation may be slightly unresolved in this 
simulation (see discussion in \S \ref{sec:jeans_cond}). 
We have repeated the above simulation with adiabatic heating initiating at 
$\rho_{0}$ = 9 x 10$^{-14}$ g cm$^{-3}$. We obtain exactly the same results as 
above, with the bar fragmenting again at $t$ $\sim$1.28 $t_{ff}$. Therefore, we conclude 
that the bar fragments produced above are not due to the Jeans condition not being obeyed.

We have repeated the simulation after changing $\rho_{0}$, the density where 
adiabatic heating initiates. We have used both smaller ($\rho_{0}$ = 
5 x 10$^{-14}$ g cm$^{-3}$ and $\rho_{0}$ = 10$^{-14}$ g cm$^{-3}$) and larger 
($\rho_{0}$ = 5 x 10$^{-13}$ g cm$^{-3}$) values than above.

\subsection{Decreasing $\rho_{0}$}{\label{sec:heating_lrho}}

The simulation with $\rho_{0}$ = 5 x 10$^{-14}$ g cm$^{-3}$ produces the expected
results ({\it cf}. the corresponding simulation of Bate \& Burkert 
\shortcite{BateMNRAS1997} with $\rho_{0}$ = 3 x 10$^{-14}$ g cm$^{-3}$). 
Since more heating is applied than before, the bar fragments later,
at $t$ $\sim$1.285 $t_{ff}$, as it becomes thin later. At the end of the 
simulation (Fig. \ref{fig:5x14_end}), at $t$ 
$\sim$1.31 $t_{ff}$, the peak density is lower than before, $\rho_{peak}$ = 
9.5 x 10$^{-12}$ g cm$^{-3}$, due to the increased amount of heating. 
The bar has fragmented to only one 
fragment at the centre. The simulation has advanced further in time and the 
binary components have 
approached closer. There are some lumps in the spiral tails. The binary 
components each have mass 0.13 M$_{\odot}$ 
and radius 33 AU. Their separation is 270 AU. The mass of the bar fragment 
is 0.03 M$_{\odot}$.

In the simulation with $\rho_{0}$ = 10$^{-14}$ g cm$^{-3}$ the bar does not 
fragment. The binary fragments have thermal support from an earlier time and 
thus become thicker. At the end of the 
simulation, $t$ $\sim$1.33 $t_{ff}$ (Fig. \ref{fig:14_end}), the peak density 
is even lower 
than before, $\rho_{peak}$ = 1.8 x 10$^{-12}$ g cm$^{-3}$. The binary 
fragments have merged and completed 3/4 of a full rotation. The merging of the 
fragments is probably due to the 
artificial shear viscosity 
being too large with our implementation of viscosity (\S \ref{sec:artvisc} -- 
see discussion in \S \ref{sec:rot_conclusi}). 
This way, we cannot reproduce the binary system followed by Klein {\it et al.} 
\shortcite{KleinTOKYO1998}.

\subsection{Increasing $\rho_{0}$}{\label{sec:heating_mrho}}

For the simulation with the least heating ($\rho_{0}$ = 
5 x 10$^{-13}$ g cm$^{-3}$) the results change in the opposite sense: 
the bar fragments earlier ($t$ $\sim$1.270 $t_{ff}$) and at the end 
of the simulation, $t$ $\sim$1.275 $t_{ff}$ (Fig. \ref{fig:5x13_end}), 
the bar has broken into 9 fragments; there are also 2 lumps in each 
spiral tail. The masses of the binary components are 0.05 M$_{\odot}$ and 
their radii are 19 AU. Their separation is 515 AU. The total mass of 
all bar fragments is 0.06 M$_{\odot}$. We cannot rule out artificial 
fragmentation for this simulation as we did above, since at 
$\rho_{0}$ = 5 x 10$^{-13}$ g cm$^{-3}$, the Jeans mass has stopped being 
resolved for half a decade in density. At $t$ $\sim$1.275 $t_{ff}$, 
the peak density has reached $\rho_{peak}$ = 1.7 x 10$^{-10}$ g cm$^{-3}$, 
a value much higher than before.

To resolve fragmentation up to this high density ($\rho_{0}$ = 
5 x 10$^{-13}$ g cm$^{-3}$) we need 185,000 particles ({\it cf.} Eqn. 
\ref{equa:Jeans_condition3}). We have conducted such a simulation and the 
results are similar to those of the simulation with 
$\rho_{0}$ = 10$^{-13}$ g cm$^{-3}$. 
In particular, the binary and a thin bar formed. The mass of each binary 
component is 0.04 M$_{\odot}$ and its radius is 12 AU. Their separation is 
495 AU. At $t$ $\sim$1.267 $t_{ff}$ the bar fragments. There are 7 fragments 
in the bar plus one lump in each spiral tail, at the end of the simulation 
($t$ $\sim$1.271 $t_{ff}$ -- Fig. \ref{fig:185_end}). The total mass of all 
bar fragments is 0.02 M$_{\odot}$. Since the Jeans mass is always resolved, 
fragmentation in this simulation is realistically modelled ($\rho_{peak}$ = 1.5 x 
10$^{-10}$ g cm$^{-3}$).

\subsection{600,000 particle simulation}{\label{sec:600kptcls}}

Finally, we have conducted a simulation with 600,000 particles where adiabatic 
heating initiates at $\rho_{0}$ = 5 x 10$^{-12}$ g cm$^{-3}$ (we use such a high
density for the switch to adiabatic heating in order to evolve the cloud 
isothermally as long as possible). At the end of the simulation ($t$ $\sim$1.245
$t_{ff}$) the bar has not fragmented. The simulation has reached a peak density 
of $\rho_{peak}$ = 1.8 x 10$^{-9}$ g cm$^{-3}$, the highest of all the 
simulations we conducted. As Truelove {\it et al}. \shortcite{TrueloveApJ1997} 
suggest, a coarse 
simulation is less evolved at the same time compared to a fine simulation. This 
happens in SPH as well, since with higher resolution the particle $h$ is smaller
and thus the density modelled becomes higher. Truelove {\it et al}. even 
suggest that between 
simulations of different resolution comparison should be made when they both 
have advanced to the same density and not to the same time.

\begin{figure}
\begin{center}
\resizebox{8cm}{!}{\includegraphics{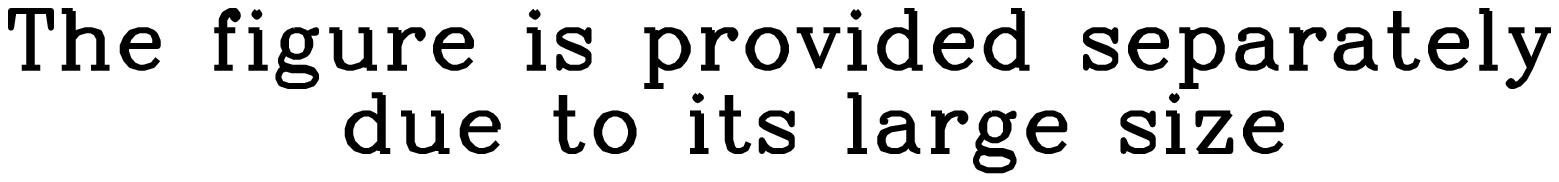}}
\end{center}    
      \caption[Column density plot for a cloud of 80,000 particles at the end 
      ($\rho_{0}$ = 5 x 10$^{-13}$ g cm$^{-3}$)]
      {Column density plot for a cloud of 80,000 particles at the end 
      ($\rho_{0}$ = 5 x 10$^{-13}$ g cm$^{-3}$). The time is $t$ = 1.275 
      $t_{ff}$. The linear size of this plot is 0.004 pc. The colour table has 
      units of 1.18 x 10$^{8}$ g cm$^{-2}$.}
      \label{fig:5x13_end}
\end{figure}

\begin{figure}
\begin{center}
\resizebox{8cm}{!}{\includegraphics{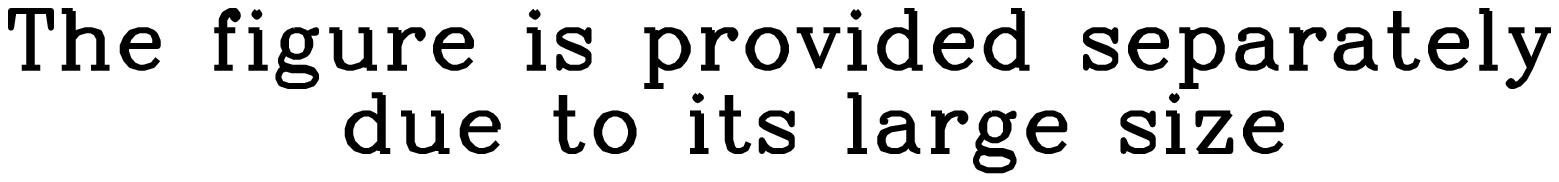}}
\end{center}    
      \caption[Column density plot for a cloud of 185,000 particles at the 
      end ($\rho_{0}$ = 5 x 10$^{-13}$ g cm$^{-3}$)]
      {Column density plot for a cloud of 185,000 particles at the end 
      ($\rho_{0}$ = 5 x 10$^{-13}$ g cm$^{-3}$). The time is $t$ = 1.271 
      $t_{ff}$. The linear size of this plot is 0.004 pc. The colour table has 
      units of 1.18 x 10$^{8}$ g cm$^{-2}$.}
      \label{fig:185_end}
\end{figure}

\begin{figure}
\begin{center}
\resizebox{8cm}{!}{\includegraphics{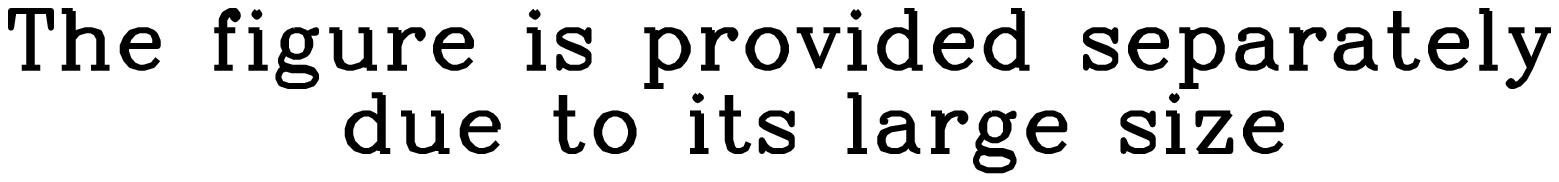}}
\end{center}
\begin{center}
\resizebox{8cm}{!}{\includegraphics{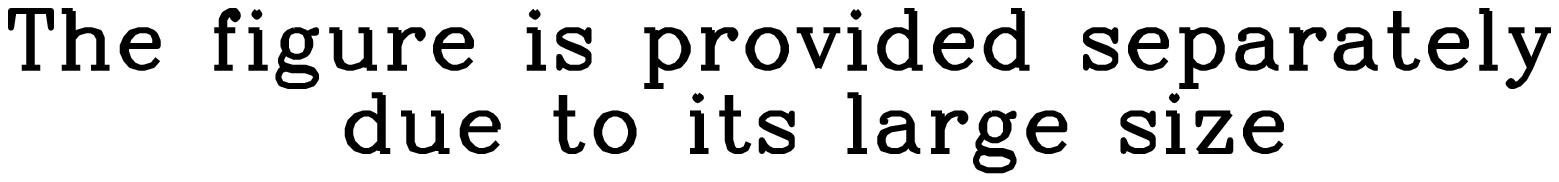}}
\end{center}    
      \caption[Column density plots for a cloud of 600,000 particles before 
      heating starts and at the end ($\rho_{0}$ = 5 x 10$^{-12}$ g cm$^{-3}$)]
      {Column density plots for a cloud of 600,000 particles before heating 
      starts and at the end ($\rho_{0}$ = 5 x 10$^{-12}$ g cm$^{-3}$). The 
      linear size of this plot is 0.004 pc. The colour table has units of 1.18 x 
      10$^{8}$ g cm$^{-2}$.
      \underline{Top}: Column density plot before heating starts ($t$ = 1.237 
      $t_{ff}$). 
      \underline{Bottom}: Column density plot at the end ($t$ = 1.245 $t_{ff}$).}
      \label{fig:600_end} 
\end{figure}

The bottom panel of Fig. \ref{fig:600_end} is a column density plot at the end 
of the simulation ($t$ $\sim$1.245 $t_{ff}$). The bar is very sparse and it 
has just started becoming self-gravitating. Until a few time-steps before, the 
binary components were still rather elongated apart from their very centres 
where a spherical core had developed. 
This core is spherical mainly due to the fact that the thickness of this 
elongated 
self-gravitating object is smaller than the $h$ of a particle in it (these are 
the particles with the smallest $h$ in the whole simulation being the 
particles with the highest density). This way, due to the spherical symmetry 
of the kernel, the centre of each elongated object settles to form a spherical 
object containing its $\sim$50 neighbours. The high effective shear viscosity 
has since then put the binary components into solid-body rotation and they 
have grown in size due to interaction with the accretion flows. At the end, the 
masses of the binary components are 0.008 M$_{\odot}$ and their radii are 3 AU. 
Their separation is 680 AU.

The bottom panel of Fig. \ref{fig:600_end} compares well with Fig. 2 of Klein 
{\it et al}. \shortcite{KleinTOKYO1998}. 
In fact, our simulation has reached a higher peak density as it has evolved 
isothermally for 2 orders of magnitude 
more than that of Klein {\it et al}. The top panel of Fig. \ref{fig:600_end} 
is a column density plot just before 
adiabatic heating starts ($t$ $\sim$1.237 $t_{ff}$). It compares very well 
with Fig. 12 of Truelove {\it et al}. 
\shortcite{TrueloveApJ1998}. Again the peak density ($\rho_{peak}$ = 5.2 x 
10$^{-12}$ g cm$^{-3}$) has reached a 
higher value than that of the simulation of Truelove {\it et al}., as 
adiabatic heating in our simulation starts 
one order of magnitude higher in density than in theirs. 

\subsection{Conclusions}{\label{sec:rot_conclusi}}

\begin{table}
\begin{center}
\begin{tabular}{|l|c|c|c|c|c|c|}\hline
 $\rho_{0}$ / g cm$^{-3}$ & Particles & $t_{bin}$ / $t_{ff}$ & $t_{end}$ / $t_{ff}$ & Bar frags. & $t_{frag}$ / $t_{ff}$ & $\rho_{peak}$ / g cm$^{-3}$\\ \hline \hline
 10$^{-14}$ & 80,000 & 1.25 & 1.33 & 0 & N/A & 1.8 x 10$^{-12}$\\ \hline
 5 x 10$^{-14}$ & 80,000 & 1.25 & 1.31 & 1 & 1.285 & 9.5 x 10$^{-12}$\\ \hline
 9 x 10$^{-14}$ & 80,000 & 1.25 & 1.30 & 3 & 1.280 & 2.1 x 10$^{-11}$\\ \hline
 10$^{-13}$ & 80,000 & 1.25 & 1.30 & 3 & 1.280 & 2.1 x 10$^{-11}$\\ \hline
 5 x 10$^{-13}$ & 80,000 & 1.25 & 1.275 & 9 & 1.270 & 1.7 x 10$^{-10}$\\ \hline
 5 x 10$^{-13}$ & 185,000 & 1.24 & 1.271 & 7 & 1.267 & 1.5 x 10$^{-10}$\\ \hline
 5 x 10$^{-12}$ & 600,000 & 1.237 & 1.245 & 0 & N/A & 1.8 x 10$^{-9}$\\ \hline
\end{tabular}
\end{center}
\caption[Summary of results for the simulations with different $\rho_{0}$]{Summary 
of results for the simulations with different $\rho_{0}$. For each simulation the 
third column gives the time of the binary formation, the fourth the final time of 
the simulation and the sixth the time of the fragmentation of the bar. All times 
are quoted in units of the free-fall time of the initial cloud, $t_{ff}$. The 
fifth column gives the number of bar fragments at the end of each simulation.
The last column gives the peak density at the end of each simulation.}
\label{tab:changerho}
\end{table}

In general, our SPH code faithfully reproduces the results of the 80,000 particle 
simulations of Bate \& Burkert 
\shortcite{BateMNRAS1997}. We, like they, conclude that with higher $\rho_{0}$ 
the bar becomes thinner and produces more fragments as the ratio of the 
length of the bar to its thickness increases, i.e. we show that fewer 
fragments are produced when more heating is applied. In addition, with the 600,000 
particle simulation our code 
faithfully reproduces the results of Truelove {\it et al}. 
\shortcite{TrueloveApJ1997,TrueloveApJ1998} and Klein {\it et al}. 
\shortcite{KleinTOKYO1998}. This makes us confident that our code models 
self-gravitating gas dynamics realistically and therefore, we can use it in our 
effort to implement a method for obtaining higher resolution (chapter 
\ref{sec:part-split}), as well as in simulations of clump-clump collisions 
(chapter \ref{sec:results}). Table \ref{tab:changerho} shows a summary of 
results for the simulations with different $\rho_{0}$.

From this series of simulations, there are several conclusions we can draw. 
Firstly, the fragmentation of the bar in the simulations of Bate \& 
Burkert \shortcite{BateMNRAS1997} and our coarse simulations (80,000 
particles) is due to poor sampling. The fact that Bate \& Burkert 
\shortcite{BateMNRAS1997} find that with increasing resolution the bar 
produces more fragments, cannot prove that the bar will always fragment, 
as they have not achieved convergence with their 
80,000 particle simulation. Our 600,000 particle simulation suggests 
the opposite, i.e. the bar should not fragment, in accordance with 
Truelove {\it et al}. \shortcite{TrueloveApJ1997,TrueloveApJ1998}. 
Unfortunately, we cannot predict bar fragmentation theoretically in 
a manner similar to the prediction of the fragmentation of a layer 
\cite{WhitworthMNRAS1994,WhitworthAnA1994}, as it is not possible to 
make a dimensional analysis in one dimension. Truelove {\it et al}. 
\shortcite{TrueloveApJ1997,TrueloveApJ1998} and Klein {\it et al}. 
\shortcite{KleinTOKYO1998} with their adaptive finite difference code 
can regulate their grid size achieving theoretically infinite resolution, 
a feature that SPH codes do not possess. One of the primary aims of the PhD 
project discussed in this thesis is to invent a method that would allow 
the resolution of an SPH simulation to be increased on-the-fly by particle 
splitting (see chapter \ref{sec:part-split}).

Secondly, there may be a way of preventing bar fragmentation even for 
simulations with 80,000 particles. This could be done with the implementation 
of a triaxial kernel, where $h$ 
is replaced by a tensor that gives different smoothing lengths in different directions. 
The values for $h$ in different directions are such that a particle still 
overlaps with $\sim$50 of its neighbours \cite{OwenApJSS1998}.

Thirdly, the times for bar fragmentation are based on our perception of a bar 
fragment to be a density enhancement on the bar twice as large as the 
underlying bar density. It has been shown that in all cases, such an 
enhancement will form 
a fragment, due to the elongated geometry of the bar and the spherical symmetry 
imposed by the kernel. This is an empirical law for this kind of simulation. 
We have considered the time that such a density enhancement needs to grow as 
the time of bar fragmentation. Our definition may differ from the relevant 
definition of Bate \& Burkert \shortcite{BateMNRAS1997} and this may be the 
reason for the small mismatch in the quoted bar fragmentation times.  

Fourthly, the fact that the fragments of the bar merge with the binary 
components in all of the simulations that the bar fragmented, is 
probably due to excess shear viscosity introduced by our application 
of artificial viscosity (\S \ref{sec:artvisc}). This is also the reason 
for the merger of the binary components in the simulation with 
$\rho_{0}$ = 10$^{-14}$ g cm$^{-3}$. In particular, our code has failed the 
test for the collapse simulation of a rotating unperturbed sphere, suggested by 
Norman, Wilson \& Barton \shortcite{NormanApJ1980b}. Shear viscosity acts as an 
agent that redistributes angular momentum and we could not obtain a perfect 
singular disc. Instead a ring was produced around the disc centre. 

An alternative implementation 
could be explored. One could use the switch introduced by Morris \& Monaghan
\shortcite{MorrisJCP1997}. 
We refer to their Eqn. 30 with the Balsara \shortcite{BalsaraJCP1995} source 
term - Eqn. 4. Another switch that calculates viscosity only for 
physically approaching streams could also be used, e.g. using 
$\nabla(\nabla \cdot v)$.

Finally, it is interesting to note that bar fragmentation happens in a symmetric
fashion, with a central object and equal number of fragments on either side of 
it, at equal distances. This can be attributed to the symmetric initial 
distribution of particles (taken from a lattice) that has removed some of the 
numerical noise and prevented the randomness in the fragment positions present 
in the simulations of Bate \& Burkert \shortcite{BateMNRAS1997} and \S 
\ref{sec:settle}.

\section{Changing the number of SPH neighbours}{\label{sec:neighbours}}

In \S \ref{sec:smoothing} and all simulations so far, we have used a fixed 
value for the number of neighbours, $N_{n} \sim$50, contained within the smoothing 
radius, $h$, of all particles. The choice of this value for $N_{n}$ was 
dictated by the need to balance accuracy -- that requires large $N_{n}$ -- 
against computational cost which is reduced 
with small values for $N_{n}$. We have shown that medium resolution (80,000 particles) 
SPH implementations of the standard test simulation cannot produce the expected 
evolution for the bar between the binary components and that we need to increase 
the number of particles, $N$, by one order of magnitude before the results 
converge to those of the finite difference code of Truelove {\it et al}. 
\shortcite{TrueloveApJ1997}. We have tried increasing not just the number of particles, 
$N$, but also the number of neighbours, $N_{n}$; this is suggested by Rasio 
\shortcite{RasioICCP51999} as a means of increasing SPH accuracy (reducing sampling 
noise). We use the limiting 
case of Rasio's suggestion by increasing $N_{n}$ and $N$ with the same rate, in an 
effort to keep the same resolution (constant $N / N_{n}$) for all simulations and 
therefore make comparison between them more meaningful.

We have conducted a series of simulations by increasing $N_{n}$ from 50 to 200 in 
steps of 50. $N$ has changed respectively from 80,000 to 320,000 in steps of 
80,000. In order to be able to identify the exact point where our simulations 
converge to the expected solution, we have also set up a series of low resolution 
simulations with $N$ ranging from 10,000 to 40,000 in steps of 10,000. All these 
simulations are evolved both with an isothermal and an adiabatic equation of state. 

We present each of the four groups of simulations separately in order to avoid 
confusion and to identify clearly the effect of increasing $N_{n}$ on the final 
state of the simulations. For consistency, we present column density plots for all 
simulations, viewed along the rotation axis. All plots have the same linear size 
as the plots of \S \ref{sec:heating_changerho}

\subsection{Isothermal Simulations}

\subsubsection{3.4.1.1 Low resolution isothermal simulations}

\begin{figure}
\resizebox{7.75cm}{!}{\includegraphics{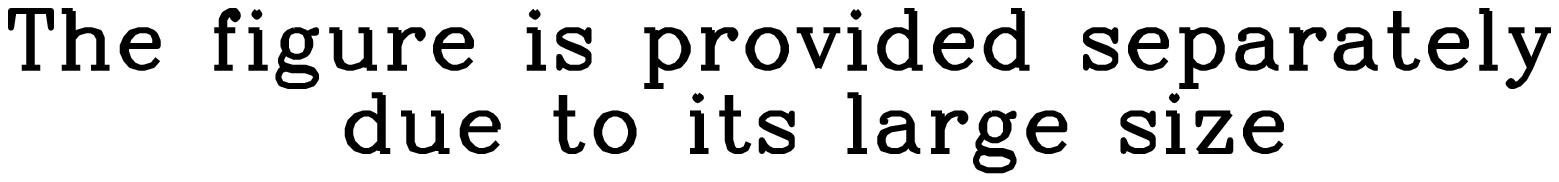}}
\resizebox{7.75cm}{!}{\includegraphics{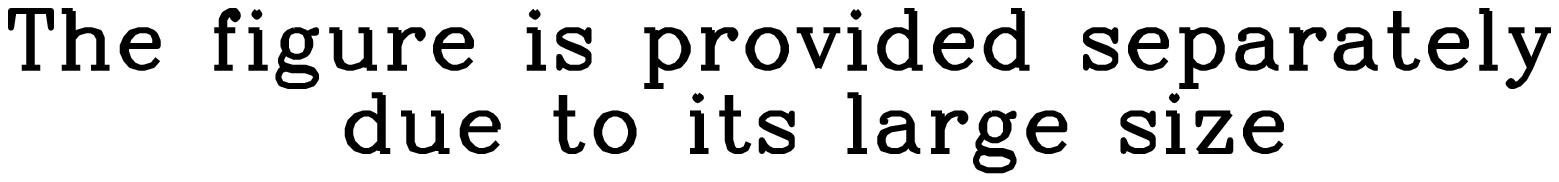}}
\resizebox{7.75cm}{!}{\includegraphics{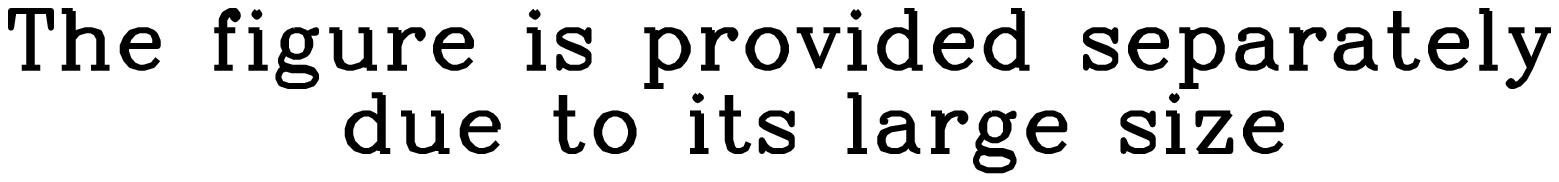}}
\resizebox{7.75cm}{!}{\includegraphics{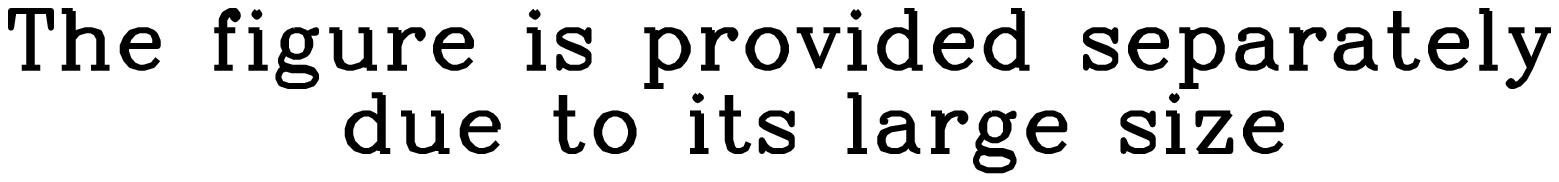}}
      \caption[Column density plots for the low-resolution isothermal simulations with 
increasing $N_{n}$]{Column density plots for the low-resolution isothermal simulations with 
increasing $N_{n}$. Final stage of the simulations with $N$=10,000 and $N_{n}$=50 (left), 
$N$=20,000 and $N_{n}$=100 (right) on the top row and $N$=30,000 and $N_{n}$=150 (left), 
$N$=40,000 and $N_{n}$=200 (right) on the bottom row. The details for each simulation are given in 
Table \ref{tab:iso_low}. The results converge after the simulation with $N$=30,000 and 
$N_{n}$=150. The linear size of all plots is 0.004 pc. The colour table has units of 1.18 x 
10$^{8}$ g cm$^{-2}$.}
      \label{fig:iso_low}
\end{figure}

\begin{figure}
\resizebox{7.75cm}{!}{\includegraphics{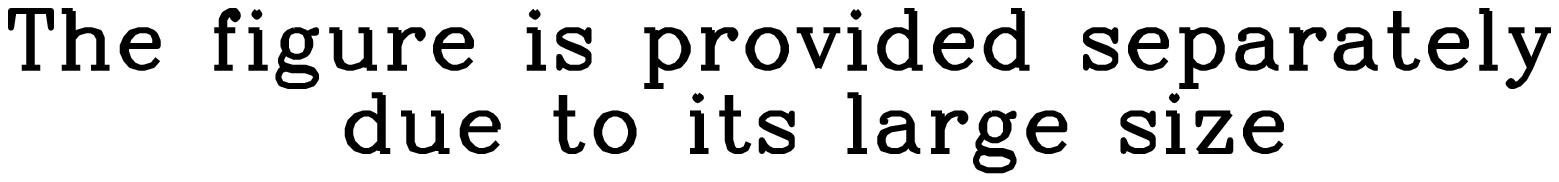}}
\resizebox{7.75cm}{!}{\includegraphics{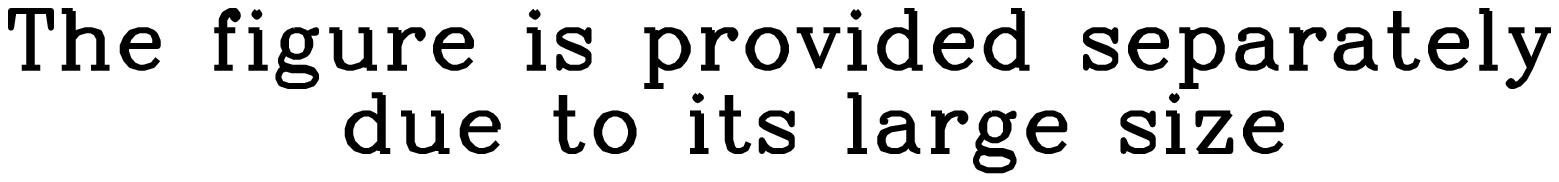}}
\resizebox{7.75cm}{!}{\includegraphics{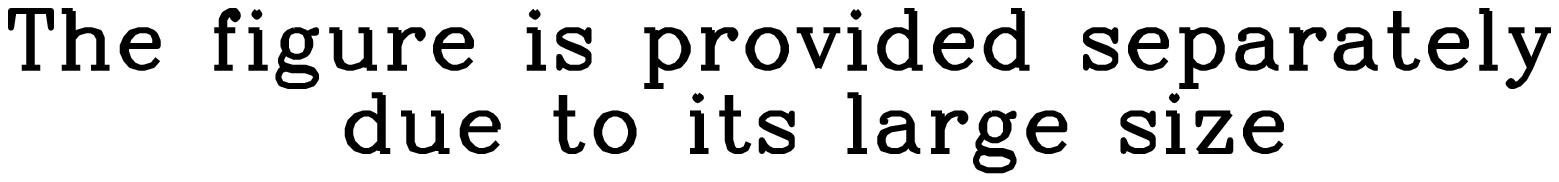}}
\resizebox{7.75cm}{!}{\includegraphics{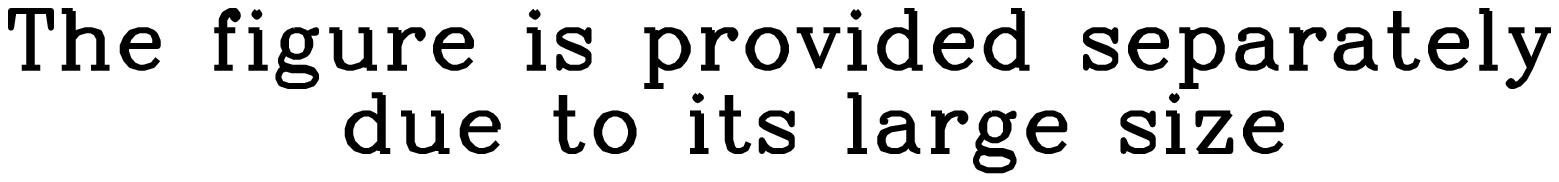}}
      \caption[Column density plots for the high-resolution isothermal simulations with 
increasing $N_{n}$]{Column density plots for the high-resolution isothermal simulations with 
increasing $N_{n}$. Final stage of the simulations with $N$=80,000 and $N_{n}$=50 (left), 
$N$=160,000 and $N_{n}$=100 (right) on the top row and $N$=240,000 and $N_{n}$=150 (left), 
$N$=320,000 and $N_{n}$=200 (right) on the bottom row. The details for each simulation are given in 
Table \ref{tab:iso_high}. The results converge after the simulation with $N$=160,000 and 
$N_{n}$=100. The linear size of all plots is 0.004 pc. The colour table has units of 1.18 x 
10$^{8}$ g cm$^{-2}$.}
      \label{fig:iso_high}
\end{figure}

\begin{table}
\begin{center}
\begin{tabular}{|l|c|c|c|c|c|c|}\hline
$N$ & $N_{n}$ & $t_{bin}$ / $t_{ff}$ & $t_{end}$ / $t_{ff}$ & Bar fragments & $t_{frag}$ / $t_{ff}$ & $\rho_{peak}$ / g cm$^{-3}$\\ \hline \hline
10,000 & 50 $\pm5$ & 1.265 & 1.274 & 2 & 1.272 & 3.2 x 10$^{-7}$\\ \hline
20,000 & 100 $\pm10$ & 1.260 & 1.278 & 0 & N/A & 2.5 x 10$^{-8}$\\ \hline
30,000 & 150 $\pm15$ & 1.259 & 1.276 & 0 & N/A & 6.0 x 10$^{-9}$\\ \hline
40,000 & 200 $\pm20$ & 1.258 & 1.274 & 0 & N/A & 2.7 x 10$^{-9}$\\ \hline
\end{tabular}
\end{center}
\caption[Summary of results for the low-resolution isothermal simulations with 
increasing $N_{n}$]{Summary of results for the low-resolution isothermal 
simulations with increasing $N_{n}$. For each simulation the 
third column gives the time of the binary formation, the fourth the final time of 
the simulation and the sixth the time of the fragmentation of the bar. All times 
are quoted in units of the free-fall time of the initial cloud, $t_{ff}$. The 
last column gives the peak density at the end of each simulation.}
{\label{tab:iso_low}}
\end{table}

A summary of the low resolution isothermal simulations is given in Table 
\ref{tab:iso_low}. All simulations can resolve fragmentation up to a 
density of $\rho_{max}$ = 1.75 x 10$^{-15}$ g cm$^{-3}$. They are all 
terminated when the multiple time-step method (\S \ref{sec:timesteps}) runs 
out of time bins. In fact, all simulations stop when the time-step becomes 
less than 2 x 10$^{-6}$ t$_{ff}$, so that it would have been computationally 
inefficient to continue (i.e. in order for time to progress by 10$^{-2}$ 
t$_{ff}$ we would need 5-6 times the run-time up to that point). The four 
simulations have shown the following:

\begin{enumerate}

\item {\bf $N$=10,000 and $N_{n}$=50} 
Due to poor sampling, both the initial expansion phase and the collapse 
that follows it are not properly modelled. This causes the binary to form 
later than expected, at $t_{bin}$ $\sim$1.265 $t_{ff}$ ({\it cf}. the values 
of the third column of Table \ref{tab:changerho}). In fact, a filament 
forms first and both ends of the filament become self-gravitating shortly 
after, forming the binary (in accordance with the evolution of the 10,000 
particle simulation of Bate \& Burkert \shortcite{BateMNRAS1997}). 
The bar fragments shortly after the binary formation, at $t_{frag}$ 
$\sim$1.272 $t_{ff}$. The top left hand panel of Fig. \ref{fig:iso_low} is a 
column density plot 
at the end of the simulation, $t_{end}$ $\sim$1.274 $t_{ff}$. The binary 
components appear to be elongated. The peak density of the 
simulation has reached a non-physical value, $\rho_{peak}$ = 3.2 x 10$^{-7}$ 
g cm$^{-3}$. This simulation does not reproduce the expected evolution.

\item {\bf $N$=20,000 and $N_{n}$=100}
The overall evolution of the cloud is similar to the that of the previous simulation. 
However, the binary forms a bit earlier than before and the binary components start collapsing 
almost simultaneously with the bar and not after the bar. At the end of the simulation, 
$t_{end}$ $\sim$1.278 $t_{ff}$ (top right hand panel of Fig. \ref{fig:iso_low}), the bar has 
not fragmented although it is very dense and it might have fragmented if we had continued 
the simulation. At this stage, the binary components are rather elongated apart from a spherical 
core that has developed at their centres. The reason for the formation of such a core is discussed 
in \S \ref{sec:600kptcls}. The peak density at the end, $\rho_{peak}$ = 2.5 x 10$^{-8}$ g cm$^{-3}$ 
is lower than before, therefore the density field is more realistically modelled, as expected 
for a simulation with less noise (more particles contribute to each SPH sum and all 
quantities are better modelled).

\item {\bf $N$=30,000 and $N_{n}$=150}
The binary forms even earlier and its components end up elongated with spherical cores formed at their 
centres. The bar is not so dense as before and clearly it has not 
fragmented by the end of the simulation, at $t_{end}$ $\sim$1.276 $t_{ff}$ (bottom left hand 
panel of Fig. \ref{fig:iso_low}). The peak density, $\rho_{peak}$ = 6.0 x 10$^{-9}$ g cm$^{-3}$, 
is even lower.

\item {\bf $N$=40,000 and $N_{n}$=200} 
The binary still forms a bit earlier and its components have the same shape as in the previous 
simulation. The peak density at the end, $\rho_{peak}$ = 2.7 x 
10$^{-9}$ g cm$^{-3}$, is even lower. The bar does not fragment and it is even less dense. The 
bottom right hand panel of Fig. \ref{fig:iso_low} is a column density plot at the end of the 
simulation, at $t_{end}$ $\sim$1.274 $t_{ff}$. The evolution of this simulation is not much 
different from that of the previous simulation. Therefore, we believe that convergence is 
achieved from the simulation with $N$=30,000 and $N_{n}$=150. 

\end{enumerate}

Obviously, this set of low 
resolution simulations cannot reproduce the exact solution. It appears that low resolution in the early stages delays binary formation.

We have also conducted a simulation with $N$=40,000 and $N_{n}$=50 (having resolution four times 
higher than that of the above simulations) in order to confirm that the above results depend 
not just on the increasing 
number of particles but also on the increasing number of neighbours. With $N$=40,000 and 
$N_{n}$=50, we have found that the binary forms earlier, at $t_{bin}$ $\sim$1.252 $t_{ff}$, 
in accordance with the corresponding simulation of Bate \& Burkert \shortcite{BateMNRAS1997}. 
The peak density at the end of the simulation with $N$=40,000 and $N_{n}$=50 is two orders of 
magnitude higher than that of the $N$=40,000 and $N_{n}$=200 simulation. This is partly due to 
the higher resolution (a finer simulation is more evolved) and partly due to the increased noise 
(less particles contribute to the SPH sums) of the $N$=40,000 and $N_{n}$=50 simulation. 
The difference in $t_{bin}$ and $\rho_{peak}$ in the $N$=40,000 and $N_{n}$=50 simulation confirms 
the dependence of the results of the above four simulations on the increasing number of neighbours.

\subsubsection{3.4.1.2 High resolution isothermal simulations}

\begin{table}
\begin{center}
\begin{tabular}{|l|c|c|c|c|}\hline
$N$ & $N_{n}$ & $t_{bin}$ / $t_{ff}$ & $t_{end}$ / $t_{ff}$ & $\rho_{peak}$ / g cm$^{-3}$\\ \hline \hline
80,000 & 50 $\pm5$ & 1.248 & 1.252 & 1.7 x 10$^{-8}$\\ \hline
160,000 & 100 $\pm10$ & 1.247 & 1.250 & 3.7 x 10$^{-9}$\\ \hline
240,000 & 150 $\pm15$ & 1.247 & 1.250 & 7.3 x 10$^{-8}$\\ \hline
320,000 & 200 $\pm20$ & 1.247 & 1.250 & 8.8 x 10$^{-8}$\\ \hline
\end{tabular}
\end{center}
\caption[Summary of results for the high-resolution isothermal simulations with 
increasing $N_{n}$]{Summary of results for the high-resolution isothermal 
simulations with increasing $N_{n}$. For each simulation the third column gives 
the time of the binary formation and the fourth the final time of the simulation. 
All times are quoted in units of the free-fall time of the initial cloud, $t_{ff}$. 
The last column gives the peak density at the end of each simulation.}
{\label{tab:iso_high}}
\end{table}

A summary of the high resolution isothermal simulations is given in Table 
\ref{tab:iso_high}. All simulations can resolve fragmentation up to a 
density of $\rho_{max}$ = 9.6 x 10$^{-14}$ g cm$^{-3}$. They are all 
terminated when the multiple time-step method (\S \ref{sec:timesteps}) runs 
out of time bins. In fact, all simulations stop when the time-step becomes 
less than 10$^{-6}$ t$_{ff}$, so that it would have been computationally 
inefficient to continue (i.e. in order for time to progress by 10$^{-2}$ 
t$_{ff}$ we would need 8-10 times the run-time up to that point). 

All four 
simulations have shown that the binary forms at the expected time ($t_{bin}$ 
$\sim$1.25 $t_{ff}$) and that its components are elongated apart from their 
centres where a spherical core develops due to the spherical symmetry of the 
kernel (see discussion in \S \ref{sec:600kptcls}). The binary components become 
thinner with increasing $N_{n}$, so that the simulations converge towards the 
expected solution of filamentary singularities with the noise being reduced. 

The bar between the binary components does not fragment, is sparse and becomes less dense with 
increasing $N_{n}$. The peak density at the end of the simulations becomes 
lower with increasing $N_{n}$, as the density field is more smoothed with 
less noise. Note that the last two simulations initially ended a bit 
earlier than the quoted values of $t_{end}$ in Table \ref{tab:iso_high}. We 
have restarted them by resetting the time bin hierarchy and continued them for 
a few time-steps. The binary components are in free-fall collapse so that the 
peak density has increased by more than one order of magnitude during this 
short time. 

Fig. \ref{fig:iso_high} shows the end state of all four 
simulations. Convergence to the expected solution starts to appear from the 
simulation with $N$=160,000 and $N_{n}$=100. This is a clear computational 
gain as we only need to double $N$ and $N_{n}$ with respect to the 80,000 
particle simulations of \S \ref{sec:heating_changerho} to obtain a solution 
similar to that of the finite difference simulations of Truelove {\it et al}. 
\shortcite{TrueloveApJ1997,TrueloveApJ1998}. Comparison with the low resolution 
isothermal simulations has shown that the increased resolution (8 times 
higher) has a clear effect on the morphology of all structures formed; the binary components 
appear more well-defined and the bar is more sparse. In the series of 
high resolution simulations convergence is achieved with a lower value for 
$N_{n}$. 

However, following the simulations 
at such high densities with an isothermal equation of state is not physical and 
in order to model the evolution of the cloud realistically we need to include 
adiabatic heating. This is presented in the following subsection.

\subsection{Simulations with Adiabatic Heating}{\label{sec:heating}}

\subsubsection{3.4.2.1 Low resolution simulations with adiabatic heating}{\label{sec:heating_low}}

A summary of the low resolution simulations with adiabatic heating is given in Table 
\ref{tab:heat_low}. All simulations can resolve fragmentation up to a 
density of $\rho_{max}$ = 1.75 x 10$^{-15}$ g cm$^{-3}$. Adiabatic heating 
starts at $\rho_{0}$ = 5 x 10$^{-13}$ g cm$^{-3}$. All simulations are
terminated at $t_{end}$ $\sim$1.29 $t_{ff}$ (Fig. \ref{fig:heat_low}). 

All four simulations have shown that the binary forms later than expected ($t_{bin}$ 
$\sim$1.26 $t_{ff}$) due to the low resolution that leads to inadequate modelling of 
the initial stages of the cloud evolution. After heating switches on, the binary 
components obtain thermal support and become spherical. The excess shear viscosity 
puts them in solid-body rotation and they eventually grow in size and come closer 
together (see discussion in \S \ref{sec:rot_conclusi}). The binary parameters are 
similar to those of the simulations of \S \ref{sec:heating_changerho}.

\begin{table}
\begin{center}
\begin{tabular}{|l|c|c|c|c|c|c|}\hline
$N$ & $N_{n}$ & $t_{bin}$ / $t_{ff}$ & $t_{end}$ / $t_{ff}$ & Bar fragments & $t_{frag}$ / $t_{ff}$ & $\rho_{peak}$ / g cm$^{-3}$\\ \hline \hline
10,000 & 50 $\pm5$ & 1.265 & 1.295 & 4 & 1.279 & 5.7 x 10$^{-10}$\\ \hline
20,000 & 100 $\pm10$ & 1.260 & 1.293 & 9 & 1.281 & 4.2 x 10$^{-10}$\\ \hline
30,000 & 150 $\pm15$ & 1.259 & 1.297 & 9 & 1.283 & 4.7 x 10$^{-10}$\\ \hline
40,000 & 200 $\pm20$ & 1.258 & 1.294 & 9 & 1.284 & 4.6 x 10$^{-10}$\\ \hline
\end{tabular}
\end{center}
\caption[Summary of results for the low-resolution simulations with adiabatic heating and 
increasing $N_{n}$]{Summary of results for the low-resolution simulations with adiabatic heating 
($\rho_{0}$ = 5 x 10$^{-13}$ g cm$^{-3}$) and 
increasing $N_{n}$. For each simulation the 
third column gives the time of the binary formation, the fourth the final time of 
the simulation and the sixth the time of the fragmentation of the bar. All times 
are quoted in units of the free-fall time of the initial cloud, $t_{ff}$. The 
last column gives the peak density at the end of each simulation.}
{\label{tab:heat_low}}
\end{table}

The whole cloud obtains thermal support and the peak densities do not reach such high 
values as in the isothermal simulations. The peak density decreases with increasing 
$N_{n}$. The last two simulations have progressed a bit more in time, therefore 
their peak density at the final stage is a bit higher than that of the simulation 
with $N$=20,000 and $N_{n}$=100.

In all simulations the bar fragments. It fragments at a later stage with less 
noise. However, decreasing the noise does not decrease the number of bar 
fragments as one might expect. The simulation with $N$=10,000 and $N_{n}$=50 produces 
4 bar fragments but the other three simulations produce 9 fragments. Some of these 
fragments merge with the binary components. In all four simulations a few lumps are 
produced in both spiral tails. 

Bate \& Burkert \shortcite{BateMNRAS1997} have argued that with increasing $\rho_{0}$ the bar 
produces more fragments as it becomes thinner and the ratio of its length to its 
thickness increases. One could argue the same for increasing $N_{n}$. Moreover, since 
the Jeans condition is not obeyed in all four simulations, lumps containing less than 
$N_{n}$ particles can start condensing out and it is easier to find such lumps with 
increasing $N_{n}$.

We conclude that from the simulation with $N$=20,000 and $N_{n}$=100 convergence is 
achieved, but clearly not to the right solution as given in Klein {\it et al}. 
\shortcite{KleinTOKYO1998}.

\begin{figure}
\resizebox{7.75cm}{!}{\includegraphics{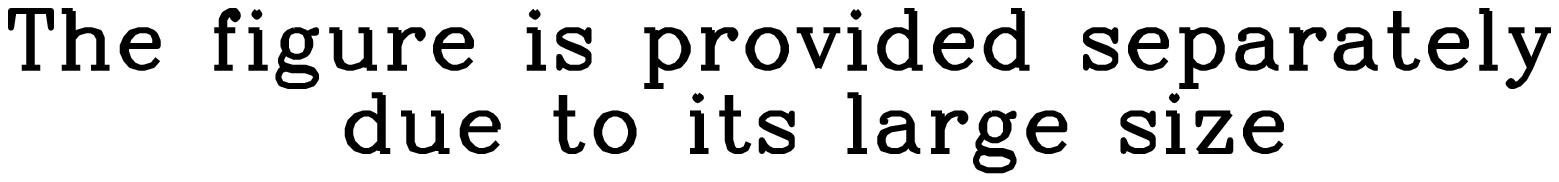}}
\resizebox{7.75cm}{!}{\includegraphics{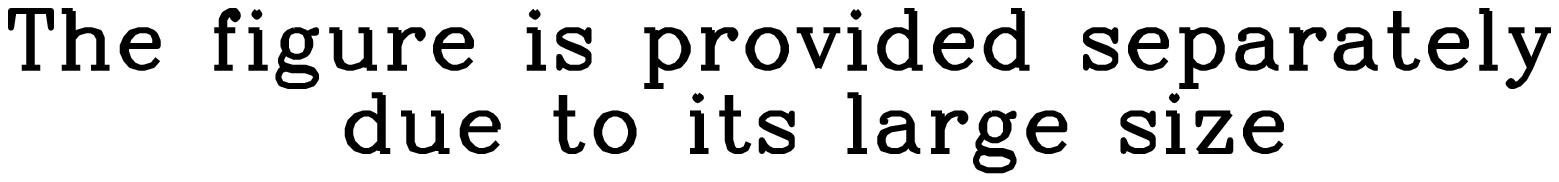}}
\resizebox{7.75cm}{!}{\includegraphics{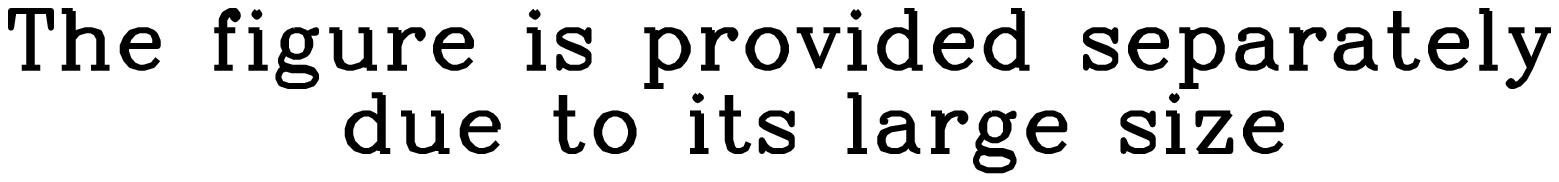}}
\resizebox{7.75cm}{!}{\includegraphics{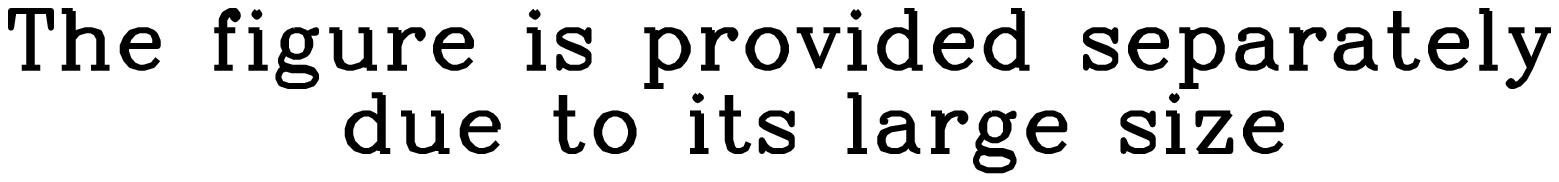}}
      \caption[Column density plots for the low-resolution simulations with adiabatic heating and 
increasing $N_{n}$]{Column density plots for the low-resolution simulations with adiabatic heating 
($\rho_{0}$ = 5 x 10$^{-13}$ g cm$^{-3}$) and 
increasing $N_{n}$. Final stage of the simulations with $N$=10,000 and $N_{n}$=50 (left), 
$N$=20,000 and $N_{n}$=100 (right) on the top row and $N$=30,000 and $N_{n}$=150 (left), 
$N$=40,000 and $N_{n}$=200 (right) on the bottom row. The details for each simulation are given in 
Table \ref{tab:heat_low}. The results converge after the simulation with $N$=20,000 and 
$N_{n}$=100. The linear size of all plots is 0.004 pc. The colour table has units of 1.18 x 
10$^{8}$ g cm$^{-2}$.}
      \label{fig:heat_low}
\end{figure}

\begin{figure}
\resizebox{7.75cm}{!}{\includegraphics{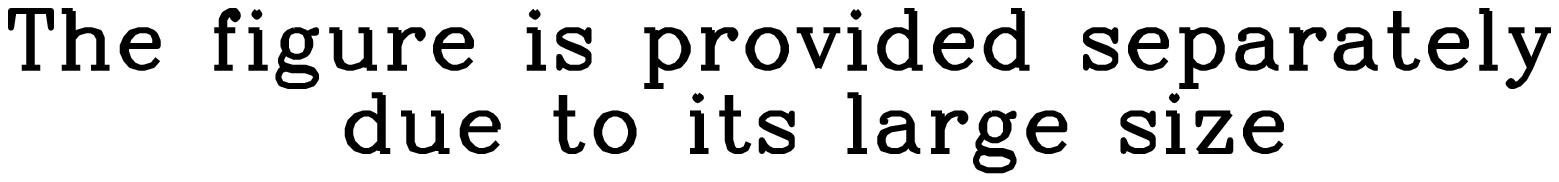}}
\resizebox{7.75cm}{!}{\includegraphics{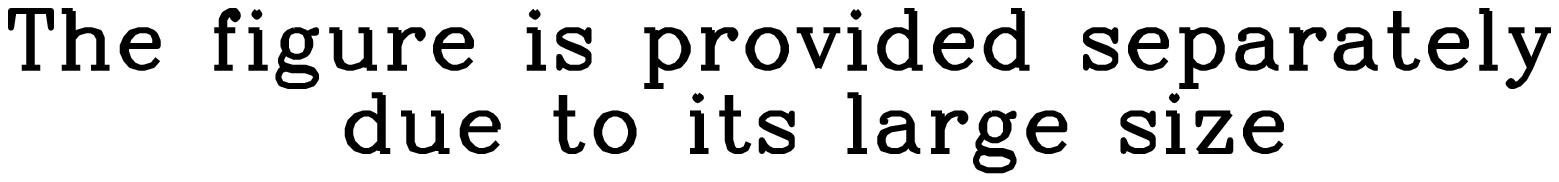}}
\resizebox{7.75cm}{!}{\includegraphics{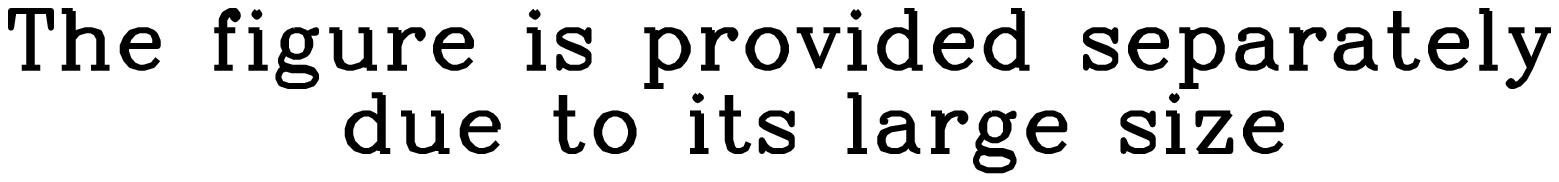}}
\resizebox{7.75cm}{!}{\includegraphics{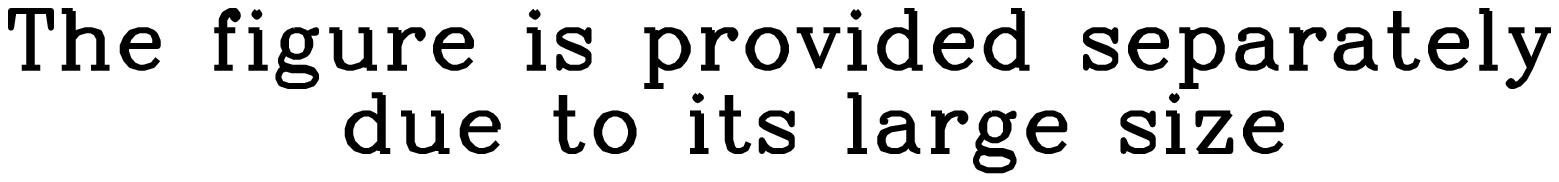}}
      \caption[Column density plots for the high-resolution simulations with adiabatic heating and 
increasing $N_{n}$]{Column density plots for the high-resolution simulations with adiabatic heating 
($\rho_{0}$ = 5 x 10$^{-14}$ g cm$^{-3}$) and increasing $N_{n}$. Final stage of the simulations with 
$N$=80,000 and $N_{n}$=50 (left), $N$=160,000 and $N_{n}$=100 (right) on the top row and $N$=240,000 
and $N_{n}$=150 left on the bottom row. The details for each simulation are given in Table 
\ref{tab:heat_high}. The results converge after the simulation with $N$=160,000 and $N_{n}$=100. 
The right panel of the bottom row is a column density plot for the high resolution simulation with 
$N$=160,000 and $N_{n}$=100 and delayed adiabatic heating ($\rho_{0}$ = 5 x 10$^{-13}$ g cm$^{-3}$). The 
linear size of all plots is 0.004 pc. The colour table has units of 1.18 x 10$^{8}$ g cm$^{-2}$.}
      \label{fig:heat_high}
\end{figure}

\subsubsection{3.4.2.2 High resolution simulations with adiabatic heating}{\label{sec:heating_high}}

\begin{table}
\begin{tabular}{|l|c|c|c|c|c|c|}\hline
$N$ & $N_{n}$ & $t_{bin}$ / $t_{ff}$ & $t_{end}$ / $t_{ff}$ & Bar fragments & $t_{frag}$ / $t_{ff}$ & $\rho_{peak}$ / g cm$^{-3}$\\ \hline \hline
80,000 & 50 $\pm5$ & 1.248 & 1.307 & 1 & 1.285 & 9.5 x 10$^{-12}$\\ \hline
160,000 & 100 $\pm10$ & 1.247 & 1.316 & 1 & 1.294 & 1.2 x 10$^{-11}$\\ \hline
240,000 & 150 $\pm15$ & 1.247 & 1.309 & 1 & 1.300 & 1.2 x 10$^{-11}$\\ \hline
\end{tabular}
\caption[Summary of results for the high-resolution simulations with adiabatic heating and 
increasing $N_{n}$]{Summary of results for the high-resolution simulations with adiabatic heating 
($\rho_{0}$ = 5 x 10$^{-14}$ g cm$^{-3}$) and 
increasing $N_{n}$. For each simulation the 
third column gives the time of the binary formation, the fourth the final time of 
the simulation and the sixth the time of the fragmentation of the bar. All times 
are quoted in units of the free-fall time of the initial cloud, $t_{ff}$. The 
last column gives the peak density at the end of each simulation.}
{\label{tab:heat_high}}
\end{table}

A summary of the high resolution simulations with adiabatic heating is given in Table 
\ref{tab:heat_high}. All simulations can resolve fragmentation up to a 
density of $\rho_{max}$ = 9.6 x 10$^{-14}$ g cm$^{-3}$. Adiabatic heating 
starts at $\rho_{0}$ = 5 x 10$^{-14}$ g cm$^{-3}$ in order for the Jeans condition to be obeyed. 
All simulations are terminated at $t_{end}$ $\sim$1.31 $t_{ff}$ (both top row panels and bottom 
left hand panel of Fig. \ref{fig:heat_high}). 

The binary forms at the expected time ($t_{bin}$ $\sim$1.25 $t_{ff}$) 
and its physical parameters are similar to those of the simulations of 
\S \ref{sec:heating_changerho}. After heating switches on, the binary 
components obtain thermal support and become spherical. The excess shear viscosity 
puts them in solid-body rotation and they eventually grow in size and come closer 
together (see discussion in \S \ref{sec:rot_conclusi}). 

The whole cloud obtains thermal support and the peak densities do not reach such high 
values as in the isothermal simulations. The peak density decreases with increasing 
$N_{n}$. The last two simulations have progressed a bit more in time, therefore, 
their peak density at the final stage is a bit higher than that of the first simulation. 

In all simulations the bar produces one fragment at the centre. The bar fragments at a 
later stage with less noise. In the simulation with $N$=80,000 and $N_{n}$=50 
each spiral arm produces a small lump. There are no lumps in the spiral tails in
the other two simulations. Therefore, we conclude that convergence is achieved from the 
simulation with $N$=160,000 and $N_{n}$=100. Since convergence is achieved, we have not 
conducted a simulation with $N$=320,000 and $N_{n}$=200 to avoid the most computationally 
expensive of this series of simulations (it requires more than 1.2 Gbytes of 
memory).

The simulations converge to a result similar to that of Klein {\it et al}. 
\shortcite{KleinTOKYO1998} apart from the one bar fragment. In all three simulations the bar 
fragment appears to form when the binary components have approached each other enough so that 
they start to spiral in towards their eventual merger (see discussion in \S \ref{sec:heating_lrho}). 
This may imply that the bar fragment is produced by a 
tidal disruption on the bar by the first binary encounter. However, SPH with reduced noise 
and a medium resolution is still not able to prevent this artificial fragment from 
forming. 

The fact that the peak densities of the high-resolution simulations are lower than those of the low-resolution simulations is due to heating being applied at a lower density in the 
former. For a comparison, we have conducted a simulation with $N$=160,000, $N_{n}$=100 and 
$\rho_{0}$ = 5 x 10$^{-13}$ g cm$^{-3}$. The peak density at the end of this simulation 
($t_{end}$ $\sim$1.277 $t_{ff}$ -- bottom right hand panel of Fig. \ref{fig:heat_high}) has 
reached $\rho_{peak}$ = 1.9 x 10$^{-10}$ g cm$^{-3}$, a value much higher than those of the 
low resolution simulations with the same amount of adiabatic heating, at the same time.

In the simulation with $N$=160,000, $N_{n}$=100 and $\rho_{0}$ = 5 x 10$^{-13}$ g cm$^{-3}$ 
the bar fragments earlier than that with $N$=80,000, $N_{n}$=50 and 
$\rho_{0}$ = 5 x 10$^{-13}$ g cm$^{-3}$ (\S \ref{sec:heating_mrho}) and it produces 9 fragments. 
Some of them merge with the binary components. Again each spiral tail contains two lumps.
Reducing the noise for the high resolution simulations with $\rho_{0}$ = 5 x 10$^{-13}$ 
g cm$^{-3}$ has not prevented the bar from fragmenting into quite a few fragments. We conclude 
that for these simulations convergence is achieved from the simulation with $N_{n}$=80,000 and 
$N$=50 (Fig. \ref{fig:5x13_end}), but clearly not to the right solution as given in 
Klein {\it et al}. \shortcite{KleinTOKYO1998}.

Again, we conclude that, with higher resolution, convergence is achieved earlier, as it was in 
the isothermal simulations. More importantly, we conclude that if the Jeans condition is not 
obeyed then by having the noise decreased we cannot obtain the 
expected result. Only when the Jeans condition was obeyed did we obtain a result close to that of 
Klein {\it et al}. \shortcite{KleinTOKYO1998}. 

With medium resolution, we have shown that we achieve convergence to the expected solution (or 
close to it) with $N_{n}$=100 both in isothermal simulations and simulations with adiabatic 
heating. To obtain the exact solution in SPH simulations with adiabatic heating, one would need to 
increase the resolution as well.

\section{Changing the kernel}{\label{sec:kernel}}

\begin{figure}
\resizebox{7.75cm}{!}{\includegraphics{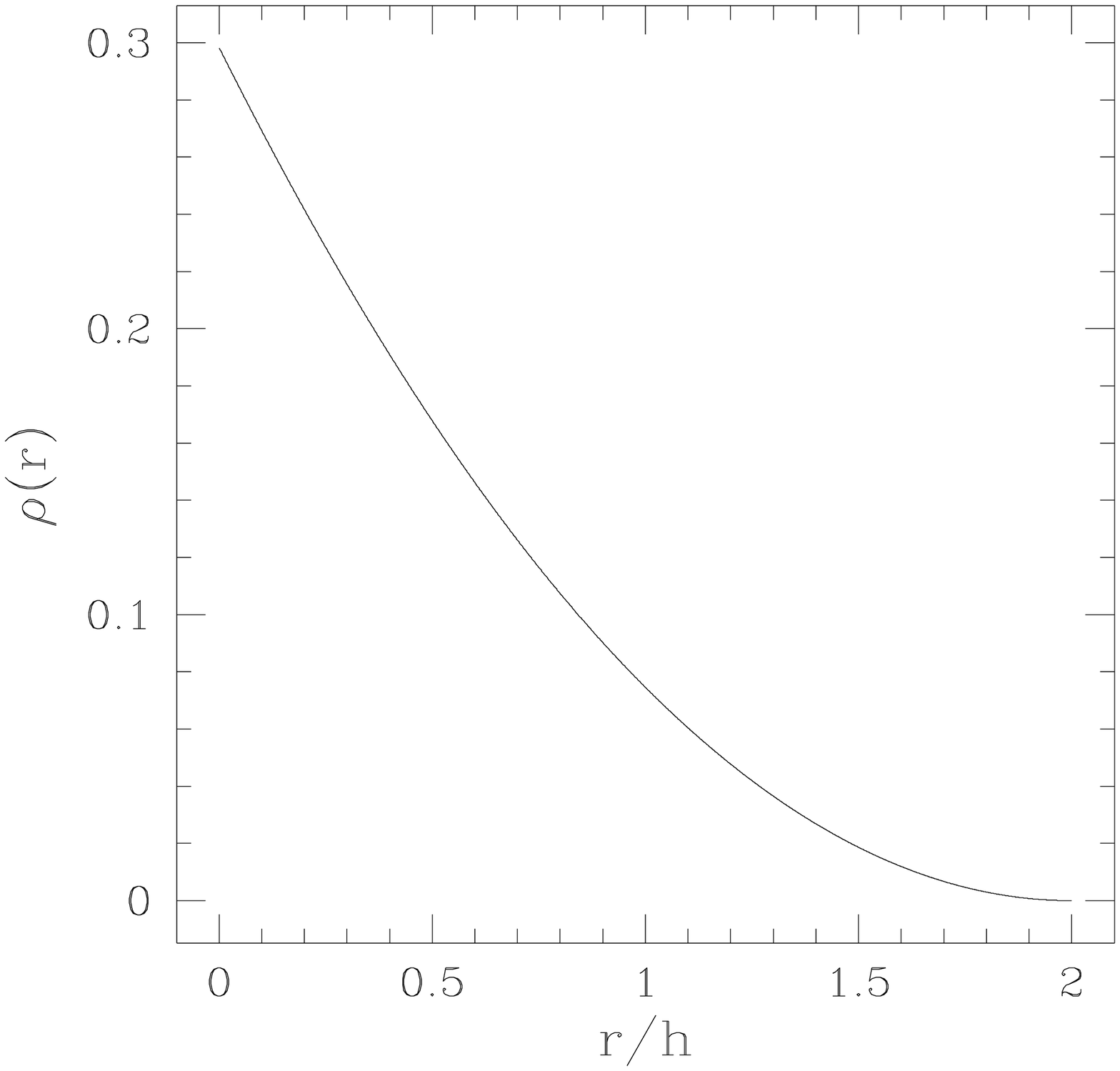}}
\resizebox{7.75cm}{!}{\includegraphics{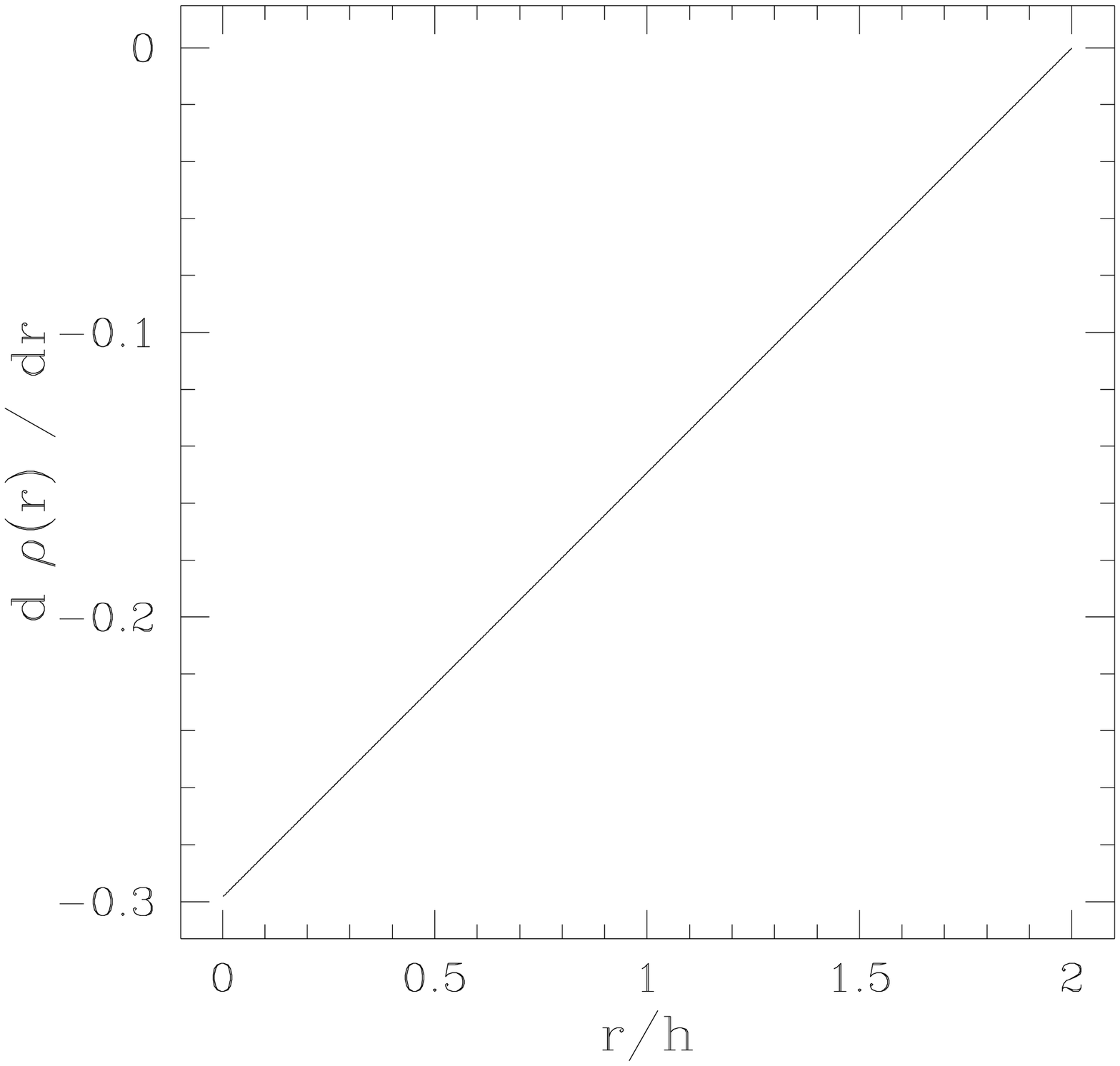}}  
      \caption[Density and its gradient for an isolated particle using the kernel of Eqn. \ref{equa:ant_kern}]
      {\underline{Left}: Radial density profile of an isolated particle of unit mass and smoothing length $h$ 
      using the kernel of Eqn. \ref{equa:ant_kern}. 
      \underline{Right}: The gradient of the radial density profile of an isolated particle of unit mass and 
      smoothing length $h$ using the kernel of Eqn. \ref{equa:ant_kern}.}
      \label{fig:ant_kern}
\end{figure}

Since in most of the above simulations with adiabatic heating the bar has fragmented, one could 
argue that the M4-kernel we have used is unable to prevent artificial clustering of particles and 
therefore artificial fragmentation (see Fig. \ref{fig:sph_profile} and the discussion on the 
gradient of the kernel in \S \ref{sec:density}). We have experimented with another kernel that 
possesses most of the properties of the M4-kernel: it has compact support and it is truncated at 
$r=2h$, it has almost the same value at $r=0$, its value at $r=h$ is one quarter of the value at 
$r=0$, it contains half of the mass in the $(0,h)$ range and the other half in the $(h,2h)$ range, 
its gradient has almost the same value at $r=0$, there are $\sim$ 47 neighbours within $h$ for a 
uniform distribution of particles. This kernel, its derivative and volume integral are given by: 

\begin{equation}
\label{equa:ant_kern}
W(s) \; = \; \left \{ 
       \begin{array}{ll}
          \frac{15}{64 \pi} (2-s)^{2}, & 0 \leqslant s \leqslant 2; \\
          0, & s \geqslant 2,
       \end{array} \right .
\end{equation}

\begin{equation}
\label{equa:dantkern}
W^{'}(s) \; = \; \left \{ 
       \begin{array}{ll}
          - \frac{15}{32 \pi} (2-s), & 0 \leqslant s \leqslant 2; \\
          0, & s \geqslant 2,
       \end{array} \right .
\end{equation}

\noindent and 

\begin{equation}
\label{equa:iantkern}
W^{*}(s) \; = \; \frac{1}{16} \left \{ 
       \begin{array}{ll}
          3s^{5} - 15s^{4} + 20s^{3}, & 0 \leqslant s \leqslant 2; \\
          16, & s \geqslant 2.
       \end{array} \right .
\end{equation}

The fact that the second derivative of the kernel is always positive, $W^{''}(s) = \frac{15}{32 \pi}$, 
ensures that the magnitude of the gradient of the kernel decreases monotonically with increasing $s$ and therefore, there is no inversion 
in the direction of the hydrostatic force. The fact that $W^{'}(0) \neq 0$ ensures that there is not 
going to be formed artificially any pairs of particles due to a vanishing hydrostatic force. The density 
profile and its gradient for an isolated particle of unit mass and smoothing length $h$, using Eqns. 
\ref{equa:ant_kern} \& \ref{equa:dantkern} respectively, are shown in Fig. \ref{fig:ant_kern}. 

\begin{figure}
\begin{center}
\resizebox{8cm}{!}{\includegraphics{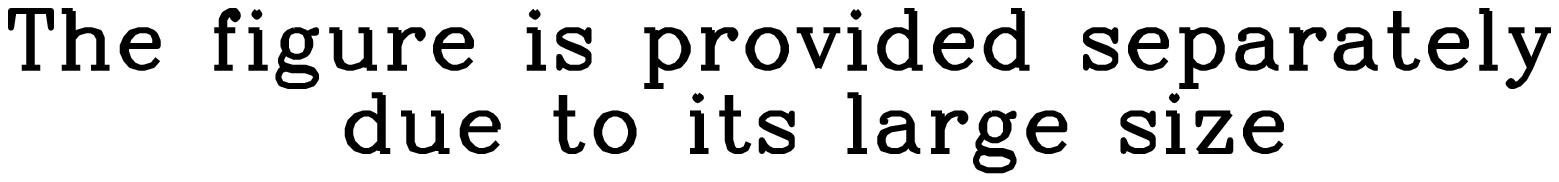}}
\end{center}  
\begin{center}
\resizebox{8cm}{!}{\includegraphics{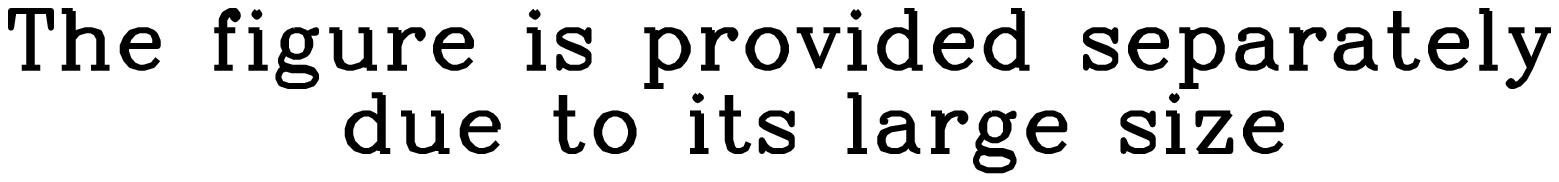}}
\end{center}
      \caption[Column density plots for a cloud of 80,000 particles using the new kernel ($\rho_{0}$ = 10$^{-13}$ g cm$^{-3}$)]
      {Column density plots for a cloud of 80,000 particles using the new kernel ($\rho_{0}$ = 10$^{-13}$ g cm$^{-3}$). 
      The linear size of these plots is 0.004 pc. The colour table has units of 1.18 x 10$^{8}$ g cm$^{-2}$. 
      \underline{Top}: Column density plot at the end ($t$ = 1.293 $t_{ff}$) using the new kernel only for hydrodynamics. 
      \underline{Bottom}: Column density plot at the end ($t$ = 1.295 $t_{ff}$) using the new kernel for both gravity and hydrodynamics.}
      \label{fig:new_kern}
\end{figure}

We have used this kernel in simulations of a rotating, spherical, uniform-density, isothermal cloud with an 
m = 2 perturbation. In fact, we have conducted two different simulations: in one we have included 
the new kernel only for calculating the hydrodynamics (we use the M4 for the gravity calculations) 
and in the other we have used the new kernel for both gravity and hydrodynamics. This way, we can 
make a clearer comparison with the performance of the M4 kernel. We have used 80,000 particles 
initially taken from a lattice. We have included adiabatic heating starting at $\rho_{0}$ = 
10$^{-13}$ g cm$^{-3}$. The results should be compared with the corresponding simulation in \S 
\ref{sec:rho0=-13} (Fig. \ref{fig:13_heat}). Both simulations were ended at $t_{end}$ $\sim$1.295 $t_{ff}$ (Fig. 
\ref{fig:new_kern}). 

The binary forms at the expected time ($t_{bin}$ $\sim$1.25 $t_{ff}$). The bar produces several 
fragments in random positions and there are also lumps in the spiral tails (one in each). 

The peak density at the end of both simulations is $\rho_{peak}$ = 3.8 x 10$^{-11}$ g cm$^{-3}$, 
almost twice the value of the simulation of \S \ref{sec:rho0=-13}. This implies that the difference 
is due to the hydrodynamics of the new kernel. In particular, the new kernel appears to produce more 
centrally condensed objects than the M4 kernel. 

This may also explain the reason for the earlier bar fragmentation ($t_{frag}$ $\sim$1.275 $t_{ff}$) 
in the simulation where the new kernel was only used for calculating hydrodynamics. In the other 
simulation, the bar fragmented at the same time as the simulation of \S \ref{sec:rho0=-13} 
($t_{frag}$ $\sim$1.280 $t_{ff}$). 

Finally, the fact that the simulation with the new kernel used only for hydrodynamics produces less fragments 
than the simulation with the new kernel used for both gravity and hydrodynamics, should be attributed 
to the effect the new kernel has on the gravity calculation. In particular, it seems that the new kernel 
emphasises density enhancements in the bar so that they become self-gravitating faster.

We conclude that a kernel that theoretically should prevent artificial clustering of particles from 
occurring appears to be more problematic than the M4. This does not prove that the M4-kernel is the ideal 
kernel for Star Formation simulations. It just indicates that the artificial fragmentation of the bar 
with the M4 is not only due to it allowing artificial clustering of particles. Further extensive comparison 
studies should be made with the use of other kernels to identify those that operate best with 
self-gravitating SPH codes. 

\section{Settled distribution}{\label{sec:settle}}

\begin{figure}
\begin{center}
\resizebox{8cm}{!}{\includegraphics{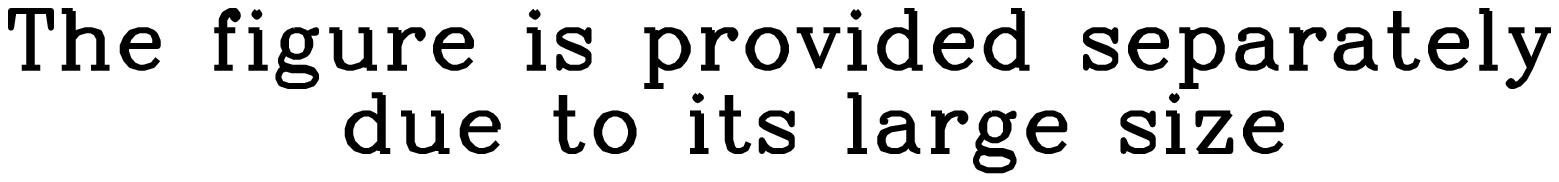}}
\end{center}    
      \caption[Column density plot for a cloud of 80,000 particles at the end with particles initially taken 
      from a settled distribution ($\rho_{0}$ = 5 x 10$^{-14}$ g cm$^{-3}$)]
      {Column density plot for a cloud of 80,000 particles at the end with particles initially taken 
      from a settled distribution ($\rho_{0}$ = 5 x 10$^{-14}$ g cm$^{-3}$). The time is $t$ = 1.31 $t_{ff}$. 
      The linear size of this plot is 0.004 pc. The colour table has units of 1.18 x 10$^{8}$ g cm$^{-2}$.}
      \label{fig:settle_run}
\end{figure}

Finally, we have conducted a simulation with 80,000 particles initially taken from a settled distribution. 
Adiabatic heating starts at $\rho_{0}$ = 5 x 10$^{-14}$ g cm$^{-3}$ in order for the Jeans condition to be obeyed. 
The results should be compared 
with the corresponding simulation in \S \ref{sec:heating_lrho} (Fig. \ref{fig:5x14_end}). Fig. \ref{fig:settle_run} is a column density plot 
at the end of the simulation (at $t_{end}$ = 1.312 $t_{ff}$). 

The binary forms at the expected time but the bar fragments into 3 objects, formed in random positions. In fact, 
one of them forms from the merger of two smaller ones, something which was never observed in the simulations where 
particles where initially taken from a lattice. Having more bar fragments and them being in random positions, 
shows that a settled distribution clearly contains more 
noise than a distribution of particles taken from a lattice. This is also suggested by the fact that the peak 
density at the end, $\rho_{peak}$ = 7.5 x 10$^{-12}$ g cm$^{-3}$, is lower than that of the corresponding simulation 
in \S \ref{sec:heating_lrho}. 

The bar fragmentation is delayed with the settled distribution ($t_{frag}$ = 1.298 $t_{ff}$). This indicates that 
it takes more time for a 
self-gravitating object to form in the bar due to the cloud containing more noise. In fact, the initial noise 
seems to provide an extra means of (turbulent) support to the bar. This is not necessarily a 
disadvantage for realistic Star Formation simulations, as the initial conditions for Star Formation in nature 
are far from being smooth and without noise. In chapter \ref{sec:results}, we shall use settled distributions 
for our initial conditions.

\chapter{Particle Splitting}{\label{sec:part-split}}

\section{Jeans condition}{\label{sec:jeans_cond}}

Several authors have recently discussed the significance of results based on 
numerical simulations that use either Eulerian or Lagrangian formulations, in 
particular the ability of these codes to prevent the non-physical growth of 
numerical perturbations, and their ability to resolve all the structure formed 
and therefore produce reliable and realistic results.

Truelove {\it et al.} \shortcite{TrueloveApJ1997,TrueloveApJ1998} describe the 
need for a condition to regulate the linear size of their grid. They 
have concluded that the linear size of their smallest grid must always 
be smaller than a quarter of the local Jeans length. By setting this Jeans condition, they 
suppress the formation and propagation of artificially induced perturbations that
could otherwise corrupt their results.

In SPH, such a condition is also needed for similar reasons. As the SPH
particles are allowed to move with the fluid, the properties of the fluid are 
kept updated. Therefore, the artificial growth of perturbations is inhibited, 
provided that the SPH calculations give accurate estimates for these properties,
i.e. there is adequate sampling of the fluid with enough particles of similar 
properties within each kernel. Bate \& Burkert \shortcite{BateMNRAS1997} have 
argued that SPH, with its current formulations, does not necessarily fulfil this
condition. They demonstrate the need for a Jeans condition for SPH, {\it viz.} 
that the local Jeans mass should be resolved at all times. By this, they mean 
that there should always be enough sampling points in a clump near to or above the 
locally defined Jeans mass.

They also give a number of complementary rules for the formulation of the 
Jeans condition in order to obtain reliable results in fragmentation 
simulations. They suggest smoothing the hydrodynamical forces at a scale 
similar to the gravity softening (i.e. $\epsilon = h$), as they find that if 
$\epsilon < h$ then artificial fragmentation is induced, while for $\epsilon > 
h$ fragmentation is inhibited.

Whitworth \shortcite{WhitworthMNRAS1998a} has also looked into the definition 
of the Jeans condition for SPH and his findings are in very good agreement 
with those of Bate \& Burkert \shortcite{BateMNRAS1997}. In his analysis, he 
introduces 3 different masses: the minimum resolvable mass by SPH, M$_{min}$, 
i.e. the mass within the radius of a kernel, with $M_{min} = N_{n} \, 
m_{ptcl}$, where $N_{n}$ is the number of neighbours within a kernel (equal 
to $\sim$50 in 3-dimensional simulations) and m$_{ptcl}$ the mass of each SPH 
particle\footnote{It is assumed that the simulation is implemented with 
particles of equal mass.}; the mass of a proto-condensation, M$_{0}$; and the 
local Jeans mass, M$_{J}$, defined for the gas confined in and around M$_{0}$. 
If M$_{0} \gg$ M$_{min}$ the proto-condensation is resolved. For M$_{0} \ll$ 
M$_{min}$, the proto-condensation is unresolved. If M$_{0} >$ M$_{J}$ it is 
unstable against collapse. For M$_{0} <$ M$_{J}$, it is stable.

He concludes that SPH is treating resolved proto-condensations (M$_{0} \gg$ 
M$_{min}$) adequately\footnote{There is a small under-estimation of the 
time scale for the growth of a possible Jeans instability, which becomes 
significant only for M$_{0} \sim$ M$_{min}$.}. Problems may arise when they are 
unresolved. In this case, unresolved proto-condensations that are Jeans 
unstable (M$_{J} <$ M$_{0} \ll$ M$_{min}$) are not allowed to collapse as 
fragmentation is inhibited, while for unresolved proto-condensations that are 
Jeans stable (M$_{0} \ll$ M$_{min}$ \& M$_{0} <$ M$_{J}$) artificial 
fragmentation is induced. He shows that artificial fragmentation is prevented 
provided that the Jeans mass is resolved, M$_{min} <$ M$_{J}$, and the 
interpolating kernel is sufficiently centrally condensed. This condition gives
a very strong constraint, as it does not allow any unresolved 
proto-condensations to form (neither stable nor unstable ones) and thus eliminates 
all possible problems that could arise in a fragmentation calculation.

Bate \& Burkert \shortcite{BateMNRAS1997} give an even stronger Jeans 
condition which states that 

\begin{equation}
\label{equa:Jeans_condition1}
2 M_{min} \lesssim M_{J}. 
\end{equation}

\noindent The proof for this is not robust, as it is based on 
qualitative evidence from weak convergence of a series of numerical 
simulations to a certain result. However, we will use this strong Jeans 
condition, as, even in the limiting case of 2 M$_{min} \sim$ M$_{J}$, the 
Whitworth \shortcite{WhitworthMNRAS1998a} condition is still satisfied.

Using Eqn. \ref{equa:jeans8}, the limiting case of the Jeans condition (Eqn. 
\ref{equa:Jeans_condition1}) becomes

\begin{equation}
\label{equa:Jeans_condition2}
2 N_{n} \frac{M_{total}}{N_{total}} = 2 N_{n} m_{ptcl} = 2 M_{min} \lesssim M_{J} = \frac{c^{3} 
\pi^{5/2}}{6 G^{3/2} \rho_{max}^{1/2}},
\end{equation}

\noindent where M$_{total}$ and N$_{total}$ are respectively the total mass 
and the total number of particles in the simulation, and $\rho_{max}$ is the 
maximum resolvable density. For a clump of given mass and temperature, the
maximum resolvable density is a function only of the total number of particles

\begin{equation}
\label{equa:Jeans_condition3}
\rho_{max} = \frac{c^{6} \pi^{5} (N_{total})^{2}}{36 G^{3} (2 N_{n} 
M_{total})^{2}} = \frac{c^{6} \pi^{5}}{144 G^{3} (N_{n} m_{ptcl})^{2}}.
\end{equation}

The objective of numerical simulations of star formation is to approach 
stellar densities, i.e. to achieve as high maximum densities as possible. Eqn.
\ref{equa:Jeans_condition3} clearly sets an obstacle against the 
implementation of this objective. For example, for resolving fully a 
simulation involving an isothermal ($c_{0}=0.17$ km s$^{-1}$) clump of 1M$_{\odot}$ to a 
density of $10^{-10}$ g cm$^{-3}$ then $\sim 1.8 \mathrm{x} 10^{7}$ particles 
are needed, which is at the limit of present-day computer capabilities.

An alternative would be to redirect resources only to regions of 
particular interest in an existing simulation. This can be achieved by 
increasing the number of particles locally in order to maintain the 
validity of Eqn. \ref{equa:Jeans_condition1} in regions approaching the 
resolution limit, whilst retaining the coarse resolution in resolved regions. 
This way, we can balance the need for higher resolution against present-day
computer capabilities. 

We have invented a method to implement this. All particles in a region of 
interest are split. We give this method the obvious name ``Particle Splitting''. 
The development and testing of this method constitute one of the primary aims of 
this work. Since we can use this method at several levels every time the resolution 
limit is reached during a simulation, we have introduced the notion of simulations 
of increasingly high resolution nested inside the original coarse simulation. From 
this, we have named all simulations to which this method is applied ``Nested 
Simulations''. 

With particle splitting we can also address another point made by Bate \& 
Burkert \shortcite{BateMNRAS1997}: we can follow the detailed evolution of all 
fragments as well as the global evolution of the simulation. The ability to do 
so is of course constrained by the time-step.

\section{Particle splitting: Concept}{\label{sec:concept}}

Our aim is to increase the number of particles in a small sub-region of the 
computational domain of an existing simulation, just before this region 
reaches its resolution limit (Eqn. \ref{equa:Jeans_condition1}). This way, we 
will be able to continue the simulation at a higher resolution, but only where 
this is really necessary. The shape of the sub-region will depend on the 
geometry of the problem (e.g. cylindrical in simulations involving flattened 
structures), but the simulation will always be fully 3-dimensional. In the 
sequel we shall refer to particles in the high resolution region as fine 
particles, and particles in all other regions as coarse particles. We shall also
refer to a simulation as a fine simulation, if it includes fine particles, and as 
a coarse simulation, if it includes no fine particles.

The method will be applied at time $t_{spl}$, when significant -- but 
always resolved -- structure has formed in the coarse simulation, just before 
this structure reaches its resolution limit. We will stop the coarse 
simulation and decide the position, shape and size of the sub-region manually. 
This involves choosing the appropriate co-ordinates for the sub-region, 
so that it contains all the significant structure. The initial 
conditions for the fine simulation will be interpolated from the coarse 
simulation. At $t_{spl}$, each coarse particle in the sub-region will be 
replaced by 13 new equal-mass fine particles distributed symmetrically round 
the coarse particle's position (\S \ref{sec:positions}). These fine particles 
will be given velocities interpolated from the velocities of the coarse particles 
they replace (\S \ref{sec:veloc}). Using the appropriate subroutines of the code
(chapter \ref{sec:SPH}), they will also be given values for all the physical 
quantities involved in the SPH equations, such as density (\S \ref{sec:density} 
\& \ref{sec:ptcl_spl}), smoothing length (\S \ref{sec:newh}), 
temperature, acceleration, time-step information (\S \ref{sec:fluid}). After 
this initialisation, the fine simulation, with only fine particles inside the 
sub-region and only coarse particles outside the sub-region, will start. 
In subsequent time-steps, if a coarse particle is found to cross the 
sub-region's boundary, then it will be split on-the-fly into 13 fine particles 
and with a procedure similar to the initialising one, these fine particles 
will be given velocities, densities, temperatures, smoothing lengths, 
accelerations, time-steps, and they will immediately become active. On the 
other hand, if a fine particle exits the nested sub-region, it will continue 
being active. The positions, velocities and accelerations of the coarse 
particles outside the sub-region will evolve together with those of the fine 
particles. In effect, the outside coarse particles will provide the boundary 
conditions to the fine simulation. This means that the boundary conditions 
will be as exact as those of the coarse simulation.

A possible problem that we will have to deal with is the interaction at the 
boundary of two different populations of particles: the massive and extended 
coarse particles, and the light and compact fine particles. This can 
cause interpenetration and mixing of the two populations and an 
artificial blurring of the boundary. We solve this problem by 
adjusting $h$ so that the kernel contains a fixed multiple of the mass of the 
central particle (\S \ref{sec:newh}), rather than a fixed number of neighbours.

An alternative implementation of particle splitting involves setting a threshold
density above which particles are automatically split. The advantage of the latter 
method is that we don't have to stop the 
simulation to decide the co-ordinates of the sub-region. All splitting 
happens on-the-fly. We call this version of the new method ``on-the-fly splitting'' 
as opposed to the former version which we call ``nested splitting''.

On-the-fly splitting is our preferred version of the new method. However, 
we will discuss both methods. We will first define the positions of 
the fine particles with respect to the positions of the coarse particles they 
replace. This will involve fine-tuning each coarse particle's $h$ and the 
distance between the 13 fine particles. Therefore, this will require adjusting 
the density profile of the configuration of the 13 fine particles to the 
density profile of the coarse particle they replace as well as trying to match 
the density profile of coarse particle distributions with the density 
profile after particle splitting. We will then illustrate the difficulty of
simulating the boundary between fine and coarse particles. Subsequently, we 
will describe the new method for finding $h$ for all particles both fine and 
coarse. The next section will finish by describing the way we assign other 
physical and numerical (e.g. velocity, acceleration, temperature and time 
step information) properties to all fine particles. 

We will then perform some tests (\S \ref{sec:split_tests}) in order to validate 
the performance of the method. We will simulate uniform 
collapse, where we will also use sink particles (\S \ref{sec:ff_collapse}). We 
will show the efficiency of the new method for calculating $h$ by applying it 
to the simulation of a stable isothermal sphere (\S \ref{sec:stable}). Finally, 
we will apply particle splitting to a collapse simulation of a rotating, uniform-density, 
isothermal cloud with an m=2 perturbation (see chapter \ref{sec:rotating}) to 
utilise the new method in a more realistic application (\S \ref{sec:rot_cloud}), 
showing its efficiency in reproducing accurate results, as well as economy in terms of computational cost.

In the next chapter, we will apply particle splitting to clump-clump collision 
simulations. Firstly, we will extend previous simulations of high mass cloud 
collisions to quantify the benefits of the new method. Secondly we will 
investigate a new part of the parameter space, looking for realistic 
fragmentation mechanisms in collisions between low-mass clumps.

\section{Particle splitting: Implementation}{\label{sec:splitting}} 

\subsection{Positions for the fine particles}{\label{sec:positions}}

\begin{figure}
\resizebox{7.75cm}{!}{\includegraphics{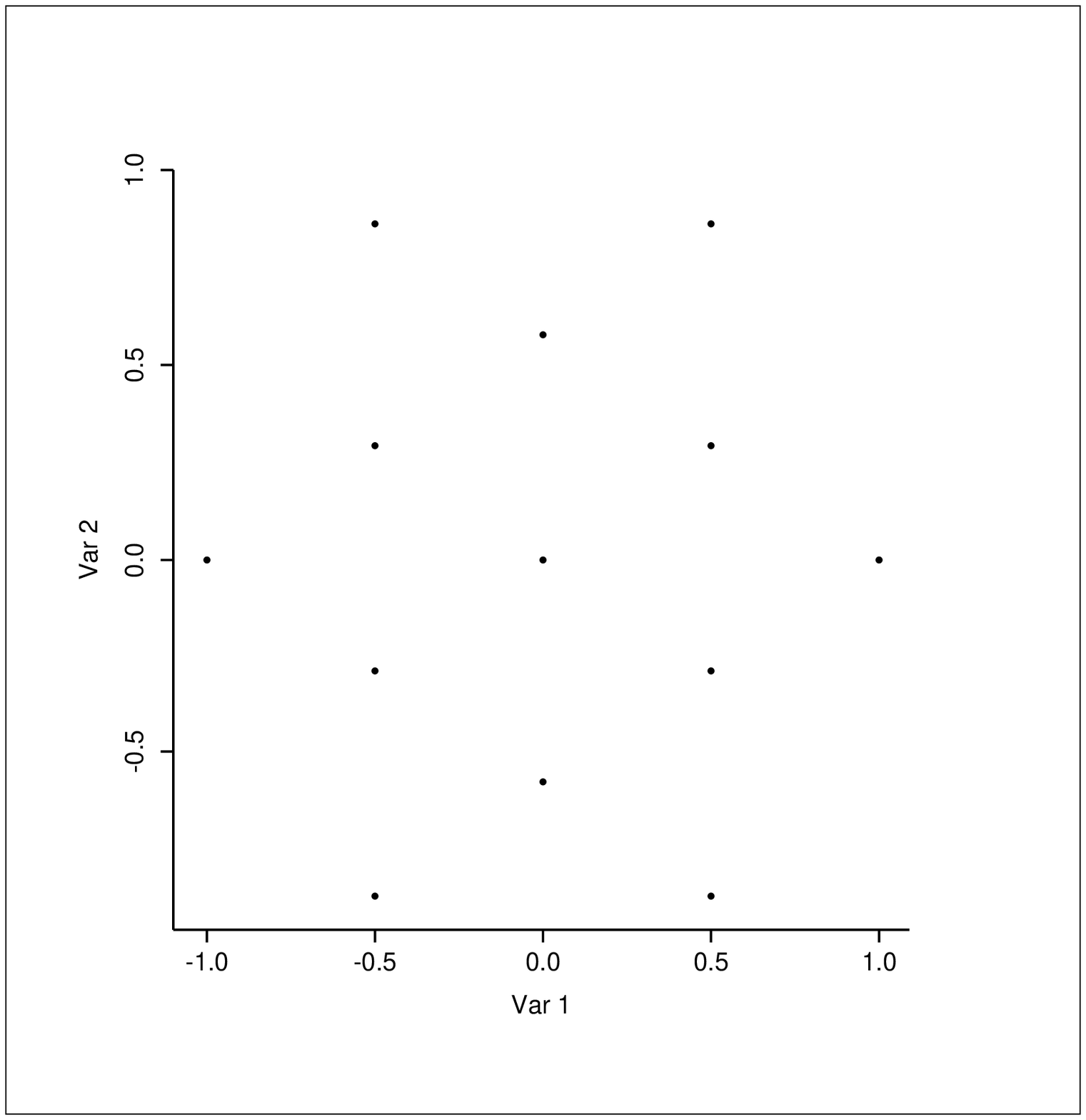}}
\resizebox{7.75cm}{!}{\includegraphics{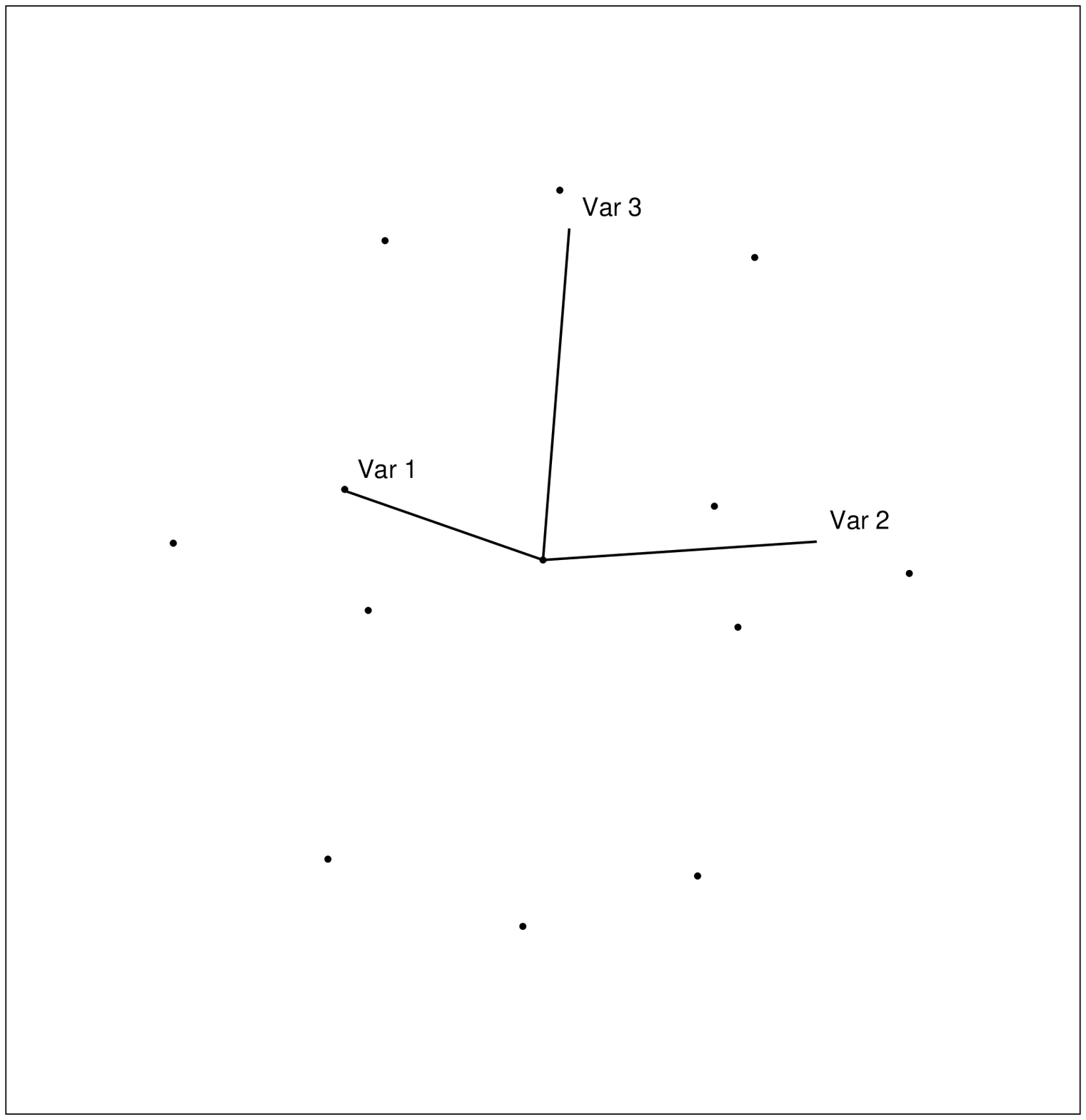}}  
      \caption[Configuration of 13 fine particles in 2 and 3 dimensions]{Graphical representation of the configuration of 13 fine particles, in two dimensional projection (left) and three dimensions (right). }
      \label{fig:particles}
\end{figure}

We replace a coarse particle with a configuration of 13 particles as shown in 
Fig. \ref{fig:particles}. A fine particle is put exactly at the position of 
the coarse particle, while 6 particles are put on the vertices of a hexagon 
centred on the position of the coarse particle. The remaining 6 particles are 
put on the vertices of two equilateral triangles parallel to the plane of the 
hexagon but on either side. All particles are put at equal distances $r_{i}$ 
from their nearest neighbours\footnote{We will deal with the value of $r_{i}$ 
shortly.}. Such a lattice is the simplest possible arrangement that has 
minimum interstitial volume \cite{KittelBOOK1962}. In fact, it is the 
primitive cell of a face-centred cubic (cubic closed-packed) structure (see 
Fig. \ref{fig:particles}, left panel). Specifically, Table \ref{tab:fcc} gives 
the co-ordinates of the 13 fine particles at unit distance away from each other, 
in the reference frame of the coarse particle they replace. 

\begin{table}
\begin{center}
\begin{tabular}{|c|c|c|c|}\hline
 & x & y & z \\ \hline
1 & 0 & 0 & 0 \\
2 & 1 & 0 & 0 \\
3 & -1 & 0 & 0 \\
4 & 0.5 & $\sqrt{3}/2$ & 0 \\
5 & 0.5 & $-\sqrt{3}/2$ & 0 \\
6 & -0.5 & $\sqrt{3}/2$ & 0 \\
7 & -0.5 & -$\sqrt{3}/2$ & 0 \\
8 & 0 & $\sqrt{3}/3$ & $-\sqrt{2/3}$ \\
9 & -0.5 & $\sqrt{3}/3$-$\sqrt{3}/2$ & $-\sqrt{2/3}$ \\
10 & 0.5 & $\sqrt{3}/3$-$\sqrt{3}/2$ & $-\sqrt{2/3}$ \\
11 & 0 & -$\sqrt{3}/3$ & $\sqrt{2/3}$ \\
12 & -0.5 & -$\sqrt{3}/3$+$\sqrt{3}/2$ & $\sqrt{2/3}$ \\
13 & 0.5 & -$\sqrt{3}/3$+$\sqrt{3}/2$ & $\sqrt{2/3}$ \\ \hline
\end{tabular}
\end{center}
      \caption[Co-ordinates of the 13 fine particles at unit distance away from each other]{Co-ordinates of the 13 fine particles at unit distance away from each other, 
in the reference frame of the coarse particle they replace.}
      \label{tab:fcc}
\end{table}

The value of $r_{i}$ is of great importance as well as the value of the 
smoothing length of the fine particles, $h_{i}$. They should be fine-tuned to 
minimise the deviation between the configuration of 13 fine particles and 
their parent coarse particle. Apart from reproducing the density profile of 
the parent coarse particle, the 13 fine particles should be positioned in such 
a way that they experience the same accelerations as if their parent coarse 
particle was present. Finally, we should take care that any density fluctuations
input to the fine region after particle splitting are kept to a minimum. We will
now calculate the density profile of the configuration of 13 fine particles in 
isolation as well as estimate any density fluctuations in a particle 
distribution due to the application of particle splitting.

\subsection{Density profile of the configuration of 13 fine particles}{\label{sec:density}}

\begin{figure}
\begin{center}
\resizebox{15.5cm}{!}{\includegraphics{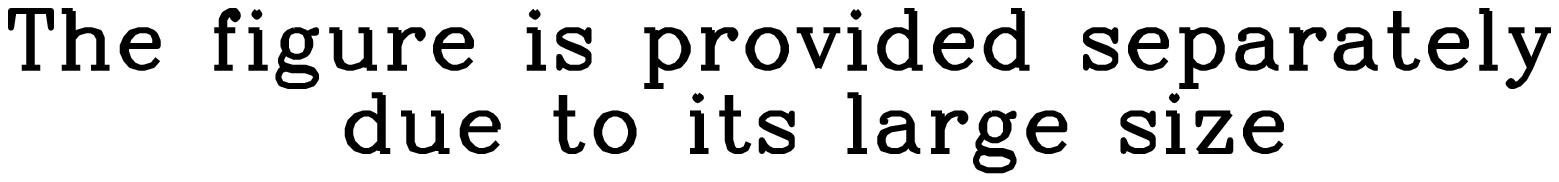}}
\end{center}
      \caption[Calculating the density profile of the configuration of 13 fine particles]{Two dimensional projection of the model used for the calculation of the mass distribution. The two concentric circles represent the one and two smoothing length spheres of an outside fine particle. The sphere with radius $r$ is centred on the central fine particle which is at a distance $r_{i}$ away from each outside fine particle.}
      \label{fig:circles}
\end{figure}

In order to calculate the mean density profile of such a configuration of 
particles, we consider a sphere of radius $r$ centred on the central fine 
particle. We define the distance $d$ from the position of any one of the 12 
outside fine particles to any point on this sphere as a function of $r$ and the 
angle $\theta$ (see Fig. \ref{fig:circles}) as

\begin{equation}
\label{equa:dist}
d \; = \; \left( r^{2}  +  r_{i}^{2}  -  2 r r_{i} cos(\theta) \right)^{\frac{1}{2}}.
\end{equation}

The mean density on the surface of the sphere is then evaluated using

\begin{eqnarray}
\label{equa:integral1}
\bar{\rho} (r) 4 \pi r^{2} \delta r & = & 12 \int_{A} m_{i} h_{i}^{-3} W_{M4} \left( \frac{d}{h_{i}} \right) \delta r \, \mathrm{d}A \; \; \Rightarrow \nonumber \\
\bar{\rho} (r) & = & 6 m_{i} h_{i}^{-3} \int_{\theta = 0}^{\theta = \theta_{max}}  W_{M4} \left( \frac{d}{h_{i}} \right) sin(\theta) \mathrm{d} \theta,
\end{eqnarray}

\noindent where the first equation gives the total mass swept up by a shell of 
radius $r$ and infinitesimal thickness $\delta r$. We have used $A  =  2 \pi r^{2} sin(\theta)$, 
the surface area of the sphere that is limited by the circle defined by angle 
$\theta$ (produced when $d$, the dashed line on Fig. \ref{fig:circles}, is 
rotated about $r_{i}$ through an angle of $\phi  =  2 \pi$). $m_{i}$ is the 
mass of one of the fine particles ($m_{i}  =  M/13$ where $M$ is the mass of 
the coarse particle) and $h_{i}$ its smoothing length ($h_{i}  =  (H/13)^{1/3}$,
where $H$ is the smoothing length of the coarse particle). We have not taken 
into account the mass of the central fine particle yet. Since we are averaging 
over all angles we have multiplied by 12 in order to account for all the 
outside fine particles.

The integral of Eqn. \ref{equa:integral1} becomes

\begin{equation}
\label{equa:integral3}
\bar{\rho} (r)  =  6 m_{i} h_{i}^{-3} \int_{\mu = \mu_{max}}^{\mu = 1}  W_{M4} \left( \left[ \frac{r^{2}  +  r_{i}^{2}  -  2 r r_{i} \mu}{h_{i}^{2}} \right]^{1/2} \right) \mathrm{d} \mu,
\end{equation}

\noindent where we have substituted $\mu = cos(\theta)$. We have used the 
minus sign of the $\mathrm{d}cos(\theta)$ term to exchange the limits of the integral.

If we write $W_{M4}(s)$ in the form 
\[ W_{M4}(s) = w_{0} + w_{1} s + w_{2} s^{2} + \dots + w_{p} s^{p} + \dots \]
then Eqn. \ref{equa:integral3} takes its final form

\begin{equation}
\label{equa:integral4}
\bar{\rho} (r)  =  6 m_{i} h_{i}^{-3} \int_{\mu = \mu_{max}}^{\mu = 1} \left\{ 
\sum_{p=0}^{p=3} w_{p}  \left( \frac{r^{2}  +  r_{i}^{2}  -  2 r r_{i} 
\mu}{h_{i}^{2}} \right)^{p/2} \right\} \mathrm{d} \mu,
\end{equation}

\noindent with the sum being the polynomial form of the M4-kernel (Eqn. 
\ref{equa:M4}).

\begin{figure} 
\resizebox{7.75cm}{!}{\includegraphics{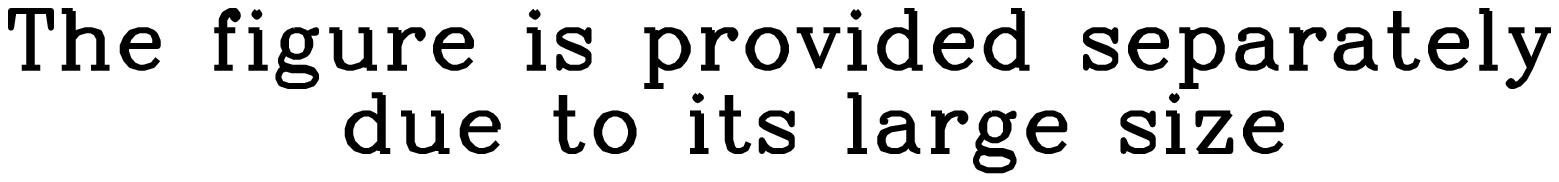}}
\resizebox{7.75cm}{!}{\includegraphics{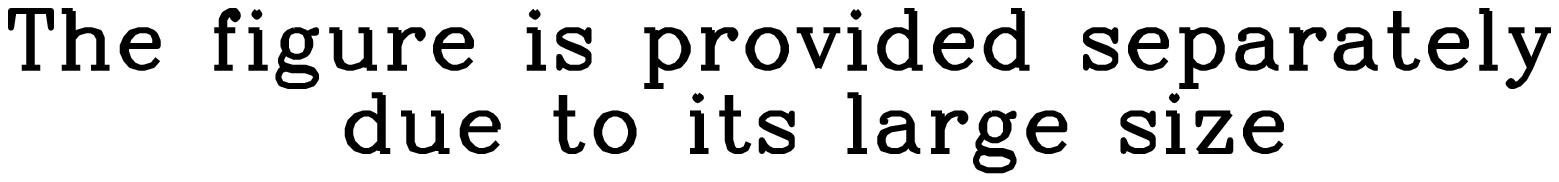}} 
      \caption[Linear density profiles of a coarse particle in isolation and of the ensemble of 13 fine particles that replace it ($r_{i} = 2h_{i}$ \& $r_{i} = 1.9h_{i}$)]{Linear density profiles of a coarse particle in isolation (solid curve) and of the ensemble of 13 fine particles that replace it (dashed curve). The distance of the 12 outside particles from the centre is set to be exactly $2h_{i}$ in the left panel and $1.9h_{i}$ in the right panel, and it is indicated by the vertical lines.}
      \label{fig:profile}
\end{figure}  

The value of $\mu_{max}$ has not been defined yet. There are four different 
cases that we must take into account for the value of $\mu_{max}$, depending on 
the value of $r$. In particular:

\begin{enumerate}

\item If the sphere centred on the central fine particle intercepts the two 
smoothing length sphere of an outside fine particle but not the one smoothing 
length sphere ($h_{i} \leqslant |r_{i} - r| \leqslant 2h_{i}$, Fig. 
\ref{fig:circles}), then $\theta_{max} = 
\theta_{2}$, where \[ \theta_{2} = cos^{-1} \left( \frac{r^{2} + r_{i}^{2} - 
4h_{i}^{2}}{2r r_{i}} \right) \] (Fig. 
\ref{fig:circles}). Therefore, in this case, \[ \mu_{2} = \frac{r^{2} + 
r_{i}^{2} - 4h_{i}^{2}}{2r r_{i}}, \] and for Eqn. \ref{equa:integral4} we 
use the second part of the M4-kernel (Eqn. \ref{equa:M4}) for $1 \leqslant s \leqslant 2$ 
and $\mu_{max} = \mu_{2}$.

\item If the central fine particle lies inside the $2h$ spheres of the outside 
fine particles and outside of the $h$ spheres without intercepting the $h$ spheres 
($h_{i}\leqslant r_{i} - r$ and $r_{i} + r \leqslant 2h_{i}$, Fig. \ref{fig:circles}), 
then $\theta_{max} = \pi$, and for Eqn. \ref{equa:integral4} we 
use the second part of the M4-kernel (Eqn. \ref{equa:M4}) for $1 \leqslant s \leqslant 2$ 
and $\mu_{max} = -1$.

\item If the sphere centred on the central fine particle intercepts both spheres ($|r_{i} - r| \leqslant h_{i}$, Fig. 
\ref{fig:circles}), then the integral of Eqn. \ref{equa:integral4} breaks into 
two parts

\begin{eqnarray}
\label{equa:integral2}
\bar{\rho} (r)  =  6 m_{i} h_{i}^{-3} \left\{ \int_{\mu = \mu_{1}}^{\mu = 1}  \left\{ \sum_{p=0}^{p=3} w_{p}  \left( \frac{r^{2}  +  r_{i}^{2}  -  2 r r_{i} \mu}{h_{i}^{2}} \right)^{p/2} \right\} \mathrm{d} \mu + \right. \nonumber \\ 
\left. \int_{\mu = \mu_{max}}^{\mu = \mu_{1}}  \left\{ \sum_{p=0}^{p=3} w_{p}  \left( \frac{r^{2}  +  r_{i}^{2}  -  2 r r_{i} \mu}{h_{i}^{2}} \right)^{p/2} \right\} \mathrm{d} \mu \right\},
\end{eqnarray}

\noindent where we use \[ \theta_{1} = cos^{-1} \left( \frac{ r^{2} + r_{i}^{2} 
- h_{i}^{2}}{2r r_{i}} \right) \] (Fig. 
\ref{fig:circles}) so that \[\mu_{1} = \frac{ r^{2} + r_{i}^{2} - h_{i}^{2}}{2r 
r_{i}} \] and $\mu_{max} = \mu_{2}$. For Eqn. \ref{equa:integral2} we take 
$w_{p}$ and $p$ from the first part of the M4-kernel (Eqn. \ref{equa:M4} for $0 
\leqslant s \leqslant 1$) for the first sum, and from the second part of the M4-kernel 
(Eqn. \ref{equa:M4} for $1 \leqslant s \leqslant 2$) for the second sum.

\item If the central fine particle lies inside the $2h$ spheres of the outside 
fine particles and outside of the $h$ spheres but it now intercepts the $h$ spheres 
($h_{i} \geqslant r_{i} - r$ and $r_{i} + r \leqslant 2h_{i}$, Fig. 
\ref{fig:circles}) then we use Eqn. \ref{equa:integral2} with $\theta_{max} = 
\pi$, thus $\mu_{max} = -1$.

\end{enumerate} 
 
The  analytical calculation of these two integrals has now 
become trivial as, in every case we have to deal with integration of 
polynomial functions. In all other cases, the integral becomes equal to 0 due to the M4-kernel having compact support.

There are two parameters of the problem which we have not dealt with yet. 
Namely, the density of the central fine particle and the value of $r_{i}$, the 
distance from the central to the outside fine particles. The calculation of the 
former is based on Eqn. \ref{equa:density} with ${\bf r}_{i} = 0$ the position 
of the central fine particle, $m_{i}  =  M/13$ the mass and $h_{i}  =  (H/13)^{1/3}$ 
the smoothing length of the fine particles. The total density 
$\bar{\rho}(r)$ for the ensemble of 13 fine particles is derived when we add 
the density of the central particle to the result of Eqns. \ref{equa:integral4} 
and/or \ref{equa:integral2}. This is illustrated with the dashed curve on the 
left panel of Fig. \ref{fig:profile}, with the outside particles at a distance 
of $r_{i} = 2 h_{i}$. 

$r_{i}$ is a free parameter in this problem and its value can be determined by 
minimising the integral

\begin{equation}
\label{equa:difference}
\int_{r=0}^{r=2H} 4 \pi r^{2} |\rho(r) - \bar{\rho}(r)| dr,
\end{equation}

\begin{figure}  
\resizebox{7.75cm}{!}{\includegraphics{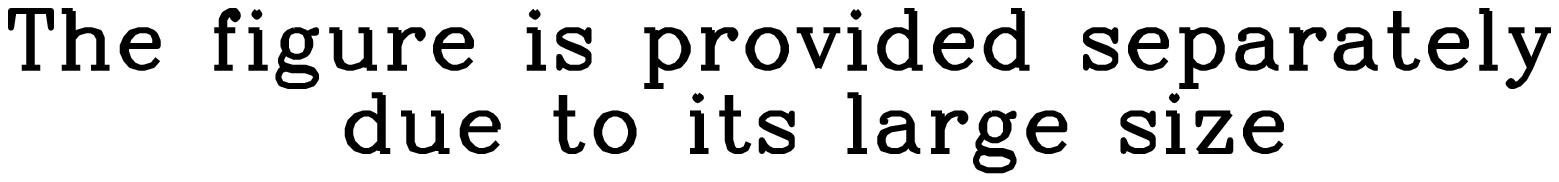}}
\resizebox{7.75cm}{!}{\includegraphics{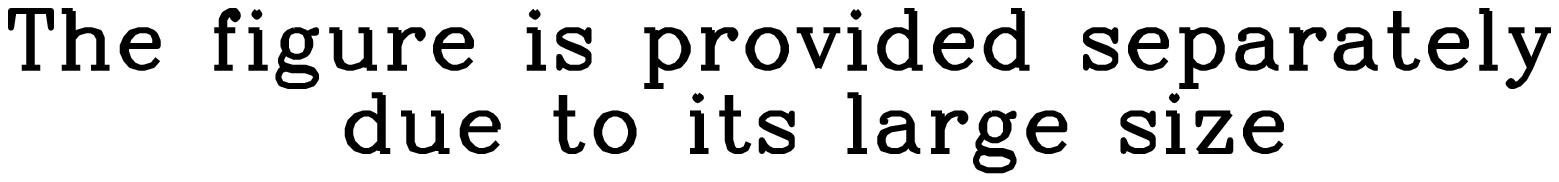}}
      \caption[Gradient of the density profiles of a coarse particle in isolation and of the ensemble of 13 fine particles that replace it ($r_{i} = 2h_{i}$ \& $r_{i} = 1.9h_{i}$)]{Gradient of the density profiles of a coarse particle in isolation (solid curve) and of the ensemble of 13 fine particles that replace it (dashed curve). The distance of the 12 outside particles from the centre is set to be exactly $2h_{i}$ in the left panel and $1.9h_{i}$ in the right panel, and it is indicated by the vertical lines.}
      \label{fig:gradient}
\end{figure}

\noindent where $\rho(r)$ is the density of the coarse particle\footnote{Its 
calculation is similar to that of the central fine particle.} and 
$\bar{\rho}(r)$ is the density of the configuration of 13 fine particles that 
replace it. This integral gives the volume average of the absolute difference in
the density estimates 
given by $\rho(r)$ and $\bar{\rho}(r)$ as defined above. It is calculated 
numerically using the {\it extended trapezoidal rule} \[ \int_{x_{1}}^{x_{N}} f(x) dx = h \left\{ \frac{1}{2} f_{1} + f_{2} + f_{3} + \dots +  f_{N-1} + \frac{1}{2} f_{N} \right\} + O (f^{\prime \prime}), \] 
where $f_{i}$ are the values of the function $f$ evaluated at $x_{i}$ ($i = 1, 2,\dots , N$) 
and $h = x_{i} - x_{i-1}$ ($i = 2, 3,\dots , N$) the constant step 
\cite{PressBOOK1990}.

The minimum of integral \ref{equa:difference} gives $r_{i} = 1.9 h_{i}$ (i.e. $95\%$ 
of $2h_{i}$). The right panel of Fig. \ref{fig:profile} shows $\rho(r)$ and 
$\bar{\rho}(r)$ (solid and dashed curves respectively) when $r_{i} = 1.9 h_{i}$.
The left panel of Fig. \ref{fig:zoom} shows a zoom on this, for $r$ between 0 
and 0.2.

Examination of Fig. \ref{fig:profile} shows that the shape of the density 
profile for the ensemble of fine particles is significantly different compared 
to the density profile of a coarse particle. The dip that appears in the 
former is due to the geometry of the fine particle configuration, i.e. the 
density drops with distance as we move away from the central particle before 
it rises again due to the outside particles. 

\begin{figure}
\resizebox{7.75cm}{!}{\includegraphics{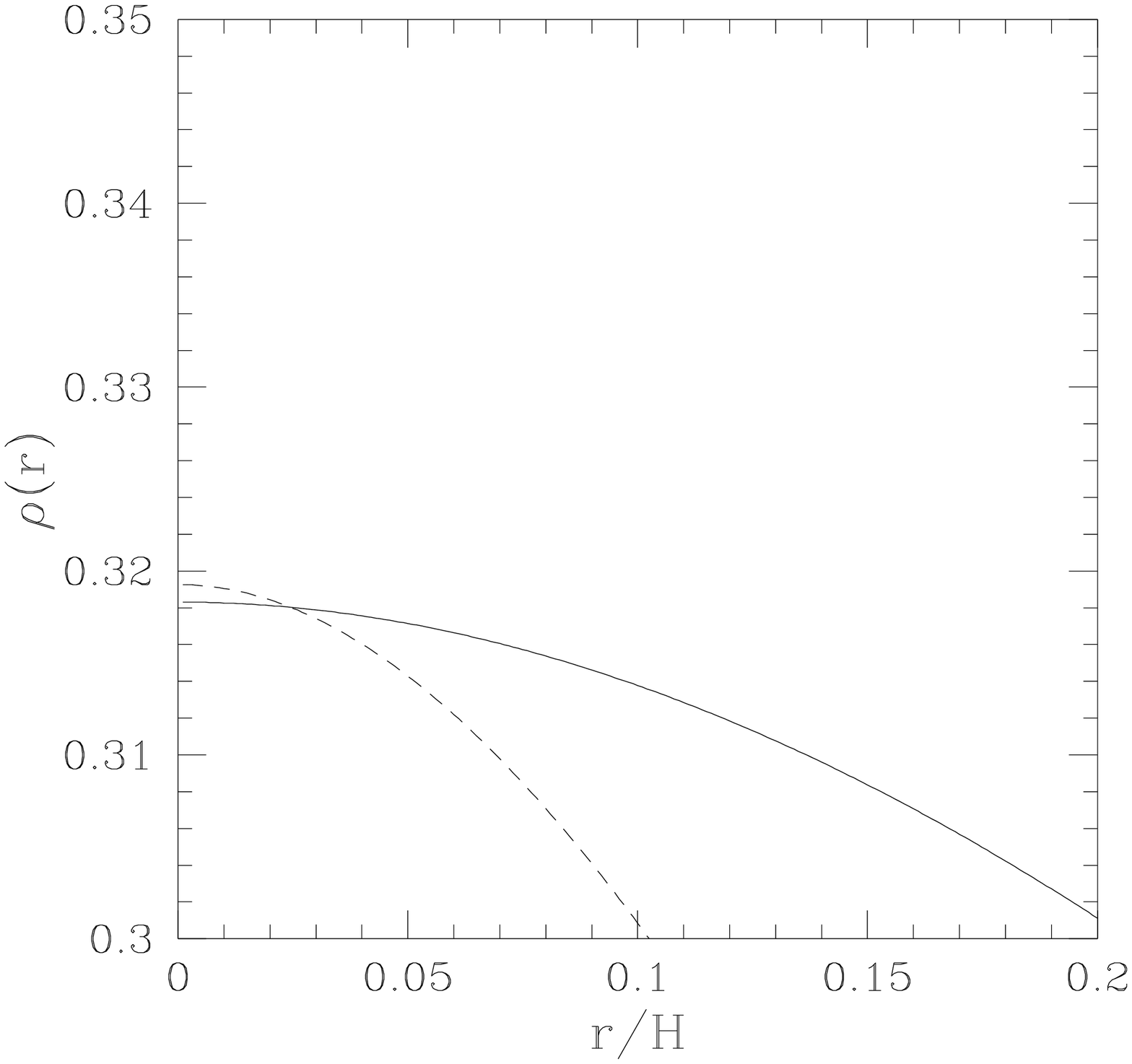}}
\resizebox{7.75cm}{!}{\includegraphics{./figs/size.eps}}  
      \caption[Zoom on the density profiles of a coarse particle in isolation and of the ensemble of 13 fine particles that replace it ($r_{i} = 1.9h_{i}$)\& gradient of the density profiles for $r_{i} = 1.3h_{i}$]{\underline{Left}: Zooming on the right panel of Fig. \ref{fig:profile}. Density profiles of a coarse particle in isolation (solid curve) and of the ensemble of 13 fine particles that replace it (dashed curve). The distance of the 12 outside particles from the centre is set to be $1.9h_{i}$. \underline{Right}: Gradient of the density profiles of a coarse particle in isolation (solid curve) and of the ensemble of 13 fine particles that replace it (dashed curve). The distance of the 12 particles from the centre is set to be $1.3h_{i}$, and it is indicated by the vertical line.}
\label{fig:zoom}
\end{figure}

We have also checked the behaviour of the density gradient. Fig. 
\ref{fig:gradient} shows the density gradient of a coarse particle (solid 
curve) and of the group of fine particles that replace it (dashed curve), 
calculated numerically\footnote{$\left[ \frac{d\rho}{dr} \right]_{i} = (\rho _{i} - \rho _{i-1})/h$, 
where $\rho _{i}$ is the value of the density evaluated at $x_{i}$ ($i = 1, 2,\dots , N$) 
and $h = x_{i} - x_{i-1}$ ($i = 2, 3,\dots , N$) the constant step.} for 
$r_{i} = 2 h_{i}$ and $r_{i} = 1.9 h_{i}$ (on the left and right panel 
respectively). One non-physical property of the M4-kernel is that the density 
gradient vanishes as $r \rightarrow 0$, and therefore if two particles are 
very close the repulsive hydrostatic force between them is very small. This 
means that there will be a length scale below which artificial clustering of 
particles may occur. We would expect the density gradient of the 13 fine 
particles that will replace a coarse particle to behave at least similarly to 
that of their parent coarse particle. However, the density gradient of the 
group of fine particles has two minima (dashed lines of Fig. \ref{fig:gradient})
which is worrying. Firstly, this means that there are going to be two length 
scales between which particles will be pulled towards the centre. Secondly, 
the smaller of these two length scales is a point of stable equilibrium, i.e. 
artificially clustered groups of particles may form and persist. Fortunately, 
it seems that there is an intrinsic geometrical constraint in the process of 
replacing coarse particles by fine ones. The fine particles are always 
positioned outside the outer of the two length scales where the hydrodynamical 
forces are reversed in direction. This means that our fine particles will 
always be positioned at a region where they will experience outward 
hydrodynamical forces. Problems may appear only if they are perturbed inwards 
and end up in the non-physical region. It appears that this does not happen in 
practice. 

\begin{figure} 
\resizebox{7.75cm}{!}{\includegraphics{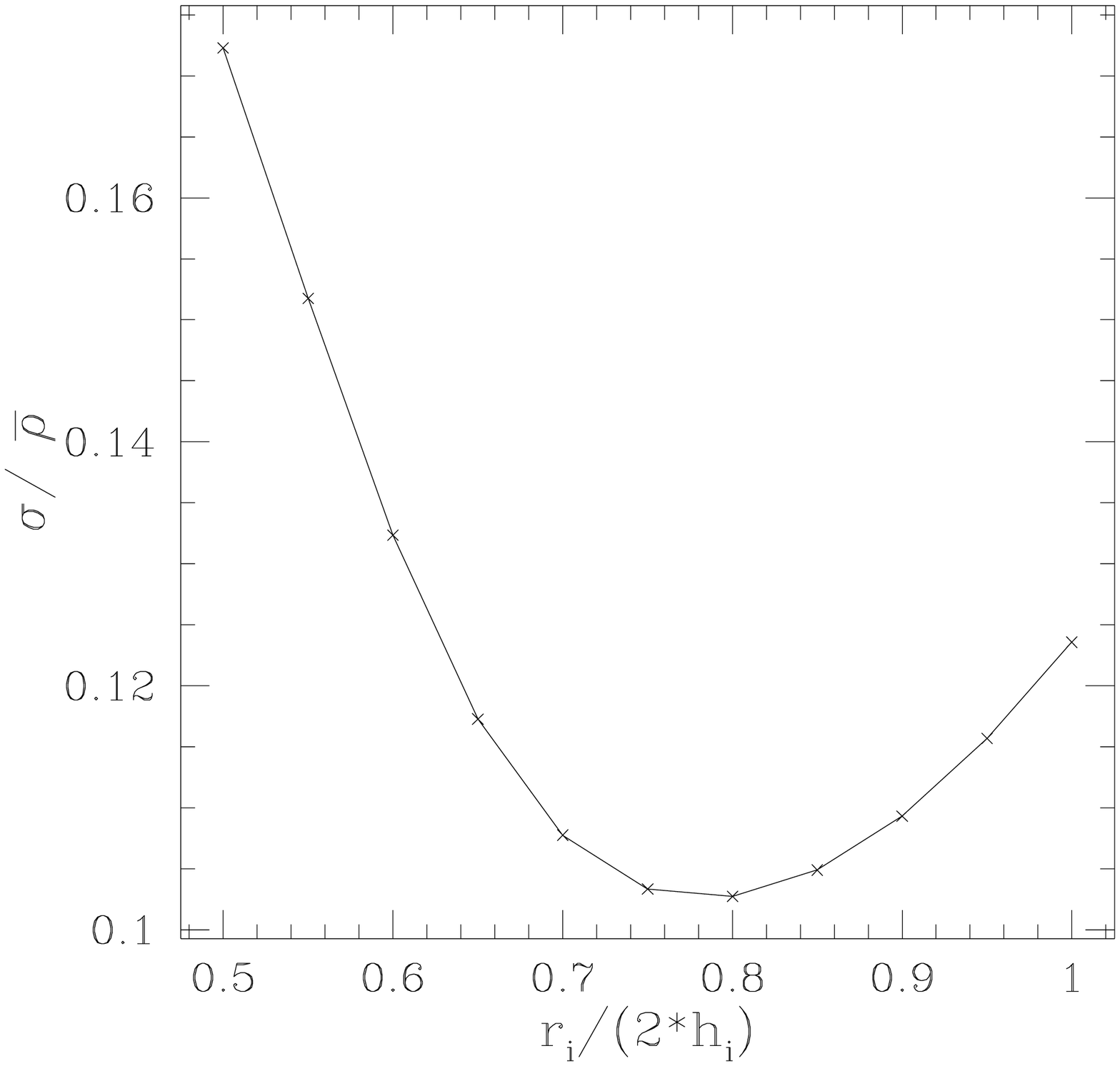}}
\resizebox{7.75cm}{!}{\includegraphics{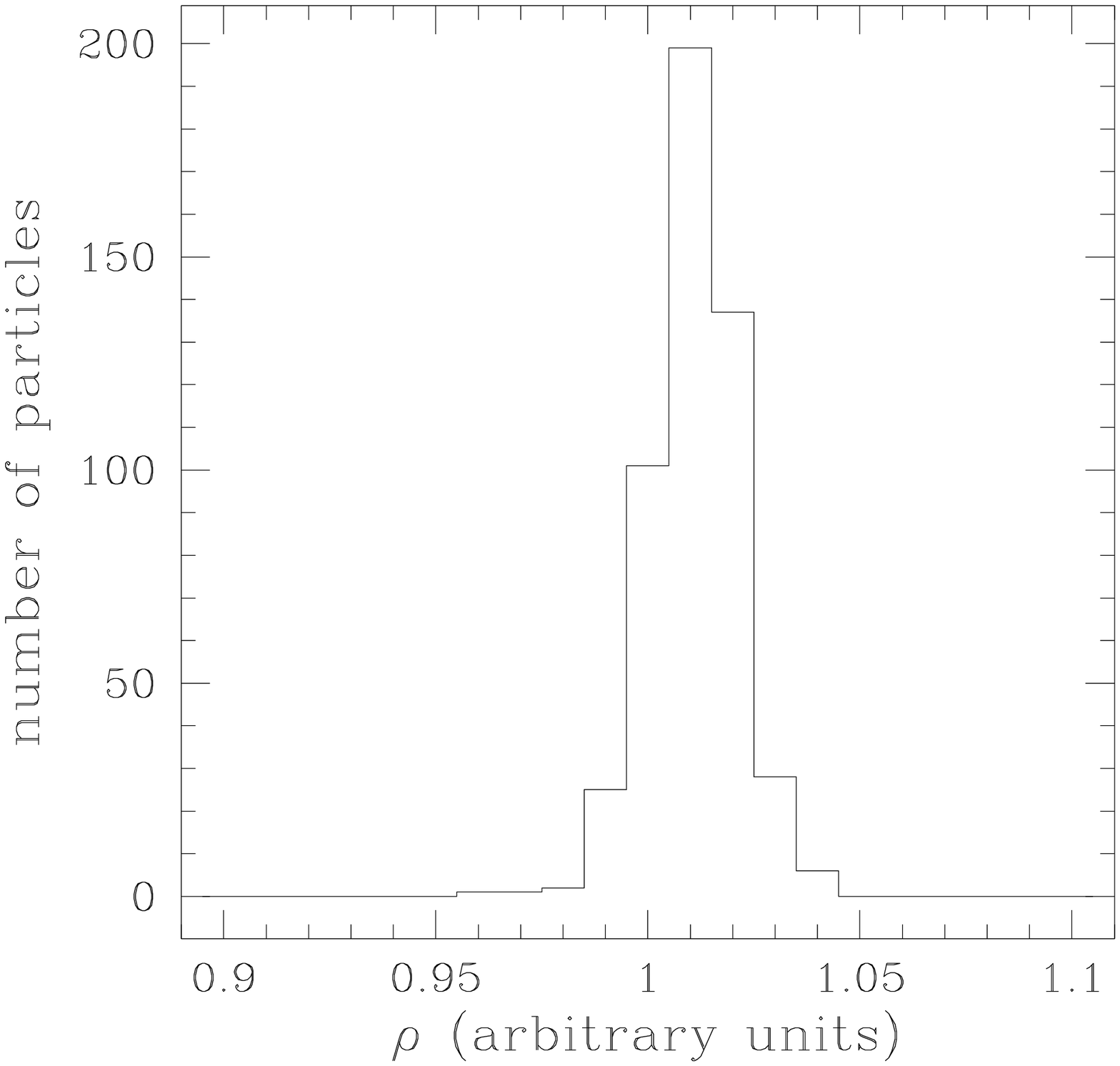}} 
      \caption[$\frac{\sigma}{\bar{\rho}}$ immediately after the splitting {\it vs.} $r_{i}$ \& the density distribution of the initial settled box (before splitting), evaluated on the particle positions]{\underline{Left}: $\frac{\sigma}{\bar{\rho}}$ immediately after the splitting {\it vs.} $r_{i}$. \underline{Right}: the density distribution of the initial settled box (before splitting), evaluated on the particle positions.}
      \label{fig:settle}
\end{figure}

In order to explore if a group of fine particles can produce a density 
gradient that behaves similarly to that of a coarse particle, we have 
conducted another short parameter search for the best value of $r_{i}$. For 
$r_{i} \leqslant 1.3 h_{i}$ (right panel of Fig. \ref{fig:zoom} for $r_{i} = 1.3 h_{i}$) 
the density gradient appears to take a shape similar to that of a coarse 
particle. However, we found that for $r_{i} = 1.3 h_{i}$, integral 
(\ref{equa:difference}) gives a value for the absolute difference in the density 
estimates twice as large as the $r_{i} = 1.9 h_{i}$ does.

\subsection{Density stability with particle splitting}{\label{sec:ptcl_spl}}

\begin{figure}
\resizebox{7.75cm}{!}{\includegraphics{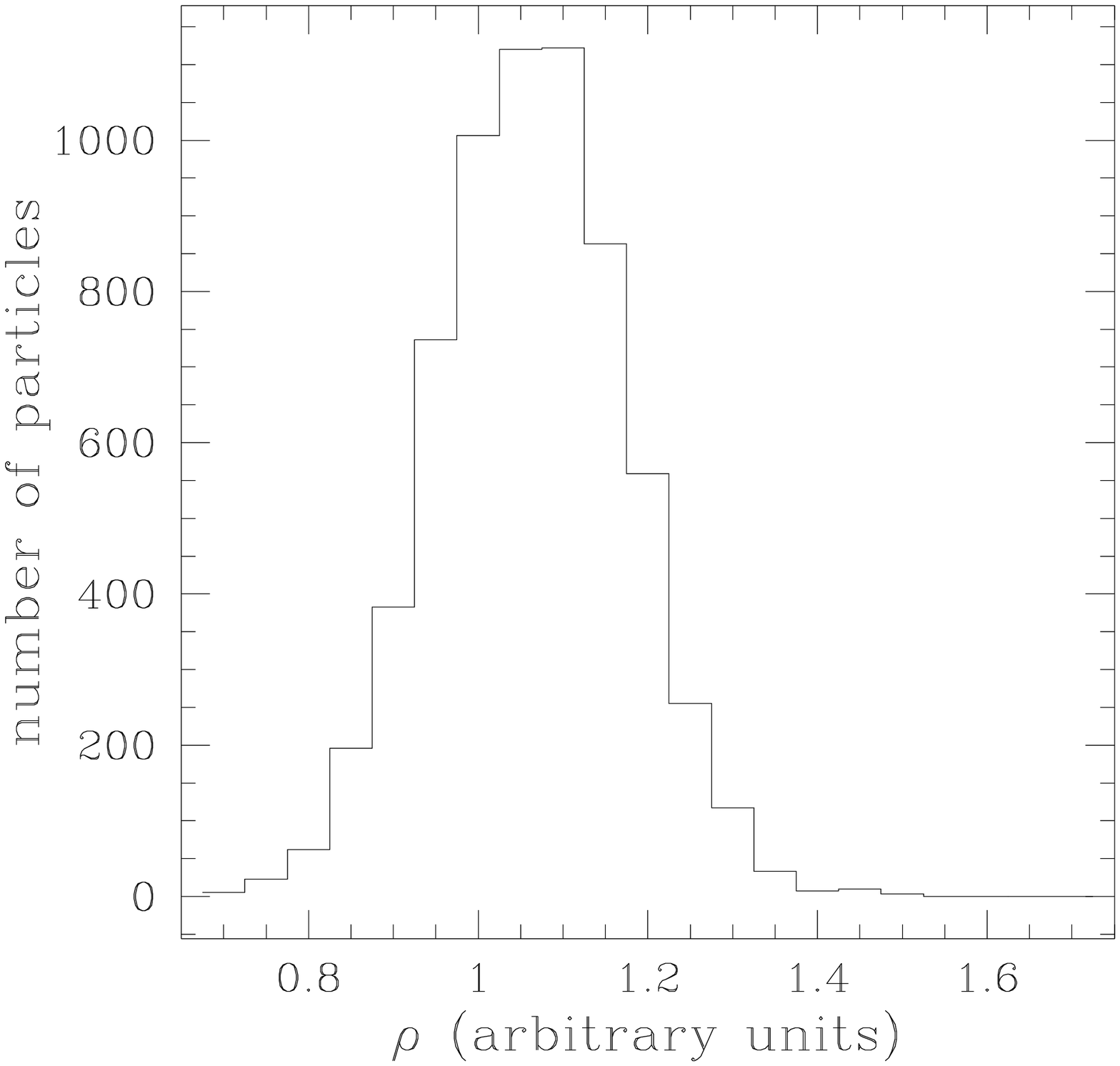}}
\resizebox{7.75cm}{!}{\includegraphics{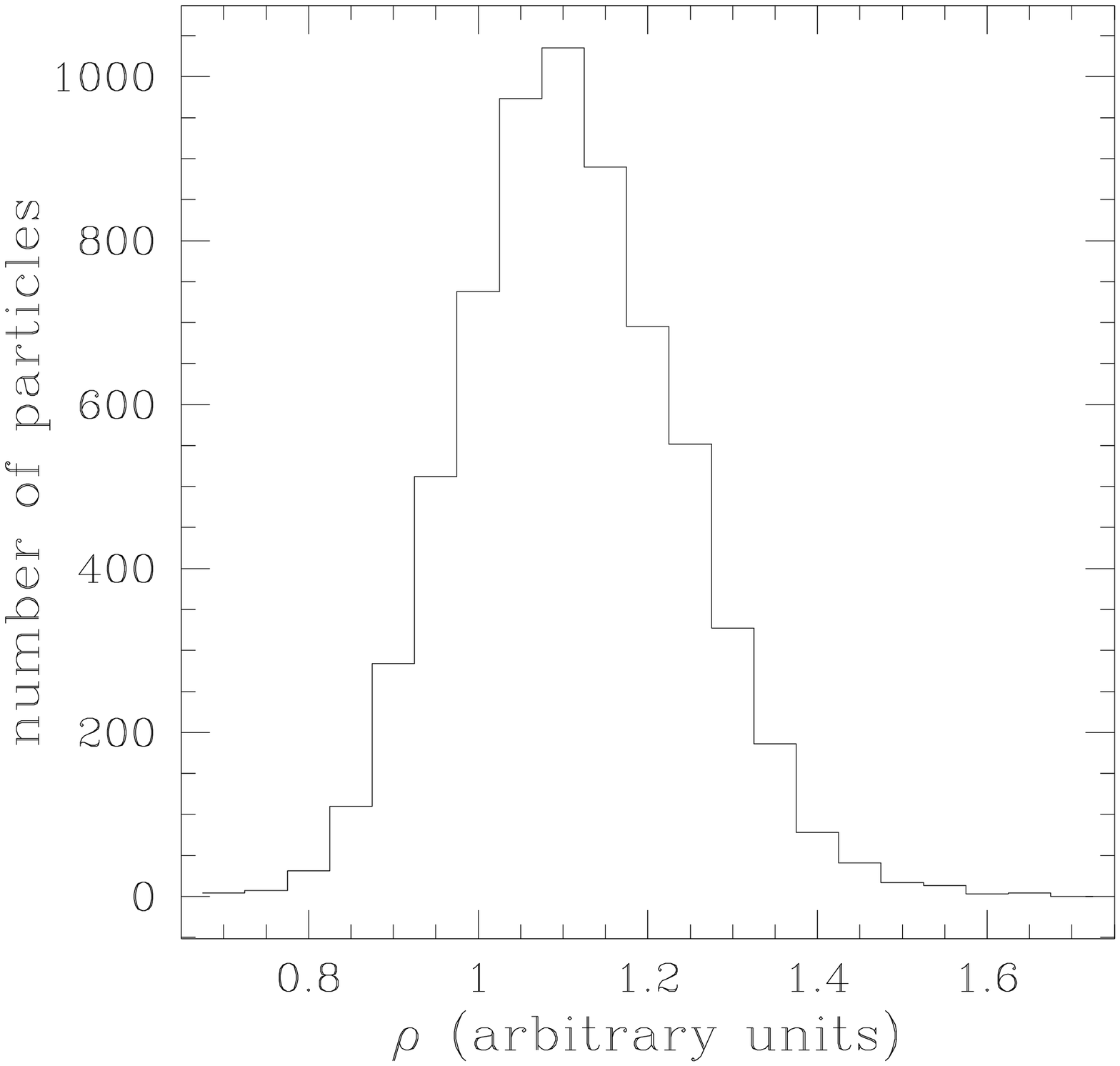}}
      \caption[Density distribution of the box after splitting for $r_{i} = 1.5h_{i}$ \& $r_{i} = 1.9h_{i}$]{Density distribution of the box after splitting for $r_{i} = 1.5h_{i}$ (left panel) and $r_{i} = 1.9h_{i}$ (right panel), evaluated on the particle positions.}
      \label{fig:distribution}
\end{figure}

In this section we demonstrate that particle splitting does not significantly affect an  
existing density distribution of coarse particles. We have produced a uniform 
distribution of 500 particles in a box using periodic boundary conditions. We 
have evaluated the density both on the particle positions and on a square grid. 
Using the method described in \S \ref{sec:positions}, we then replaced each 
particle with 13 fine ones randomly oriented and we tried to reproduce the 
same uniform density allowing for a short settling of the fine particles. We 
have discovered that the value of $r_{i}$, the distance of the 12 outside fine 
particle from the central fine particle, can influence our results 
significantly. 

In particular, the amount of settling required after particle splitting for 
the density distribution to be identical to the initial settled box depended 
on this distance $r_{i}$. We have used the ratio of the variance over the mean 
of the density distribution, $\frac{\sigma}{\bar{\rho}}$, as the quantity that 
determines if our distribution is settled or not. We have settled until 
$\frac{\sigma}{\bar{\rho}} < 0.01$. The left panel of Fig. \ref{fig:settle} 
shows the dependence of $\frac{\sigma}{\bar{\rho}}$ on $r_{i}$ immediately 
after the splitting. The right panel illustrates the density distribution of 
the initial settled box (before splitting). The minimum value of 
$\frac{\sigma}{\bar{\rho}}$ is for $r_{i}$ 
between $1.5h_{i}$ and $1.6h_{i}$, with $1.5h_{i}$ having the absolute minimum 
in both variance and mean of the density. Fig. \ref{fig:distribution} shows 
the density distribution of the box after splitting for $r_{i} = 1.5h_{i}$ 
(left panel) and $r_{i} = 1.9h_{i}$ (right panel). Thus, for $r_{i} = 1.5h_{i}$ 
the least settling is required. This applies for both the evaluation of 
density on particle positions and on the square grid. 

Therefore, we finally 
decided that $r_{i}$ should neither be equal to $1.9h_{i}$ nor $1.3h_{i}$ 
({\it cf}. \S \ref{sec:density}) but $1.5h_{i}$. This is because it is the 
distribution of an ensemble of particles that we are particularly interested 
in, and not the density profile of a single particle. Any 
over-evaluation of the density around the position of the coarse particle 
({\it cf.} Fig. \ref{fig:ri_75} left panel) is expected to be treated 
appropriately by the hydrodynamical forces (dispersal of the ensemble of fine 
particles that replace it). The shape of the density gradient reassures 
us that this will happen (right panel of Fig. \ref{fig:ri_75}).

In \S \ref{sec:rot_cloud} we will discuss the effects of particle splitting on 
a more realistic particle distribution.

\begin{figure}
\resizebox{7.75cm}{!}{\includegraphics{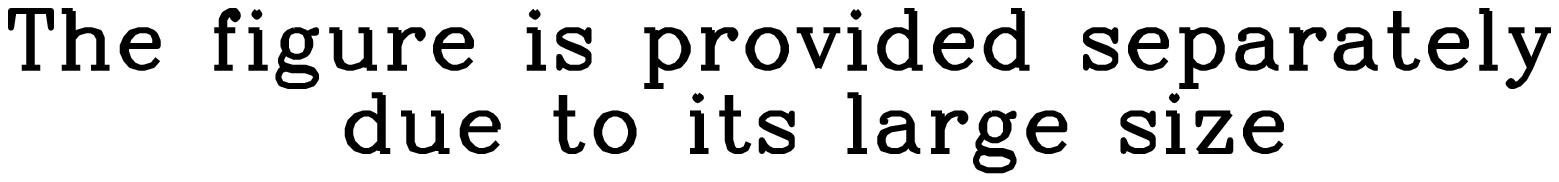}}
\resizebox{7.75cm}{!}{\includegraphics{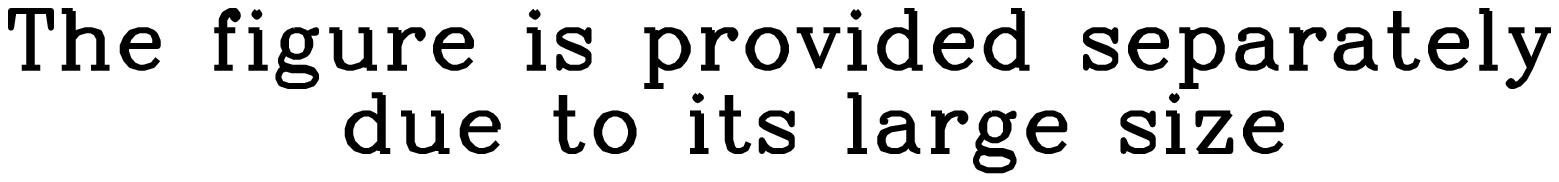}}
      \caption[Density profile and density gradient of the ensemble of 13 fine particles for $r_{i} = 1.5h_{i}$]{Density profile (left) and density gradient (right) of the ensemble of 13 fine particles (dashed lines) that replaced a coarse particle (solid lines). The outside fine particles are at $r_{i} = 1.5h_{i}$ from the centre. The vertical lines indicate the position of $r_{i}$.}
      \label{fig:ri_75}
\end{figure}

\subsection{Smoothing lengths for the fine particles}{\label{sec:newh}}

As shown in Fig. \ref{fig:settle} \& \ref{fig:distribution}, there is an 
increase in the mean density of a uniform distribution of 500 particles after 
particle splitting is applied, even for the case of $r_{i} = 1.5h_{i}$ when 
this increase is minimum (i.e. minimum settling is required for the 
fine distribution to obtain again unit mean density). In particular, for this 
case the mean density is increased by 5.04\%. Furthermore, the 
calculation of the number of neighbours for each fine particle does not always 
give the right result ($\sim 50$).

In order to explore this further we have conducted another similar test. We 
have simulated a stable isothermal sphere in isolation and have applied 
particle splitting (in particular the nested splitting version) to its central 
region (the initial conditions are given in 
\S \ref{sec:stable} and shown in Fig. \ref{fig:stable}). We have found that it 
is not only 
the number of neighbours that gives wrong results, but also the density inside 
and outside the fine sub-region is incorrectly modelled after splitting is 
applied. The reason for this is that there now exist particles of different 
mass in close contact. In particular, the method for finding the smoothing 
length for all particles is based on counting the number of their neighbours 
(\S \ref{sec:smoothing}). This makes fine particles just inside the sub-region 
boundary, which have coarse particles as neighbours, look for some of their 
neighbours in a region of lower resolution. Therefore, their smoothing length 
is over-estimated and their density under-estimated. The coarse particles just 
outside the sub-region boundary have fine particles as neighbours and look for 
their neighbours in a region of higher resolution. Thus, their smoothing 
length is under-estimated and their density over-estimated (top panel of Fig.
\ref{fig:stable_nested_old} {\it cf}. with top panel of Fig. 
\ref{fig:stable}). However, there is 
an even more severe boundary effect disturbing the evolution of the fine simulation. Since the 
sub-region boundary is fixed in space, the fine particles should always be 
inside the boundary. Nevertheless, with the implementation of nested splitting 
fine particles penetrate through the non-moving coarse particles and end up at 
the other side of the boundary (bottom panel of Fig. 
\ref{fig:stable_nested_old} {\it cf}. with bottom panel of Fig. 
\ref{fig:stable}). The gravitational field around the boundary 
fine particles is not balanced as they are in contact with more massive 
particles from one side. This make them move to the other side of the boundary. 
In particular, they move through the gaps between the boundary coarse 
particles, occupying the potential energy gaps between those particles.
This creates an expanding shell of fine particles similar to an artificial 
rarefraction wave which will eventually corrupt the evolution of the 
isothermal sphere even if the sphere may finally evolve to another stable 
equilibrium state.

Therefore, the new method clearly needs a special feature to eliminate these 
boundary effects. A possible solution would be to define a region on either
side of the boundary, where both fine and coarse particles would have their 
smoothing length evolution constrained in order to create a smooth transition 
from the low to the high resolution and vice versa. However, this idea would 
be rather complicated to implement as the position and velocity of particles 
with respect to the boundary would need to be calculated at every time-step.

We have found a simpler method. It involves calculating the smoothing length of 
a particle by specifying the total mass of neighbours, rather than the 
number of neighbours. Specifically, the smoothing length of each particle is 
such that it contains $\sim$50 times its mass. We think this is more 
appropriate in our case where there exists mixing of particles with different 
mass. Away from the boundary, finding $h$ is exactly the same as before: 
there are $\sim 50$ equal mass neighbours. Close to the boundary, though, a coarse 
particle appears as 13 fine ones when $h$ is calculated for a fine particle, 
while it takes 13 fine particles to add one effective neighbour to a coarse particle. 
This method is valid as the volume of a coarse particle ($\propto H^{3}$) is not
significantly altered when the coarse particle is split with $r_{i}=1.5 h_{i}$ 
(\S \ref{sec:density})\footnote{For the first trial value of $h_{i} = (H/13)^{1/3}$, 
the configuration of 13 fine particles has a radius of $\sim 1.06 H$.}. Close 
to  the boundary, coarse particles have more than 50 neighbours and fine particles have 
less than 50, but both their smoothing lengths and densities are 
correctly modelled. In its numerical details, the method is implemented 
exactly the same way as before (\S \ref{sec:smoothing}), but the variable that 
we now use as a criterion for the acceptance of a trial smoothing length is 
the enclosed mass and not the number of enclosed neighbours. The notion of a 
particle always tracing constant mass is maintained and SPH 
with different mass particles preserves its Lagrangian character. The test 
simulation for the evolution of a stable isothermal sphere is repeated (\S 
\ref{sec:stable}) and shows that the new method for calculating $h$ greatly 
improves the treatment of the boundary as well as preventing the growth of 
numerical perturbations that may disturb the evolution of the stable 
isothermal sphere. 

\begin{figure}                  
\begin{center}
\resizebox{9cm}{!}{\includegraphics{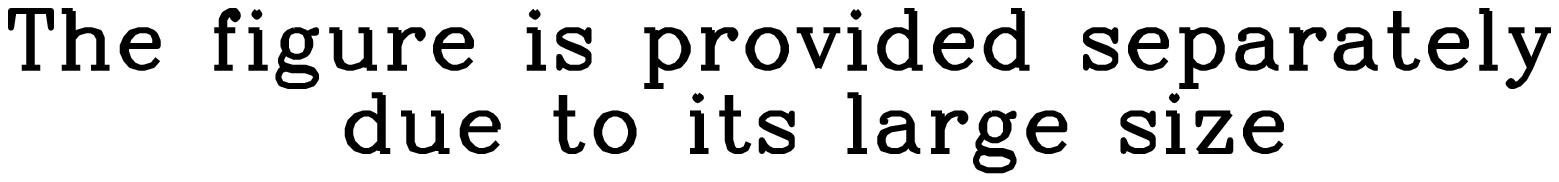}}
\end{center}  
\begin{center}
\resizebox{9cm}{!}{\includegraphics{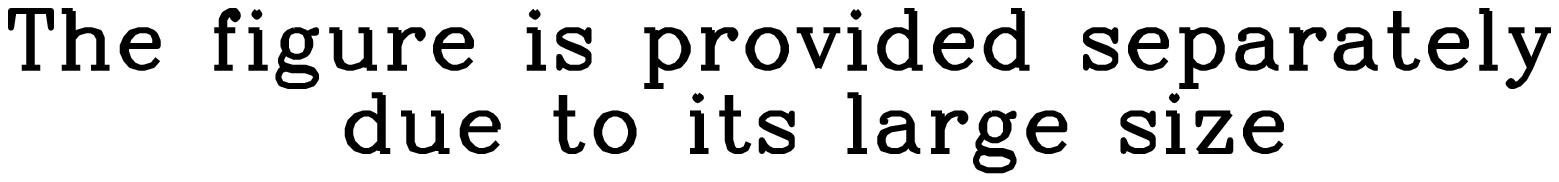}}
\end{center}  
      \caption[Evolution of a stable isothermal sphere when nested splitting 
      was applied, without implementation of the new method for 
      calculating $h$]
      {Evolution of a stable isothermal sphere after $t$ = 9.07 $t_{ff}$, when 
      nested 
      splitting was applied within a radius of 3 x 10$^{-2}$ pc, without 
      implementation of the new 
      method for calculating $h$. The green points show fine 
      particles and the 
      black points coarse particles. \underline{Top}: Radial 
      density profile of the isothermal sphere. The red line 
      indicates the solution of Eqn. \ref{equa:hydrobalancedimeless}. 
      \underline{Bottom}: Thin equatorial slice ($\Delta z = 4 \times 10^{-3}$ 
      pc) of the isothermal sphere showing the boundary between the coarse and 
      the fine particles.}
      \label{fig:stable_nested_old}
\end{figure}

\begin{figure}
\resizebox{15cm}{!}{\includegraphics{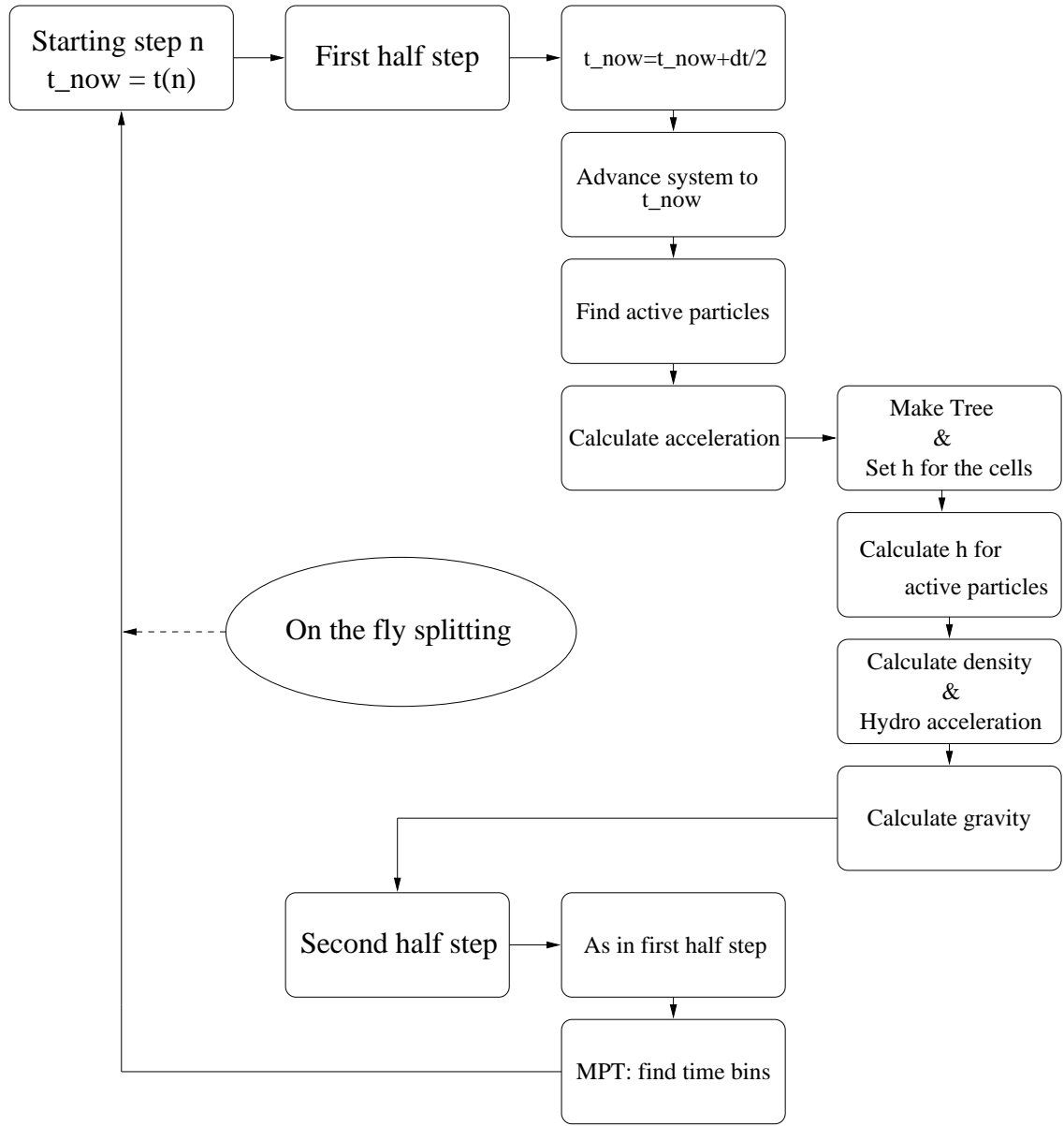}}  
      \caption[The code `step by step' including particle splitting]{The code `step by step': Flow-chart of the algorithm that dictates the evolution of the fluid in time, after the on-the-fly splitting subroutines have been added. It represents the cycle, $n$, of the integration scheme, when the system advances with the time-step in the minimum time bin, $dt = \Delta t_{min}$.}
      \label{fig:step-by-step1}
\end{figure}

\subsection{Velocities for the fine particles}{\label{sec:veloc}}

To evaluate the velocity of each fine particle we use Eqn. \ref{equa:alpha} 

\begin{equation}
\label{equa:alpha1}
{\bf v}({\bf r}_{j}) \; = \; \sum_{i} m_{i} \frac{{\bf v}_{i}}{\rho_{i}}H_{i}^{-3} W_{M4} \left( \frac{|{\bf r}_{j} - {\bf r}_{i}|}{H_{i}} \right),
\end{equation}

\noindent where ${\bf v}_{i}$, ${\bf r}_{i}$ and $H_{i}$ are the velocity, 
position and smoothing length, respectively, of coarse particle $i$, and 
${\bf r}_{j}$ is the position of fine particle $j$. We do not sum contributions 
from coarse particles which are more than $2H_{i}$ away from fine particle $j$ (i.e. 
$\frac{|{\bf r}_{j} - {\bf r}_{i}|}{H_{i}} > 2$). Since each fine particle is 
contained within the $2H$ radius of its parent coarse particle, the sum of Eqn. 
\ref{equa:alpha1} is over the $\sim$50 neighbours of its parent particle. 
Obviously, the largest contribution comes from this parent particle, but Eqn. 
\ref{equa:alpha1} guarantees that the velocity field around each coarse 
particle is accurately passed to the fine particles that replace it.

In all the tests to which particle splitting has been applied (\S \ref{sec:split_tests}), 
the velocity field is accurately represented by the fine particles after 
splitting occurs. The new method also conserves global linear and angular 
momenta. 

\subsection{Updating other fluid properties \& numerical parameters}{\label{sec:fluid}}

In ``nested splitting'', the exact steps we follow in implementing particle splitting are: 

\begin{itemize}

\item We decide on the dimensions of the fine region. We count all the coarse 
particles that are inside this region at time $t_{spl}$ when the method is 
initiated.

\item The fine particles that are produced are put in position by randomly 
rotating the co-ordinates given in Table \ref{tab:fcc} around the y- and z-axes.

\item The velocities of the fine particles are calculated using Eqn. 
\ref{equa:alpha1}. Their mass is 13 times smaller than that of a coarse 
particle. Their $h$ is calculated with the new method described in the 
previous section (\S \ref{sec:newh}) with a first trial value of 
$h_{i}=(H/13)^{1/3}$.

\item The fine particles are given the same temperature as their parent coarse 
particle. Taking into account that splitting is necessary in order for a 
simulation to be properly resolved up to the density where adiabatic heating 
initiates (chapter \ref{sec:SPH}), this assumption of isothermality is valid for the 
fine particles, as all other particles remain isothermal as well. If particle 
splitting were necessary at later stages in regions with temperature gradients, 
temperatures could be calculated by interpolation in the same way as velocities. 

\item All fine particles are put in the minimum time bin to make sure that the 
method sustains its accuracy. They are given the minimum time-step. If the 
multiple time-step method decides that they can be moved to a higher time 
bin, then this will happen as soon as it is allowed (\S \ref{sec:timesteps}).
The extra computational cost of keeping some fine particles in the minimum 
time bin -- even though this may not really be needed -- is minimal. Nonetheless, our 
decision for the initial choice of time-step for the fine particles is 
dictated by the need to obtain the highest possible accuracy. Moving all the 
fine particles and updating their properties during the first time-steps after 
their introduction to the simulation also gives the code a cushion for 
correcting any values that do not match the local fluid properties around the 
fine particles.

\item The fine particles are given a splitting identifier indicating the 
level of splitting they represent, as the method may be applied several times 
in a simulation creating a series of nested simulations each at a higher 
resolution.

\item The total number of particles is updated to contain the fine particles. 
For book keeping purposes only, the central fine particle replaces its parent 
coarse particle in all variable arrays, while all other fine particles are put 
at the end of the arrays.

\item Finally, after the density of all fine particles is calculated, we can 
calculate the acceleration for these particles. For these calculations we use
the usual SPH routines (\S \ref{sec:final}). The fine particles are now ready
to take part in the next cycle of the integration scheme and then the fine 
simulation starts.

\item At the end of all subsequent time-steps, the code checks for any coarse 
particles that may have crossed the sub-region boundary during the previous 
time-step. Then these coarse particles are split on-the-fly, without stopping 
the fine simulation. The code just repeats all the above steps for the coarse
particles that have just entered the fine region. Fig. \ref{fig:step-by-step1} 
shows the flow-chart of the algorithm that dictates the evolution of the fluid 
in time, after the on-the-fly splitting subroutines have been added.

\end{itemize}

In ``on-the-fly splitting'' we don't have to stop the coarse simulation to apply 
particle splitting, but instead we use an automated test in the particle splitting 
subroutine. We calculate initially the density threshold, $\rho_{max}$, above which 
the simulation stops resolving the Jeans mass. $\rho_{max}$ is given by Eqn. 
\ref{equa:Jeans_condition3}. Then we start the coarse simulation and particle 
splitting automatically initiates on-the-fly when particles exceed this density. 

This way, we don't have to apply splitting to particles unnecessarily, just 
because they are in spatial proximity to developing proto-condensations. 
Therefore, we can achieve even greater economy in computational cost, that can 
make the new method even more attractive and efficient. This is our preferred 
version of particle splitting and this is the method we apply to clump-clump collisions 
in the next chapter. In the following section we describe the application of both 
versions of particle splitting to standard test simulations. 

\section{Tests}{\label{sec:split_tests}}

We apply both versions of particle splitting to the central 
region of a collapse simulation, to see if there is propagation of boundary 
effects. We do the same for the central region of a simulation of a stable 
isothermal sphere, to show that the boundary effects mentioned in \S \ref{sec:newh} 
have been eliminated by the new calculation of smoothing lengths. Finally, we 
apply both versions of particle splitting to the collapse simulation of a 
rotating, uniform-density, isothermal cloud with an m=2 perturbation ({\it cf}. 
chapter \ref{sec:rotating}). 
We demonstrate that SPH with particle splitting gives very good results to 
this problem. We show that the Jeans condition for fragmentation provides a 
very strong test for the significance of numerical results. We also show
the efficiency of the new method in terms of computational economy and in 
particular the superiority of ``on-the-fly splitting''.

In quantifying the results of our tests, we need to be able to associate these results only 
with the performance of our numerical code. In order to decrease the numerical noise input by 
the initial distribution of particles, we perform all tests using clouds whose particles are taken 
initially to be on a lattice. In the sequel, we will refer to the particles of such simulations 
as ``particles initially taken on a lattice''. To verify that the results of tests with particles 
initially taken on a lattice are not biased due to some preferred orientation of
the initial lattice, we will also perform one simulation for each test where particles are taken initially 
from a ``settled'' distribution. Such a distribution of particles is produced when the particles 
are taken in random positions and then they are relaxed to uniform density, using the SPH 
formulation described in Whitworth {\it et al}. \shortcite{WhitworthAnA1995}.

\subsection{Collapse Simulation}{\label{sec:ff_collapse}} 

We test both versions of particle splitting at the centre of a collapse 
simulation. This way we test the method under conditions of homogeneous inflow. 
Both implementations show good results. There is very little dispersion of fine 
particles out of the fine region and there is a well-defined boundary between 
the fine and the coarse regions.

The initial conditions consist of a spherical cloud of mass M = 1 M$_{\odot}$ and 
radius R = 0.016 pc. The cloud has uniform density $\rho$ = 3.74 x 10$^{-18}$ 
g cm$^{-3}$ and uniform temperature T = 7.9 K. The ratio of thermal to 
gravitational energy is $\alpha$ = 0.26 and the cloud has a Jeans mass of 0.2 
M$_{\odot}$, so that it is unstable and it collapses. There is no rotation. We 
have used 10,185 particles to simulate the cloud, initially taken on a lattice. 
For this test we use our complete self-gravitating SPH code (chapter \ref{sec:SPH}). 
We also include adiabatic heating to slow down collapse. Adiabatic 
heating initiates at $\rho_{0}$ = 10$^{-13}$ g cm$^{-3}$. It is included in 
order to prevent small time-steps from occurring. Adiabatic heating is necessary at the very centre of the simulation, 
where matter accretes to form a central object. Outside the centre, isothermal 
collapse occurs. In this region, the collapsing gas evolves with a self-similar 
form, $\rho \propto r^{-2}$, as predicted by Bodenheimer \& Sweigart 
\shortcite{BodenheimerApJ1968}. 

Nevertheless, the time-step does get very small 
after some time. To prevent this, we have repeated the simulations using a sink 
particle to simulate the core of the accreted central object. With this, we 
allow accretion of more matter to the central object and we finally end up at 
a state where most coarse particles have entered the fine region and have been 
split.

A sink particle is a particle that accretes all matter that enters its radius. 
In this case the sink particles have a radius of 2 x 10$^{-4}$ pc. There is only 
one sink particle at the centre of each simulation. The particles that lie 
inside this radius stop being active (i.e. their properties are not followed 
any further). The accreted particles have their mass decreased to 
zero and their mass is added to the mass of the sink. Therefore, the region 
just outside the sink is not very well evolved, as the active particles just outside the 
sink radius do not ``see'' the accreted particles inside the sink radius. 
As a result, these active particles look for neighbours only outside the sink. 
Their smoothing length is over-estimated and their densities under-estimated.

We stop all simulations when the time-step becomes less than 2 x 10$^{-6}$ t$_{ff}$, so that it 
is computationally inefficient to continue (i.e. in order for time to progress by 10$^{-2}$ t$_{ff}$ we 
would need 5 times the run-time up to that point).

We present five different simulations: collapse using nested splitting without a sink, 
collapse using nested splitting with a sink, collapse using on-the-fly splitting without a sink, 
collapse using on-the-fly splitting with a sink, and collapse using nested splitting with particles 
initially taken from a settled distribution.

\begin{figure}
\begin{center}
\resizebox{9cm}{!}{\includegraphics{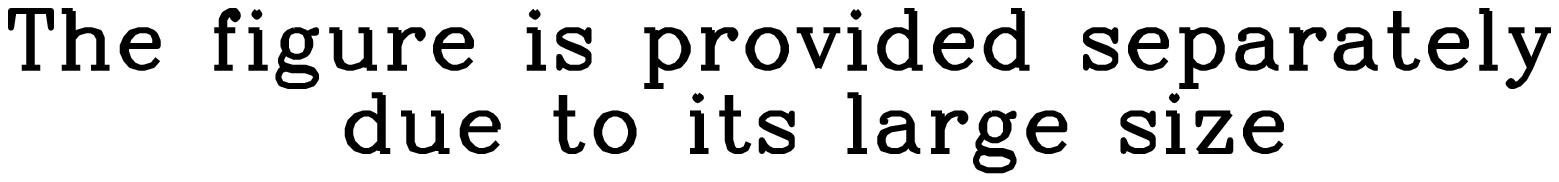}}
\end{center}  
\begin{center}
\resizebox{9cm}{!}{\includegraphics{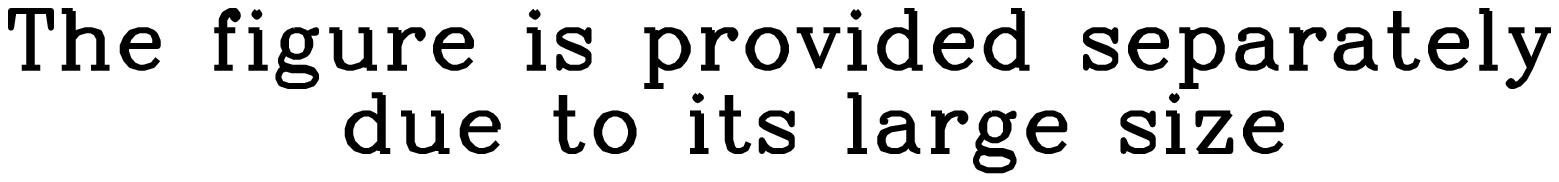}}
\end{center}  
      \caption[Radial density profile of a collapse simulation when nested 
      splitting was applied]
      {Radial density profile of a collapse simulation when nested splitting 
      was applied within a radius of 2 x 10$^{-3}$ pc. The green points 
      show fine particles and the black points coarse particles. The red line 
      indicates the $\rho \propto r^{-2}$ profile. 
      \underline{Top}: Radial density profile after $t$ = 1.01 $t_{ff}$ for a 
      collapse simulation without a sink. \underline{Bottom}: Radial 
      density profile after $t$ = 1.05 $t_{ff}$ for a collapse simulation with 
      a sink of radius 2 x 10$^{-4}$ pc simulating the accreted central 
      object.}
      \label{fig:collapse_nest}
\end{figure}

\begin{figure}
\begin{center}
\resizebox{9cm}{!}{\includegraphics{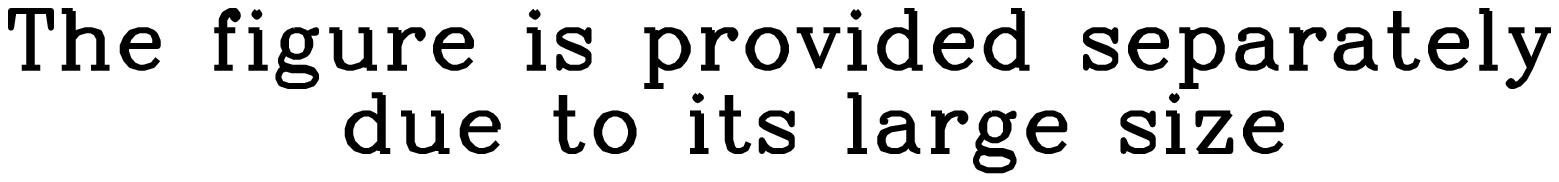}}
\end{center}  
\begin{center}
\resizebox{9cm}{!}{\includegraphics{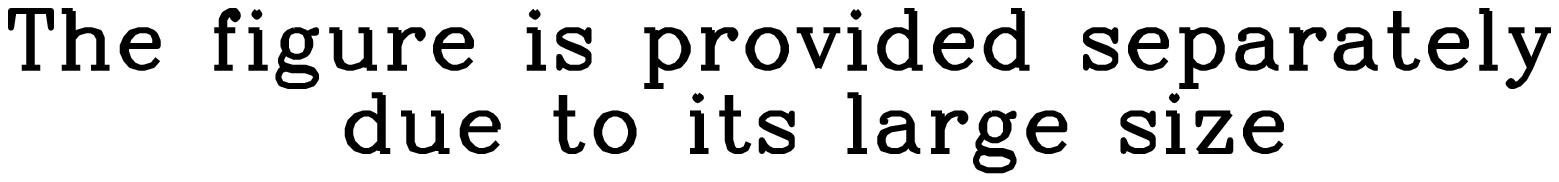}}
\end{center}  
      \caption[Radial density profile of a collapse simulation when on-the-fly 
      splitting was applied]
      {Radial density profile of a collapse simulation when on-the-fly splitting
      was applied. The density threshold is taken to be $\rho_{max}$ =  8.55 x 
      10$^{-16}$ g cm$^{-3}$. The green points show fine particles and the black
      points coarse particles. The red line indicates the $\rho \propto r^{-2}$ 
      profile. \underline{Top}: Radial density profile after $t$ = 1.01 
      $t_{ff}$ for a collapse simulation without a sink. \underline{Bottom}: 
      Radial density profile after $t$ = 1.05 
      $t_{ff}$ for a collapse simulation with a sink of radius 2 x 10$^{-4}$ pc 
      simulating the accreted central object.}
      \label{fig:collapse_fly}
\end{figure}

\subsubsection{4.4.1.1. Isothermal collapse using nested splitting without a sink}{\label{sec:collapse_nest}}

The top panel of Fig. \ref{fig:collapse_nest} shows the radial density profile 
of the sphere when nested splitting was applied with the fine region having a 
radius of 2 x 10$^{-3}$ pc. The green points show fine particles and the black 
points coarse particles. The red line indicates the $\rho \propto r^{-2}$ 
profile. The top panel of Fig. \ref{fig:collapse_nest} shows the end of the 
simulation at time $t$ = 1.01 $t_{ff}$. For about 2 orders of magnitude in 
radius, we find that density complies well to the predicted profile, 
$\rho \propto r^{-2}$. Inside 3 x 10$^{-5}$ pc the density increases rapidly 
due to the formation of a central core. The cloud has contracted to a radius 
of 6.3 x 10$^{-3}$ pc. The maximum density at the centre of the cloud is 
$\rho_{peak}$ = 4.8 x 10$^{-11}$ g cm$^{-3}$. There are 69,861 particles in 
total, with 4,973 coarse particles having been split. Note that there are two 
boundary effects. Some fine particles have their density over-estimated 
immediately after their parent coarse particle is split (i.e. their first 
smoothing radius is smaller than it should be). Some other particles are 
attracted by the heavier coarse particles and temporarily exit the fine region. 
Both these effects are transient, as particle identification has proven that, 
although the effects are static with time, they are produced by different 
(new) fine particles at each time-step. This test shows that the boundary 
effects due to nested splitting are not significant as they do not alter the 
predicted density profile. SPH with the new method for calculating $h$ (\S 
\ref{sec:newh}) eventually remove any fluctuations input by the application of 
nested splitting.

\subsubsection{4.4.1.2. Isothermal collapse using nested splitting with a sink}

The bottom panel of Fig. \ref{fig:collapse_nest} shows the radial density 
profile of the sphere when nested splitting was applied with the same radius 
for the fine region, but now with a sink particle having radius 2.0 x 10$^{-4}$ pc, 
simulating the central 
accreted object. It is the end of the simulation at time $t$ = 1.05 $t_{ff}$. 
The radius of the sphere is 5.4 x 10$^{-3}$ pc. Accretion to the 
centre of the cloud is not modelled properly due to the sink. The time-step does not 
decrease so rapidly and at the end of the simulation more matter 
has entered the fine region. In particular, there are 99,837 particles in total, with 7,471 
coarse particles having been split. The radial density profile obeys the $\rho 
\propto r^{-2}$ profile for one order of magnitude. A small density under-estimation is evident 
outside the sink radius due to the effects described at the end of \S 
\ref{sec:ff_collapse}. This test extends the conclusion that the boundary 
effects due to nested splitting are not significant in a case where most 
coarse particles ($\sim 75\%$) are split.

\subsubsection{4.4.1.3. Isothermal collapse using on-the-fly splitting without a sink}{\label{sec:collapse_fly}}

The top panel of Fig. \ref{fig:collapse_fly} shows the radial density profile 
of the sphere when on-the-fly splitting was applied. The density threshold is 
taken to be $\rho_{max}$ = 8.55 x 10$^{-16}$ g cm$^{-3}$. The green points 
show fine particles and the black points coarse particles. The red line 
indicates the $\rho \propto r^{-2}$ profile. The top panel of Fig. 
\ref{fig:collapse_fly} shows the end of the simulation at time $t$ = 1.01 
$t_{ff}$. For about 2 orders of magnitude in radius, we find that density 
complies well to the predicted profile, $\rho \propto r^{-2}$. Inside 3 x 
10$^{-5}$ pc the density increases rapidly due to the formation of the central 
core. The density profile compares very well with the density profile of the 
nested splitting simulation (top panel of Fig. \ref{fig:collapse_nest}). The 
fine region has smaller radius than in the nested splitting simulation, but 
splitting happens again within the isothermal region (i.e. along the $\rho 
\propto r^{-2}$ line), so that comparison between the two methods is 
legitimate. Again, the cloud has contracted to a radius of 6.3 x 10$^{-3}$ pc, 
and the maximum density at the centre of the cloud is $\rho_{peak}$ = 4.8 x 
10$^{-11}$ g cm$^{-3}$. In the on-the-fly splitting simulation there are only 
46,917 particles in total, with  3,061 coarse particles having been split. Therefore, 
on-the-fly splitting, involving $\sim$20,000 particles less than nested 
splitting, advanced to the same state as nested splitting a lot faster, 
requiring less computational effort (the required memory and the size of the 
output files are proportional to the number of particles). Note that the two 
boundary effects are still present. Some fine particles have their density 
over-estimated immediately after their parent coarse particle is split (i.e. 
their first smoothing radius is smaller than it should be). Some other 
particles are attracted by the heavier coarse particles and temporarily exit 
the fine region moving to a region of lower resolution. Both these effects are 
again transient, as particle identification has proven that, although the 
effects are static with time, they are produced by different (new) fine 
particles at each time-step. This test shows that the boundary effects due to 
on-the-fly splitting are not significant as they do not alter the predicted 
density profile. SPH with the new method for calculating $h$ (\S \ref{sec:newh}) 
eventually removes any fluctuations input by the application of on-the-fly 
splitting. Comparison with nested splitting demonstrates that on-the-fly 
splitting produces the same results as nested splitting, much more efficiently.

\subsubsection{4.4.1.4. Isothermal collapse using on-the-fly splitting with a sink}

The bottom panel of Fig. \ref{fig:collapse_fly} shows the radial density 
profile of the sphere when on-the-fly splitting was applied with the same 
density threshold, but now with a sink particle having radius 2.0 x 10$^{-4}$ pc, 
simulating the central 
accreted object. It is the end of the simulation at time $t$ = 1.05 $t_{ff}$. 
The radius of the sphere is 5.4 x 10$^{-3}$ pc. Accretion to the 
centre of the cloud is not modelled properly due to the sink. The time-step does not 
decrease so rapidly and at the end of the simulation more matter 
has entered the fine region. In particular, there are 86,817 particles in total, with 6,386 
coarse particles having been split. The radial density profile obeys the $\rho 
\propto r^{-2}$ profile for one order of magnitude. A small density under-estimation is evident 
outside the sink radius due to the effects described at the end of \S 
\ref{sec:ff_collapse}. The density profile of this test compares very well 
with that of the corresponding test of nesting splitting (i.e. nested splitting with a sink particle). 
This test extends the conclusion that the boundary effects due to on-the-fly 
splitting are not significant in a case where most coarse particles ($\sim 65\%$) 
are split.

\begin{figure}
\begin{center}
\resizebox{9cm}{!}{\includegraphics{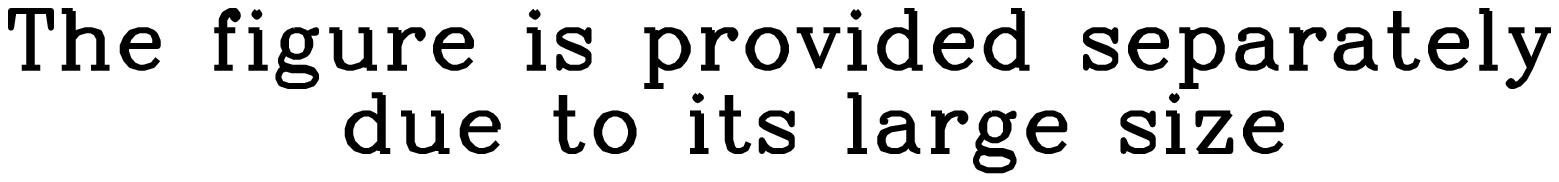}}
\end{center}
      \caption[Radial density profile of a collapse simulation when nested 
      splitting was applied with 
      particles initially taken from a settled distribution]
      {Radial density profile of a collapse simulation when nested 
      splitting was applied within a radius of 2 x 10$^{-3}$ pc, with 
      particles initially taken from a settled distribution. 
      The green points show fine particles and the black points coarse 
      particles. The red line indicates the $\rho \propto r^{-2}$ profile.
      This is the end of the simulation at $t$ = 1.01 $t_{ff}$.}
      \label{fig:collapse_nest_settle}
\end{figure}

\begin{figure}
\begin{center}
\resizebox{9cm}{!}{\includegraphics{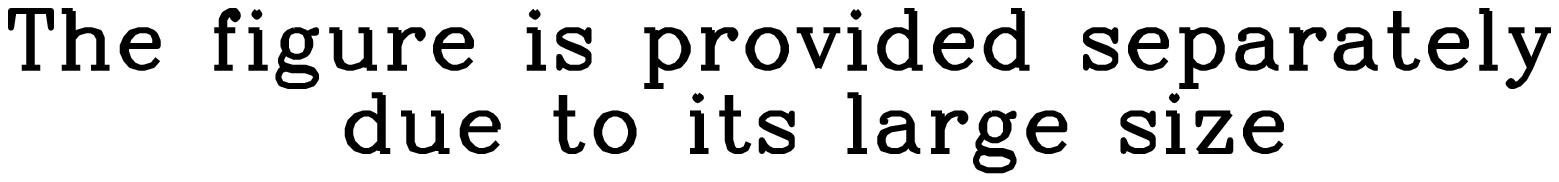}}
\end{center}  
\begin{center}
\resizebox{9cm}{!}{\includegraphics{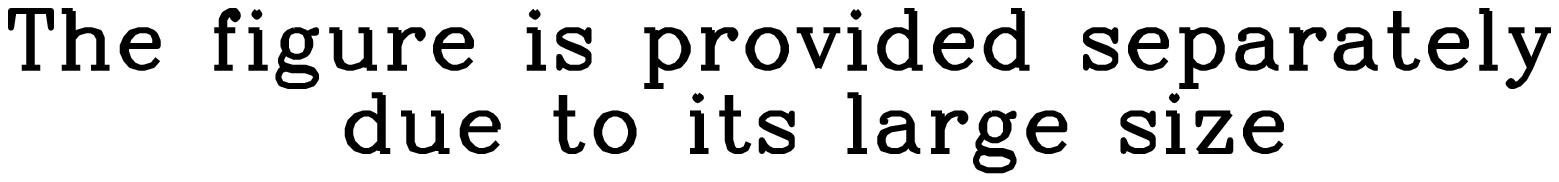}}
\end{center}  
      \caption[Initial state of a stable isothermal sphere]
      {Initial state of a stable isothermal sphere. 
      \underline{Top}: Radial density profile of the isothermal sphere. The red 
      line indicates the solution of Eqn. \ref{equa:hydrobalancedimeless}. 
      \underline{Bottom}: Thin equatorial slice ($\Delta z = 4 \times 10^{-3}$ 
     pc) of the isothermal sphere showing the boundary between the coarse (black
      points) and the fine (green points) particles, immediately after nested 
      splitting was applied within a radius of 3 x 10$^{-2}$ pc.}
      \label{fig:stable}
\end{figure}

\subsubsection{4.4.1.5. Isothermal collapse using nested splitting with particles initially taken from a settled distribution}

We have repeated the collapse simulation applying nested splitting to the 
centre of a cloud of 10,482 particles initially taken from a settled distribution with a uniform 
density of $\rho$ = 3.74 x 10$^{-18}$ g cm$^{-3}$. We have used the same 
initial conditions. We have simulated collapse with the same code as above 
(\S \ref{sec:collapse_nest}.1). 
The fine region has the same radius of 2 x 10$^{-3}$ pc. There is no sink 
particle. Fig. \ref{fig:collapse_nest_settle} shows the 
radial density profile 
of the cloud at $t$ = 1.01 $t_{ff}$. The density complies well to the 
predicted profile, $\rho \propto r^{-2}$. It compares very well with the 
simulation of \S \ref{sec:collapse_nest}.1, when particles were initially 
taken on a lattice (top panel of Fig. \ref{fig:collapse_nest}). Note that in Fig. 
\ref{fig:collapse_nest_settle} there is more noise that in the
top panel of Fig. \ref{fig:collapse_nest}. 
The cloud has contracted to a radius of 6.5 x 
10$^{-3}$ pc. The maximum density at the centre of the cloud is $\rho_{peak}$ 
= 1.8 x 10$^{-10}$ g cm$^{-3}$. There are 74,394 particles in total, with 
5,326 coarse particles having been split. This test rules out the possibility that 
the results presented above are biased due to some preferred orientation of the 
initial lattice.

\subsection{Stable Isothermal Sphere}{\label{sec:stable}} 

We test particle splitting on a simulation of a stable isothermal sphere. Both
methods show again good results. The new calculation of $h$ (\S \ref{sec:newh}) 
has clearly eliminated previous inefficiencies in treating the fine region 
boundary.
Fine particles still move towards the gaps between coarse particles at the 
boundary, but there is little penetration through the boundary. In particular, 
as the fine region slowly expands, coarse particles trapped in the expanding
fine region are automatically split. Any fluctuations input by the application of 
particle splitting are being removed and the sphere settles towards the predicted density 
profile after a few free-fall times.

The initial conditions consist of a spherical cloud of mass M = 1 M$_{\odot}$ and 
radius R = 0.1 pc. The cloud has central density $\rho_{peak}$ = 3.04 x 
10$^{-20}$ g cm$^{-3}$ and uniform temperature T = 7.9 K. The density at the 
outer edge of the cloud is $\rho_{edge}$ = 1.05 x 10$^{-20}$ g cm$^{-3}$. The 
cloud is stable as it has $\Xi$ = 3 (\S \ref{sec:test_stable}). There is no 
rotation. We have used 10,185 particles to simulate the cloud, initially taken 
on a lattice. Their positions are calculated according to the scheme given in 
\S \ref{sec:test_stable}. For this test we use our complete self-gravitating 
SPH code (chapter \ref{sec:SPH}). The top panel of Fig. \ref{fig:stable} shows 
the initial radial density profile of the cloud. We use crosses as the particles
initially are on a lattice and the radial profile therefore consists of very few 
different points. The red line indicates the solution of Eqn. 
\ref{equa:hydrobalancedimeless}. The bottom panel of Fig. \ref{fig:stable} is a 
thin equatorial slice ($\Delta z = 4 \times 10^{-3}$ 
pc) of the isothermal sphere showing the boundary between the coarse (black 
points) and the fine (green points) particles, immediately after nested 
splitting was applied within a radius of 3 x 10$^{-2}$ pc.

We present three different simulations for the evolution of a stable isothermal sphere: 
using nested splitting, using on-the-fly splitting and using nested splitting with particles 
initially taken from a settled distribution.

\subsubsection{4.4.2.1. Stable isothermal sphere using nested splitting}{\label{sec:stable_nested}}

The top panel of Fig. \ref{fig:stable_nested} shows the radial density profile 
of the isothermal sphere after $t$ = 8.17 $t_{ff}$, when nested splitting was 
applied with the fine region having a radius of 3 x 10$^{-2}$ pc. The 
green points show fine particles and the black points coarse particles. The red 
line indicates the solution of Eqn. \ref{equa:hydrobalancedimeless}. The 
density profile compares well with the initial one (top panel of Fig. 
\ref{fig:stable}). There are 22,833 particles in total with 1,054 coarse 
particles having been split. There are still some particles on either side of the 
boundary between the coarse and the fine regions having their density 
incorrectly modelled. These are basically new fine particles (split a few 
time-steps before) that have taken inappropriate initial values for their 
smoothing length. Their smoothing length and density will very soon settle to the expected values. 
However, the number of these particles is significantly 
reduced due to the application of the new method for calculating $h$. Moreover, 
the deviation from the expected density profile for these particles is greatly 
reduced ({\it cf}. with the top panel of Fig. \ref{fig:stable_nested_old}). The 
introduction of new fine particles into the simulation creates the opposite 
effect to neighbouring coarse particles: their smoothing length is 
over-estimated for a few time-steps and thus their density is under-estimated.

\begin{figure}
\begin{center}
\resizebox{9cm}{!}{\includegraphics{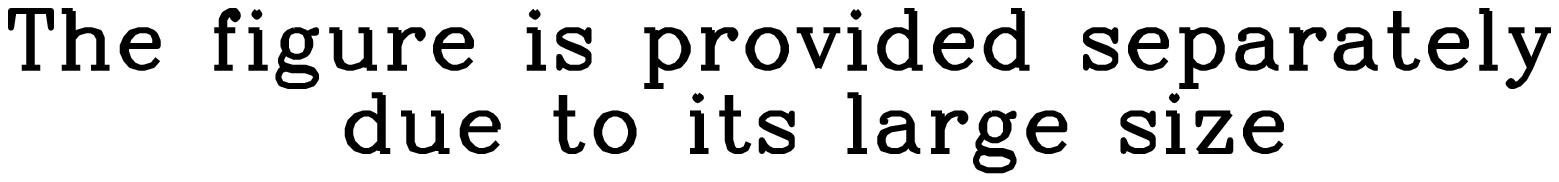}}
\end{center}  
\begin{center}
\resizebox{9cm}{!}{\includegraphics{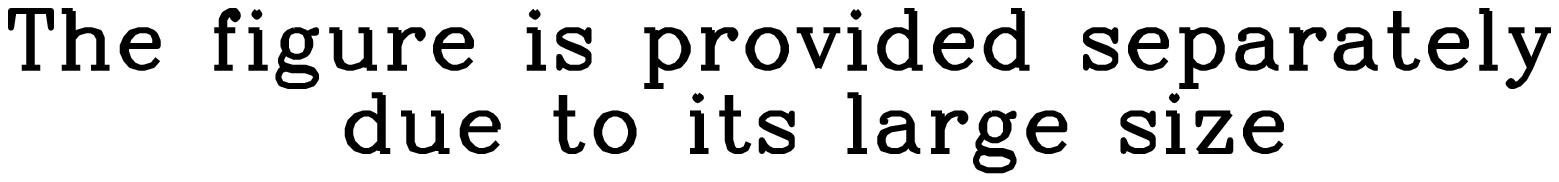}}
\end{center}  
      \caption[Evolution of a stable isothermal sphere when nested splitting 
      was applied]
      {Evolution of a stable isothermal sphere after $t$ = 8.17 $t_{ff}$, when 
      nested splitting was applied within a radius of 3 x 10$^{-2}$ pc. The 
      green points show fine particles and the black points coarse particles. 
      \underline{Top}: Radial density profile of the isothermal sphere. The red 
      line indicates the solution of Eqn. \ref{equa:hydrobalancedimeless}. 
      \underline{Bottom}: Thin equatorial slice ($\Delta z = 4 \times 10^{-3}$ 
      pc) of the isothermal sphere showing the 
      boundary between the coarse and the fine particles.}
      \label{fig:stable_nested}
\end{figure}

\begin{figure}
\begin{center}
\resizebox{9cm}{!}{\includegraphics{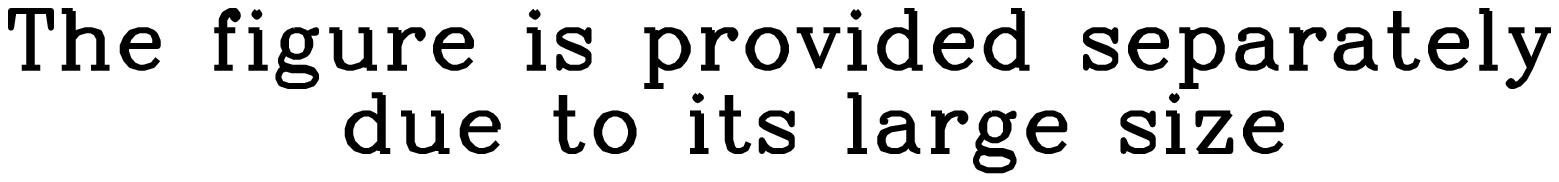}}
\end{center}  
\begin{center}
\resizebox{9cm}{!}{\includegraphics{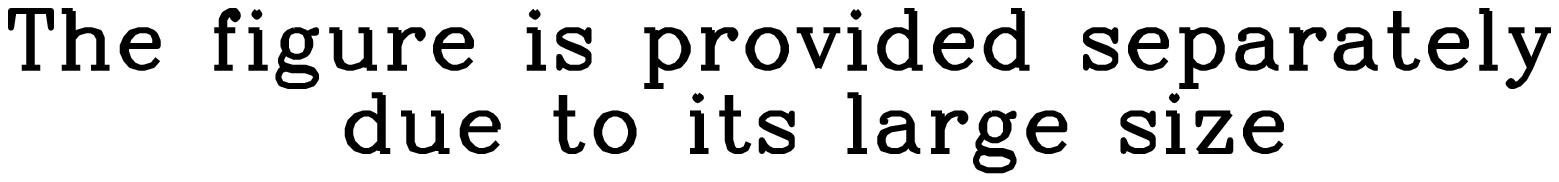}}
\end{center}  
      \caption[Evolution of a stable isothermal sphere when on-the-fly splitting
      was applied]
      {Evolution of a stable isothermal sphere after $t$ = 8.59 $t_{ff}$, when 
      on-the-fly splitting was applied. The density threshold is taken to be 
      $\rho_{max}$ = 2.48 x 10$^{-20}$ g cm$^{-3}$. The green points 
      show fine particles and the black points coarse particles. 
      \underline{Top}: Radial density profile of the isothermal sphere. The red 
      line indicates the solution of Eqn. \ref{equa:hydrobalancedimeless}. 
      \underline{Bottom}: Thin equatorial slice ($\Delta z = 4 \times 10^{-3}$ 
      pc) of the isothermal sphere showing the 
      boundary between the coarse and the fine particles.}
      \label{fig:stable_fly}
\end{figure}

The fine particles inside the fine region boundary are still attracted by the 
coarse particles just outside the fine region. The fine region appears to 
expand slowly. However, there is now a clear boundary between the coarse and 
fine regions (bottom panel of Fig. \ref{fig:stable_nested} {\it cf}. with the 
bottom panel of Fig. \ref{fig:stable_nested_old}). Fine particles do  
penetrate through the coarse particles. However, the majority of the latter are 
always kept outside the fine region. In fact, as fine particles slowly move to a
region of lower density some coarse particles move inside the fine region radius
and are automatically split. The coarse particles surrounded by fine particles 
in the bottom panel of Fig. \ref{fig:stable_nested} are split in the subsequent 
time-steps. This way, the number of coarse particles being split 
increases from 459 initially to 1,054 at the end of the simulation. The voids in 
the fine region shown in the equatorial slice of the bottom panel of Fig. 
\ref{fig:stable_nested} are projection effects produced when coarse particles 
exit the thin equatorial slice. Note that the thickness of the equatorial slice 
is less than, or of the order of the smoothing length of the coarse particles 
($\Delta z = 4 \times 10^{-3}$ pc). Coarse particles exit temporarily the 
equatorial slice due to numerical noise in their inward motion (this motion is caused by the 
slowly expanding fine region). There is no way one can damp this noise at a 
scale smaller than the smoothing length of the particles.

We conclude that SPH with nested splitting constrains the cloud to remain at the
same overall equilibrium state at all times, while the fine region boundary is 
permitted to evolve slowly within this global profile. In fact, the 
particles whose density deviates form the expected profile quickly have 
their density settled to the predicted values. This test shows that the 
boundary effects introduced by the application of nested splitting are not 
significant as the stable isothermal sphere retains its overall properties. 
It also demonstrates the efficiency of the new method for calculating 
$h$ (\S \ref{sec:newh}).

\begin{figure}
\begin{center}
\resizebox{9cm}{!}{\includegraphics{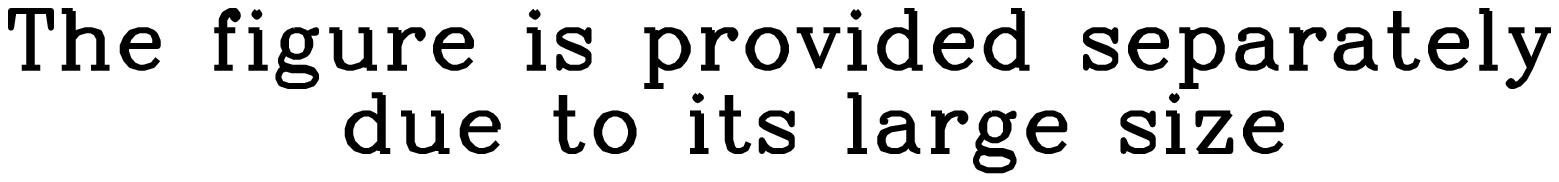}}
\end{center}  
      \caption[Evolution of a stable isothermal sphere when nested splitting 
      was applied with 
      particles initially taken from a settled distribution]
      {Evolution of a stable isothermal sphere when nested splitting 
      was applied within a radius of 3 x 10$^{-2}$ pc, with 
      particles initially taken from a settled distribution. 
      The green points show fine particles and the black points coarse 
      particles. The red line indicates the solution of Eqn. 
      \ref{equa:hydrobalancedimeless}. Radial density profile of the 
      simulation after $t$ = 6.63 $t_{ff}$.}
      \label{fig:stable_nest_settle}
\end{figure}

\subsubsection{4.4.2.2. Stable isothermal sphere using on-the-fly splitting}{\label{sec:stable_fly}}

The top panel of Fig. \ref{fig:stable_fly} shows the radial density profile 
of the isothermal sphere after $t$ = 8.59 $t_{ff}$, when on-the-fly splitting 
was applied with density threshold $\rho_{max}$ =  2.48 x 10$^{-20}$ g cm$^{-3}$. 
The green points show fine particles and the black points coarse particles. The 
red line indicates the solution of Eqn. \ref{equa:hydrobalancedimeless}. 
The density profile compares very well with the initial one (top panel of Fig. 
\ref{fig:stable}). There are 27,057 particles in total with 1,406 coarse 
particles having been split. The density profile is similar to that of the nested 
splitting simulation (top panel of Fig. \ref{fig:stable_nested}). Similar 
boundary effects can be observed in this test. There are some particles on 
either side of the boundary between the coarse and the fine regions having 
their density incorrectly modelled. These are basically new fine particles 
(split a few time-steps before) that have taken inappropriate initial values 
for their smoothing length. Their smoothing length and density will very soon settle to  the expected values. The 
introduction of new fine particles into the simulation creates the opposite 
effect on neighbouring coarse particles: their smoothing length is 
over-estimated for a few time-steps and thus their density is under-estimated.

The fine particles inside the fine region 
are attracted by the coarse particles just outside the fine region. The 
fine region appears to expand slowly. However, there is again a clear boundary 
between the coarse and fine regions (bottom panel of Fig. \ref{fig:stable_fly} 
{\it cf}. with the bottom panel of Fig. \ref{fig:stable_nested_old}; note that 
in on-the-fly splitting the fine region's radius is larger). Fine particles do  
penetrate through the coarse particles. However, the majority of the latter are 
always kept outside the fine region. In fact, as fine particles slowly move to a
region of lower density some coarse particles have their density estimates 
increased above the density threshold and are automatically split. The coarse 
particles surrounded by fine particles in the bottom panel of Fig. 
\ref{fig:stable_fly} are split in the subsequent time-steps. This way, 
the number of coarse particles being split increases from 959 initially to 
1,406 at the end of the simulation. Therefore, a smaller 
fraction of coarse particles is split compared to the nested splitting 
simulation. The voids in the fine region shown in the equatorial slice of the 
bottom panel of Fig. \ref{fig:stable_fly} are again projection effects produced 
when coarse particles exit the thin equatorial slice. Note that the thickness 
of the equatorial slice is less than, or of the order of the smoothing length 
of the coarse particles ($\Delta z = 4 \times 10^{-3}$ pc). As in the previous 
simulation, coarse particles exit temporarily the equatorial slice due to 
numerical noise in their inward motion. There is no way one can 
damp this noise at a scale smaller than the smoothing length of the particles.

We conclude that SPH with on-the-fly 
splitting constrains the cloud to remain at the same overall equilibrium state 
at all times, while the fine region boundary is permitted to evolve slowly 
within this global profile. In fact, the particles whose 
density deviates form the expected profile quickly have their density settled 
to the predicted values. This test shows that the boundary effects introduced 
by the application of on-the-fly splitting are not significant as the stable 
isothermal sphere retains its overall properties. In this test, on-the-fly 
splitting produces similar results to nested splitting, again more efficiently.

\subsubsection{4.4.2.3. Stable isothermal sphere using nested splitting with particles initially taken from a settled distribution}

We have repeated the simulation for the evolution of the stable isothermal 
sphere applying nested splitting to the centre of a cloud of 10,482 particles 
initially taken from a settled distribution of uniform density. 
The particles are then positioned according to the scheme given in \S 
\ref{sec:test_stable}. We have used the same initial conditions and the same 
code as above (\S \ref{sec:stable_nested}.1). The fine region has the same 
radius of 3 x 10$^{-2}$ pc. 
Fig. \ref{fig:stable_nest_settle} shows the radial 
density profile of the cloud at $t$ = 6.63 $t_{ff}$. The density complies 
well to the initial profile (top panel of Fig. \ref{fig:stable}). It compares 
very well with the simulation of \S \ref{sec:stable_nested}.1, when particles 
were initially taken on a lattice 
(top panel of Fig. \ref{fig:stable_nested}). There are 22,602 particles 
in total, with 1,010 coarse particles having been split. The boundary effects 
are similar to those discussed above. In the simulation of \S 
\ref{sec:stable_nested}.1, 
at $t$ = 6.63 $t_{ff}$, roughly the same number of coarse particles had been 
split. This test rules out the possibility that the results presented above are 
biased due to some preferred orientation of the initial lattice.

\subsection{Rotating Cloud}{\label{sec:rot_cloud}}

Having tested the new method against a problem of homogeneous inflow as well as the 
evolution of a stable cloud, we now apply it to the standard test simulation, 
discussed in chapter \ref{sec:rotating}. Application of the new method to a 
more realistic problem with known solution may reveal disadvantages of the new method 
that we have not calculated or thought of. 

We apply the initial conditions used in the simulations of the previous 
chapter (\S \ref{sec:rot-initial}), i.e. we use uniform-density, isothermal 
($c_{0}=0.17$ km s$^{-1}$; T = 7.9 K; $\alpha \approx$ 0.26), rotating ($\Omega$ = 7.2 x 10$^{-13}$ 
rad s$^{-1}$; $\beta \approx$ 0.16), spherical clouds of mass M = 1 M$_{\odot}$ 
and radius R $\approx$ 0.02 pc, with particles cut initially from a face-centred 
cubic lattice and then given an m = 2 azimuthal perturbation by adjusting their 
spherical polar azimuthal coordinate, $\phi$, to a value $\phi^{*}$ given by 
\[ \phi \, = \, \phi^{*} + \frac{A \sin(m \phi^{*})}{m}, \; \mathrm{with} \; A = 10\% \; \mathrm{amplitude}. \] 
We use our complete self-gravitating SPH code with 50 neighbours (chapter 
\ref{sec:SPH}), including adiabatic heating. Our findings in the tests 
of the previous chapter, and in the tests of particle splitting in the present chapter, assist in fine-tuning the 
values of the initial number of particles and the density above which adiabatic 
heating switches on. This way, the simulations always obey the Jeans condition 
with minimum computational cost. 

In chapter \ref{sec:rotating}, we have shown that convergence to the results of 
Truelove {\it et al}. \shortcite{TrueloveApJ1997,TrueloveApJ1998} and Klein 
{\it et al}. \shortcite{KleinTOKYO1998} can be achieved only with high resolution 
simulations. We have also shown that filamentary singularities can be obtained only if 
the simulations are evolved isothermally for sufficient time. The new method allows 
us to start a coarse simulation with low- to medium-resolution until it reaches the 
maximum resolvable density, $\rho_{max}$ (Eqn. \ref{equa:Jeans_condition3}), when 
particle splitting can be applied to increase the resolution 13-fold. If adiabatic 
heating starts before the fine simulation reaches its resolution limit, then the 
simulation obeys the Jeans condition at all times. Specifically, in the adiabatic heating regime 
the Jeans mass increases with increasing density ({\it cf}. Eqns. \ref{equa:jeans8} 
\& \ref{equa:eosfinal}) so that Eqn. \ref{equa:Jeans_condition1} is always valid as 
long as adiabatic heating switches on before the fine simulation reaches its resolution 
limit. The higher the initial number of particles the longer the simulation can 
evolve isothermally, allowing for the use of a higher value for $\rho_{0}$, the 
density above which adiabatic heating starts. 

In order to use a high density for the switch to the 
adiabatic regime ($\rho_{0}$=10$^{-12}$ g cm$^{-3}$), we have to start the simulation 
with no less than 40,000 particles. For comparison with the simulation of \S 
\ref{sec:rho0=-13}, we also present a simulation with $\rho_{0}$=10$^{-13}$ 
g cm$^{-3}$. In order for such a simulation to obey the Jeans condition at all times, 
a cloud of as low as 10,000 particles initially can be used. In our effort to evolve both 
simulations with minimum computational cost, we use the minimum initial number of particles 
that allows the Jeans condition to be obeyed at all times for the above values of $\rho_{0}$. 
In particular, the simulation with $\rho_{0}$=10$^{-12}$ g cm$^{-3}$ starts with 40,000 
particles and the simulation with $\rho_{0}$=10$^{-13}$ g cm$^{-3}$ starts with 10,000 
particles. We apply both versions of the new method to each simulation. 

Bate \& Burkert 
\shortcite{BateMNRAS1997} have shown that clouds of 10,000 particles do not 
resolve the binary formation. In contrast, clouds of 40,000 particles do resolve 
the formation of the binary and, in general, produce results very similar to 
those obtained with isothermal simulations of higher resolution. 

The particle splitting simulation with 40,000 particles initially gives results consistent with those 
of Truelove {\it et al}. and Klein {\it et al}. as well as those presented in 
\S \ref{sec:600kptcls} for the 600,000 particle simulation, but with fewer 
particles, demonstrating the economy in computation achieved with particle 
splitting. The new method reduces the computational cost 
on two counts: we do not have to use high resolution from the beginning of the 
simulation and when we do increase the resolution, this does not have to happen 
everywhere but only in regions of interest.

Using particle splitting just before the 10,000 particle simulation violates 
the Jeans condition, proves 
to be sufficient to reproduce results of higher resolution 
simulations (Bate \& Burkert \shortcite{BateMNRAS1997}, \S \ref{sec:rho0=-13},  
\ref{sec:heating_high}.2), but not the results of Truelove {\it et al}. 
\shortcite{TrueloveApJ1997,TrueloveApJ1998}, Klein {\it et al}. 
\shortcite{KleinTOKYO1998} and \S \ref{sec:600kptcls}. This happens because the 
initial stages of the cloud evolution are not properly modelled due to low 
resolution. 
We present a simulation that cannot give results consistent with those of Truelove 
{\it et al}. and Klein {\it et al}. despite the fact that it obeys the Jeans 
condition at all times, to demonstrate that particle splitting is a necessary, 
but not a sufficient, condition for the reliability of a simulation.  

Both versions of the new method produce similar results for each simulation. 
On-the-fly splitting is again more efficient than nested splitting in terms of 
computational cost. Tables \ref{tab:nest_rot} \& \ref{tab:nest_rot2} show a summary of our results. 
For comparison purposes, 
they also list the corresponding results of Truelove {\it et al}. 
\shortcite{TrueloveApJ1998} and Klein {\it et al}. \shortcite{KleinTOKYO1998}. 

We do not need to present the simulation 
with particles initially taken from a settled distribution. The only case that the bar fragments is the 
nested splitting simulation for a cloud of 10,000 particles initially and the results of such a simulation with 
particles initially taken from a settled distribution are similar to those of \S \ref{sec:settle}. Therefore, 
we refer to the discussion of the results of \S \ref{sec:settle}.

All figures presented here are column density 
plots viewed along the rotation axis. The captions indicate the units of the 
colour tables. They also give the linear size of the figure and the time 
of the simulation.

\subsubsection{4.4.3.1. Nested Splitting for a Cloud of 40,000 Particles}{\label{sec:nest40}}

The top panel of Fig. \ref{fig:40_nest_1} is a column density plot viewed along 
the rotation axis and shows the 
density projected on the x-y plane initially (at $t=0$), where the density 
enhancements indicate the m=2 perturbation. In a medium to high resolution simulation of a rotating, 
spherical, uniform, isothermal cloud with an m = 2 perturbation, the binary 
separation is expected to be $\sim$0.004 pc (\S \ref{sec:heating_changerho}). 
Taking 
into account the flattened shape that the cloud takes after $\sim$1 $t_{ff}$, 
the fine region in nested splitting takes a cylindrical shape and its radius 
is large enough to contain the binary formed and any other structure 
associated with it (e.g. spiral tails). The cylinder's height is as large as 
the thickness of the disc formed ($\sim$1/4 of the disc radius). Specifically, 
the fine region radius is 0.003 pc. 

\begin{figure}
\begin{center}
\resizebox{8cm}{!}{\includegraphics{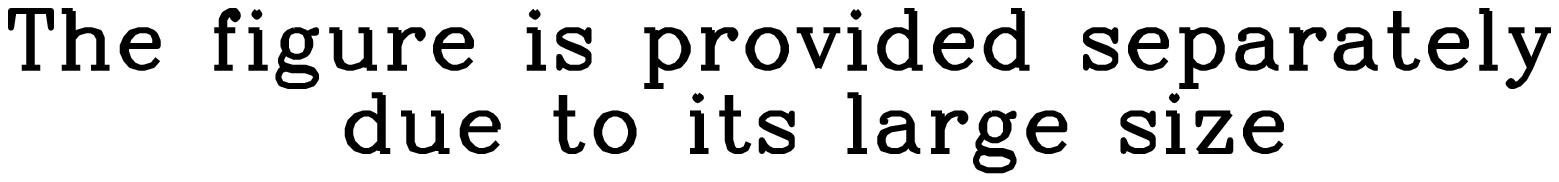}}
\end{center}  
\begin{center}
\resizebox{8cm}{!}{\includegraphics{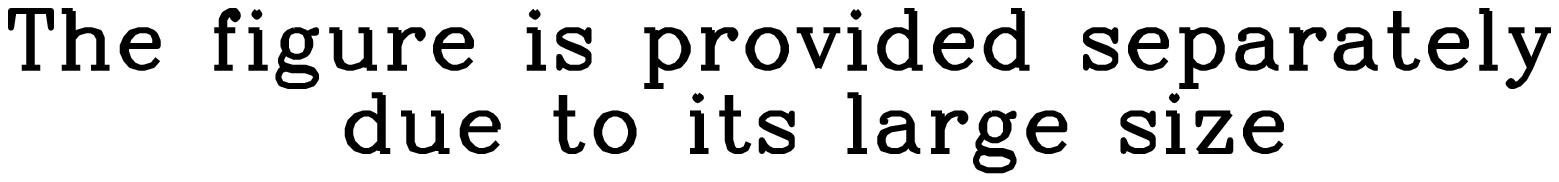}}
\end{center}  
      \caption[Nested Splitting for a Cloud of 40,000 Particles: Column density plots of the initial sphere and the cloud after splitting is applied]
      {Nested Splitting for a Cloud of 40,000 Particles: Column density plots of the initial sphere and the cloud after splitting is applied. 
      \underline{Top}: Column density plot of the cloud initially ($t$ = 0). The linear size of this plot is 0.04 pc. The colour table has units of 1.18 x 10$^{6}$ g cm$^{-2}$.
      \underline{Bottom}: Column density plot of the cloud after splitting is applied ($t$ = 1.244 $t_{ff}$). The linear size of this plot is 0.008 pc. The colour table has units of 2.95 x 10$^{7}$ g cm$^{-2}$.}
      \label{fig:40_nest_1}
\end{figure}

\begin{figure}
\begin{center}
\resizebox{8cm}{!}{\includegraphics{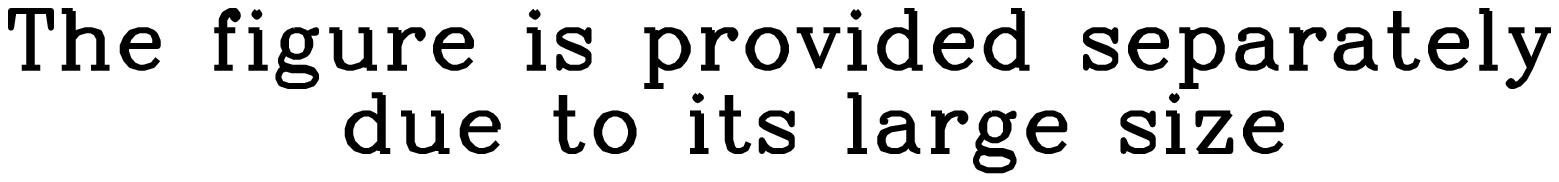}}
\end{center}  
\begin{center}
\resizebox{8cm}{!}{\includegraphics{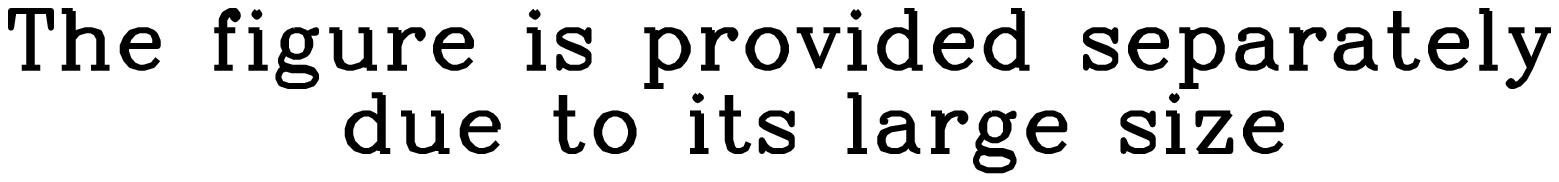}}
\end{center}  
      \caption[Nested Splitting for a Cloud of 40,000 Particles: Column density plots of the cloud before heating is applied and at the end]
      {Nested Splitting for a Cloud of 40,000 Particles: Column density plots of the cloud before heating is applied and at the end.
      \underline{Top}: Column density plot of the cloud before heating is applied ($t$ = 1.254 $t_{ff}$). The linear size of this plot is 0.004 pc. The colour table has units of 1.18 x 10$^{8}$ g cm$^{-2}$. 
      \underline{Bottom}: Column density plot of the cloud at the end ($t$ = 1.265 $t_{ff}$). The linear size of this plot is 0.004 pc. The colour table has units of 1.18 x 10$^{8}$ g cm$^{-2}$.}
      \label{fig:40_nest_2}
\end{figure}

\begin{figure}
\begin{center}
\resizebox{8cm}{!}{\includegraphics{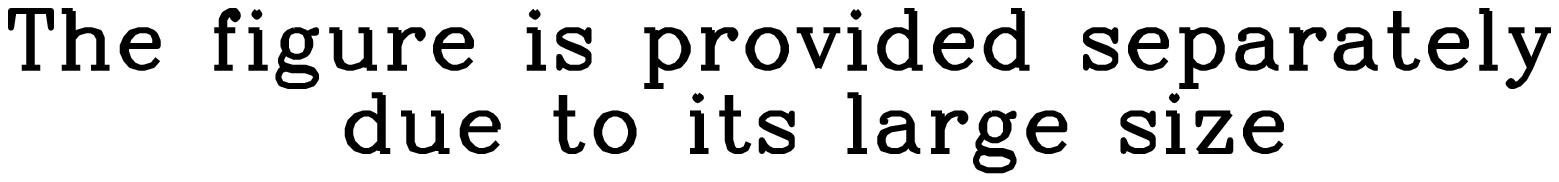}}
\end{center}  
\begin{center}
\resizebox{8cm}{!}{\includegraphics{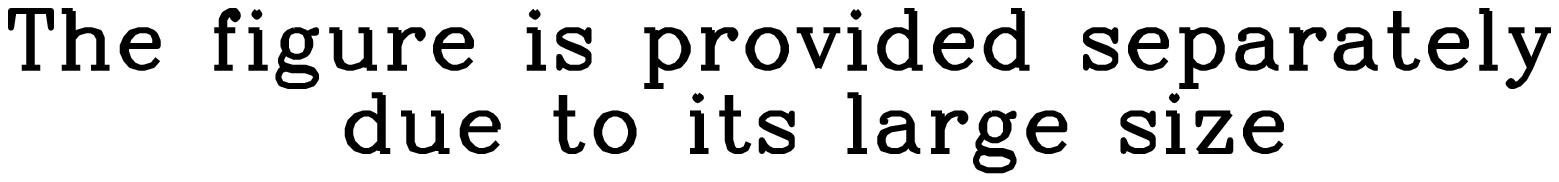}}
\end{center}  
      \caption[On-the-fly Splitting for a Cloud of 40,000 Particles: Column density plots after splitting is applied and before heating starts]
      {On-the-fly Splitting for a Cloud of 40,000 Particles: Column density plots after splitting is applied and before heating starts. 
      \underline{Top}: Column density plot of the cloud after splitting is applied ($t$ = 1.251 $t_{ff}$). The linear size of this plot is 0.008 pc. The colour table has units of 2.95 x 10$^{7}$ g cm$^{-2}$.
      \underline{Bottom}: Column density plot of the cloud before heating starts ($t$ = 1.258 $t_{ff}$). The linear size of this plot is 0.004 pc. The colour table has units of 1.18 x 10$^{8}$ g cm$^{-2}$.}
      \label{fig:40_fly_1}
\end{figure}

\begin{figure}
\begin{center}
\resizebox{8cm}{!}{\includegraphics{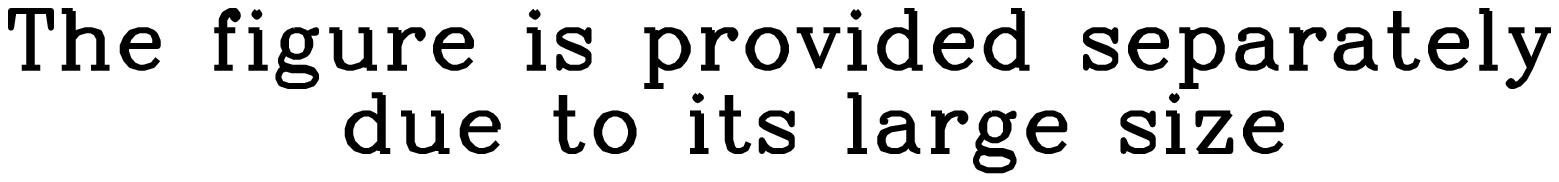}}
\end{center}    
      \caption[On-the-fly Splitting for a Cloud of 40,000 Particles: Column density plot of the cloud at the end]
      {On-the-fly Splitting for a Cloud of 40,000 Particles: Column density plot of the cloud at the end ($t$ = 1.277 $t_{ff}$). The linear size of this plot is 0.004 pc. The colour table has units of 1.18 x 10$^{8}$ g cm$^{-2}$.} 
      \label{fig:40_fly_2}
\end{figure}

Nested splitting 
is applied after $t=$1.244 $t_{ff}$, when the binary has started forming and the
maximum density is about to exceed the density threshold, $\rho_{max}$ = 2.4 x 
10$^{-14}$ g cm$^{-3}$, above which the simulation stops resolving the Jeans 
mass. Initially, 
17,400 coarse particles lie inside the fine region radius and are split. The 
bottom panel of Fig. \ref{fig:40_nest_1} 
is a column density plot immediately after the application of nested splitting. 
Note the smooth transition of the column density through the fine region boundary 
(its diameter is 0.006 pc while the linear size of the 
bottom panel of Fig. \ref{fig:40_nest_1} is 0.008 pc). 

In 
subsequent time-steps, all particles crossing the fine region radius are split 
on-the-fly. A well-defined binary forms and between the binary components a bar 
grows. The fine simulation would reach its resolution limit at $\rho_{max}$ = 
4 x 10$^{-12}$ g cm$^{-3}$, but adiabatic heating starts at $\rho_{0}$ = 10$^{-12}$ 
g cm$^{-3}$ and the simulation obeys the Jeans condition at all times. 
The top panel of Fig. \ref{fig:40_nest_2} shows the 
column density just before adiabatic heating initiates at $t=$1.254 $t_{ff}$. 

The top panel of Fig. \ref{fig:40_nest_2} 
compares well with Fig. 12 of Truelove {\it et al}. \shortcite{TrueloveApJ1998};
there are two elongated objects and a thin bar connecting them ($\rho_{peak}$ = 
7.14 x 10$^{-13}$ g cm$^{-3}$). 

The bottom panel of Fig. \ref{fig:40_nest_2} shows the column density at the end
of the simulation, $t=$1.265 $t_{ff}$. There are 281,331 particles in total with
20,020 coarse particles having been split. The mass of each fragment 
is 0.02 M$_{\odot}$ and its radius $\sim$7 AU. Their separation is 515 AU. The 
peak density of the simulation has risen to $\rho_{peak}$ = 1.97 x 10$^{-10}$ g 
cm$^{-3}$. The bar 
has not fragmented. There is a well-defined contrast between the density of the 
binary components and the bar that connects them (2 orders of magnitude). 

The final results compare well with those of Klein {\it et al}. 
\shortcite{KleinTOKYO1998} (their Fig. 2 - note that in Klein {\it et al}. 
adiabatic heating starts about one order of magnitude earlier). 

The resolution of the simulation is sufficient that we can trust the result that
the bar does not fragment up to $t=$1.265 $t_{ff}$. In fact, the resolution of this simulation is
equivalent to the resolution of a 520,000 
particle simulation and the results are indeed similar to those of the 
simulation with the highest resolution we conducted without splitting in \S 
\ref{sec:600kptcls} (600,000). In particular, the peak 
density of the simulation is comparable to that of the simulation with 600,000 
particles (see discussion in \S \ref{sec:600kptcls}). This indicates the efficiency of nested 
splitting as only 281,331 particles have been used.

\subsubsection{4.4.3.2. On-the-fly Splitting for a Cloud of 40,000 Particles}

On-the-fly splitting is applied after $t=$1.244 $t_{ff}$, when the maximum 
density is about to exceed the density threshold, $\rho_{max}$ = 2.4 x 
10$^{-14}$ g cm$^{-3}$, over which the simulation stops resolving the Jeans mass. Initially, 
2,955 coarse particles are split. The top panel of Fig. \ref{fig:40_fly_1} is 
the first column density plot after the application of on-the-fly splitting 
($t$= 1.251 $t_{ff}$). Note the smooth transition of the
column density through the fine region boundary (within the red coloured area). 

In subsequent time-steps, all particles whose density exceeds the density 
threshold are split on-the-fly. A well-defined binary forms and between the binary 
components a bar grows. The fine simulation would reach its resolution limit at $\rho_{max}$ = 
4 x 10$^{-12}$ g cm$^{-3}$, but adiabatic heating starts at $\rho_{0}$ = 10$^{-12}$ 
g cm$^{-3}$ and the simulation obeys the Jeans condition at all times. 
The bottom panel of Fig. \ref{fig:40_fly_1} shows the column density just before 
adiabatic heating starts at $t=$1.258 $t_{ff}$. 

The bottom panel of Fig. 
\ref{fig:40_fly_1} compares well with Fig. 12 of Truelove {\it et al}. 
\shortcite{TrueloveApJ1998}; there are two elongated objects and a thin bar 
connecting them ($\rho_{peak}$ = 6.86 x 10$^{-13}$ g cm$^{-3}$). 

Fig. \ref{fig:40_fly_2} shows the column density at the end of 
the simulation, $t=$1.277 $t_{ff}$. There are 150,135 particles in total with 
9,087 coarse particles having been split. The mass of each fragment is 0.03 
M$_{\odot}$ and its radius $\sim$10 AU. Their separation is 475 AU. The 
peak density of the simulation has risen to $\rho_{peak}$ = 1.85 x 10$^{-10}$ 
g cm$^{-3}$. The bar has not fragmented. There is a well-defined contrast 
between the density of the binary components and the bar that connects them 
(2 orders of magnitude). 

The final
results compare well with those of Klein {\it et al}. 
\shortcite{KleinTOKYO1998} (their Fig. 2 - note that in Klein {\it et al}. 
adiabatic heating starts about one order of magnitude earlier). 

The resolution of the simulation is sufficient that we can trust the result that
the bar does not fragment up to $t=$1.277 $t_{ff}$. In fact, the resolution of this simulation is
equivalent to the resolution of a 520,000 
particle simulation and the results are indeed similar to those of the 
simulation with the highest resolution we conducted without splitting in \S 
\ref{sec:600kptcls} (600,000). In particular, the peak 
density of the simulation is comparable to that of the simulation with 600,000 
particles (see discussion in \S \ref{sec:600kptcls}). This indicates the efficiency of on-the-fly
splitting as only 150,135 particles have been used. 

Comparison between the results of the on-the-fly splitting 
and the nested splitting simulations clearly indicates that the results are similar (see Tables 
\ref{tab:nest_rot} \& \ref{tab:nest_rot2}), although the on-the-fly splitting simulation has progressed more in 
time (the binary 
components have accreted more matter, they are closer and the bar is more inclined). 
However, on-the-fly splitting is much more efficient in obtaining them (using 
131,196 less particles in total).

\subsubsection{4.4.3.3. Nested Splitting for a Cloud of 10,000 Particles}

The top panel of Fig. \ref{fig:10_nest_1} is a column density plot viewed along the rotation axis and shows the density 
projected on the x-y plane initially (at $t=0$) for a cloud of 10,000 particles,
where the density enhancements indicate the m=2 perturbation. The coarser 
resolution used here does not change the initial appearance of the cloud ({\it cf}. the top panel of Fig. 
\ref{fig:40_nest_1}). Following the same reasoning as in \S \ref{sec:nest40}.1, the fine region is cylindrical with the
radius now taken to be 0.002 pc (the expected bar length is extracted from Fig. 7 of Bate \& Burkert 
\shortcite{BateMNRAS1997}). 

\begin{figure}
\begin{center}
\resizebox{8cm}{!}{\includegraphics{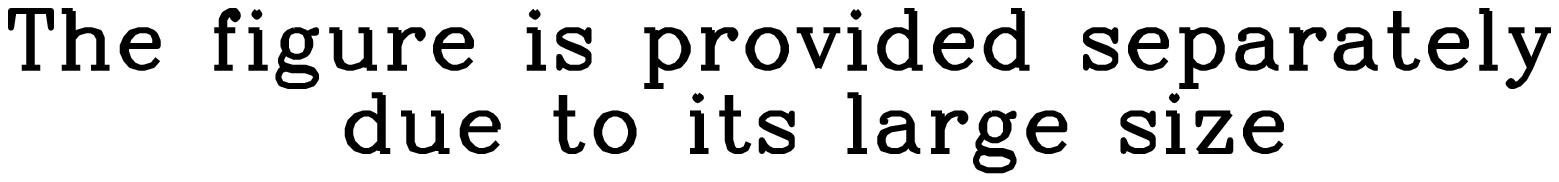}}
\end{center}  
\begin{center}
\resizebox{8cm}{!}{\includegraphics{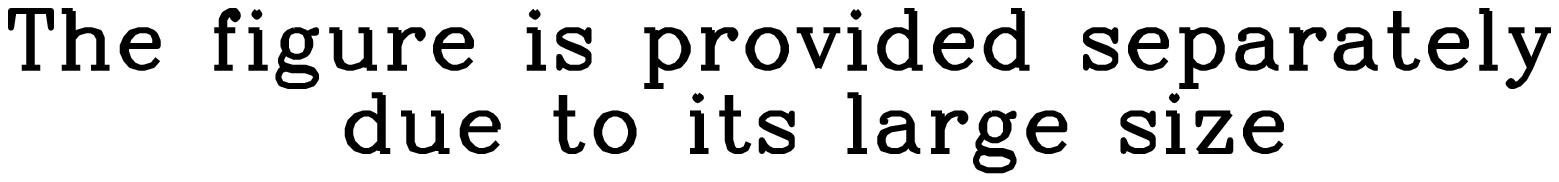}}
\end{center}  
      \caption[Nested Splitting for a Cloud of 10,000 Particles: Column density plots of the initial sphere and the cloud after splitting is applied]
      {Nested Splitting for a Cloud of 10,000 Particles: Column density plots of the initial sphere and the cloud after splitting is applied. 
      \underline{Top}: Column density plot of the cloud initially ($t$ = 0). The linear size of this plot is 0.04 pc. The colour table has units of 1.18 x 10$^{6}$ g cm$^{-2}$.
      \underline{Bottom}: Column density plot of the cloud after splitting is applied ($t$ = 1.105 $t_{ff}$). The linear size of this plot is 0.008 pc. The colour table has units of 2.95 x 10$^{7}$ g cm$^{-2}$.}
      \label{fig:10_nest_1}
\end{figure}

\begin{figure}
\begin{center}
\resizebox{8cm}{!}{\includegraphics{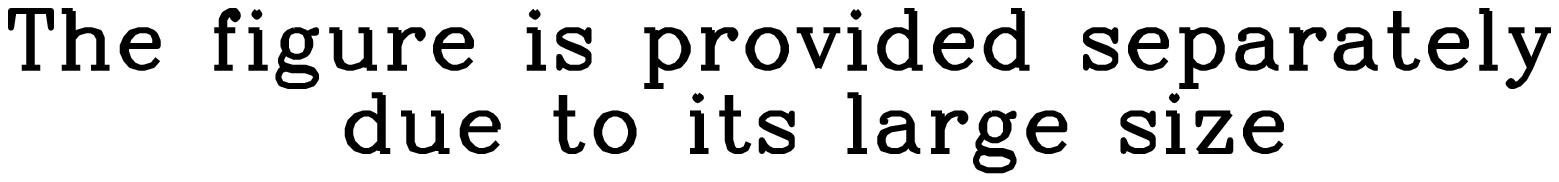}}
\end{center}  
\begin{center}
\resizebox{8cm}{!}{\includegraphics{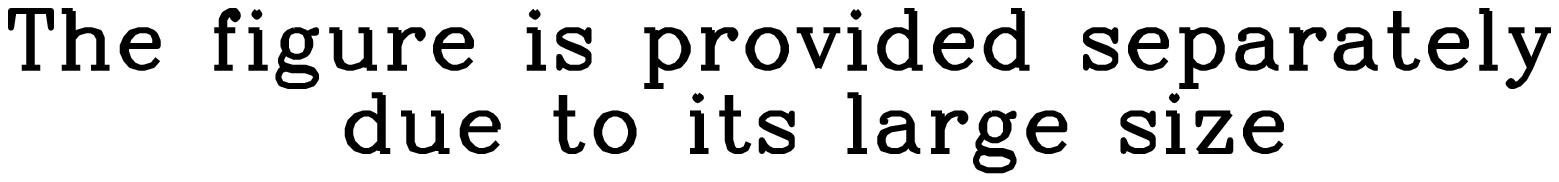}}
\end{center}  
      \caption[Nested Splitting for a Cloud of 10,000 Particles: Column density plots of the cloud before heating is applied and at the end]
      {Nested Splitting for a Cloud of 10,000 Particles: Column density plots of the cloud before heating is applied and at the end.
      \underline{Top}: Column density plot of the cloud before heating is applied ($t$ = 1.265 $t_{ff}$). The linear size of this plot is 0.004 pc. The colour table has units of 1.18 x 10$^{8}$ g cm$^{-2}$. 
      \underline{Bottom}: Column density plot of the cloud at the end ($t$ = 1.301 $t_{ff}$). The linear size of this plot is 0.004 pc. The colour table has units of 1.18 x 10$^{8}$ g cm$^{-2}$.}
      \label{fig:10_nest_2}
\end{figure}

Nested splitting is applied after $t=$1.105 $t_{ff}$, 
when the maximum density is about to exceed the density threshold, $\rho_{max}$ 
= 1.5 x 10$^{-15}$ g cm$^{-3}$, over which the simulation stops resolving the 
Jeans mass. At this 
time, the binary has not formed yet and the centre of the cloud has just started
collapsing after the initial expansion phase \cite{BateMNRAS1997}. Initially, 
705 coarse particles lie inside the fine region radius and are split. The bottom
panel of Fig. \ref{fig:10_nest_1} is a 
column density plot immediately after the application of nested splitting. Note 
that the lattice which the particles were initially taken on has not broken yet.
Also note the smooth transition of the column density through the fine region 
boundary (its diameter is 0.004 pc while the size of the bottom
panel of Fig. \ref{fig:10_nest_1} is 0.008 pc). The density of the fine particles is 
emphasised due to effects similar to those discussed in \S 
\ref{sec:stable_nested}.1 and the small dynamic range of the bottom panel of Fig. \ref{fig:10_nest_1}. 

In subsequent 
time-steps, all particles crossing the fine region radius are split on-the-fly. 
A central density enhancement grows and later forms a bar. 
On either side of the bar spiral tails form due to rotation. At both ends of the bar two objects 
finally form at $t$ = 1.265 $t_{ff}$.
The fine simulation would reach its resolution limit at $\rho_{max}$ = 
2.53 x 10$^{-13}$ g cm$^{-3}$, but adiabatic heating starts at $\rho_{0}$ = 10$^{-13}$ 
g cm$^{-3}$ and the simulation obeys the Jeans condition at all times. 
The top panel of Fig. \ref{fig:10_nest_2} shows the column density 
just before adiabatic heating starts at $t=$1.265 $t_{ff}$. Subsequently, the bar fragments 
producing three fragments, one at the centre and two at equal distances from the centre. 

This is in accordance with the findings of the low 
resolution run of Bate \& Burkert \shortcite{BateMNRAS1997}, where the binary 
formation is not resolved and just a bar forms that later fragments into multiple fragments. However, 
in this simulation, with application of nested splitting, the binary forms, although at a later stage than expected. 

The bottom panel of Fig. \ref{fig:10_nest_2} shows the 
column density at the end of the simulation, $t=$1.301 $t_{ff}$. There are 55,125 
particles in total with 3,794 coarse particles having been split. The mass of the binary fragments 
is 0.06 M$_{\odot}$ and of the bar fragments 0.01 M$_{\odot}$. The binary fragments have radii of $\sim$25 AU. 
The bar is 410 AU long. The peak density 
of the simulation has risen to $\rho_{peak}$ = 2.14 x 10$^{-11}$ g cm$^{-3}$. The bar 
fragments as it does in the corresponding runs of Bate \& Burkert \shortcite{BateMNRAS1997} and those of \S 
\ref{sec:rho0=-13}, \ref{sec:heating_high}.2. 

Clearly, the evolution of this simulation does not imitate the evolution predicted by the high resolution 
simulations of Truelove {\it et al}. \shortcite{TrueloveApJ1997,TrueloveApJ1998}, Klein {\it et al}. 
\shortcite{KleinTOKYO1998} and \S \ref{sec:600kptcls} (see Tables \ref{tab:nest_rot} \& \ref{tab:nest_rot2}).

\begin{figure}
\begin{center}
\resizebox{8cm}{!}{\includegraphics{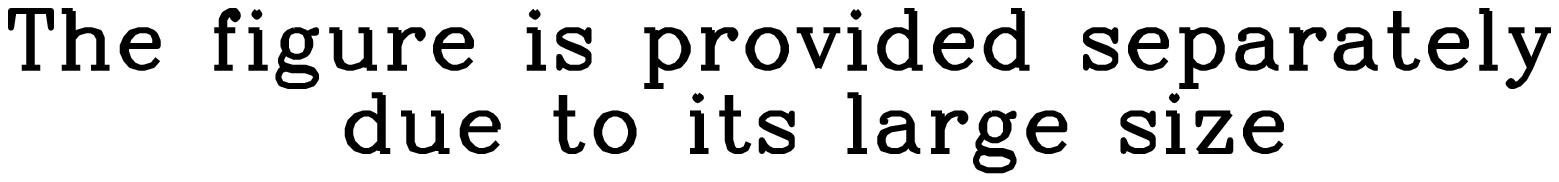}}
\end{center}
      \caption[Simulation of a Cloud of 130,000 Particles without Particle Splitting: Column density plot of the cloud at the end]
      {Simulation of a Cloud of 130,000 Particles without Particle Splitting: Column density plot of the cloud at the end ($t=$1.291 $t_{ff}$). The linear size of this plot is 0.004 pc. The colour table has units of 1.18 x 10$^{8}$ g cm$^{-2}$.}
      \label{fig:130_nest_1}
\end{figure}

\begin{figure}  
\begin{center}
\resizebox{8cm}{!}{\includegraphics{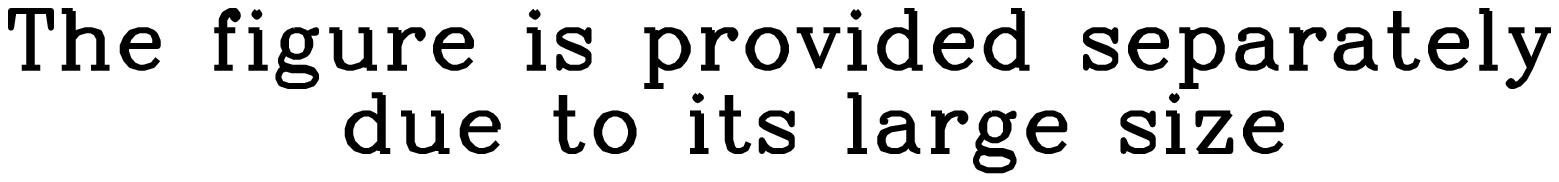}}
\end{center}
\begin{center}
\resizebox{8cm}{!}{\includegraphics{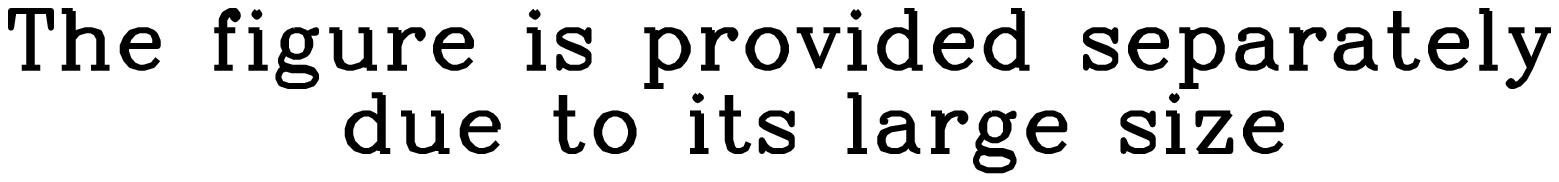}}
\end{center}
      \caption[On-the-fly Splitting for a Cloud of 10,000 Particles: Column density plots of the cloud after splitting is applied and before heating starts]
      {On-the-fly Splitting for a Cloud of 10,000 Particles: Column density plots of the cloud after splitting is applied and before heating starts. 
      \underline{Top}: Column density plot of the cloud after splitting is applied ($t$ = 1.142 $t_{ff}$). The linear size of this plot is 0.008 pc. The colour table has units of 2.95 x 10$^{7}$ g cm$^{-2}$.
      \underline{Bottom}: Column density plot of the cloud before heating starts ($t$ = 1.293 $t_{ff}$). The linear size of this plot is 0.004 pc. The colour table has units of 1.18 x 10$^{8}$ g cm$^{-2}$.
}
      \label{fig:10_fly_1}
\end{figure}

\begin{figure}  
\begin{center}
\resizebox{8cm}{!}{\includegraphics{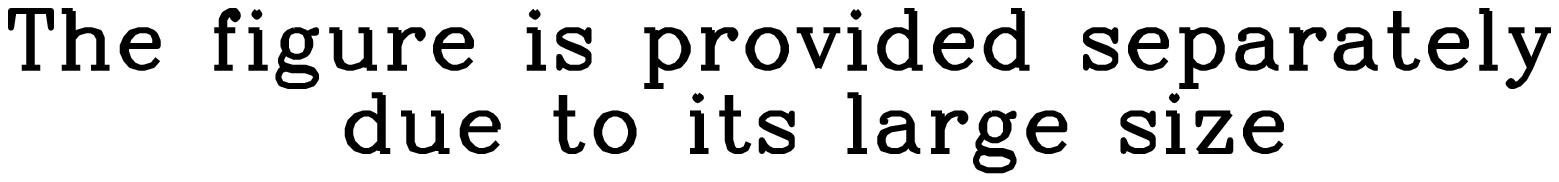}}
\end{center}  
      \caption[On-the-fly Splitting for a Cloud of 10,000 Particles: Column density plot of the cloud at the end]
      {On-the-fly Splitting for a Cloud of 10,000 Particles: Column density plot of the cloud at the end ($t$ = 1.326 $t_{ff}$). The linear size of this plot is 0.004 pc. The colour table has units of 1.18 x 10$^{8}$ g cm$^{-2}$.}
      \label{fig:10_fly_2}
\end{figure}

\paragraph{The equivalent resolution from the beginning of the simulation}

At the centre of this simulation the resolution is equivalent to that of a simulation with 130,000 coarse particles. We 
have constructed such a test to compare with the above simulation. We have used 
the same initial conditions with 130,000 particles initially taken on a lattice.
Adiabatic heating initiates at $\rho_{0}$ = 10$^{-13}$ g cm$^{-3}$. Particle splitting is \underline{not} applied. 

Fig. \ref{fig:130_nest_1} shows 
the end of the 130,000 particles simulation, at $t=$1.291 $t_{ff}$. In this case,
the binary formation is resolved (forms at $t=$1.239 $t_{ff}$). The bar between the binary components does 
fragment, i.e. the results are similar to the simulations of Bate \& Burkert \shortcite{BateMNRAS1997} and those of 
\S \ref{sec:rho0=-13},  \ref{sec:heating_high}.2. 

The difference between 
the results of this run and the results of the above simulation with only 
10,000 particles initially is due to the low resolution initial phase of the 
latter simulation. In particular, in the latter simulation the initial expansion and the 
subsequent collapse that leads to the formation of the binary 
\cite{BateMNRAS1997} are not properly modelled (due to poor sampling) and this causes the 
delayed formation of the binary. The application of nested splitting and the fact that 
the Jeans condition is obeyed clearly assists in the eventual formation of the binary 
({\it cf}. the result of the 10,000 particle simulation of Bate \& Burkert \shortcite{BateMNRAS1997}). 

This conclusion is confirmed by the results of 
simulations of 20,000 and 30,000 particles initially, with application of on-the-fly 
splitting. In particular, we find that in these two runs the binary formation is resolved. Moreover, 
the evolution of the 30,000 particle simulation is closer to the converged solution of Bate \& Burkert \shortcite{BateMNRAS1997}. 
However, application of particle splitting, despite assisting this simulation in creating the 
binary that would not be formed otherwise, can not guarantee solving all problems, due to poor 
sampling. 

\subsubsection{4.4.3.4. On-the-fly Splitting for a Cloud of 10,000 Particles}

On-the-fly splitting is applied to the cloud of 10,000 particles after $t=$1.105
$t_{ff}$, when the maximum density is about to exceed the density threshold, 
$\rho_{max}$ = 1.5 x 10$^{-15}$ g cm$^{-3}$, over which the simulation stops resolving the 
Jeans mass. Initially, 3,146 coarse particles are split. The top panel of Fig. 
\ref{fig:10_fly_1} is the first column density plot after the application of on-the-fly splitting 
($t$=1.142 $t_{ff}$). Note 
the smooth transition of the column density through the fine region boundary 
(within the green coloured area). Note 
that the lattice which the particles were initially taken on has not broken yet.
Within the few time-steps elapsed after the first coarse particles were split, a rugged 
appearance similar to the bottom panel of Fig. \ref{fig:10_nest_1} has been smoothed out as it did in 
\S \ref {sec:stable_fly}.2. Due to the low density threshold, many coarse particles are split on the 
fly within these few time-steps. At $t$=1.142 $t_{ff}$, the binary has not formed yet and 
the centre of the cloud has just started collapsing after the initial expansion 
phase \cite{BateMNRAS1997}.

In subsequent time-steps, all particles whose 
density exceeds the density threshold are split on-the-fly.
A central density enhancement grows and later forms a bar. 
On either side of the bar spiral tails form due to rotation. 
The fine simulation would reach its resolution limit at $\rho_{max}$ = 
2.53 x 10$^{-13}$ g cm$^{-3}$, but adiabatic heating starts at $\rho_{0}$ = 10$^{-13}$ 
g cm$^{-3}$ and the simulation obeys the Jeans condition at all times. 

The bottom panel of Fig. \ref{fig:10_fly_1} shows the column density just before
adiabatic heating starts at $t=$1.293 $t_{ff}$. At both ends of the bar two objects finally 
form at $t$ = 1.317 $t_{ff}$. Subsequently, the bar does not fragment. This result is in accordance with 
the findings of the high 
resolution simulations of Truelove {\it et al}. \shortcite{TrueloveApJ1997,TrueloveApJ1998}, Klein {\it et al}. 
\shortcite{KleinTOKYO1998} and \S \ref{sec:600kptcls}, where the bar does not fragment. However, in this simulation, due to the initial coarse resolution phase, 
the binary forms at a much later stage than expected. Fig. \ref{fig:10_fly_2} shows the column 
density at the end of the simulation ($t=$1.326 $t_{ff}$). The density contrast between the binary components 
and the bar that connects them is greater than one order of magnitude. There are 96,255 particles in 
total with 7,219 coarse particles having been split. The mass of each binary fragment 
is 0.06 M$_{\odot}$ and its radius $\sim$40 AU. The bar is 270 AU long. The peak density of the 
simulation has risen to $\rho_{peak}$ = 5.04 x 10$^{-12}$ g cm$^{-3}$.  

In this simulation the initial expansion and the 
subsequent collapse that leads to the formation of the binary 
\cite{BateMNRAS1997} are not properly modelled (due to poor sampling) and this causes 
the delayed formation of the binary. The application of on-the-fly splitting and the 
fact the Jeans condition is obeyed clearly assists in the eventual formation of 
the binary ({\it cf}. the result of the 10,000 particle simulation of Bate \& Burkert 
\shortcite{BateMNRAS1997}). The fact that the binary forms later than in all other low 
resolution simulations of \S \ref{sec:neighbours} as well as the above 
simulation (nested splitting with 10,000 particles initially) can be attributed to the large number of 
coarse particles being split almost simultaneously at the initial stages. During the 
collapse phase following the initial expansion, many coarse particles obtain density 
larger than the low density threshold we have used. We have used such a low value for 
$\rho_{max}$ (1.5 x 10$^{-15}$ g cm$^{-3}$) due to the small initial number of particles. 
The simultaneous splitting of few thousand particles introduced excess density 
perturbations to the simulation. SPH and the new method for calculating $h$ have been shown 
to be efficient in eliminating such perturbations. This is why we finally obtain the 
right result. 

Comparison between the results 
of the on-the-fly splitting and the nested splitting simulations with 10,000 
particles initially, shows that only with on-the-fly splitting we can reproduce the results of the high 
resolution simulations of Truelove {\it et al}. \shortcite{TrueloveApJ1997,TrueloveApJ1998}, 
Klein {\it et al}. \shortcite{KleinTOKYO1998} and \S \ref{sec:600kptcls} (see Tables \ref{tab:nest_rot} 
\& \ref{tab:nest_rot2}). However, with the low density threshold set due to the small initial number 
of particles the computational cost of the on-the-fly splitting simulation grows considerably. 
Therefore, accuracy has been achieved at the expense of computational efficiency. This is against one of 
the primary aims of the new method that is designed to gain both in accuracy and computational efficiency. 
This confirms our previous conclusion that particle splitting, the way it has been formulated in this work, is a 
necessary, but not a sufficient, condition for the reliability and efficiency of a simulation. 

\subsection{Conclusions}

\begin{sidewaystable}
\begin{center}
\begin{tabular}{|l|l|c|c|c|c|c|c|}\hline
Simulation & Particles & $\rho_{0}$ / g cm$^{-3}$ & $\rho_{max}$ / g cm$^{-3}$ & $t_{spl}$ / $t_{ff}$ & $t_{bin}$ / $t_{ff}$ & $t_{heat}$ / $t_{ff}$ & $\rho_{peak}^{iso}$ / g cm$^{-3}$ \\ \hline \hline
Nested & 40,000 & 10$^{-12}$ & 2.4 x 10$^{-14}$ & 1.244 & 1.244 & 1.254 & 7.14 x 10$^{-13}$ \\ \hline
On-the-fly & 40,000 & 10$^{-12}$ & 2.4 x 10$^{-14}$ & 1.251 & 1.244 & 1.258 & 6.86 x 10$^{-13}$ \\ \hline
Nested & 10,000 & 10$^{-13}$ & 1.5 x 10$^{-15}$ & 1.105 & 1.265 & 1.265 & 8.52 x 10$^{-14}$ \\ \hline
On-the-fly & 10,000 & 10$^{-13}$ & 1.5 x 10$^{-15}$ & 1.142 & 1.317 & 1.293 & 7.81 x 10$^{-14}$ \\ \hline \hline
Truelove (1998) & R$_{32}$ & -- & -- & -- & -- & 1.317 & 3.91 x 10$^{-13}$ \\ \hline
Klein (1998) & R$_{32}$ & -- & -- & -- & -- & 1.304 & 8.97 x 10$^{-15}$ \\ \hline
\end{tabular}
\end{center}
\caption[Summary of results for the standard test simulations using particle splitting (first part)]{Summary of results for the standard test simulations using particle splitting (first part). 
For each simulation the second column gives the initial number of particles and the third column lists the density above which adiabatic heating starts. 
$\rho_{max}$ is the density at which the coarse simulation reaches its resolution limit. $\rho_{peak}^{iso}$ is the peak density at the end of the 
isothermal regime. $t_{spl}$ is the time at which particle splitting is applied, $t_{heat}$ is the 
time at which heating starts and $t_{bin}$ is the time the binary forms. The last two rows list the relevant results of the finite 
difference simulations of Truelove {\it et al}. \shortcite{TrueloveApJ1998} and Klein {\it et al}. \shortcite{KleinTOKYO1998}. For these two rows, we have used the 
initial number of cells instead of the number of particles in column 2.}
\label{tab:nest_rot}
\end{sidewaystable}

\begin{sidewaystable}
\begin{center}
\begin{tabular}{|l|l|c|c|c|c|c|c|}\hline
Simulation & Particles & $\rho_{0}$ / g cm$^{-3}$ & $t_{end}$ / $t_{ff}$ & Bar Fragm. & $t_{frag}$ / $t_{ff}$ & $\rho_{peak}^{heat}$ / g cm$^{-3}$ & Particles \\ \hline \hline
Nested & 40,000 & 10$^{-12}$ & 1.265 & 0 & N/A & 1.97 x 10$^{-10}$ & 281,331\\ \hline
On-the-fly & 40,000 & 10$^{-12}$ & 1.277 & 0 & N/A & 1.85 x 10$^{-10}$ & 150,135\\ \hline
Nested & 10,000 & 10$^{-13}$ & 1.301 & 3 & 1.281 & 2.14 x 10$^{-11}$ & 55,125\\ \hline
On-the-fly & 10,000 & 10$^{-13}$ & 1.326 & 0 & N/A & 5.04 x 10$^{-12}$ & 96,255\\ \hline \hline
Truelove (1998) & R$_{32}$ & -- & 1.319 & 0 & N/A & 5 x 10$^{-10}$ & R$_{131,072}$\\ \hline
Klein (1998) & R$_{32}$ & -- & 1.351 & 0 & N/A & 1.88 x 10$^{-11}$ & R$_{4,096}$\\ \hline
\end{tabular}
\end{center}
\caption[Summary of results for the standard test simulations using particle splitting (second part)]{Summary of results for the standard test simulations using particle splitting (second part). 
For each simulation the second column gives the initial number of particles and the last column the final number of particles. The third column lists the density above which adiabatic heating starts. 
$\rho_{peak}^{heat}$ is the peak density at the end of the simulation. $t_{end}$ is the time at the end of the simulation and $t_{frag}$ is the 
time the bar fragments. The fifth column gives the number of fragments produced in the bar. The last two rows list the relevant results of the finite 
difference simulations of Truelove {\it et al}. \shortcite{TrueloveApJ1998} and Klein {\it et al}. \shortcite{KleinTOKYO1998}. For these two rows, we have used the 
initial and the final number of cells instead of the number of particles in columns 2 and 13 respectively.}
\label{tab:nest_rot2}
\end{sidewaystable}

The first two tests (\S \ref{sec:ff_collapse} \& \ref{sec:stable}) have proven that 
the new method for calculating $h$ (\S \ref{sec:newh}) has greatly reduced the 
boundary perturbations introduced by the application of particle splitting to the 
regions where fine and coarse particles are in contact. In particular, the collapse 
simulations (\S \ref{sec:ff_collapse}) show that there is no outward propagation of 
the fine region boundary in cases of inflow, which are the flows usually investigated in 
simulations of Star Formation. It also shows that the fine region reproduces the
expected density profiles. 

The simulation of the evolution of a stable isothermal 
sphere (\S \ref{sec:stable}) shows that there is a distinct boundary between the fine 
and the coarse regions. The boundary may expand in cases of quiescent evolution of 
stable spheres, but the overall density profile accurately obeys the predicted one, 
irrespectively of the size of the fine region. Both tests have revealed that the on the 
fly version of particle splitting is much more efficient in terms of computational cost. 
In principle, nested splitting could be superior to on-the-fly splitting if the boundary 
perturbations mentioned in \S \ref{sec:newh} persisted after the implementation of the new 
method for calculating $h$, as in nested splitting the fine region is larger than it 
really needs to be. This could prevent the boundary effects from propagating inwards and 
corrupting the simulation. However, both tests show that the boundary effects 
of both versions of the new method have been eliminated to a large extent and that the 
insignificant errors produced by particle splitting are similar in both versions of the 
method. Therefore, on-the-fly splitting can be safely used as it is superior in terms of 
computational cost.

We have also applied particle splitting to collapse simulations like those of 
the previous chapter. We conclude that the simulations with 40,000 particles initially, 
that use a sufficiently high density for the switch to adiabatic heating ($\rho_{0}$=10$^{-12}$ 
g cm$^{-3}$) while obeying the Jeans condition at all times, 
reproduce the results of Truelove {\it et al.} \shortcite{TrueloveApJ1998},
of Klein {\it et al}. \shortcite{KleinTOKYO1998}, and of our highest resolution 
simulation of \S \ref{sec:600kptcls} (600,000), but with only $\sim$200,000 
particles in total at the final stages. The resolution of these simulations 
is equivalent to the resolution of a simulation with 520,000 particles, 
which clearly indicates the efficiency of particle splitting. In particular, with particle splitting we achieved $\sim 60\%$ economy in memory and $\sim 70-75\%$ economy in CPU. A summary 
of our results is given in Tables \ref{tab:nest_rot} \& \ref{tab:nest_rot2}. 

With particle 
splitting the simulation obeys the Jeans condition at all times with great 
computational gain. Our new method has succeeded in meeting the objectives 
we have set in \S \ref{sec:concept}. We can now apply it to simulations of 
clump-clump collisions. 

On-the-fly splitting is more efficient than nested splitting, as even 
less particles are required. This is due to the fact that in on-the-fly splitting no 
particles are split unnecessarily. On-the-fly splitting is our preferred version of 
the new method. We will use on-the-fly splitting in simulations of clump-clump collisions in the next chapter.

In all simulations presented here, the Jeans condition is always obeyed and therefore 
there are no artificial fragments. This demonstrates the Jeans condition for 
fragmentation provides a very strong test for the significance of numerical results. 

However, with the simulations with 10,000 initially and $\rho_{0}$=10$^{-13}$ g cm$^{-3}$, 
we have shown that particle splitting in response to the imminent violation of the Jeans condition is 
a necessary, but not a sufficient, condition for the reliability and efficiency of a simulation. 
In particular, the low resolution of the initial stages of the cloud evolution 
has prevented the nested splitting simulation from reproducing the result of 
finite difference simulations of the same problem. It has also prevented the 
on-the-fly splitting simulation from obtaining this result efficiently.

\chapter[Clump-Clump Collisions]{On-the-fly Splitting Simulations of Clump-Clump Collisions}{\label{sec:results}}

Simulations of cloud-cloud collisions 
\cite{ChapmanNATURE1992,PongracicMNRAS1992,TurnerMNRAS1995,WhitworthMNRAS1995,BhattalMNRAS1998} evolve over several 
orders of magnitude in density and involve fragmentation to produce protostellar objects. Therefore, it 
is essential that fragmentation is properly modelled, i.e. the resolution is sufficient that the 
Jeans condition is always obeyed. The particle splitting simulations of the standard test with only 
10,000 particles initially (\S \ref{sec:rot_cloud}) have shown that the Jeans condition is a necessary but not sufficient condition for reliable simulations; in that case the initial expansion 
and the subsequent collapse of the cloud were not properly modelled. 

The cloud-cloud collision simulations of Bhattal {\it et al}. \shortcite{BhattalMNRAS1998} were conducted with 
small numbers of particles (2,000-8,000 per cloud) and implemented with a constant gravity 
softening, $\epsilon$. However, small numbers of particles lead to violation of the Jeans condition and inaccurate modelling of the density field (\S \ref{sec:rot_conclusi}, 
\ref{sec:neighbours} \& \ref{sec:jeans_cond}). Also Bate \& Burkert 
\shortcite{BateMNRAS1997} have demonstrated that the implementation of constant $\epsilon$-softening can be very 
misleading for SPH calculations with gravity, as it can both artificially induce fragmentation of 
Jeans stable lumps and/or stabilise Jeans unstable lumps against collapse, depending on the 
$\epsilon / h$ ratio (\S \ref{sec:jeans_cond}).

Therefore, it is worth conducting 
new high resolution simulations in which the Jeans condition is explicitly monitored and obeyed, in order to obtain a more detailed understanding of the processes involved. 
In this chapter, we perform such high resolution simulations of clump-clump collisions by applying 
particle splitting, and in particular the on-the-fly splitting version. 

In chapter \ref{sec:part-split}, we have shown that on-the-fly 
splitting is both a reliable and a computationally efficient method. 
Therefore, with on-the-fly splitting, we can overcome the problem of insufficient resolution for modelling
fragmentation and so we can obtain credible results. 
We use $h$-softening, i.e. we set 
$\epsilon=h$. All particles have an individual time-varying $h$. 

We perform three simulations starting from the initial conditions used by Bhattal {\it et al}. 
\shortcite{BhattalMNRAS1998} in order to make direct comparisons with their results, and hence to
quantify the effects of low resolution and $\epsilon$-softening on those results (\S \ref{sec:amar}). 
We also investigate fragmentation induced by collisions 
between low-mass clumps by performing a series of simulations with various combinations of relative velocity and impact parameter 
(\S \ref{sec:clump}). We note that collisions of clumps from the low-mass end 
of the clump mass function have not been investigated before. 

Previous collision simulations have shown that all fragmentation events happen in 
the shocked gas layer formed at the interface of the collision. It is therefore very important 
that the formation of this layer is properly modelled. In the following section, we define the 
appropriate minimum initial number of particles for our simulations to obey the Jeans condition 
at all times, and to resolve the formation of this shock. We also define the physical and 
numerical initial conditions for our simulations. 

\section{Initial conditions}{\label{sec:clump_initial}}

\begin{sidewaystable}
\begin{center}
\begin{tabular}{|c|c|c|c|c|c|c|c|c|}\hline
$\mathcal{M}$ & $M_{0}/\mathrm{M}_{\odot}$ & $N_{frag}^{min}$ & $N_{\rho_{0}}^{min}$ & $M_{frag}/\mathrm{M}_{\odot}$ & $a_{0}$/km s$^{-1}$ & $T$/K & $v_{coll}$/km s$^{-1}$ & $v_{cloud}$/km s$^{-1}$\\ \hline \hline
5 & 1 & 224 & 1,410 & 0.45 & 0.2 & 10 & 1.0 & 0.5\\ \hline
  & 10 & 1,258 & 14,099 & 0.77 & 0.36 & 32 & 1.0 & 0.5\\ \hline
  & 75 & 5,699 & 105,739 & 1.31 & 0.59 & 87 & 1.0 & 0.5\\ \hline
  & 100 & 7,072 & 140,985 & 1.43 & 0.63 & 100 & 1.0 & 0.5\\ \hline
9 & 1 & 300 & 1,410 & 0.33 & 0.2 & 10 & 1.8 & 0.9\\ \hline
  & 10 & 1,688 & 14,099 & 0.57 & 0.36 & 32 & 1.8 & 0.9\\ \hline
  & 75 & 7,646 & 105,739 & 0.97 & 0.59 & 87 & 1.8 & 0.9\\ \hline
  & 100 & 9,487 & 140,985 & 1.07 & 0.63 & 100 & 1.8 & 0.9\\ \hline
10 & 1 & 317 & 1,410 & 0.32 & 0.2 & 10 & 2.0 & 1.0\\ \hline
   & 10 & 1,779 & 14,099 & 0.54 & 0.36 & 32 & 2.0 & 1.0\\ \hline
   & 75 & 8,060 & 105,739 & 0.92 & 0.59 & 87 & 2.0 & 1.0\\ \hline
   & 100 & 10,000 & 140,985 & 1.01 & 0.63 & 100 & 2.0 & 1.0\\ \hline
15 & 1 & 388 & 1,410 & 0.26 & 0.2 & 10 & 1.0 & 0.5\\ \hline
   & 10 & 2,178 & 14,099 & 0.44 & 0.36 & 32 & 1.0 & 0.5\\ \hline
   & 75 & 9,871 & 105,739 & 0.75 & 0.59 & 87 & 1.0 & 0.5\\ \hline
   & 100 & 12,248 & 140,985 & 0.83 & 0.63 & 100 & 1.0 & 0.5\\ \hline
\end{tabular}
\end{center}
\caption[Minimum initial numbers of particles $N_{frag}^{min}$ \& $N_{\rho_{0}}^{min}$ for different 
        values of $M_{0}$ and $\mathcal{M}$]{Minimum initial numbers of particles $N_{frag}^{min}$ 
        \& $N_{\rho_{0}}^{min}$ for different values of $M_{0}$ and $\mathcal{M}$ so that the Jeans 
        condition is obeyed at all times. In all cases, we have used $N_{n}$=50. The fifth column lists 
        the mass of the fragments produced in the layer.}
\label{tab:clump_initial}
\end{sidewaystable}

When two clouds collide, a layer of shocked gas forms at the interface between the clouds 
\cite{ChapmanNATURE1992,PongracicMNRAS1992,TurnerMNRAS1995,WhitworthMNRAS1995,BhattalMNRAS1998}. The layer eventually 
fragments producing groups of protostars. Whitworth {\it et al}. \shortcite{WhitworthAnA1994} have 
shown that the mass of a fragment, $M_{frag}$, produced in such a shocked layer is given by

\begin{equation}
\label{equa:fragmentmass}
M_{frag} \sim \frac{a_{s}^{3}}{(G^{3} \rho_{cloud} \mathcal{M})^{1/2}},
\end{equation}

\noindent where $a_{s}$ is the effective sound speed of the gas in the shocked layer, $\rho_{cloud}$ is the initial 
pre-shock density of each one of the two clouds involved in the collision and $\mathcal{M}$ is the Mach 
number of the collision, defined as the ratio of the relative velocity of the clouds, $v_{coll}$, 
to $a_{s}$ ($v_{coll}=2 v_{cloud}$, where $v_{cloud}$ is the bulk velocity of each cloud). 

The pre-shock density of each cloud, $\rho_{cloud}$, is given by 

\begin{equation}
\label{equa:cloudensity}
\rho_{cloud} \sim \frac{a_{0}^{6}}{G^{3} M_{0}^{2}},
\end{equation}

\noindent if we make the assumption that the clouds are in hydrostatic balance (i.e. in an equilibrium state). $M_{0}$ is the cloud mass and $a_{0}$ is the 
effective pre-shock sound speed, i.e. a sound speed which represents thermal and turbulent pressure. 

Combining Eqns. \ref{equa:fragmentmass} \& \ref{equa:cloudensity} we obtain

\begin{equation}
\label{equa:fragmentmass2}
M_{frag} \sim \left( \frac{a_{s}}{a_{0}} \right)^{3} M_{0} \mathcal{M}^{-1/2}.
\end{equation}

In order for the formation of these fragments to be resolved, their mass must be larger, or of the order of, 
the minimum mass resolvable by SPH, 2$M_{min}$. Substituting $M_{min}$ = $N_{n} m_{ptcl}$ we obtain,

\begin{equation}
\label{equa:fragmentmass3}
\left( \frac{a_{s}}{a_{0}} \right)^{3} M_{0} \mathcal{M}^{-1/2} \gtrsim 2 N_{n} \frac{M_{0}}{N_{total}},
\end{equation}

\noindent and hence

\begin{equation}
\label{equa:fragmentmass4}
N_{total} \gtrsim 2 N_{n} \left( \frac{a_{0}}{a_{s}} \right)^{3} \mathcal{M}^{1/2}.
\end{equation}

Using the scaling relation, 
\[ a_{0} \sim 0.2 \; \mathrm{km} \; \mathrm{s}^{-1} \left( \frac{M_{0}}{\mathrm{M}_{\odot}} \right)^{1/4} \] 
({\it cf}. Larson \shortcite{LarsonMNRAS1981}), we finally obtain

\begin{equation}
\label{equa:numberofptcls}
N_{total} \gtrsim 2 N_{n} \left( \frac{M_{0}}{\mathrm{M}_{\odot}} \right)^{3/4} \mathcal{M}^{1/2},
\end{equation}

\noindent where we have used $a_{s}$=0.2 km s$^{-1}$ for the post-shock sound speed, or equivalently 
$T_{s}$=10 K for the post-shock temperature. This is a reasonable assumption since for typical post-shock 
densities, the gas cools very rapidly and efficiently down to 
10 K. The third column of Table \ref{tab:clump_initial} lists the minimum initial numbers of particles, $N_{frag}^{min}$, 
needed to resolve fragmentation in the shocked layer for the different values of 
$M_{0}$ and $\mathcal{M}$. In \S \ref{sec:tube}, we have shown that the shock is adequately modelled with our code for $v_{cloud}$=0.5 km s$^{-1}$ ($\mathcal{M}$=5).

Above $\rho_{0} = 10^{-14}$ g cm$^{-3}$ the gas heats up adiabatically, as described in chapter \ref{sec:SPH}. In chapter 
\ref{sec:rotating}, we have used different values for $\rho_{0}$ in order to explore the performance 
of our code under different circumstances. The 
value of $\rho_{0}$ above which adiabatic heating switches on is defined by where heating due to gravitational collapse exceeds the rate of cooling by dust -- because the dust becomes optically thick to its own 
radiation (e.g. Whitworth {\it et al}. \shortcite{WhitworthMNRAS1998b}). $\rho_{0}$ is 
given by \[ \rho_{0} \sim \dfrac{\bar{m}^{3/2} (kT)^{5/2}}{h^{3} c^{2}}, \] where $h$ and $k$ are 
the Planck and Boltzmann constants respectively and $c$ is the speed of light. If we substitute the values 
of the constants, $T$=10 K and assume that the gas is molecular, then we arrive to a 
value for $\rho_{0} \sim 10^{-14}$ g cm$^{-3}$.

In order for the Jeans condition to be obeyed not just for the first fragments formed in the 
layer but also up to $\rho_{0}$ = 10$^{-14}$ 
g cm$^{-3}$, we need 

\begin{equation}
\label{equa:numberofptcls2}
N_{total} \gtrsim \dfrac{12 N_{n} M_{0} G^{3/2} \rho_{0}^{1/2}}{13 \pi^{5/2} a_{s}^{3}}.
\end{equation}

\noindent (Eqn. \ref{equa:Jeans_condition2}). To find the minimum initial number of particles, $N_{\rho_{0}}^{min}$, needed for 
the Jeans condition to be obeyed in the layer up to $\rho_{0}$ = 10$^{-14}$ g cm$^{-3}$, we 
divide by 13 the number of particles given by Eqn. \ref{equa:Jeans_condition2}, as we can increase the number 
of particles 13-fold by using particle splitting once. The fourth column of Table \ref{tab:clump_initial} lists 
the values of $N_{\rho_{0}}^{min}$ for the different values of $M_{0}$.

From Table \ref{tab:clump_initial} we observe that $N_{\rho_{0}}^{min}$ is always larger than $N_{frag}^{min}$ 
so that we always use $N_{\rho_{0}}^{min}$ as the initial number of particles in our simulations. In all cases, 
we use $N_{n}$=50. 

Collisions between low-mass clumps ($M_{0}$=10 M$_{\odot}$) seem to be the most computationally efficient to 
explore as they only need $\sim$15,000 particles initially for the Jeans condition to be obeyed at all times. Collisions between such clumps 
have not been investigated before. In all cases, two equal mass clumps of 10 M$_{\odot}$ collide with 
opposite bulk velocity vectors. In molecular clouds with hierarchical substructure, collisions between clumps 
at the same level of the hierarchy (i.e. with comparable mass) are the most probable \cite{ScaloBLACK1985}. 

We have conducted simulations with $\mathcal{M}$=5, 10, and 15 and with $b$, the ratio of the impact parameter to the radius of the cloud, satisfying $b$=0.0, 0.2, 0.4, 0.6 and 0.8. Whitworth {\it et al}. 
\shortcite{WhitworthMNRAS1994} have argued that the equations used to estimate  $M_{frag}$ are approximately valid not only for head-on 
collisions ($b$=0.0) but also for collisions with $b<$0.7. The $b$=0.8 collisions are conducted only for 
completeness, as the clumps do not interact strongly, even in the slowest collision ($\mathcal{M}$=5). 

We also repeat three simulations using the initial conditions of Bhattal {\it et al}. 
\shortcite{BhattalMNRAS1998} with the Jeans condition obeyed at all times (seventh row in Table 
\ref{tab:clump_initial}). Our aim is to make a direct comparison of our results with theirs, in order to be 
able to quantify the effect of fragmentation being properly modelled and the 
benefit of using particle splitting. We perform the simulations with $b$=0.2, 0.4 and 
0.5, as the first and the third ones are categorised by Bhattal {\it et al}. as having different mechanisms in 
play for the formation of binary/multiple protostellar systems (rotational instability in a disc {\it vs}. 
fragmentation of the layer). We also perform the second one as Bhattal {\it et al}. claim that it combines both mechanisms.

From Table \ref{tab:clump_initial} we see that for 10 
M$_{\odot}$ clumps, the pre-shock sound speed $a_{0} \sim$0.35 km s$^{-1}$ so that the pre-shock temperature 
of the clumps is $T_{0} \sim$35 K. For typical post-shock densities cooling of the gas is very rapid and efficient so that it cools down to 
10 K in the shocked layer very quickly. In order to be able to model this behaviour, we have slightly altered our barotropic equation of state (Eqn. \ref{equa:state}) so that

\begin{equation}
\label{equa:statecool}
\frac{P({\bf r})}{\rho({\bf r})} \; = \; \left \{ 
   \begin{array}{ll}
      a_{0}^{2}, & \rho \leqslant \rho_{1};\\
      \left( (a_{0}^{2} - a_{s}^{2}) \left( \frac{\rho}{\rho_{1}} \right)^{-2/3} + a_{s}^{2} \right) \left[1 + \left( \frac{\rho({\bf r})}{\rho_{0}} \right) ^{4/3} \right]^{1/2}, & \rho > \rho_{1},
   \end{array} \right .
\end{equation}

\noindent where $a_{0} \sim$0.35 km s$^{-1}$, $a_{s} \sim$0.2 km s$^{-1}$ and $\rho_{1}$=1.2 x 10$^{-20}$ 
g cm$^{-3}$. $\rho_{1}$ is the density at which cooling initiates \cite{TohlineCANUTO1982}. This equation 
of state can be divided 
into four regimes. An isothermal regime with $T_{0}$=35 K for densities below $\rho_{1}$. A cooling stage, 
approximating to $T \propto \rho^{-2/3}$ for densities just above $\rho_{1}$, until the temperature reaches 
$T_{s}$=10 K. The temperature then remains constant at $T_{s}$=10 K for densities below $\rho_{0}$. Finally, 
the gas heats up adiabatically above $\rho_{0} = 10^{-14}$ g cm$^{-3}$.

We can summarise our initial conditions as follows. Two equal mass clumps collide with equal but opposite 
velocities. In repeating the simulations of Bhattal {\it et al}. we use $M_{0}$=75 M$_{\odot}$, $a_{0} \sim$0.6 km s$^{-1}$ (corresponding to $T_{0}$=100 K); $\mathcal{M}$=9 (corresponding to $v_{cloud}$=0.9 
km s$^{-1}$), $b$=0.2, 0.4 and 0.5, and 110,000 particles 
per clump. In the second set of simulations, which we call low-mass clump collisions, we use $M_{0}$=10 M$_{\odot}$, $a_{0} \sim$0.35 km s$^{-1}$ (corresponding to $T_{0}$=35 K); $\mathcal{M}$=5, 10, 15 (corresponding to $v_{cloud}$=0.5, 1.0 and 1.5 km s$^{-1}$), $b$=0.0, 0.2, 0.4, 0.6 and 0.8, and 15,000 particles 
per clump.  

\begin{figure}
\begin{center}
\resizebox{4.55cm}{!}{\includegraphics{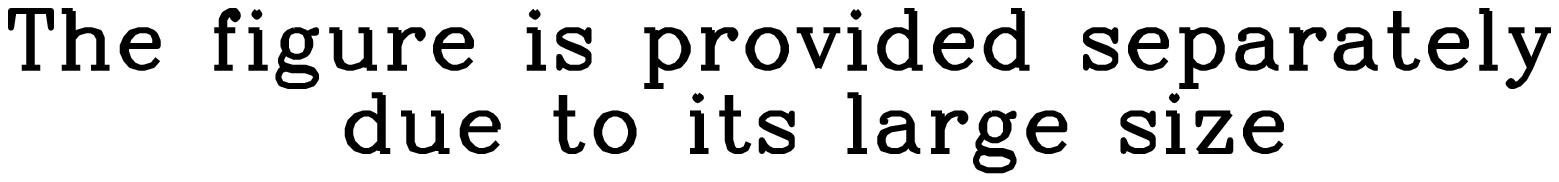}}
\resizebox{8cm}{!}{\includegraphics{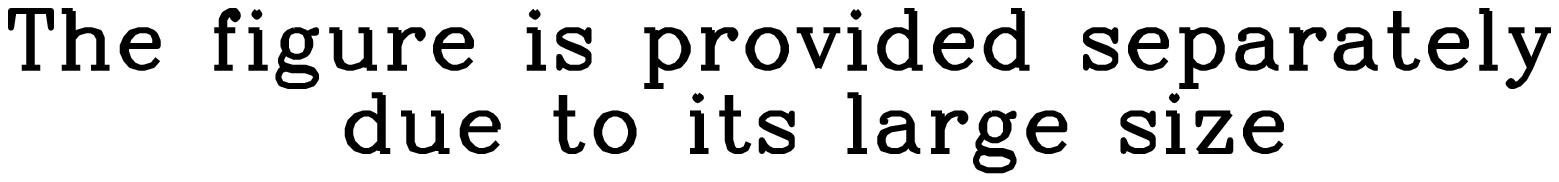}}
\end{center}
      \caption[Column density plots for the $b$=0.2 collision of two 75 M$_{\odot}$ clumps]
      {Column density plots for the $b$=0.2 collision of two 75 M$_{\odot}$ clumps at the end 
      of the simulation, $t$ $\sim$0.64 Myr. 
      \underline{Left}: Column density plot viewed along the z-axis. The linear 
      size of this plot is 0.04 pc. The colour table has units of 1.18 x 10$^{6}$ g cm$^{-2}$. 
      \underline{Right}: Column density plot viewed along the y-axis. The linear 
      size of this plot is 0.028 pc. The colour table has units of 2.41 x 10$^{6}$ g cm$^{-2}$.}
      \label{fig:amar2}
\end{figure}

\begin{figure}
\begin{center}
\resizebox{8cm}{!}{\includegraphics{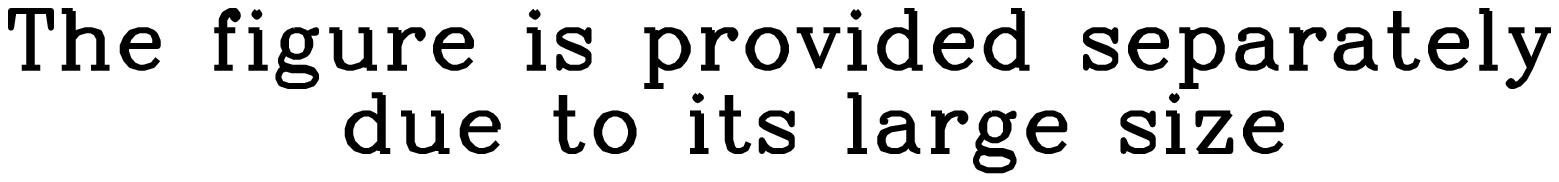}}
\end{center}  
      \caption[Column density plot for the $b$=0.2 collision of two 75 M$_{\odot}$ clumps showing 
      the network of filaments produced]
      {Column density plot for the $b$=0.2 collision of two 75 M$_{\odot}$ clumps viewed along the 
      y-axis, at the end of the simulation, $t$ $\sim$0.64 Myr, showing the network of filaments 
      produced. The linear size of this plot is 0.2 pc. The colour table has units of 
      4.73 x 10$^{4}$ g cm$^{-2}$.}
      \label{fig:amar2fil}
\end{figure}

In all cases, the clumps are stable isothermal spheres made with a procedure similar to that of \S 
\ref{sec:test_stable}. The particles, before they are moved to their final positions, are relaxed to uniform 
density, as they were initially taken from a random distribution (this relaxed distribution is what we called in previous chapters a 
``settled'' distribution). The colliding clumps are taken to be touching before the simulations start, to save 
computational time. During the preceding approach of the colliding clouds, mutual tidal distortion will be small because $\mathcal{M} \gtrsim 5$. We use our full self-gravitating 
SPH code (chapter \ref{sec:SPH}) with 
the above barotropic equation of state (Eqn. \ref{equa:statecool}); we use 
$h$-softening with our TCG method 
(\S \ref{sec:TCG}); we use $N_{n}$=50; and we 
use on-the-fly splitting at $\rho_{max}$ = $\rho_{0} / 13^{2}$ = 6.0 x 10$^{-17}$ g cm$^{-3}$, as at this stage the Jeans condition would 
otherwise stop being obeyed. After application of particle splitting, adiabatic heating starts before the fine simulation reaches 
its resolution limit. Therefore, the Jeans condition is obeyed all the way up to the highest densities we can achieve.

We use false-colour column density plots to present our results. All structure formed is contained within a single 
layer and therefore such plots are not greatly confused by projection effects. Column density plots are 
preferred to particle plots as the former give a more accurate representation of
the total density 
field, and of what would be seen in optically thin molecular-line (or dust-continuum) radiation, assuming a uniform excitation temperature (or dust temperature). 
We do not use contour plots, as we believe that colour column density plots are easier to read and give more information. The figure captions give the linear size 
of each plot and the plane onto which mass is projected. They also give the time of the snapshot they represent and the units of the colour 
table. We have taken the clouds always to collide along 
the x-axis, with offset parallel to the y-axis. 

In all cases, when a fragment forms, its linear density profile appears like a normal distribution around the 
peak density. At the end of our simulations, this profile is very steep. To infer its mass, we take the fragment to extend out to 
3-$\sigma$, where $\sigma$ is the FWHM, and we measure the mass within this radius. This method gives good results for the mass and the radius of spherical as well as disc-like objects.

\section{Repeating the Simulations of Bhattal {\it et al}.}{\label{sec:amar}}

\subsection{Simulation with $M_{0}$=75 M$_{\odot}$ and $b$=0.2}{\label{sec:amar2}}

Bhattal {\it et al}. \shortcite{BhattalMNRAS1998} found that for low-$b$ collisions a single spherical 
protostellar object formed at the centre of the collision and then accreted material from the spindle-shaped 
filament. The protostellar object soon became disc-like and grew in mass and angular momentum as the offset between the opposing accretion flows along the spindle increased. Several accretion-induced rotational instability events took place as the primary 
protostellar disc ejected material and angular momentum through spiral arms. The ejected material 
interacted with the accretion flows and produced multiple lower mass companions to the massive primary. 
The primary had mass between 20-60 M$_{\odot}$. The low-mass companions formed were not always resolved 
properly, as in some cases they contained less than 50 particles. Bhattal {\it et al}. used a constant 
$\epsilon$-softening with $\epsilon \sim$500 AU, so that they could not resolve the formation of an 
object until its size became $\gtrsim$500 AU. 

Our simulation evolves with very small time-step due to the large number of particles. After about 10,000 
time-steps, it becomes extremely inefficient to continue the simulation, as even if the time-step did not 
decrease any further, after another 10,000 time-steps the time would only progress by about 1\% of the total 
time that far. Therefore, we have terminated our simulation at that point, after 0.64 Myr. There are 
$\sim$260,000 particles at the end.

The fact that our simulation has to stop very early prevents us from presenting figures with the time 
evolution of properties such as the mass, angular momentum and radius of fragments or the binding energies of 
groups of protostars. 

After $\sim$0.61 Myr, a network of filaments forms within the shocked layer. Protostellar objects start 
forming at the intercepts of the filaments. On-the-fly splitting is applied at $\sim$0.63 Myr. In total, four 
objects have formed via fragmentation of the filaments when the simulation terminates. Two of them quickly merge. At the end of the 
simulation ($\sim$0.64 Myr), three disc-like objects are still accreting matter from the filaments. They are already centrally condensed. They 
are rotating fast and are attended by thin spiral arms. An offset develops between the opposing accretion flows and 
the inflowing material steadily increases the specific angular momentum of the discs. They are spinning increasingly fast. At this early stage of the evolution there is no sign of interaction between the spiral arms and the accretion flows. 
We believe that secondary companions to the discs may eventually form in our simulation via interaction between the spiral arms and the accretion flows. However, we note that the 
accretion flows in our simulation are extremely homogeneous, and not lumpy as they are in the simulation of Bhattal {\it et al}. 

The three discs are still in the sub-solar mass range, in accordance with the mass evolution of Fig. 3 of 
Bhattal {\it et al}. Specifically, the masses of the three disc-like objects are $m_{1}$=0.35 M$_{\odot}$, 
$m_{2}$=0.30 M$_{\odot}$ and $m_{3}$=0.19 M$_{\odot}$. They all extend out to $\sim$120 AU. Their central 
densities have reached $\rho_{1}^{peak}$=2.7 x 10$^{-12}$ g cm$^{-3}$, $\rho_{2}^{peak}$=5.5 x 10$^{-13}$ 
g cm$^{-3}$ and $\rho_{3}^{peak}$=6.8 x 10$^{-14}$ g cm$^{-3}$, which implies that the temperature at their centres is increasing. Their separation is of the order of 10$^{4}$ AU and they appear to be a weakly bound system at this stage. The left panel of Fig. \ref{fig:amar2} is 
a column density plot viewed along the z-axis and the right panel of Fig. \ref{fig:amar2} is a column 
density plot viewed along the y-axis showing the three protostellar discs at the end of the simulation.

Although it is possible that rotational instabilities may subsequently produce secondaries to the objects formed in our 
simulations, the initial formation of multiple protostars is due to {\em layer 
fragmentation}; this is not present in the simulation of Bhattal {\it et al}. The total mass of the three 
fragments of our simulation is of the same order of magnitude as the mass of the central object in the simulation of 
Bhattal {\it et al}., at corresponding times. 

Moreover, a network of 3-4 filaments is produced in the shocked layer of our simulation; Bhattal {\it et al}. 
do not produce a network. At the end of the simulation, the minimum density of the filaments is 
$\rho_{fil}$=3.0 x 10$^{-17}$ g cm$^{-3}$, which corresponds to $n_{H_{2}} \gtrsim$5 x 10$^{6}$ cm$^{-3}$ (Fig. \ref{fig:amar2fil}). 

We presume that the evolution of the density field 
is followed better in our simulation. The fact that the Jeans condition is obeyed, prevents a central object from 
forming artificially before the filaments. Since with on-the-fly splitting there is no preferred length scale, 
we can trust the detailed evolution of the discs themselves as well as their dynamical interaction. However, 
because of the large number of particles, the time-step decreases rapidly and we have to stop the simulations 
at a very early stage before the group of protostellar discs can evolve into a bound system.

\subsection{Simulation with $M_{0}$=75 M$_{\odot}$ and $b$=0.4}{\label{sec:amar4}}

For this simulation, Bhattal {\it et al}. found that a multiple system formed both via fragmentation of a single 
filament in the shocked layer and via rotational instability in the central, and more massive, fragment 
formed in the filament. The fragments formed a bit later than the central object in their $b$=0.2 collision. 

Their findings are repeated in our simulation with the exception of the rotational instability event. 

As in the $b$=0.2 collision, we terminate this simulation after $\sim$0.69 Myr, since the time-step has 
decreased considerably and continuing becomes computationally inefficient. The number of particles has 
increased to 260,000 at the end of the simulation.

\begin{figure}
\begin{center}
\resizebox{8cm}{!}{\includegraphics{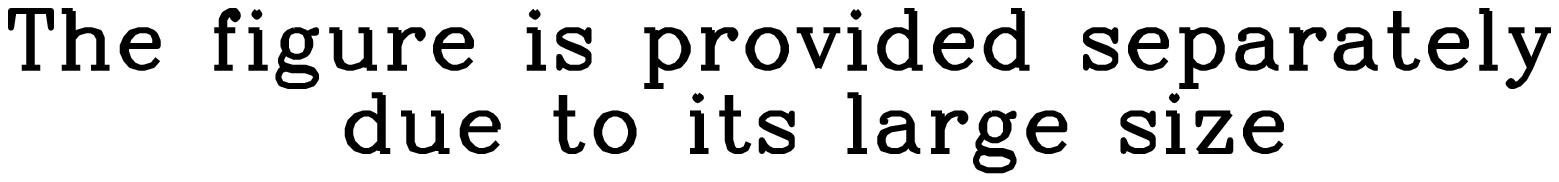}}
\end{center}  
      \caption[Column density plot for the $b$=0.4 collision of two 75 M$_{\odot}$ clumps]
      {Column density plot for the $b$=0.4 collision of two 75 M$_{\odot}$ clumps viewed along the 
      z-axis at the end of the simulation, $t$ $\sim$0.69 Myr. The linear size of this plot is 0.008 pc. 
      The colour table has units of 2.95 x 10$^{7}$ g cm$^{-2}$.}
      \label{fig:amar4}
\end{figure}

\begin{figure}
\begin{center}
\resizebox{8cm}{!}{\includegraphics{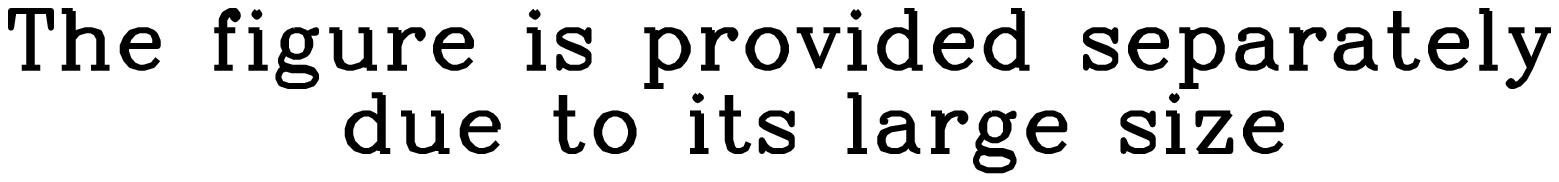}}
\end{center}  
      \caption[Column density plot for the $b$=0.5 collision of two 75 M$_{\odot}$ clumps]
      {Column density plot for the $b$=0.5 collision of two 75 M$_{\odot}$ clumps viewed along the 
      z-axis at the end of the simulation, $t$ $\sim$0.75 Myr. The linear size of this plot is 0.02 pc. 
      The colour table has units of 4.73 x 10$^{6}$ g cm$^{-2}$.}
      \label{fig:amar5}
\end{figure}

A single filament forms perpendicular to the collision axis, at $\sim$0.66 Myr. At the same time, on-the-fly splitting initiates. Four objects have formed in the filament by 
$t$ $\sim$0.69 Myr and two more 
have started condensing out of the filament. The masses of the four objects are $m_{1}$=0.33 M$_{\odot}$, 
$m_{2}$=$m_{3}$=$m_{4}$=0.02 M$_{\odot}$. The more massive object is disc-like with a radius of $\sim$170 AU. 
The three smaller ones are still spherical with radii of $\sim$45 AU. Their central 
densities have reached $\rho_{1}^{peak}$=2.0 x 10$^{-12}$ g cm$^{-3}$, $\rho_{2}^{peak}$=3.3 x 10$^{-13}$ 
g cm$^{-3}$, $\rho_{3}^{peak}$=1.3 x 10$^{-13}$ g cm$^{-3}$ and $\rho_{4}^{peak}$=8.2 x 10$^{-14}$ g cm$^{-3}$. 

The four objects are formed in random positions along the filament and are falling together. They 
would presumably interact if the simulation were continued further.  

The more massive object rotates 
fast and might well become rotationally unstable at a later stage. It forms first and, after becoming self-gravitating, it evolves with the minimum time-step. 
This is why the other fragments seem to evolve rather slowly compared to it. 

Fig. \ref{fig:amar4} shows the four objects within the filament at the end of the simulation. At this stage, 
the minimum density of the filament is $\rho_{fil}$=2.0 x 10$^{-17}$ g cm$^{-3}$, corresponding to $n_{H_{2}} \gtrsim$5 x 10$^{6}$ cm$^{-3}$.

\subsection{Simulation with $M_{0}$=75 M$_{\odot}$ and $b$=0.5}{\label{sec:amar5}}

For this collision, Bhattal {\it et al}. found that a single filament formed diagonally in the x-y plane. The 
filament subsequently produced several fragments that merged. After $\sim$1 Myr, a binary with equal 
mass components was spiraling in towards a possible merger. The filament fragmented later than in their $b$=0.4 collision. When they repeated their simulation with higher resolution, they resolved each binary component to 
be a binary itself, i.e. an hierarchical quadruple. 

With our simulation, we confirm their results. We stop the simulation 
at $\sim$0.75 Myr, when it involves $\sim$342,000 particles. The time-step decreases so much that it is inefficient to continue any further. Particle splitting initiates at 
$\sim$0.72 Myr. At $\sim$0.73 Myr the filament forms. It soon starts fragmenting into several objects.
Two of them, at the bottom right hand corner of Fig. \ref{fig:amar5}, condense out first and eventually 
collapse faster, thereby dominating the smallest time-step bins. At $\sim$0.746 Myr they merge. At the end of the 
simulation, another three objects appear to be forming (the two green lumps at the centre of Fig. \ref{fig:amar5}, plus another one out of Fig. 
\ref{fig:amar5}, above and to the left).

All fragments remain spherical and rotate but considerably slower than in the two previous simulations. In particular, they become very centrally condensed. Some 
other density enhancements appear in the filament suggesting that more fragments may be produced (via fragmentation of the filament). The masses of 
the four objects at the final stage are $m_{1}$=0.46 M$_{\odot}$, $m_{2}$=0.07 M$_{\odot}$, $m_{3}$=0.06 
M$_{\odot}$ and $m_{4}$=0.04 M$_{\odot}$. The most massive one has radius $\sim$35 AU and the radius of the other 
three is $\sim$100 AU. Their central densities are $\rho_{1}^{peak}$=1.6 x 10$^{-11}$ g cm$^{-3}$, 
$\rho_{2}^{peak}$=7.7 x 10$^{-13}$ g cm$^{-3}$, $\rho_{3}^{peak}$=6.0 x 10$^{-13}$ g cm$^{-3}$ and 
$\rho_{4}^{peak}$=4.6 x 10$^{-14}$ g cm$^{-3}$. The minimum density of the filament is $\rho_{fil}$=2.0 x 
10$^{-18}$ g cm$^{-3}$, corresponding to $n_{H_{2}} \gtrsim$5 x 10$^{5}$ cm$^{-3}$.

\subsection{Conclusions}

Apart from the $b$=0.2 simulation, our results are in general agreement with those of Bhattal {\it et al}. 
\shortcite{BhattalMNRAS1998}. In the $b$=0.2 collision, we do not obtain a central object. Instead, a
network of filaments forms, the filaments fragment, and a group of protostellar discs forms at the 
intercepts of the filaments. We believe that with our simulation, where fragmentation is properly modelled, 
we obtain a more realistic evolution for the shocked layer. 

We conclude that in all cases {\em fragmentation of the filaments} produces the seed for a group of protostars. 
Accretion-induced rotational instability can then produce secondaries to the protostars formed. Our simulations have to stop at an early stage due to the small time-step with which 
the simulation progresses; the time-step drops due to the large number of particles. Thus, we can not 
confirm the development of rotational instabilities. We expect that such instabilities may develop at a later 
stage when the angular momentum of the discs increases further. We anticipate that accretion-induced 
rotational instabilities are more probable with small $b$ ($b < $0.5). 

The filaments fragment later with increasing $b$, as the leading edge of each colliding clump has to 
travel more into the other clump before the shocked layer becomes massive enough. 

Bhattal {\it et al}. found that 
the primary mass was larger with decreasing $b$. One might argue that this is evident in our simulations 
as well, since the total mass of the primaries is largest in the $b$=0.2 collision ($\sim$0.85 
M$_{\odot}$). However, our evidence is somewhat inconclusive since only the fragments that collapse 
first are properly modelled; they dominate the small time bins and are evolved with the smallest 
time-steps while all other fragments are in suspended animation. 

The simulations of Bhattal {\it et al}. were of particularly low resolution and may have suffered 
from spurious fragmentation. Furthermore, we believe that since they could not properly model the 
density field (they could not resolve the filament formation in the $b$=0.2 collision), the primary 
masses they inferred were over-estimated. The primary protostars of our simulations are only a few ten thousand 
years old, since they form simultaneously with, or shortly after, the filaments. From our simulations a mass accretion rate $\gtrsim 10^{-5}$ M$_{\odot}$ yr$^{-1}$ can be inferred. 
With average mass accretion 
rates of 10$^{-5}$ M$_{\odot}$ yr$^{-1}$ for the first few hundred 
thousand years and 10$^{-6}$ M$_{\odot}$ yr$^{-1}$ subsequently until they become a million years old, 
the primaries will end up as 3-6 M$_{\odot}$ stars if they are not disturbed by rotational instability 
and/or mergers. If we consider as typical the 4-6 primary fragments that seem to be produced in our 
simulations, then 20-30 M$_{\odot}$ in total will end up in stars in such collisions. This 
would imply that the efficiency of star formation in such clump-clump collisions is less than, or of 
the order of, 20\% which is in accordance with the findings of Hunter {\it et al}. 
\shortcite{HunterApJ1986} from simulations of colliding gas flows. In fact, they suggest a star formation 
efficiency slightly higher than this, but since their flows move faster than our clumps, their shocks are 
stronger, and thus fragmentation of the resulting filaments may produce more primaries. The star formation efficiency estimated from our simulations is similar to that of most molecular 
clouds \cite{RengarajanApJ1985}. The efficiency inferred from the simulations of Bhattal {\it et al}. is 
of the order 30-40\% which is unrealistically high. 

Furthermore, we believe that the conclusion of Bhattal {\it et al}. that in low-$b$ collisions unequal 
mass systems are produced whereas in medium- to high-$b$ collisions equal mass systems occur, is not 
necessarily sound since they arrive at this conclusion having not resolved the filament formation in low-$b$ 
collisions or the subsequent filament fragmentation. 

In all collisions, our 
simulations end before we can have conclusive evidence on the protostellar systems formed being bound or 
not. In all cases, the accretion flows bring them closer together steadily. 

It is also interesting to note that in all cases the filaments produced have 
$n_{H_{2}} \gtrsim$10$^{6}$ cm$^{-3}$, which means that such filaments could be observed in molecular line radiation using 
tracers that are excited at such densities, e.g. NH$_{3}$, CS. 

Finally, we conclude that in low $b$ collisions, a network of filaments develops instead of the single spindle-like filament formed in medium to high $b$ collisions ($b \geqslant$0.4). In low $b$ collisions, more mass from the two clouds ends up in the shocked layer, as they collide almost head-on. This increases the surface area of the layer. 

\section{Low-mass clump collisions}{\label{sec:clump}}

\begin{sidewaystable}
\begin{center}
\begin{tabular}{|lr|p{35mm}|p{50mm}|p{45mm}|p{45mm}|p{17mm}|}\hline
 & $b$ & 0.0 & 0.2 & 0.4 & 0.6 & 0.8 \\ \hline
$\mathcal{M}$ & & & & & & \\ \hline \hline
5 & & One spherical rotating object. 0.33 M$_{\odot}$. No filaments. & One disc-like object. 0.7 M$_{\odot}$. Spiral arms. No companions. A single filament. & Two disc-like objects. 1.11 M$_{\odot}$ in total. Only more massive with spiral arms. Possible companions. Single filament. & Two well-separated rotating objects. 0.68 M$_{\odot}$ in total. Only more massive with spiral arms. Possible companions. No filaments. & No shock. \\ \hline 
10 & & One spherical rotating object. 0.35 M$_{\odot}$. Two filaments. & Two disc-like objects (+ a third forming). 1.0 M$_{\odot}$ in total. Both spiral arms. No companions. Network of filaments. & Single disc-like object (+ a second forming). 0.48  M$_{\odot}$. Spiral arms. Possible companions. Single filament. & No shock. & No shock. \\ \hline 
15 & & One disc-like rotating object. 0.47 M$_{\odot}$. No spiral arms. Network of filaments. & Two disc-like objects (+ a third forming). 0.96 M$_{\odot}$ in total. Only more massive with spiral arms. Possible companions. Well-defined network of filaments. & Single disc-like object. 0.31 M$_{\odot}$. Spiral arms. No companions. No filaments. & No shock. & No shock. \\ \hline 
\end{tabular}
\end{center}
\caption[Summary of simulations and most important results for the low-clump collisions ($M_{0}$=10 M$_{\odot}$) 
        for the different values of $b$ and $\mathcal{M}$]{Summary of simulations and most important results for the 
        low-clump collisions ($M_{0}$=10 M$_{\odot}$) for the different values of $b$ and $\mathcal{M}$.}
\label{tab:lowclump}
\end{sidewaystable}

In our low-mass clump collision simulations, two equal mass clumps, with $M_{0}$=10 M$_{\odot}$, collide 
with equal but opposite velocities. We use $M_{0}$=10 M$_{\odot}$, $a_{0} \sim$0.35 km s$^{-1}$ (corresponding to $T_{0}$=35 K); $\mathcal{M}$=5, 10, 15 (corresponding to $v_{cloud}$=0.5, 1.0 and 1.5 km s$^{-1}$), $b$=0.0, 0.2, 0.4, 0.6 and 0.8, and 15,000 particles 
per clump. Table \ref{tab:lowclump}
gives a summary of all simulations conducted as well as the most important results.

All collisions were conducted with and without application of particle splitting in order to be able to 
quantify the effect of particle splitting -- i.e. the Jeans condition being obeyed -- on the final result. All simulations 
evolve with very small time-step. Especially after the application of particle splitting, the time-step 
decreases due to the increased number of particles (decreased inter-particle distance; see Eqn. 
\ref{equa:deltatime2}). This happened in the higher mass clump collisions as well (\S \ref{sec:amar}). We stop the 
particle splitting simulations when it becomes inefficient for them to be continued any further (see 
discussion in \S \ref{sec:amar2}). We also stop the simulations without particle splitting at a time close 
to that of the corresponding particle splitting simulation only for comparison purposes. 

In the sequel all figures refer to the particle splitting simulations only.

\subsection{Simulation with $M_{0}$=10 M$_{\odot}$, $\mathcal{M}$=5 and $b$=0.0}{\label{sec:ccb0m5}}

The initial conditions for the head-on collisions are shown in the top panel of Fig. \ref{fig:b0m15yz}.

This is a head-on collision with the clumps moving with relatively slow velocities ($v_{cloud}$=0.5 
km s$^{-1}$). On-the-fly splitting initiates at $t_{spl}$ $\sim$0.455 Myr. The simulation stops at $\sim$0.476 Myr. At the end of the simulation there are 42,700 active particles. 

Only a single spherical fragment has formed. 
It forms at $\sim$0.471 Myr. The fragment forms at 
the centre of the simulation and it accretes material from the shocked layer. No filament forms in this 
collision. At the end of the simulation, the fragment mass is 0.33 M$_{\odot}$ and its radius $\sim$50 AU. 
Its central density has reached $\rho_{peak}$=4.75 x 10$^{-12}$ g cm$^{-3}$. It rotates but not as 
fast as the fragments in the simulations of Bhattal {\it et al}. In this simulation, there seems to be (cylindrically) isotropic 
inflow of matter onto the fragment as opposed to accretion along filaments in the 
simulations of Bhattal {\it et al}. 

The simulation without on-the-fly splitting produces similar results. The protostar forms at the same time as above 
($\sim$0.471 Myr). At the end of the simulation 
($\sim$0.476 Myr), its mass is 0.47 M$_{\odot}$. The Jeans condition is not obeyed above $\rho_{max}$=6 x 10$^{-17}$ g cm$^{-3}$. The protostar's radius is smaller ($\sim$32 AU) as it has become more 
centrally condensed ($\rho_{peak}$=1.75 x 10$^{-11}$ g cm$^{-3}$). 
We presume that with on-the-fly 
splitting we obtain a more realistic picture for this simulation even for such an early time 
in its evolution.

\subsection{Simulation with $M_{0}$=10 M$_{\odot}$, $\mathcal{M}$=10 and $b$=0.0}{\label{sec:ccb0m10}}

For this head-on collision the shock is stronger as the clumps move with higher velocities 
($v_{cloud}$=1.0 km s$^{-1}$) than in the previous simulation. On-the-fly splitting initiates at $t_{spl}$ $\sim$0.343 Myr. The simulation stops 
at $\sim$0.370 Myr. At the end 
of the simulation there are 45,500 active particles.
 
Two filaments form at $\sim$0.360 Myr almost perpendicular to each other. 
At the intercept of the two filaments a single object forms and it accretes 
material from the two filaments. At the end of the 
simulation, the object is spherical and rotating. Its mass is 0.35 M$_{\odot}$ and its 
radius $\sim$45 AU. Its central density has reached $\rho_{peak}$=4.9 x 10$^{-12}$ g cm$^{-3}$. 

The 
mass and the radius of the fragment are similar to those of the fragment of the previous simulation. The 
formation of the fragment happens faster in this simulation due to the fact that the two clumps 
are moving with higher velocities (this also forces the time-step of the simulation to decrease 
considerably at an earlier time).  

The simulation without on-the-fly splitting evolves in a similar way. In particular, 
two filaments form at $\sim$0.360 Myr and the single object, formed on their intercept, is accreting
material from them. It 
eventually becomes disc-like. At the end of the simulation 
($\sim$0.367 Myr), its mass is 0.27 M$_{\odot}$ and its radius $\sim$38 AU. However, it has become more 
centrally condensed ($\rho_{peak}$=7.1 x 10$^{-12}$ g cm$^{-3}$). 

\subsection{Simulation with $M_{0}$=10 M$_{\odot}$, $\mathcal{M}$=15 and $b$=0.0}{\label{sec:ccb0m15}}

This is the head-on collision involving the highest clump velocities ($v_{cloud}$=1.5 km s$^{-1}$). 
On-the-fly splitting 
starts at $t_{spl}$ $\sim$0.310 Myr. The simulation stops at $\sim$0.332 Myr. At the end of the simulation there are 44,100 active particles. 

A network of filaments forms at $\sim$0.320 Myr and a 
single rotating object forms at the intercept of the filaments almost simultaneously with them. The central protostellar object is surrounded by an 
extended disc. Material from the filaments swirls onto the disc. Due to the high clump velocities and the fact that accretion to the object 
happens in a non-axisymmetric fashion, the angular momentum grows faster than in the two previous 
simulations. The angular momentum 
increases with time but, at the end of the simulation, it is not sufficient for the growth of 
rotational instabilities. 

At this time ($t$ $\sim$0.332 Myr -- bottom panel of Fig. \ref{fig:b0m15yz}), the 
mass of the protostellar object is 0.47 M$_{\odot}$ and its radius $\sim$145 AU. Its central density has 
reached $\rho_{peak}$=2.9 x 10$^{-12}$ g cm$^{-3}$. The mass of the fragment is again similar 
to the masses of the protostars formed in the previous simulations.  
The fragment again forms faster than in the previous simulations, as the shocked layer forms faster due to 
the clumps colliding at a higher relative velocity. 

\begin{figure}
\begin{center}
\resizebox{8cm}{!}{\includegraphics{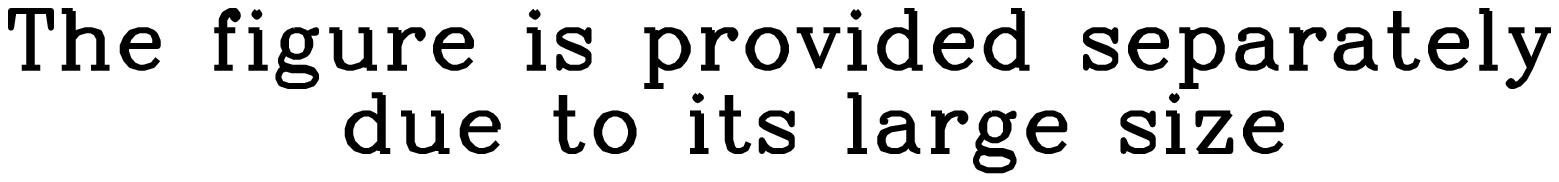}}
\end{center}
\begin{center}
\resizebox{8cm}{!}{\includegraphics{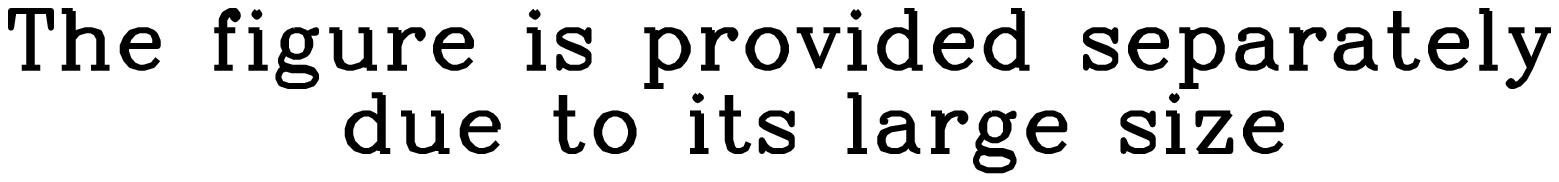}}
\end{center}  
      \caption[Column density plots for the $b$=0.0 and $\mathcal{M}$=15 collision of two 10 M$_{\odot}$ clumps]
      {Column density plots for the $b$=0.0 and $\mathcal{M}$=15 collision of two 10 M$_{\odot}$ clumps. \underline{Top}: Initial conditions viewed along the 
      z-axis. The linear size of this plot is 0.48 pc. 
      The colour table has units of 8.20 x 10$^{3}$ g cm$^{-2}$. \underline{Bottom}: Column density plot viewed along the 
      x-axis at the end of the simulation, $t$ $\sim$0.332 Myr. The linear size of this plot is 0.035 pc. 
      The colour table has units of 1.54 x 10$^{6}$ g cm$^{-2}$.}
      \label{fig:b0m15yz}
\end{figure}

\begin{figure}
\begin{center}
\resizebox{8.2cm}{!}{\includegraphics{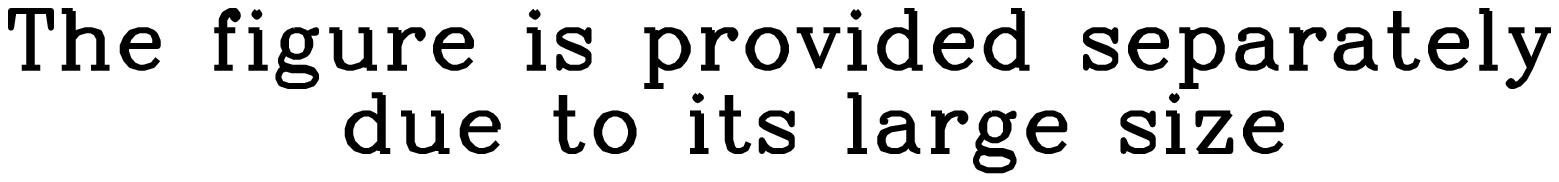}}
\end{center}
\begin{center}
\resizebox{8cm}{!}{\includegraphics{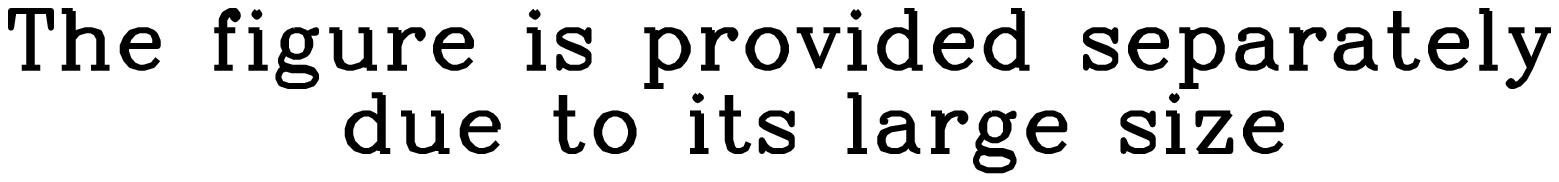}}
\end{center}  
      \caption[Column density plots for the $b$=0.2 and $\mathcal{M}$=5 collision of two 10 M$_{\odot}$ clumps]
      {Column density plots for the $b$=0.2 and $\mathcal{M}$=5 collision of two 10 M$_{\odot}$ clumps. \underline{Top}: Initial conditions viewed along the 
      z-axis. The linear size of this plot is 0.56 pc. 
      The colour table has units of 6.03 x 10$^{3}$ g cm$^{-2}$. \underline{Bottom}: Column density plot viewed along the 
      z-axis at the end of the simulation, $t$ $\sim$0.496 Myr. The linear size of this plot is 0.016 pc. 
      The colour table has units of 7.38 x 10$^{6}$ g cm$^{-2}$.}
      \label{fig:b2m5xy}
\end{figure}

In the simulation without particle splitting, a network of filaments forms and the 
rotating protostellar disc-like object forms almost simultaneously with the filaments, at $\sim$0.324 Myr. 
The protostellar disc rotates faster with time. After a short period of time, spiral arms appear in the disc. Up to the end of the simulation ($\sim$0.334 Myr), no significant interaction between the spiral 
arms and the accretion flows is observed. At this time, the mass of the protostar is 0.60 
M$_{\odot}$, its radius $\sim$100 AU, and it has become more centrally condensed 
($\rho_{peak}$=1.2 x 10$^{-11}$ g cm$^{-3}$). The protostar grows in mass considerably faster since the Jeans 
condition is not obeyed above $\rho_{max}$=6 x 10$^{-17}$ g cm$^{-3}$. Again, comparison of the the simulation 
without particle splitting and the on-the-fly splitting simulation demonstrates that the results obtained with 
particle splitting are different, and presumably more realistic.  

\subsection{Effect of changing $\mathcal{M}$ with constant 
$b$=0.0}{\label{sec:compare1}}

Comparison between the three head-on collision simulations with different clump velocities, 
shows that protostars with similar mass form. The simulations stop at a very 
early stage of the accretion process, only a few thousand years after the protostars form. 
At these early stages of protostellar collapse, protostars appear to be accreting mass with an accretion rate of $\sim$5 x 10$^{-5}$ M$_{\odot}$ yr$^{-1}$. This value is similar to that found in the 75 M$_{\odot}$ clump collisions (\S \ref{sec:amar}). The protostars formed are accreting with a rate similar to that of Class 0 objects. 

We also note that the angular momentum of the protostars increases with increasing clump velocity, as the protostar in the $\mathcal{M}$=15 collision forms a 
well-defined disc, in contrast to the protostars in the other two simulations that remain spherical. There is no evidence for the growth of rotationally induced instabilities 
up to the time the collisions end. It is more likely for these kind of instabilities to occur in 
the simulation with $v_{cloud}$=1.5 km s$^{-1}$, at a later time. Subsequent interaction of the spiral arms with the accretion flows may form companions to the primary protostar. 

The filaments become more well-defined 
with increasing clump velocity (i.e. as the shocks become stronger), and they start forming from a lower density. This implies that filaments are the mechanism by which material accretes onto the protostars in high compression shocks. We only present a column density 
plot for the simulation with $v_{cloud}$=1.5 km s$^{-1}$ as this is the most interesting simulation having the 
densest filaments (with minimum density of $\rho_{fil}$=2.5 x 10$^{-17}$ g cm$^{-3}$, i.e. $n_{H_{2}} \gtrsim$5 x 10$^{6}$ cm$^{-3}$.).

\subsection{Simulation with $M_{0}$=10 M$_{\odot}$, $\mathcal{M}$=5 and $b$=0.2}{\label{sec:ccb2m5}}

The initial conditions for the $b$=0.2 collisions are shown in the top panel of Fig. \ref{fig:b2m5xy}.

In the particle splitting simulation of the $b$=0.2 and $\mathcal{M}$=5 collision, on-the-fly splitting 
starts at $t_{spl}$ $\sim$0.463 Myr. A single filament forms perpendicular 
to the collision axis at $\sim$0.480 Myr. The filament feeds a single protostellar disc. The disc rotates fast. There is spiral structure in the disc. At the end of the 
simulation ($\sim$0.496 Myr), the spiral arms are very well-defined but there is no evidence 
for them interacting with the uniform accretion flows. We can not rule out the formation of secondaries at a later stage. 

The mass of the protostellar object is 0.70 
M$_{\odot}$ and its radius $\sim$90 AU (bottom panel of Fig. \ref{fig:b2m5xy}). Its central density has reached 
$\rho_{peak}$=7.1 x 10$^{-12}$ g cm$^{-3}$. At the end of the simulation there are 55,300 active particles. 

The simulation without particle splitting produces very similar results. The 
single rotating disc-like fragment is again within a single filament perpendicular to 
the collision axis (the filament forms at $\sim$0.480 Myr). Spiral arms develop in the disc. The fragment is not evolved so much as in the particle 
splitting simulation since the simulation without particle splitting ends a few 
thousand years earlier than the particle splitting simulation (at 
$\sim$0.490 Myr). The mass of the fragment at the end of the simulation is
0.45 M$_{\odot}$ and its radius is $\sim$84 AU. Its central 
density has reached $\rho_{peak}$=5.4 x 10$^{-12}$ g cm$^{-3}$. We note that, although there is no evidence for interaction between the accretion flows and the spiral arms, the accretion flows in the simulation without particle splitting are lumpy. This is the main difference with the particle 
splitting simulation where the accretion through the filament is very smooth 
and without any lumps.

Comparison between the particle splitting simulations with $b$=0.0 and 
$b$=0.2 (both with $\mathcal{M}$=5) shows that the single protostar rotates 
faster with larger $b$; the angular momentum is larger due to the increased 
impact parameter of the collision. However, both simulations end rather soon after 
the protostar formation and this prevents the angular momentum from becoming very large. The simulation with $b$=0.2 ends at a later time as 
it takes more time for the shocked 
layer to become massive enough to fragment. 

\begin{figure}
\begin{center}
\resizebox{8cm}{!}{\includegraphics{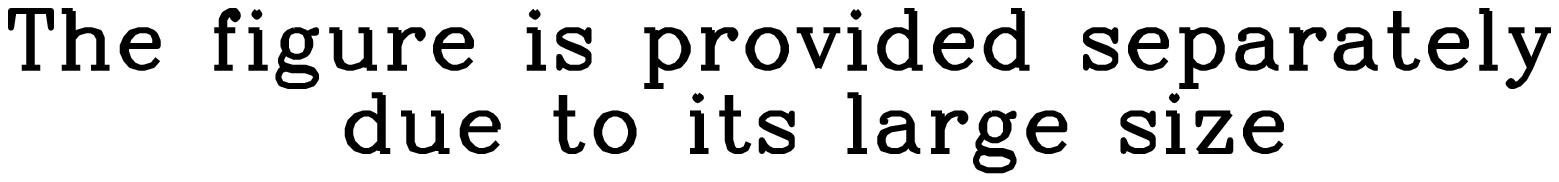}}
\end{center}
\begin{center}
\resizebox{8cm}{!}{\includegraphics{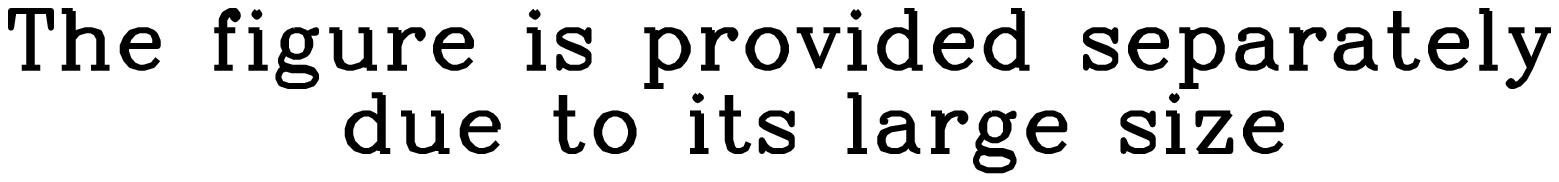}}
\end{center} 
      \caption[Column density plots for the $b$=0.2 and $\mathcal{M}$=10 collision of two 10 M$_{\odot}$ clumps]
      {Column density plots for the $b$=0.2 and $\mathcal{M}$=10 collision of two 10 M$_{\odot}$ clumps at the end of the simulation, $t$ $\sim$0.396 Myr. \underline{Top}: Column density plot viewed along the 
      z-axis. The linear size of this plot is 0.016 pc. 
      The colour table has units of 7.38 x 10$^{6}$ g cm$^{-2}$. 
      \underline{Bottom}: Column density plot viewed along the 
      y-axis. The linear size of this plot is 0.02 pc. 
      The colour table has units of 4.72 x 10$^{6}$ g cm$^{-2}$.}
      \label{fig:b2m10xy}
\end{figure}

\begin{figure}
\begin{center}
\resizebox{8cm}{!}{\includegraphics{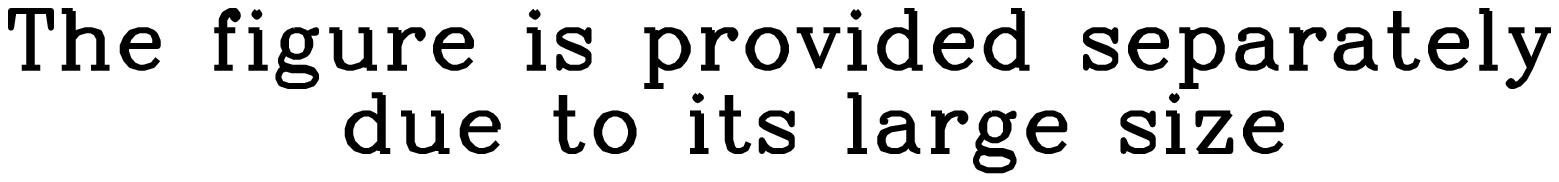}}
\end{center}
\begin{center}
\resizebox{8cm}{!}{\includegraphics{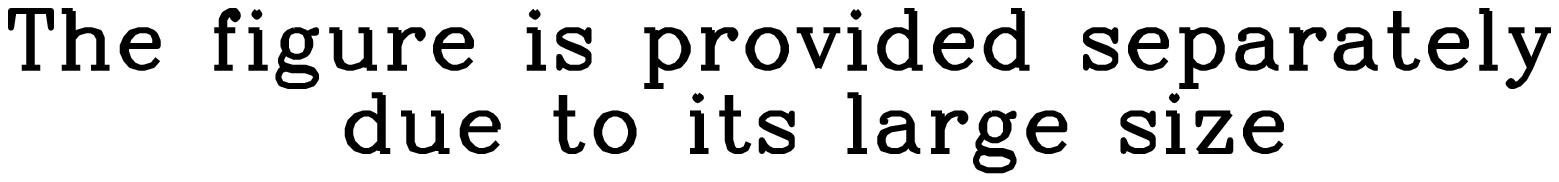}}
\end{center} 
      \caption[Column density plots for the $b$=0.2 and $\mathcal{M}$=15 collision of two 10 M$_{\odot}$ clumps]
      {Column density plots for the $b$=0.2 and $\mathcal{M}$=15 collision of two 10 M$_{\odot}$ clumps viewed along the 
      x-axis at the end of the simulation, $t$ $\sim$0.368 Myr. \underline{Top}: The linear size of this plot is 0.032 pc. 
      The colour table has units of 1.85 x 10$^{6}$ g cm$^{-2}$. 
      \underline{Bottom}: Zooming on the protostar on the left of the top panel. The linear size of this plot is 0.008 pc. 
      The colour table has units of 2.95 x 10$^{7}$ g cm$^{-2}$.}
      \label{fig:b2m15yz}
\end{figure}

\subsection{Simulation with $M_{0}$=10 M$_{\odot}$, $\mathcal{M}$=10 and $b$=0.2}{\label{sec:ccb2m10}}

In the particle splitting simulation of the $b$=0.2 and $\mathcal{M}$=10 collision, on-the-fly splitting starts at $t_{spl}$ $\sim$0.339 Myr. A network of filaments 
forms at this time. At $\sim$0.360 Myr, two objects form at the intercepts of the filaments. The 
protostars rotate fast and accretion discs form around them shortly after their formation. At the end 
of the simulation ($\sim$0.396 Myr), the two protostellar discs are attended by spiral arms. There is no evidence for interaction between the spiral arms and the accretion flows. The discs are approaching each other, towards a possible capture or 
merger. The separation at periastron and alignment of the two approaching protostars, are the major factors which will determine whether the two objects merge or are captured into a binary orbit. 
It is too early to determine the evolution of this binary. 

The masses of 
the two objects at the final stage are $m_{1}$=0.59 M$_{\odot}$ and $m_{2}$=0.41 M$_{\odot}$ (top panel of Fig. \ref{fig:b2m10xy}). The 
more massive protostar has radius $\sim$76 AU and the radius of the other object is $\sim$103 AU. 
The former is more centrally condensed (i.e. its disc is smaller). Their central densities are 
$\rho_{1}^{peak}$=4.4 x 10$^{-12}$ g cm$^{-3}$ and $\rho_{2}^{peak}$=1.8 x 10$^{-12}$ g cm$^{-3}$. 
The minimum density of the filaments is $\rho_{fil}$=2.8 x 10$^{-17}$ g cm$^{-3}$, i.e. $n_{H_{2}} \gtrsim$5 x 10$^{6}$ cm$^{-3}$ (bottom panel of Fig. \ref{fig:b2m10xy}). There is a possible 
third object starting to form a few time-steps before the end of the simulation. There are 64,300 active 
particles at the end of the simulation. 

The simulation without particle splitting produces similar results but can be followed only up to 
$\sim$0.382 Myr. In particular, a network of filaments forms at $\sim$0.339 Myr and at $\sim$0.360 Myr a single object forms within the filaments. It rotates fast and becomes disc-like 
followed by spiral arms. There is no evidence for the formation of secondaries in the disc. 
The protostar is evolving with tiny time-step. This is the reason for the final time being shorter 
than that of the particle splitting simulation. The mass of the protostar at the end of the simulation is 
0.44 M$_{\odot}$ and its radius is $\sim$110 AU. Its central density has reached 
$\rho_{peak}$=3.25 x 10$^{-12}$ g cm$^{-3}$. Towards the end of the simulation, two more objects start condensing out. We presume that the results 
of the on-the-fly splitting simulation are more realistic on two counts: two objects are formed 
instead of one, and the time-step is not so small allowing the simulation to evolve a bit longer. 

Comparison between the particle splitting simulations with $b$=0.0 and $b=0.2$ 
(both with $\mathcal{M}$=10) shows that both protostars in the $b$=0.2 collision are more massive than the single protostar in the $b$=0.0 collision, as they are evolved for a longer time. In the 
former simulation both protostars rotate faster than the single object of the latter simulation. It appears that the angular momenta of the protostars increase with 
increasing $b$. The $b$=0.2 simulation ends later than the $b$=0.0 as it takes more time for the clumps to move through 
each other before the shocked layer becomes massive enough to fragment.

In our $b$=0.2 collision involving 75 M$_{\odot}$ clumps, the Mach number is 9. Comparison of the $b$=0.2 collisions with different mass clumps, shows that in both cases filament fragmentation is the mechanism for the formation of the primaries. Fewer primaries and filaments form in the low-mass clump collision as the shocked layer has lower surface density, $\Sigma_{s}$. $\Sigma_{s}$ is proportional to the time elapsed from the beginning of the collision and in the 75 M$_{\odot}$ clump collision the filaments form long after the end of the 10 M$_{\odot}$ clump collision. Higher surface density gives smaller Jeans length in the layer ($\lambda_{J} \propto \Sigma_{s}^{-1}$, Whitworth {\it et al}. \shortcite{WhitworthAnA1994}). The mean separation between filaments in the layer and fragments in the filaments is of the order 
of the Jeans length. This is why this mean separation is smaller in the 75 M$_{\odot}$ clump collision and hence more filaments form in the layer and more fragments in the filaments. 

\subsection{Simulation with $M_{0}$=10 M$_{\odot}$, $\mathcal{M}$=15 and $b$=0.2}{\label{sec:ccb2m15}}

In the particle splitting simulation of the $b$=0.2 and $\mathcal{M}$=15 collision a network of filaments forms at $\sim$0.306 Myr. The filaments are more well-defined than before and they form on the whole surface of the shocked layer, not just at its centre as in previous simulations. On-the-fly splitting starts at $\sim$0.336 Myr. Two objects form within the central filaments at $\sim$0.348 Myr. The 
protostars rotate fast and accretion discs form around them shortly after their formation. At the end 
of the simulation ($\sim$0.368 Myr), the disc of the more massive protostar is attended by spiral arms. There are density enhancements at the points where the spiral arms interact with the accretion flows (bottom panel of Fig. \ref{fig:b2m15yz}). The enhancements in density may subsequently form one or two companions to the primary. 

It is too early to determine whether the two primaries form a bound or an unbound system. The masses of 
the two objects at the final stage are $m_{1}$=0.53 M$_{\odot}$ and $m_{2}$=0.42 M$_{\odot}$ (top panel of Fig. \ref{fig:b2m15yz}). There is also another 0.1 M$_{\odot}$ associated with the spiral arms that will be the upper limit for the mass of the secondaries forming via rotational instability. The 
more massive protostar has radius $\sim$42 AU and the radius of the other object is $\sim$35 AU. Their central densities are 
$\rho_{1}^{peak}$=1.2 x 10$^{-11}$ g cm$^{-3}$ and $\rho_{2}^{peak}$=3.1 x 10$^{-12}$ g cm$^{-3}$. 
The minimum density of the filaments is $\rho_{fil} \sim$10$^{-16}$ g cm$^{-3}$, i.e. $n_{H_{2}} \sim$10$^{7}$ cm$^{-3}$. There is a possible 
third object starting to condense in another filament. There are 70,200 active 
particles at the end of the simulation. 

\begin{figure}
\begin{center}
\resizebox{8cm}{!}{\includegraphics{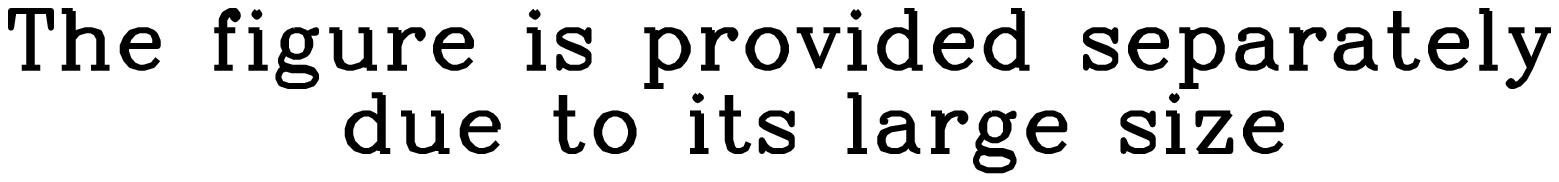}}
\end{center}
\begin{center}
\resizebox{8cm}{!}{\includegraphics{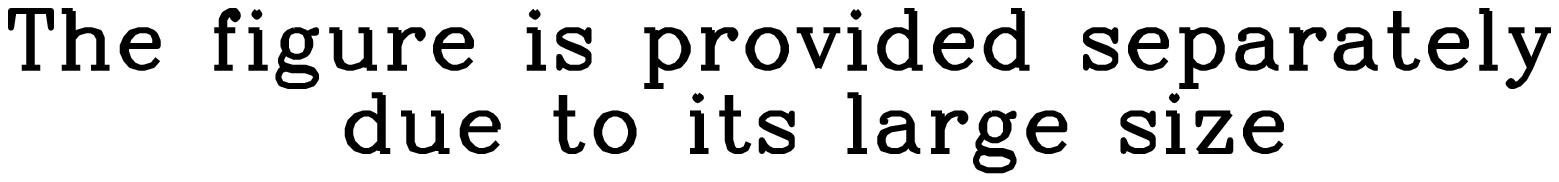}}
\end{center} 
      \caption[Column density plots for the $b$=0.4 and $\mathcal{M}$=5 collision of two 10 M$_{\odot}$ clumps]
      {Column density plots for the $b$=0.4 and $\mathcal{M}$=5 collision of two 10 M$_{\odot}$ clumps viewed along the 
      z-axis at the end of the simulation, $t$ $\sim$0.557 Myr. \underline{Top}: The linear size of this plot is 0.024 pc. 
      The colour table has units of 3.28 x 10$^{6}$ g cm$^{-2}$. 
      \underline{Bottom}: Zooming on the protostar on the bottom right hand corner of the top panel. The linear size of this plot is 0.01 pc. 
      The colour table has units of 1.89 x 10$^{7}$ g cm$^{-2}$.}
      \label{fig:b4m5xy}
\end{figure}

\begin{figure}
\begin{center}
\resizebox{8cm}{!}{\includegraphics{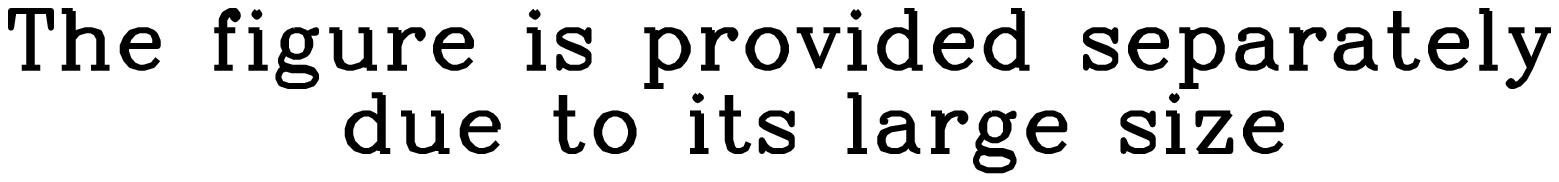}}
\end{center}
\begin{center}
\resizebox{8cm}{!}{\includegraphics{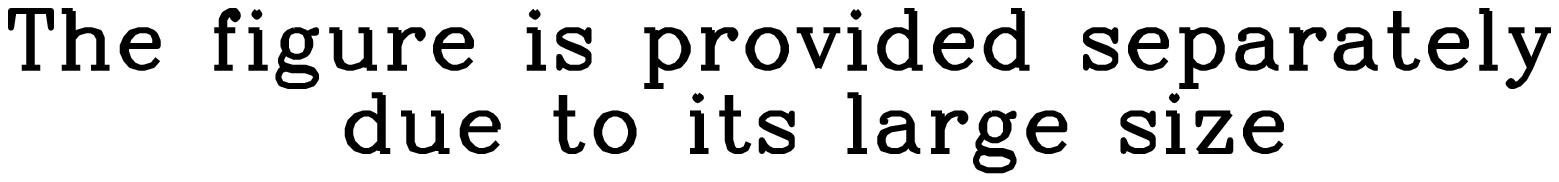}}
\end{center} 
      \caption[Column density plots for the $b$=0.4 and $\mathcal{M}$=10 collision of two 10 M$_{\odot}$ clumps]
      {Column density plots for the $b$=0.4 and $\mathcal{M}$=10 collision of two 10 M$_{\odot}$ clumps viewed along the 
      z-axis at the end of the simulation, $t$ $\sim$0.507 Myr. \underline{Top}: The linear size of this plot is 0.048 pc. 
      The colour table has units of 8.20 x 10$^{5}$ g cm$^{-2}$. 
      \underline{Bottom}: Zooming on the protostar on the bottom right hand corner of the top panel. The linear size of this plot is 0.008 pc. 
      The colour table has units of 2.95 x 10$^{7}$ g cm$^{-2}$.}
      \label{fig:b4m10xy}
\end{figure}

The simulation without particle splitting produces similar results. It is followed up to $\sim$0.360 Myr. In particular, a network of filaments forms at $\sim$0.306 Myr and at $\sim$0.348 Myr a single object forms within the filaments. It rotates fast and becomes disc-like 
with spiral arms. There is no evidence for interactions between the accretion flows and the spiral arms. However, the protostar rotates faster than all other objects formed in simulations without particle splitting.
Again, the protostar is evolving with very small time-step. This is the reason for the final time being shorter 
than that of the particle splitting simulation. The mass of the fragment at the end of the simulation is 
0.45 M$_{\odot}$ and its radius is $\sim$92 AU. Its central density has reached 
$\rho_{peak}$=2.2 x 10$^{-12}$ g cm$^{-3}$. Towards the end of the simulation, two more objects start condensing out. Again, we presume that with on-the-fly splitting we obtain more realistic results.

When we compare the simulations with $b$=0.0 and $b$=0.2 (both with $\mathcal{M}$=15), we conclude that the simulation evolution is delayed, and the angular momentum of the discs is increased with larger $b$. A similar conclusion was reached for $\mathcal{M}$=10.

\subsection{Effect of changing $\mathcal{M}$ with constant 
$b$=0.2}{\label{sec:compare2}}

By comparing the three simulations with $b$=0.2, we arrive at similar conclusions to those of \S \ref{sec:compare1}. The protostars appear to be accreting mass with an accretion rate of $\sim$5 x 10$^{-5}$ M$_{\odot}$ yr$^{-1}$.   

We also note that the angular momenta of the protostars increases with increasing clump velocity, as the protostars form more 
well-defined discs and rotate faster. A secondary companion may be starting to form at the end of the $v_{cloud}$=1.5 km s$^{-1}$ simulation. 

The filaments increase in number and become more well-defined 
with increasing clump velocity. Since $\Sigma_{s} \propto v_{cloud}$ and $\lambda_{J} \propto \Sigma_{s}^{-1}$, the Jeans length in the layer decreases with increasing Mach number of the collision (see discussion in \S \ref{sec:ccb2m10}).

The accretion rate does not increase with increasing $v_{cloud}$, which means that the filaments are somehow holding up temporarily material from falling on to the protostars to release it at a later stage. The densest filaments were formed in the $\mathcal{M}$=15 simulation (with minimum density $\rho_{fil} \sim$10$^{-16}$ g cm$^{-3}$, i.e. $n_{H_{2}} \sim$10$^{7}$ cm$^{-3}$).

\subsection{Simulation with $M_{0}$=10 M$_{\odot}$, $\mathcal{M}$=5 and $b$=0.4}{\label{sec:ccb4m5}}

The initial conditions for the $b$=0.4 collisions are shown in the top panel of Fig. \ref{fig:b4m15xy}.

In the particle splitting simulation of the $b$=0.4 and $\mathcal{M}$=5 collision on-the-fly splitting starts at $\sim$0.525 Myr. A single filament forms diagonally on the x-y plane at $\sim$0.532 Myr. Two objects form towards the ends of the filament at $\sim$0.535 Myr. The 
protostars rotate fast and accretion discs form around them shortly after their formation. At the end 
of the simulation ($\sim$0.557 Myr), one of the two protostellar discs is attended by strong spiral arms (bottom panel of Fig. \ref{fig:b4m5xy}). There are density enhancements at the points where the spiral arms interact with the accretion flows. A secondary companion to the primary protostar might subsequently form from these enhancements. 

The masses of 
the two objects at the final stage are $m_{1}$=0.67 M$_{\odot}$ and $m_{2}$=0.44 M$_{\odot}$ (top panel of Fig. \ref{fig:b4m5xy}). The 
more massive protostar has radius $\sim$77 AU and the radius of the other object is $\sim$52 AU. Their central densities are 
$\rho_{1}^{peak}$=5.25 x 10$^{-12}$ g cm$^{-3}$ and $\rho_{2}^{peak}$=7.94 x 10$^{-12}$ g cm$^{-3}$ (the second object is more centrally condensed). It can not be determined whether the protostellar system is bound or not; the separation is $\sim$3500 AU. The minimum density of the filaments is $\rho_{fil}$=2.43 x 10$^{-17}$ g cm$^{-3}$, i.e. $n_{H_{2}} \geqslant$5 x 10$^{6}$ cm$^{-3}$. There are 79,400 active 
particles at the end of the simulation. 

The simulation without particle splitting produces different results. It is followed to 
$\sim$0.544 Myr. In particular, a single thin filament forms at $\sim$0.528 Myr and at $\sim$0.533 Myr the filament starts fragmenting. It produces 10 objects. Two of them merge. They are all disc-like and they rotate fast, but there is no evidence for rotational instabilities. 
Again, the protostars are evolving with very small time-step. The total mass of the fragments at the end of the simulation is 
1.17 M$_{\odot}$. The objects have radii of $\sim$30 AU. Their central densities have reached 
$\rho_{peak}$=1.0-7.68 x 10$^{-12}$ g cm$^{-3}$. We presume again that with on-the-fly splitting we obtain more realistic results, as fragmentation of the filament into so many objects is suspicious.

\subsection{Simulation with $M_{0}$=10 M$_{\odot}$, $\mathcal{M}$=10 and $b$=0.4}{\label{sec:ccb4m10}}

\begin{figure}
\begin{center}
\resizebox{8cm}{!}{\includegraphics{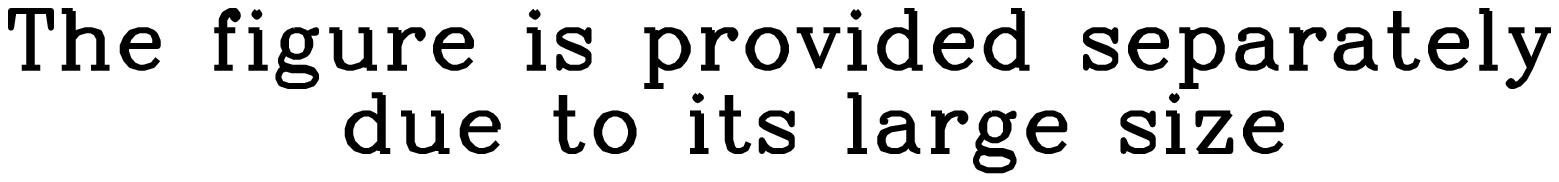}}
\end{center}
\begin{center}
\resizebox{8cm}{!}{\includegraphics{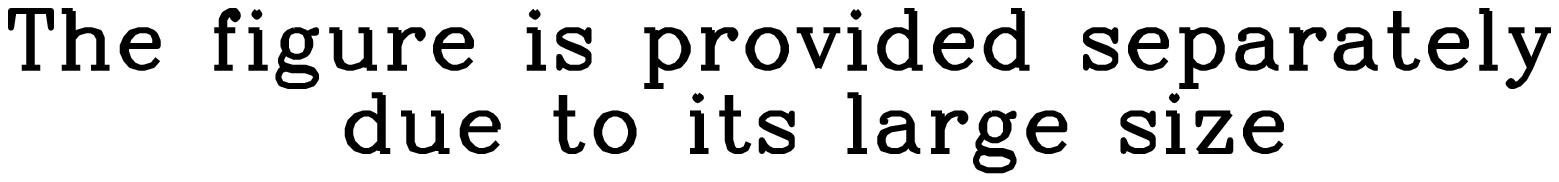}}
\end{center} 
      \caption[Column density plots for the $b$=0.4 and $\mathcal{M}$=15 collision of two 10 M$_{\odot}$ clumps]
      {Column density plots for the $b$=0.4 and $\mathcal{M}$=15 collision of two 10 M$_{\odot}$ clumps. \underline{Top}: Initial conditions viewed along the 
      z-axis. The linear size of this plot is 0.024 pc. 
      The colour table has units of 3.28 x 10$^{6}$ g cm$^{-2}$. 
      \underline{Bottom}: Column density plot viewed along the 
      z-axis at the end of the simulation, $t$ $\sim$0.453 Myr. The linear size of this plot is 0.032 pc. 
      The colour table has units of 1.85 x 10$^{6}$ g cm$^{-2}$.}
      \label{fig:b4m15xy}
\end{figure}

\begin{figure}
\begin{center}
\resizebox{8cm}{!}{\includegraphics{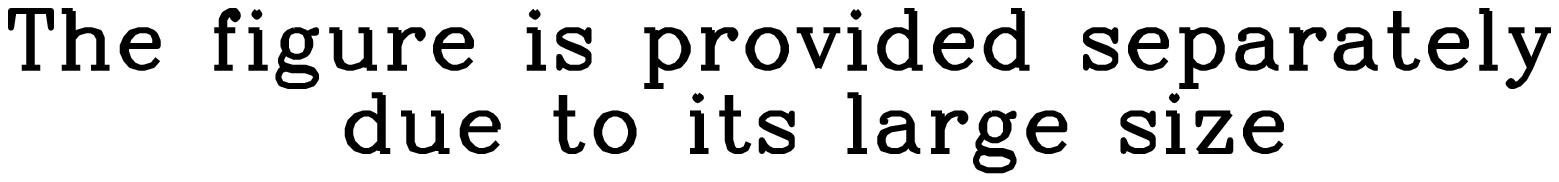}}
\end{center}
\begin{center}
\resizebox{8cm}{!}{\includegraphics{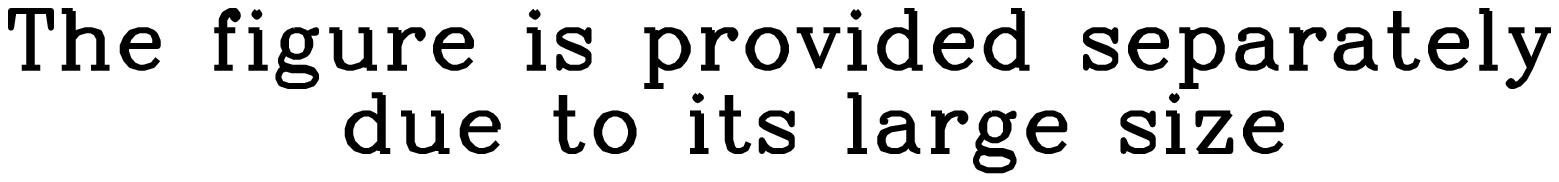}}
\end{center}  
      \caption[Column density plots for the $b$=0.6 and $\mathcal{M}$=5 collision of two 10 M$_{\odot}$ clumps]
      {Column density plots for the $b$=0.6 and $\mathcal{M}$=5 collision of two 10 M$_{\odot}$ clumps. \underline{Top}: Initial conditions viewed along the 
      z-axis. The linear size of this plot is 0.74 pc. 
      The colour table has units of 3.45 x 10$^{3}$ g cm$^{-2}$. \underline{Bottom}: Column density plot viewed along the 
      z-axis at the end of the simulation, $t$ $\sim$0.7 Myr. The linear size of this plot is 0.12 pc. 
      The colour table has units of 1.31 x 10$^{5}$ g cm$^{-2}$.}
      \label{fig:b6m5xy}
\end{figure}

In the particle splitting simulation of the $b$=0.4 and $\mathcal{M}$=10 collision on-the-fly splitting starts at $\sim$0.432 Myr. A single filament forms at $\sim$0.452 Myr. The filament is not as thin as in the previous simulation. One object forms at $\sim$0.485 Myr. The 
protostar rotates fast and an accretion disc forms around it shortly after its formation. At the end 
of the simulation ($\sim$0.507 Myr), the disc is attended by spiral arms. Due to the density enhancements at the points where the spiral arms interact with the accretion flows, we conclude that it is likely that secondaries to the protostar would be formed subsequently (bottom panel of Fig. \ref{fig:b4m10xy}). The mass of 
the  object at the final stage is 0.48 M$_{\odot}$ and its radius $\sim$48 AU. Its central density is 
$\rho_{peak}$=3.4 x 10$^{-12}$ g cm$^{-3}$. 
The minimum density of the filaments is $\rho_{fil}$=2 x 10$^{-17}$ g cm$^{-3}$, i.e. $n_{H_{2}} \geqslant$5 x 10$^{6}$ cm$^{-3}$. There is a possible 
second object starting to condense at the other end of the filament (top panel of Fig. \ref{fig:b4m10xy}). There are 59,300 active 
particles at the end of the simulation. 

The simulation without particle splitting produces different results, similar to those of the simulation without particle splitting for $b$=0.4 and $\mathcal{M}$=5. It is followed up to 
$\sim$0.477 Myr. A single filament forms at $\sim$0.458 Myr and at $\sim$0.466 Myr the filament fragments. It produces 5 objects. Two of them merge. They are all disc-like and they rotate fast. There is no evidence for rotational instabilities growing in these protostars. Again, the protostars are evolving with very small time-step. The total mass of the fragments at the end of the simulation is 
0.54 M$_{\odot}$. The objects have radii of 25-35 AU. Their central densities have reached 
$\rho_{peak}$=0.45-2.0 x 10$^{-12}$ g cm$^{-3}$. We presume that with on-the-fly splitting we obtain more realistic results.

\subsection{Simulation with $M_{0}$=10 M$_{\odot}$, $\mathcal{M}$=15 and $b$=0.4}{\label{sec:ccb4m15}}

In the particle splitting simulation of the $b$=0.4 and $\mathcal{M}$=15 collision no filaments are formed. On-the-fly splitting starts at $\sim$0.397 Myr. One object forms at the centre of the domain at $\sim$0.433 Myr. The 
protostar rotates fast and an accretion disc forms around it. At the end 
of the simulation ($\sim$0.453 Myr), the protostellar disc is attended by weak spiral arms. 
There is no evidence for interaction between the spiral arms and the accretion flows. 
The mass of 
the object at the final stage is 0.31 M$_{\odot}$ (bottom panel of Fig. \ref{fig:b4m15xy}). Its radius is $\sim$132 AU and its central density  
$\rho_{peak}$=3.3 x 10$^{-13}$ g cm$^{-3}$. 
There are 39,700 active 
particles at the end of the simulation. 

The simulation without particle splitting produces similar results. It was followed to 
$\sim$0.437 Myr. A single object forms at $\sim$0.428 Myr at the centre of the domain. It rotates fast. The protostar is evolving with very small time-step. The mass of the fragment at the end of the simulation is 
0.32 M$_{\odot}$ and its radius is $\sim$97 AU. Its central density has reached 
$\rho_{peak}$=2.43 x 10$^{-12}$ g cm$^{-3}$. We presume that with on-the-fly splitting we obtain more realistic results.

\subsection{Effect of changing $\mathcal{M}$ with constant 
$b$=0.4}{\label{sec:compare3}}

Increasing $\mathcal{M}$ in simulations with $b$=0.4 does not produce stronger shocks, nor does it increase the angular momentum of the protostars. The colliding clumps fail to form a strong shock in the $\mathcal{M}$=15 
collision. This conclusion contrasts with our previous conclusion that increasing $\mathcal{M}$ gives stronger shocks with higher global angular momentum when $b \leqslant$0.2. It appears that the $b$=0.4 collisions are the borderline, as with $b$ higher than 0.4 we shall see that the clump collisions do not form strong shocks.

\subsection{Simulations with $M_{0}$=10 M$_{\odot}$ and $b$=0.6}{\label{sec:ccb6m5}}

The initial conditions for the $b$=0.6 collisions are shown in the top panel of Fig. \ref{fig:b6m5xy}.

In the particle splitting simulation of the $b$=0.6 and $\mathcal{M}$=5 collision on-the-fly splitting starts at $\sim$0.644 Myr. The two clumps move a long way into each other before the density increases significantly. Two single well-separated objects form at $\sim$0.678 Myr. No filaments are formed in this collision. The 
protostars rotate fast and accretion discs form around them shortly after their formation. At the end 
of the simulation ($\sim$0.701 Myr), one of the two protostellar discs has strong spiral arms (bottom right protostar in the bottom panel of Fig. \ref{fig:b6m5xy}). It is possible that secondaries to the primary protostar will form  shortly. The system of the two protostars is unbound as they are at a large distance ($>$ 30,000 AU) and moving apart rapidly. 

The masses of 
the two objects at the final stage are $m_{1}$=0.40 M$_{\odot}$ and $m_{2}$=0.28 M$_{\odot}$ (bottom panel of Fig. \ref{fig:b6m5xy}). The 
more massive protostar has radius $\sim$53 AU and the radius of the other object is $\sim$85 AU. Their central densities are 
$\rho_{1}^{peak}$=3.4 x 10$^{-12}$ g cm$^{-3}$ and $\rho_{2}^{peak}$=9.6 x 10$^{-13}$ g cm$^{-3}$. There are 50,000 active 
particles at the end of the simulation. 

The simulation without particle splitting produces only one protostar (apparently corresponding to the bottom right protostar in the simulation with particle splitting; see bottom panel of Fig. \ref{fig:b6m5xy}). It forms at $\sim$0.671 Myr. It rotates fast and becomes disc-like 
with spiral arms. 
There is no evidence for interaction between the spiral arms and the accretion flows. 
Again, the protostar is evolving with very small time-step. The mass of the fragment at the end of the simulation ($\sim$0.684 Myr) is 
0.30 M$_{\odot}$ and its radius is $\sim$80 AU. Its central density has reached 
$\rho_{peak}$=2.3 x 10$^{-12}$ g cm$^{-3}$. 

In the $b$=0.6 simulations with $\mathcal{M}$=10 and $\mathcal{M}$=15 the clumps do not interact significantly to induce fragmentation, and this is also the case in all the $b$=0.8 simulations (\S \ref{sec:ccb8m5}). 

\begin{figure}
\begin{center}
\resizebox{8cm}{!}{\includegraphics{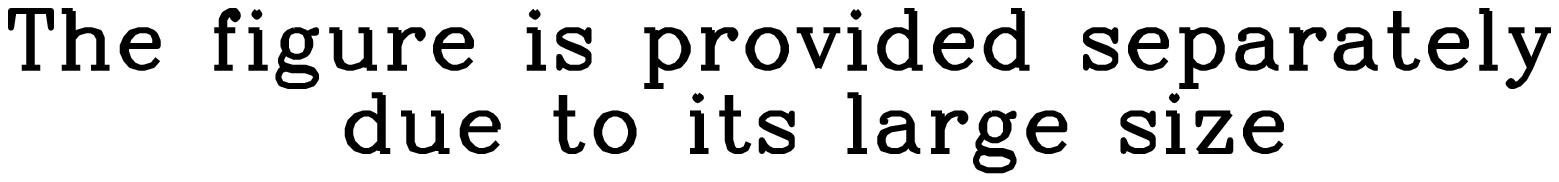}}
\end{center}
\begin{center}
\resizebox{8cm}{!}{\includegraphics{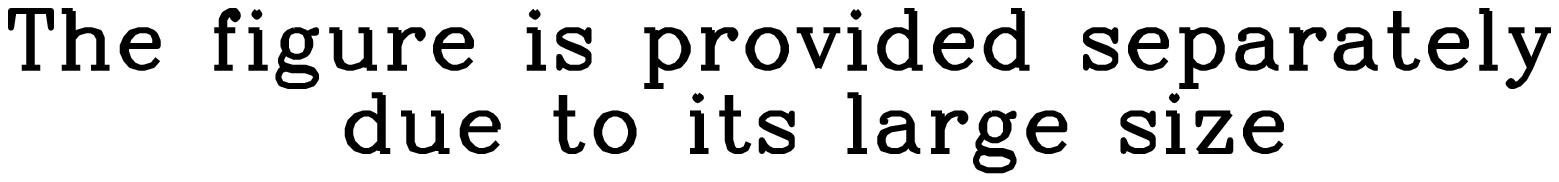}}
\end{center}  
      \caption[Column density plots for the $b$=0.8 and $\mathcal{M}$=5 collision of two 10 M$_{\odot}$ clumps]
      {Column density plots for the $b$=0.8 and $\mathcal{M}$=5 collision of two 10 M$_{\odot}$ clumps. \underline{Top}: Initial conditions viewed along the 
      z-axis. The linear size of this plot is 0.72 pc. 
      The colour table has units of 3.63 x 10$^{3}$ g cm$^{-2}$. \underline{Bottom}: Column density plot viewed along the 
      z-axis at the end of the simulation, $t$ $\sim$1.4 Myr. The linear size of this plot is 0.56 pc. 
      The colour table has units of 6.03 x 10$^{3}$ g cm$^{-2}$.}
      \label{fig:b8m5xy}
\end{figure}

\subsection{Simulations with $M_{0}$=10 M$_{\odot}$ and $b$=0.8}{\label{sec:ccb8m5}}

The initial conditions for the $b$=0.8 collisions are shown in the top panel of Fig. \ref{fig:b8m5xy}. There is no shocked layer formed in any of these collisions. The overlap of the two clumps is very small. We have conducted these simulations only for completeness. Here we present the end of the simulation with $\mathcal{M}$=5 and $b$=0.8 (bottom panel of Fig. \ref{fig:b8m5xy}, $t$ $\sim$1.4 Myr). We have chosen to present this simulation because it is the one with the slowest moving clouds and, hence, the highest {\it a priori} probability that the collision could create a shock, due to gravitational focussing (as in \S \ref{sec:ccb6m5}). Particle splitting was not applied as the density in the two clumps did not exceed $\rho_{max}$=6 x 10$^{-17}$ g cm$^{-3}$.

\section{Discussion}{\label{sec:collision_conl}}

\newcommand{\DoT}[1]{\begin{turn}{-90}#1\end{turn}}

\begin{table}
\begin{center}
\begin{tabular}{|p{8mm}p{17mm}p{25mm}|p{7mm}|p{15mm}p{25mm}|}\hline
$b$ & 0.0 & 0.2 & 0.4 & 0.6 & 0.8 \\ \hline
$\Longrightarrow$& $\Longrightarrow$ $\Longrightarrow$ & $\Longrightarrow$ $\Longrightarrow$ $\Longrightarrow$ & & $\Longrightarrow$ $\Longrightarrow$ & $\Longrightarrow$ $\Longrightarrow$ \\
 & Angular & Momentum $\uparrow$ & & Clump & Interaction $\downarrow$ \\ 
 & & $t_{frag}$ $\uparrow$ & & & \\ \hline \hline
$\mathcal{M}$ $\Downarrow$ 5 $\Downarrow$ 10 $\Downarrow$ 15 $\Downarrow$ &  Angular Filament Filament & momentum $\uparrow$ Numbers $\; \; \; \;$ $\uparrow$ Density $\; \; \; \;$ $\uparrow$ $t_{frag}$ $\downarrow$ & \DoT{\fbox{\LARGE{Transition}}} & \normalsize{No shock} & \\ 
&  &  &  &  &  \\ \hline 
\end{tabular}
\end{center}
\caption[Dependence of different quantities and phenomena on the increasing values of $b$ and $\mathcal{M}$ for simulations of cloud-cloud collisions]{Dependence of different quantities and phenomena on the increasing values of $b$ and $\mathcal{M}$ for simulations of cloud-cloud collisions. Note that the parameter space is divided in two sections: low $b$ collisions produce stronger shocks. Large $b$ collisions reduce the cloud interaction. The transition happens at $b$=0.4.}
\label{tab:lowclump1}
\end{table}

We have conducted a series of cloud-cloud collision simulations using particle splitting. Two different sets of initial conditions were used (\S \ref{sec:clump_initial}). In the set used previously by Bhattal {\it et al}., the mass of the colliding clumps was 75 M$_{\odot}$ (\S \ref{sec:amar}). In the second set of initial conditions the clump mass was 10 M$_{\odot}$ (\S \ref{sec:clump}). 

With both sets of initial conditions, we have found that simulations with $b \leqslant$0.5 produce shocked layers. Filaments or spindles form in the layers, with densities $n_{H_{2}} \gtrsim$ 10$^{5}$ cm$^{-3}$. Groups of protostellar discs form by fragmentation of the filaments. We have identified {\em fragmentation of the filaments} as the common mechanism for Star Formation in these collisions. We have also found that the 
filaments act as the accretion channel that feeds the protostars with material. 

All protostars formed, have mass accretion rates of $\sim$ 5 x 10$^{-5}$ M$_{\odot}$ yr$^{-1}$ for the first 10-20 thousand years of their evolution. Their ages and mass accretion rates are comparable to those of Class 0 objects. 

The simulations involving 75 M$_{\odot}$ clumps, indicate that a group of 4-6 stars with masses between 3 and 6 M$_{\odot}$ forms as a result of these collisions. The inferred Star Formation efficiency is $\sim$15-20\%. To derive the star masses and the Star Formation efficiency, we assume the observed mass accretion rates of protostars and a total pre-main-sequence lifetime of about a million years. 

The 10 M$_{\odot}$ clumps have smaller radii than the 75 M$_{\odot}$ clumps. Therefore, in 10 M$_{\odot}$ clump collisions, all the mass of the clumps enters the shocked layer in a shorter time than in 75 M$_{\odot}$ clump collisions. Thus, accretion to the protostars through the filaments finishes earlier. Specifically, in low-mass clump collisions, 1-2 M$_{\odot}$ stars would form. Our simulations have shown that 1-2 protostars form in each collision. Thus, the Star Formation efficiency derived from low-mass clump collisions is 10-15\%.

Bhattal {\it et al}. have found that accretion-induced rotational instabilities in protostellar discs can produce secondary companions to the protostars. In some of our simulations spiral arms form in the discs. We found evidence suggesting that the spiral arms are interacting with the accretion flows. However, our simulations stop at an early stage of the disc evolution due to time-step constraints. Thus, we can not confirm that such interactions are efficient in forming companions to the protostars. Formation of secondaries by accretion-induced rotational instabilities and/or disc-disc interactions would increase the Star Formation efficiency. 

Simulations with $b \leqslant$0.5 suggest that the protostars are falling together along the filaments. The protostars could form bound systems. On smaller scales, they could merge or form binary systems by capture. To investigate their dynamic evolution, we need to use sink particles (\S \ref{sec:collapse_nest}.1) to replace the protostars, in order to prevent small time-steps form occuring. We have repeated the simulation with $M_{0}$=10 M$_{\odot}$, $\mathcal{M}$=10 and $b$=0.2 by replacing the most massive protostar formed by a sink. The evolution of this simulation was similar to that of \S \ref{sec:ccb2m10}, as the other protostar soon became dense enough to require small time-steps. We should develop an automated algorithm to replace all protostellar objects with densities above a certain threshold (e.g. Bonnell {\it et al}. \shortcite{BonnellMNRAS1997}). 

To investigate the evolution of the system of protostars, we would also need to regulate the shear viscosity, for instance by using the time-dependent formulation of Morris \& Monaghan \shortcite{MorrisJCP1997} and/or the Balsara 
\shortcite{BalsaraJCP1995} switch.

The detailed evolution of each simulation indicated a dependence on the values of the clump mass, clump velocity (i.e. collision Mach number) and impact parameter. Table \ref{tab:lowclump1} summarises the dependence on the impact parameter, $b$, and the Mach number, $\mathcal{M}$. With $\Rightarrow$, we indicate the increase in the values of these parameters. $\uparrow$ indicates an increasing quantity or a phenomenon that becomes stronger and more frequent. $\downarrow$ marks a decreasing quantity or a phenomenon that becomes less frequent. 

In particular, we find that with larger $b$, the angular momentum of the discs and the time for filament fragmentation are also larger. This makes rotational instabilities more probable for collisions with larger $b$.

Increasing the Mach number of the collision, produces more fragments and networks of filaments, as the Jeans length in the shocked layer decreases. The filaments become more well-defined as the shock becomes stronger, and the time for filament fragmentation decreases. The angular momentum of the discs increases also with increasing $\mathcal{M}$, indicating that in high Mach number collisions, rotational instabilities are likely to be more effective for the formation of secondary companions.

Increasing the clump mass increases the number of filaments in the layer and the number of fragments in the filaments. It also creates stronger shocks and more well-defined filaments. The reason for this is again the decrease in the Jeans length of the layer.

With $b > 0.5$ the colliding clumps interact less. For $b > 0.5$ and $\mathcal{M} > 5$ the collisions do not create shocks. There seems to be a geometrical constraint on the strength of the shocked layers produced by cloud collisions: for low $b$ collisions, the shocks are strongest as more mass enters in them. Collisions with $b$=0.4 appear to produce less strong shocks. Therefore, we have concluded that the transition occurs at $b$=0.4. From Table \ref{tab:lowclump1} we conclude that collisions with $b$=0.2 and $\mathcal{M}$=10 or 15 are the most efficient. 

With particle splitting the Jeans condition remains obeyed at all times. The higher resolution achieved with particle splitting decreases the values of the time-step and the simulations had to end at an early time. 

We presume that the particle splitting simulations produce more realistic results than the simulations of Bhattal {\it et al}. for the 75 M$_{\odot}$ clump collisions and our 10 M$_{\odot}$ clump collisions without particle splitting. Our presumption is supported by the fact that when the simulation with 10 M$_{\odot}$, $b$=0.2 and $\mathcal{M}$=10 was repeated with fine particles from the beginning (i.e. with 195,000 particles per cloud) and without particle splitting, the results were in very good agreement at comparable times. 
Repeating the above particle splitting simulation with clouds of 15,000 particles but with particles initially taken from a lattice, has given again results very similar to those of the particle splitting simulation of \S \ref{sec:ccb2m10}. The simulation with particles initially taken from a lattice starts with less particle noise and it produces more symmetric results, i.e fragments in symmetric positions. However, the fragments have similar masses, radii and separations with those formed in the particle splitting simulation of \S \ref{sec:ccb2m10}. In the future, we aim to test carefully the convergence of our results.

Finally, our simulations have shown the formation of dense filaments with $n_{H_{2}} \gtrsim$ 10$^{5}$ cm$^{-3}$. We predict that filaments, and in some cases networks of filaments, could be observed in sites of dynamical star formation. In such sites, more than 2 or 3 Class 0 objects would form almost simultaneously within a radius of few thousand AU. The filaments could be observed in NH$_{3}$ molecular line observations. With arc-second resolution, filaments of the sizes inferred by our simulations could be observed in SFR regions at distances $\lesssim$ 1 kpc.

\chapter{Conclusions}{\label{sec:conclusions}}

In this thesis we have studied Star Formation triggered by low-mass cloud collisions by means of numerical simulations. Many previous simulations of cloud-cloud collisions did not obey the Jeans condition for fragmentation due to their lack of sufficient numerical resolution. To overcome this limitation without using very large numbers of particles, we have developed particle splitting. This is a method that increases the number of particles only in places where higher resolution is required in order to satisfy the Jeans condition.

The method has been applied to cloud-cloud collision simulations using clouds of different masses and examining collisions with different impact parameters and different relative cloud velocities. 

In this chapter, we summarise our conclusions. We have divided them in two sections: those on the numerical method and those on the results of cloud-cloud collisions. 

\section{SPH and Particle Splitting}

For the numerical simulations conducted in this thesis we have used Smoothed Particle Hydrodynamics, and Tree-Code-Gravity for the calculation of the gravitational forces. We have used a the second order Runge-Kutta time-integration scheme and multiple particle time steps and the Monaghan prescription for artificial viscosity ($\alpha=\beta=1$). 

The numerical code has been extensively tested in previous work, producing good results to most tests. We have shown that a) it is able to model collapse and b) it prevents numerical perturbations from developing in simulations of equilibrium spheres. The code can also model adequately shocks created by $\mathcal{M} \sim$ 5 collisions of gas flows.

The formulation of artificial viscosity in our code produces a large effective shear viscosity. As a result we cannot trust the detailed evolution of the discs around the protostars formed in our simulations. Future improvements to the code should seek to regulate the shear viscosity, for instance by using the time-dependent formulation of Morris \& Monaghan \shortcite{MorrisJCP1997} and/or the Balsara 
\shortcite{BalsaraJCP1995} switch. 

In this thesis, we have developed an algorithm for particle splitting, which 
increases the number of particles in a simulation, but only in regions where the resolution is not sufficient to continue obeying the Jeans condition. With the new method, a coarse particle in such a region is replaced by 13 particles of smaller mass. Our algorithm puts the 13 fine particles on the vertices of an 
equilateral lattice (fcc). This is a lattice with minimum interstitial volume. It is appropriate for the 3-dimensional problems we model in this thesis. 
The density above which the Jeans condition stops being resolved is inversely proportional to the particle mass squared. By decreasing the particle mass 13-fold, the density at which the simulation reaches its resolution limit increases by 13$^{2}$. In all the fine simulations performed in this thesis, adiabatic heating switched on at densities lower than the critical density. Therefore, the Jeans condition was obeyed throughout these simulations. In principle, we could split particles repeatedly, and thereby increase the resolution indefinitely.

Identifying the coarse particles that need to be split is achieved with two separate versions of the new method. In nested splitting we manually decide the volume and the position of the region where the Jeans condition stops being obeyed, and we split all particles in this region. Subsequently, all coarse particles entering this region are split on-the-fly. In on-the-fly splitting we calculate the density above which the Jeans condition will be violated and all particles that obtain densities higher than this are split on-the-fly. On-the-fly splitting is preferred as particles are not split unnecessarily nor do we have to 
stop the coarse simulation.

Particle splitting was found to introduce errors into the calculation of particle smoothing lengths. We had to revise the method for calculating particle smoothing lengths, so that it complied with the mixing of different mass particles. In calculating particle smoothing lengths we now take into account the amount of mass contained in a smoothing kernel, and not the number of neighbours.

Both versions of particle splitting have been tested and give good results. SPH with particle splitting was found to model adequetly isothermal collapse. In simulations of equilibrium spheres the errors introduced by particle splitting were not sufficient for perturbations to be produced and propagated . 

Application of particle splitting to simulations of rotating clouds with m=2 density perturbations gives results that agree with the results of high resolution finite difference simulations. SPH simulations without particle splitting required at least twice as many particles to reproduce the same results. 
Therefore, the new method is both efficient and economic, in terms of computational cost. Specifically, for the particle splitting simulations we have used only $\sim 40\%$ of the memory and $\sim 25-30\%$ of the CPU used in the 600,000 simulation without particle splitting. The simulations of rotating clouds with m=2 density perturbations are computationally demanding as a large fraction of the total mass ends up in the protostars formed. We expect particle splitting to be even more efficient in problems where a large fraction of the total mass is not evolved in fragmentation. On-the-fly splitting is more economic than nested splitting and it is the version we use in simulations of cloud-cloud collisions.

However, simulations of rotating clouds with m=2 density perturbations which start with a small number of particles, have shown that particle splitting, in response to the imminent violation of the Jeans condition, is 
only a necessary, and not a sufficient, condition for the reliability and efficiency of a simulation.  

Particle splitting can be applied not just to cloud-cloud collision simulations but also to other SPH simulations that also suffer from insufficient resolution. In particular, particle splitting could be used in collapse simulations, disc fragmentation calculations, simulations of disc interactions, simulations of stellar winds, modelling of mass transfer discs in cataclysmic variables and accretion discs in super massive black holes.

\section{Cloud-cloud collisions}

We have applied on-the-fly splitting to low-mass clump collision simulations. With particle splitting artificial fragmentation has been eliminated. Two sets of initial conditions have been used in our simulations. The first set repeat previous simulations which were performed by Bhattal {\it et al}. without particle splitting; the parameters are $M_{0}$=75 M$_{\odot}$, $\mathcal{M}$=9 and  $b$=0.2, 0.4 and 0.5, with 110,000 particles per clump. The second set explore a new set of parameters; $M_{0}$=10 M$_{\odot}$; $\mathcal{M}$=5, 10, 15; $b$=0.0, 0.2, 0.4, 0.6 and 0.8, and 15,000 particles per clump. 

The simulations with $b \leqslant$0.5 produce shocked layers. Filaments or spindles form in the shocked layers, with densities $n_{H_{2}} \gtrsim$ 10$^{5}$ cm$^{-3}$. 

Systems of protostars are produced by fragmentation of the filaments. We identify fragmentation of filaments as the common mechanism for Star Formation in these collisions.

Most of the protostars are surrounded by discs. Rotational instabilities in these discs may produce secondaries to the primary protostars. Spiral arms are formed in almost all discs, but interaction between the spiral arms and the accretion flows is observed only in a few cases. However, our simulations end early due to the decreased time-step. 
Spiral arms in discs become more frequent with increasing $b$ and $\mathcal{M}$.

The protostars formed in the $b \leqslant$0.5 simulations are falling together along the filaments. At the end of the simulations, they are still at large separations, and we can not estimate if the system they form is bound and/or if they will merge at a later time. 

All protostars formed show mass accretion rates of $\sim$ 5 x 10$^{-5}$ M$_{\odot}$ yr$^{-1}$ for the first 10-20 thousand years of their evolution. 
Their ages and mass accretion rates are comparable to those of Class 0 objects. 
The inferred Star Formation efficiency is $\sim$15-20\% for the 75 M$_{\odot}$ clump collisions and $\sim$10-15\% for the 10 M$_{\odot}$ clump collisions. The difference arises from the number of protostars formed which is larger in the former collisions. Disc instabilities and disc-disc interactions can create low-mass companions to the primaries formed in our simulations, and thus slightly increase the inferred values of Star Formation efficiency.

In high Mach number collisions ($\mathcal{M} \geqslant$ 9) with $b < 0.4$, a network of filaments forms. The filaments have higher column densities in the 75 M$_{\odot}$ clump collisions.

We predict that filaments, and in some cases networks of filaments, could be observed in sites of dynamical star formation. In such sites, more than 2 or 3 Class 0 objects would form almost simultaneously within a radius of few thousand AU. The filaments could be observed in NH$_{3}$ molecular line radiation. With arc-second resolution, filaments of the sizes inferred by our simulations could be observed in SFR regions at distances $\lesssim$ 1 kpc.

Due to time-step constraints, our simulations can only be followed for a few thousand years after protostar formation. Sink particles could replace the protostars as soon as they reach a certain density. An automated algorithm should be implemented for this purpose. With such an algorithm, we would not need to follow the detailed evolution of the individual discs and therefore the simulations could be followed further. We would not be able to monitor the efficiency of accretion-induced rotational instabilities and disc-disc interactions as mechanisms for secondary formation, but we could obtain more specific values for the Star Formation efficiency of the filament fragmentation mechanism induced by cloud-cloud collisions. We could also determine whether the protostellar systems formed are bound. 

To obtain the detailed evolution of a disc and its interaction with the accretion flows from the filaments, we need to continue the simulations for a much longer time. Parallelisation of the code and increasing speed of computation achieved by super computers may make this possible.

\appendix \chapter{Jeans criterion of stability}{\label{append:appendia}}

The fact that signatures of collapse (\S \ref{sec:collapse}.2) only appear at 
certain sites in the interstellar medium, indicates that, in general, 
interstellar gas is in a quasi-static state, where the self-gravity is 
balanced by hydrostatic pressure, turbulence and possibly magnetic fields.
We would like to know the point where this balance breaks, as at this point, 
collapse initiates and the gas is no longer efficiently supported.

Jeans \shortcite{JeansPTRS1902} dealt with the simple case of an infinite
homogeneous gas at rest, supported only by its pressure. He concentrated on 
the velocity of wave propagation of a small fluctuation in density, $\delta 
\rho$,

\begin{equation}
\label{equa:jeans1}
\rho \frac{\partial \mbox{\bf \em v}}{\partial t} \; = \; - \nabla \delta P + \rho \nabla \delta \Phi,
\end{equation}

\noindent with

\begin{equation}
\label{equa:jeans2}
\frac{\partial \delta \rho}{\partial t} \; = \; - \rho \nabla \cdot \mbox{\bf \em v},
\end{equation}

\begin{equation}
\label{equa:jeans3}
\nabla^{2} \delta \Phi \; = \; - 4 \pi G \rho.
\end{equation}

\noindent (taken from Chandrasekhar \shortcite{ChandraBOOK1939}). For 
adiabatic gas $(\delta P = c^{2} \delta \rho)$, Eqn. \ref{equa:jeans1} becomes

\begin{equation}
\label{equa:jeans4}
\rho \frac{\partial}{\partial t} \nabla \cdot \mbox{\bf \em v} \; = \; - c^{2} \nabla^{2} \delta \rho + \rho \nabla^{2} \delta \Phi,
\end{equation}

\noindent or

\begin{equation}
\label{equa:jeans5}
\frac{\partial^{2}}{\partial t^{2}} \delta \rho \; = \; c^{2} \nabla^{2} \delta \rho + 4 \pi G \rho \delta \rho.
\end{equation}

This is a typical wave equation, with solution of the form

\[
\delta \rho \propto \mathrm{e}^{i ({\bf k} \cdot {\bf x} + \sigma t)},
\]

\noindent as long as $\sigma^{2} = c^{2} k^{2} - 4 \pi G \rho$. The
velocity of propagation, or Jeans velocity, $V_{J}$, is therefore

\[
V_{J} = \frac{\sigma}{k} = c \left( 1 - \frac{4 \pi G \rho}{c^{2} k^{2}} \right)^{1/2}.
\]

\noindent The gas is unstable for all wave numbers $k < k_{J}$, where

\begin{equation}
\label{equa:jeans6}
k_{J} = \frac{1}{c} (4 \pi G \rho)^{1/2}.
\end{equation}

If we assume that the corresponding minimum wavelength

\begin{equation}
\label{equa:jeans7}
\lambda_{J} = \frac{\pi}{k_{J}} = \frac{c \pi^{1/2}}{2 (G \rho)^{1/2}}
\end{equation}

\noindent corresponds to the radius of the smallest unstable spherical fragment, the mass of this fragment is

\begin{equation}
\label{equa:jeans8}
M_{J} = \frac{c^{3} \pi^{5/2}}{6 G^{3/2} \rho^{1/2}},
\end{equation}

\noindent which is called the Jeans mass\footnote{A similar formulation is 
given in Jeans \shortcite{JeansBOOK1929}. It is based on the excess of 
gravitational energy when collapse initiates. The arithmetic coefficients in 
both cases are of the same order of magnitude.}.

\backmatter

\chapter{References}
}
\bibliography{../../Bibliography/SFGBIBLnew,SPECIALS}

\end{document}